%% file: main.tex
\documentclass[aps,prl,twocolumn,amsmath,amssymb,superscriptaddress]{revtex4-2}

\makeatletter
\def\maketitle{
\@author@finish
\title@column\titleblock@produce
\suppressfloats[t]}
\makeatother

\usepackage{orcidlink} 
\input{definitions}
\usepackage{graphicx} 
\usepackage{xcolor}%
\usepackage{ifthen}  
\usepackage{cleveref}
\usepackage{placeins}
\usepackage{float}

\newboolean{wordcount}
\setboolean{wordcount}{false} 

\ifthenelse{\boolean{wordcount}}%
{
\usepackage{bibentry} 
\usepackage{comment} 
\excludecomment{align*} 
\excludecomment{align} 
\excludecomment{equation*} 
\excludecomment{equation} 
\excludecomment{eqnarray*} 
\excludecomment{eqnarray} 
\excludecomment{acknowledgments} 
\def\maketitle{} 
}{}

\usepackage[absolute]{textpos}

\begin{document}

\title{Search for a Dark Higgs Boson Produced in Association with Inelastic Dark Matter at the \belletwo Experiment}

\ifthenelse{\boolean{wordcount}}{}{
\input{authors}
}

\begin{abstract}
\input{abstract}
\end{abstract}

\maketitle

\input{body}

\ifthenelse{\boolean{wordcount}}{} 
{
\begin{acknowledgments}
\input{acknowledgements}
\end{acknowledgments}
 }

\ifthenelse{\boolean{wordcount}}%
{ 
\nobibliography{references}
}
{ 
\bibliography{references} 
}

\ifthenelse{\boolean{wordcount}}{}
{
\clearpage
\input{endmatter}
}

\ifthenelse{\boolean{wordcount}}{}{
\clearpage
\input{supplemental_content}
}

\end{document}

%% file: authors.tex
  \author{I.~Adachi\,\orcidlink{0000-0003-2287-0173}} 
  \author{L.~Aggarwal\,\orcidlink{0000-0002-0909-7537}} 
  \author{H.~Ahmed\,\orcidlink{0000-0003-3976-7498}} 
  \author{H.~Aihara\,\orcidlink{0000-0002-1907-5964}} 
  \author{N.~Akopov\,\orcidlink{0000-0002-4425-2096}} 
  \author{S.~Alghamdi\,\orcidlink{0000-0001-7609-112X}} 
  \author{M.~Alhakami\,\orcidlink{0000-0002-2234-8628}} 
  \author{A.~Aloisio\,\orcidlink{0000-0002-3883-6693}} 
  \author{N.~Althubiti\,\orcidlink{0000-0003-1513-0409}} 
  \author{K.~Amos\,\orcidlink{0000-0003-1757-5620}} 
% ...
\collaboration{The Belle II Collaboration}

%% file: abstract.tex
Inelastic dark matter models that have two dark matter particles and a massive dark photon can reproduce the observed relic dark matter density without violating cosmological limits. 
The mass splitting between the two dark matter particles \chione and \chitwo, with $m(\chitwo) > m(\chione)$, is induced by a dark Higgs field and a corresponding dark Higgs boson \dh.
We present a search for dark matter in events with two vertices, at least one of which must be displaced from the interaction region, and missing energy.
Using a $\lumi\,\invfb$ data sample collected at \belletwo, which operates at the SuperKEKB \epem collider, we observe no evidence for a signal.
We set upper limits on the \prodbf, where $x^+x^-$ indicates $\mu^+\mu^-, \pi^+\pi^-$, or $K^+K^-$, as functions of \dh{} mass and lifetime at the level of $10^{-1}\,\fb$.
We set model-dependent upper limits on the dark Higgs mixing angle at the level of $10^{-5}$ and on the dark photon kinetic mixing parameter at the level of $10^{-3}$. 
This is the first search for dark Higgs bosons in association with inelastic dark matter.

%% file: body.tex
Despite clear observations of the gravitational effects of dark matter~(DM)\,\cite{Planck:2018vyg,Arbey:2021gdg}, the mass of dark matter particles is unknown, and they have
not yet been shown to interact with standard model~(SM) particles.
Direct detection experiments are sensitive to elastically scattering DM particles with masses typically in the GeV to TeV range~\cite{LZ:2024zvo,XENON:2023cxc,PandaX:2024qfu,DarkSide-50:2022qzh}.
At colliders, searches for particles mediating interactions with dark sector particles are sensitive to mediator masses in the MeV to TeV range~\cite{CMS:2024zqs,ATLAS:2024fdw}.
These searches cover both promptly decaying and long-lived mediators decaying after a macroscopically large distance.
A nonminimal class of models introduces inelastic DM~(iDM), where DM couples inelastically to SM states, depending on the mass difference $\Delta m = m(\chitwo)-m(\chione)$ between the two DM mass eigenstates $\chi_1$ and $\chi_2$\,\cite{Tucker-Smith:2001myb}.
The simplest iDM models introduce the inelastic coupling via a massive dark photon \ap\, that couples off diagonally to the two DM states.
The \ap{} kinetically mixes with SM photons via a mixing parameter $\epsilon$\,\cite{Izaguirre:2015zva,Izaguirre:2017bqb,Duerr:2019dmv} and decays predominantly via $\ap\to\chione\chitwo$.
A small $\Delta m$ or small coupling to the \ap\, makes the heavier state $\chitwo$ long lived before it decays into a pair of SM particles and the lighter state \chione{}.
The relic DM candidate \chione{}, which contributes to the DM abundance observed today, is stable and escapes detection~\cite{Izaguirre:2015zva}.
These models can be extended to explain the mass splitting $\Delta m$ and the \ap\, mass by introducing an additional dark Higgs boson \dh\,\cite{Duerr:2020muu}.
The \dh\, would mix with the SM Higgs boson through a mixing angle $\theta$~\cite{OConnell:2006rsp,Beacham:2019nyx}.
In total, the model has seven free parameters: the masses $m(\dh$), $m(\ap$), $m(\chione$), and the mass splitting $\Delta m$; the mixing angle $\theta$; the kinetic mixing parameter $\epsilon$; and the coupling $g_{D}=\sqrt{4\pi\alpha_D}$ between DM and the \ap{}. 
The coupling $k\approx g_D\Delta m/m(\ap)$ between DM and the \dh\, is fixed by the other parameters~\cite{Duerr:2020muu}.
We restrict the search to parameter combinations that correspond to the perturbative regime and that evade existing constraints from observations of the cosmic microwave background by Planck\,\cite{Planck:2018vyg} by requiring $m(\dh) < m(\chione) < m(\ap)$.
In this scenario, the DM relic density would be predominantly determined by the process $\chione\chione\to\dh\dh$.
The \dh{} lifetime increases for decreasing values of $\theta$, making the \dh{} long-lived at small $\theta$.

This model is already constrained by searches for \dh{} or \ap{} mediators without specific assumptions about an iDM model.
Searches for scalars exclude $\sin\theta$ larger than $10^{-3}$ to $10^{-4}$ for \dh\, masses up to about 5\,\gevcc\,(see Ref.\,\cite{Ferber:2023iso} for a review) for $m(\dh) < 2m(\chione)$, while for higher \dh\, masses the limits are considerably weaker.
For the direct production of an \ap\ through kinetic mixing with a photon and subsequent decay into iDM, the CMS\,experiment excludes $y=\epsilon^2\alpha_D(m(\chione)/m(\ap))^4$ larger than $10^{-7}$ to $10^{-8}$ for $\mchione \gtrsim 3\,\gevcc$\,\cite{CMS:2023bay}.
Reinterpretations\,\cite{Duerr:2019dmv,Tsai:2019buq,Mongillo:2023hbs} of searches for invisible decays of \ap\, at BABAR\,\cite{BaBar:2017tiz}, and of searches for long-lived \ap\, decays at NuCal\,\cite{Blumlein:2011mv,Blumlein:2013cua}, CHARM\,\cite{Gninenko:2012eq}, and NA64\,\cite{NA64:2019auh} exclude $y$ larger than about $10^{-9}$ for $\mchione \lesssim 3\,\gevcc$ and $y$ larger than $10^{-12}$ below 1\,\gevcc. 

\begin{figure}[ht]
\centering
\includegraphics[width=0.45\textwidth]{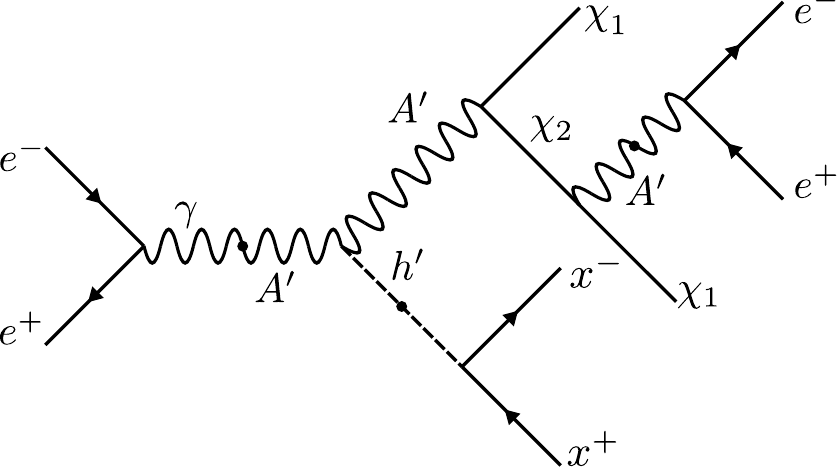}
 \caption{Feynman diagram depicting the search channel for \ap{} production in association with
a \dh{} with subsequent decays into both visible and dark sector states. 
Here $x^+x^-$ indicates $\mu^+\mu^-$, $\pi^+\pi^-$, or $K^+K^-$.
Mixing between dark sector and visible states is indicated by black dots.
\label{fig:feynman}}
\end{figure}

We present the first search for a dark Higgs boson in association with iDM.
We use events with up to two displaced vertices and missing energy, produced in $\epem$ collisions via $\epem\to\dh(\to x^+x^-)\ap\left[\to\chione\chitwo(\to\chione\eplus\eminus)\right]$, where $x^+x^-$ indicates $\mu^+\mu^-$, $\pi^+\pi^-$, or $K^+K^-$.
The corresponding Feynman diagram is shown in \cref{fig:feynman}.
We search for the signal as a narrow enhancement in the $m(x^+x^-)$ distribution.
We present our results as model-independent limits on the product of the production cross section $\sigma_\mathrm{prod} = \sigma\left(\epem \to\dh \chione \chitwo\right)$ and the branching fractions $\mathcal{B}\left(\dh\to x^+x^-\right)\times\mathcal{B}\left(\chitwo\to\chione\eplus\eminus\right)$.
In addition to the model-independent search, we interpret our results as a model-dependent limit on the mixing angle $\theta$ as a function of the \dh{} mass, and as a limits on $y$ as a function of the  \chione{} mass.

We use a $\lumi~\invfb$ data sample\,\cite{ref:lumi_paper} collected at a center-of-mass~(c.m.) energy of $\sqrt{s}=10.58\gev$ by the \belletwo experiment\,\cite{Abe:2010gxa} at the SuperKEKB \epem collider\,\cite{Akai:2018mbz}.
The beam energies are $7\gev$ for $e^-$ and $4\gev$ for $e^+$, resulting in a boost $\beta\gamma=0.28$ of the c.m.~frame relative to the laboratory frame.

The \belletwo detector consists of a variety of subdetectors surrounding the interaction point~(IP) in a cylindrical manner.
The trajectories of charged particles~(tracks) are reconstructed by a combination of a two-layer silicon-pixel detector, a four-layer silicon-strip detector, and a central drift chamber~(CDC).
The tracking detectors are surrounded by time-of-propagation and aerogel ring-imaging Cherenkov detectors used for particle identification~(PID).
The PID detectors cover an angular region of $14^\circ < \theta_{\rm{polar}} < 124^\circ$.
Photons are reconstructed by an electromagnetic calorimeter~(ECL) that also 
serves in the identification of electrons covering $12^\circ < \theta_{\rm{polar}} < 155^\circ$. 
The ECL is surrounded by a $1.5\,\mathrm{T}$ superconducting solenoid.
The outermost subdetector is a $K^0_L$ and muon detector~(KLM) which is installed in the iron flux return of the solenoid.
The longitudinal direction, the transverse plane, and the polar angle $\theta_{\rm{polar}}$ are defined with respect to the detector’s solenoidal axis in the direction of the electron beam. 
In the following, quantities are defined in the laboratory frame unless specified otherwise.

We use simulated events to optimize the event selection, and to determine efficiencies and signal resolutions.
Signal events are generated using a combination of {MadGraph5@NLO}\,\cite{Alwall:2014hca} and EVTGEN\,\cite{Lange:2001uf} taking into account effects of initial state radiation~(ISR)\,\cite{Frixione:2021zdp}.
Furthermore, we consider effects from electromagnetic final state radiation in the decay of the \dh{} using PHOTOS\,\cite{Barberio:1990ms,Barberio:1993qi}.
To correct for efficiency differences caused by different beam background conditions, we generate signal simulations for a variety of different data-taking conditions using beam-induced backgrounds sampled from data overlaid with simulated signal events and find an approximately linear correlation between background level and signal efficiency.
We use the efficiency obtained from a linear fit at the luminosity-weighted average beam background level of our dataset.
Motivated by Ref.\,\cite{Duerr:2020muu} and a previous search for a similar model\,\cite{CMS:2023bay}, we consider values of \hbox{$m(\ap) = 3\,\mchione$ and $4\,\mchione$}, \hbox{$\Delta m = 0.2\,\mchione, 0.4\,\mchione$ and $1.0\,\mchione$}, and $\alpha_D = 0.1$ and  $0.5$.
For all possible combinations of these values, we generate events for \dh\, masses between 0.2\,\gevcc and 3.0\,\gevcc in about 45 steps of varying size and various lifetimes $0.1 < c\tau(\dh) < 10\,000$\,\cm in steps that are approximately equidistant on a logarithmic scale; we generate events for \mchione between 0.2\,\gevcc and 3.0\,\gevcc in 30 steps of 0.1\,\gevcc, and various lifetimes $0.01 < c\tau(\chitwo) < 1000$\,\cm in variable steps.
Since $\map > \mchione + \mchitwo$, the \ap{} in the \chitwo decay is always off shell, while the \ap{} produced in association with the \dh{} can be either on shell or off shell with \ap{} masses up to 12\,\gevcc.
The lifetime of the \ap\, is negligible for all values of $\epsilon$ probed in this analysis.

We simulate the following background processes: $e^+e^-\to\FourS\to B\overline{B}$ with EvtGen; 
$e^+e^-\to q\bar{q} (\gamma)$, where $q=u,d,s,c$ with KKMC\,\cite{Jadach:1999vf} interfaced with PYTHIA8\,\cite{Sjostrand:2014zea} and EvtGen; 
$e^+e^-\to\mu^+\mu^-(\gamma)$ and $e^+e^-\to\tau^+\tau^-(\gamma)$ with KKMC; 
$e^+e^-\to e^+e^-e^+e^-$, $e^+e^-\to e^+e^-\mu^+\mu^-$,  $e^+e^-\to \mu^+\mu^-\mu^+\mu^-$, $e^+e^-\to e^+e^-\tau^+\tau^-$, and $e^+e^-\to \mu^+\mu^-\tau^+\tau^-$ with AAFH\,\cite{Berends:1984gf}; 
$e^+e^-\to\tau^+\tau^-\tau^+\tau^-$ with KoralW\,\cite{Jadach:1998gi}; 
$e^+e^-\to e^+e^-\pi^+\pi^-$, $e^+e^-\to e^+e^-K^+K^-$, and $e^+e^-\to e^+e^-p\bar{p}$ with TREPS\,\cite{Uehara:1996bgt}; 
$e^+e^-\to e^+e^-(\gamma)$ and $e^+e^-\to \gamma\gamma(\gamma)$ with Babayaga.NLO\,\cite{Balossini:2008xr}; 
$e^+e^-\to \KS\KL\gamma$, $e^+e^-\to \pi^+\pi^-\gamma$, $e^+e^-\to K^+K^-\gamma$, and $e^+e^-\to\pi^+\pi^-\pi^0\gamma$ with PHOKHARA\,\cite{Campanario:2019mjh}.
Decays of $\tau$ leptons are simulated with TAUOLA\,\cite{Jadach:1990mz} for KKMC, and using PYTHIA8 for all other event generators.
The detector geometry and interactions of final-state particles with detector material are simulated using GEANT4\,\cite{Agostinelli:2002hh}. 
Both experimental and simulated events are reconstructed and analyzed using the \belletwo software~\cite{Kuhr:2018lps, basf2-zenodo}.
To avoid experimenter’s bias, we examine the experimental data only after finalizing the analysis selection.
All selection criteria are chosen by iteratively optimizing the figure of merit for a discovery with a significance of five standard deviations\,\cite{Punzi:2003bu}.
To avoid additional complexity of the analysis we chose a single set of selections for all model parameter combinations.

We use events selected by a calorimeter-only trigger, which requires the sum of energy depositions in the polar angle region $22^{\circ} < \theta_{\rm{polar}} < 128^{\circ}$ of the ECL to be above 1\,\gev.
We require that the total deposited energy in this angular region exceeds 1.5\,\gev which ensures nearly 100\% trigger efficiency for events that pass this selection, to avoid systematic uncertainties introduced by the trigger requirement.

We reconstruct \dh\ and \chitwo candidates by combining pairs of oppositely charged particles reconstructed from tracks.
We require each track to have at least 20 tracking detector hits.
The track pairs are separately constrained by fits to originate from common vertices.
The opening angle between the two tracks must be greater than 0.1\,rad to suppress background from photon conversions, and the \dh\, pointing angle $\Delta\alpha_{\vec x, \vec p}$ between the \dh\, momentum $\vec{p}$ and vertex position vectors $\vec{x}$ must have $\Delta\alpha_{\vec x, \vec p} < 3.1^\circ$, so the \dh{} vertex points back to the IP.
To reduce promptly decaying SM backgrounds, at least one of the two vertices must have a transverse displacement with respect to the IP $d_{\text{v}}$ of at least $0.2$\,\cm.
To reject contributions from track-pairs produced in hadronic interactions in the ECL, both vertices must have $d_{\text{v}} < 110$\,\cm.

PID information from all relevant subdetectors is combined to classify final states~\cite{Kou:2018nap}.
At least one track from the \dh\, decay must have an extrapolated polar angle $37^{\circ} < \theta_{\rm{polar}}^{\rm{ext}} < 122^{\circ}$, calculated by extrapolating the track helix to the KLM inner surface to ensure high muon identification efficiency.
To further reduce the backgrounds in the final state with $\dh\to\pi^+\pi^-$ decays, all four tracks must be in the range $18^{\circ} < \theta_{\rm{polar}}^{\rm{ext}} < 155^{\circ}$.
To ensure high electron PID purity, we require the ratio between the energy deposition in the calorimeter and the momentum of the corresponding track to be larger than 0.6.

We reject events with \dh{} candidates with {$0.467 < M(\pi^+\pi^-) < 0.529$\,\gevcc} to reduce background from \KS decays; we remove events in the $\dh\to\pi^+\pi^-$ final state with $1.06 < M(p\pi^-) < 1.15$\,\gevcc to reduce background from $\Lambda$-baryon decays; we also remove events in the $\dh\to K^+ K^-$ final state with {$0.977 < M(K^+ K^-) < 1.061$\,\gevcc} to reduce background from $\phi$ decays produced in $\epem\to\phi(\to K^+K^-)\gamma(\to \epem)$.
The total missing energy in the c.m.\,frame, calculated from the momenta of the four charged particles and the known initial \epem kinematics, must be greater than 0.4\,\gev which is twice the minimal \chione{} mass we consider.
The missing momentum direction must be separated from any energy deposition in the KLM detector by at least 0.5\,rad to reject neutral hadron backgrounds.
To reduce backgrounds from nonreconstructed particles, the missing momentum direction must point into a more restrictive tracking region ($23^\circ< \theta^{\mathrm{miss}}_{\mathrm{polar}}< 149^\circ$) to avoid the CDC edges where data-simulation agreement is less reliable.
The reconstructed electron pair mass from the $\chitwo\to\chione\eplus\eminus$ decay must be less than 2.5\,\gevcc which corresponds to the maximal $\Delta m$ for which we provide model-dependent interpretations.
We require that no other tracks are reconstructed, and that the total deposited energy in the calorimeter not matched to tracks satisfies $E_{\mathrm{extra}} < 1.0$\,\gev.
We require that the opening angle of the tracks from the \dh{} vertex be less than 3.0\,rad to suppress background from cosmic muons crossing the detector that are incorrectly reconstructed as two back-to-back tracks.

If multiple signal candidates in the same event pass the selections, which occurs in less than 3\% of events, we choose the candidate with the smallest \dh\, pointing angle.

The overall signal selection efficiency is typically a few percent up to 20\%. 
It is generally higher for large $\Delta m$ and small displacements of the \dh.

We determine the signal mass resolution by fitting a double-sided Crystal Ball~(DSCB) function\,\cite{Gaiser:Phd, Skwarnicki:1986xj} to simulated \mdh distributions.
The resolution $\sigma_{\mathrm{sig}}^{\mathrm{DSCB}}$ increases smoothly from about 1\,\mevcc for a light \dh\, to about 7\,\mevcc for a heavy \dh\, and depends only slightly on the \dh{} lifetime or final state.
Mass hypotheses that lack a simulation sample are interpolated from adjacent simulated samples.

We extract the signal yield by counting events in narrow windows of $M(x^+x^-)$ with a width of $\pm 2\sigma_{\mathrm{sig}}^{\mathrm{DSCB}}$ in steps of $\sigma_{\mathrm{sig}}^{\mathrm{DSCB}}/2$.
Based on studies with simulated samples, we assume a uniform background as a reasonable approximation given the size of our sample.
We determine the background level from data by counting all events in $M(x^+x^-)$ sidebands~(SBs). 
In the $K^+K^-$ and $\mu^+\mu^-$ final state the SB is the full mass range excluding the respective signal window; for the $\pi^+\pi^-$ final state we split the mass region at 1\,\gevcc and determine different uniform background levels below and above this value.

We evaluate systematic uncertainties affecting selection efficiency, integrated luminosity, the limited number of simulated events, and the background model. The dominant systematic uncertainties are associated with the signal efficiency, and depend on combinations of the \dh{} and \chitwo{} masses and lifetimes. 
Relative uncertainties are typically around 4\% for most parameter configurations, but can reach 10\% for the lightest \dh{} masses and large displacements.
For large displacements, the dominant systematic uncertainty on the signal efficiency arises from data-simulation differences in track finding for displaced tracks.
We correct for this with an auxiliary measurement using $K_S^0$ decays from the process $D^{*+}\rightarrow D^0 (\rightarrow K_S^0 \pi^+ \pi^-) \pi^+$, and assign an uncertainty obtained by varying the nominal correction within the total uncertainty on the correction.
For tracks close to the IP we instead correct for momentum-dependent efficiency differences between data and simulation, resulting in uncertainties at the level of 0.5\%\,\cite{PhysRevD.111.032012}.
The uncertainties arising from PID are evaluated using the processes $e^+e^-\to e^+e^-e^+e^-$, $e^+e^-\to e^+e^-\mu^+\mu^-$, and $e^+e^-\to\mu^+\mu^-\gamma$, as well as decays of $K_S^0$ or $J/\Psi$.
For electrons these uncertainties are typically at the level of 3\%, while uncertainties for $\mu$, $\pi$, and $K$ are below 1\%.
We account for a lifetime-dependent effect on PID by introducing an additional systematic uncertainty, evaluated using $K_S^0$ and $\Lambda$ decays. 
For very displaced vertices, these uncertainties can reach up to 10\%.
The uncertainty on the luminosity is 0.47\%\,\cite{ref:lumi_paper}.
The limited number of simulated events for each signal configuration introduces systematic uncertainties at the level of 1--2\% for most parameter configurations but can reach up to 10\% for very long lifetimes.
We verify that our interpolation procedure between simulated mass points does not introduce a significant additional uncertainty.
We estimate the uncertainty introduced by splitting the mass region in the $\pi^+\pi^-$ final state by varying the split point to 0.9\,\gevcc and 1.2\,\gevcc, respectively, and take the maximum deviation from the nominal background level as the uncertainty $\delta$.

We find no events in the $\mu^+\mu^-$ final state, 8 events in the $\pi^+\pi^-$ final state, and one event in the $K^+K^-$ final state. 
The $M(\pi^+\pi^-)$ distribution in the $\dh\to\pi^+\pi^-$ final state is shown in \cref{fig:events_pion}, while the distributions for $\dh\to\mu^+\mu^-$ and $\dh\to K^+K^-$ are shown in the Supplemental Material~\cite{aux:2024}.
The statistical model used to compute the signal significances and $p$ values is discussed in Appendix~\ref{sec:bkg_only}.
The largest local significance for the model-independent search is $2.9\,\sigma$, including systematic uncertainties, found near $\mdh = 0.531\,\gevcc$ for the $\pi^+\pi^-$ final state for a lifetime of $c\tau=1.0\cm$.
Taking into account the look-elsewhere effect\,\cite{Gross:2010qma}, this excess has a global significance of $1.1\,\sigma$.

\begin{figure}[ht]
\centering
\includegraphics[width=0.45\textwidth]{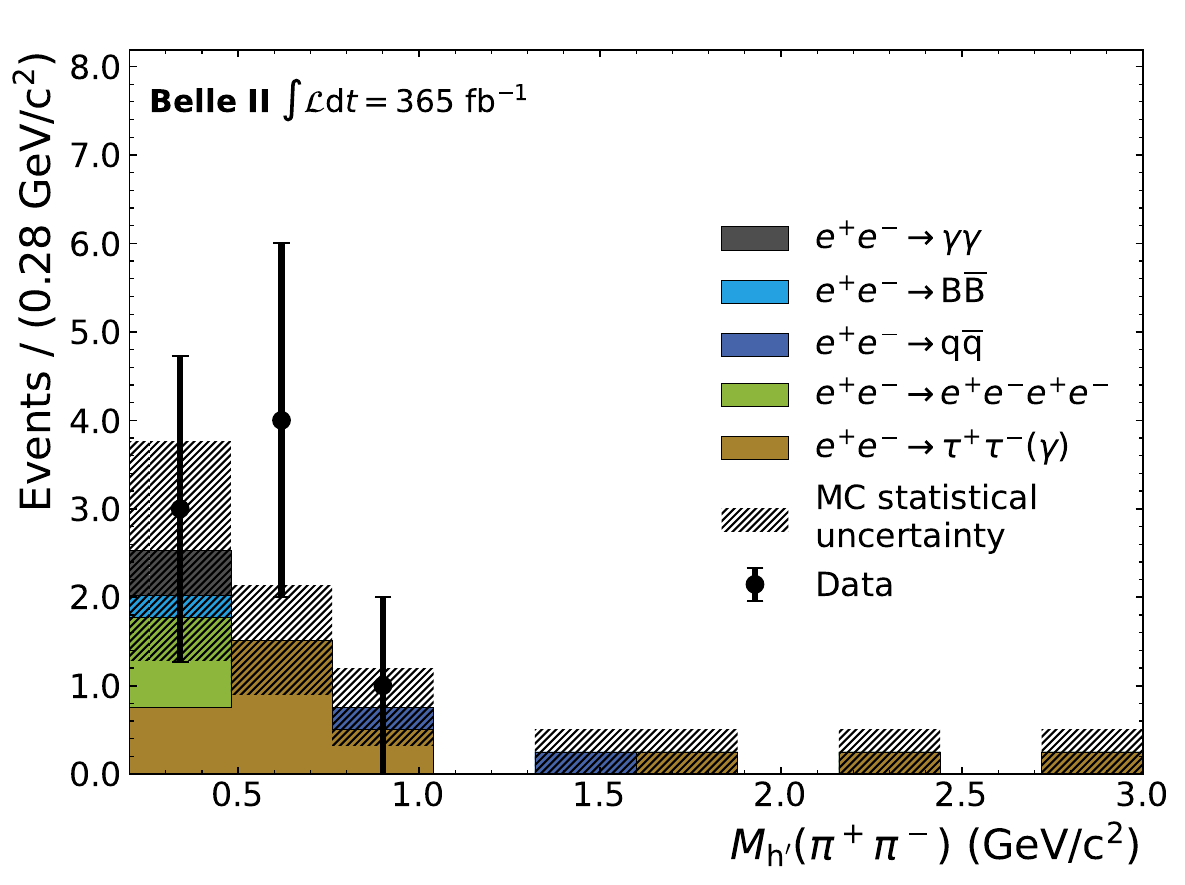}
 \caption{Distribution of $M(\pi^+\pi^-)$ together with the stacked contributions from the various simulated SM background samples for $\dh\to\pi^+\pi^-$ candidates. Simulation is normalized to a luminosity of \lumiinvfb{}. }\label{fig:events_pion}
\end{figure}

With the method described in Appendix~\ref{sec:upper_limit}, we compute 95\% Bayesian credibility level upper limits on $\sigma_\mathrm{sig} = \mathprodbfshort{}$ using the Bayesian Analysis Toolkit software package\,\cite{Caldwell:2008fw, bat-zenodo}.
The observed upper limits, including systematic uncertainties, are shown in the Supplemental Material~\cite{aux:2024}.
Using a Jeffreys prior\,\cite{Jeffreys:1946} would decrease the upper limits on $\sigma_\mathrm{sig}$ by up to 30\% with respect to the uniform prior.
The systematic uncertainties weaken the limits, with the largest increase of 2.5\% occurring for heavy \dh{} with small lifetimes.

For the model-dependent interpretations, we multiply the $p$ values in all relevant and kinematically accessible analysis channels, again separately for various lifetimes.

\begin{figure}[ht]
\centering
\includegraphics[width=0.45\textwidth]{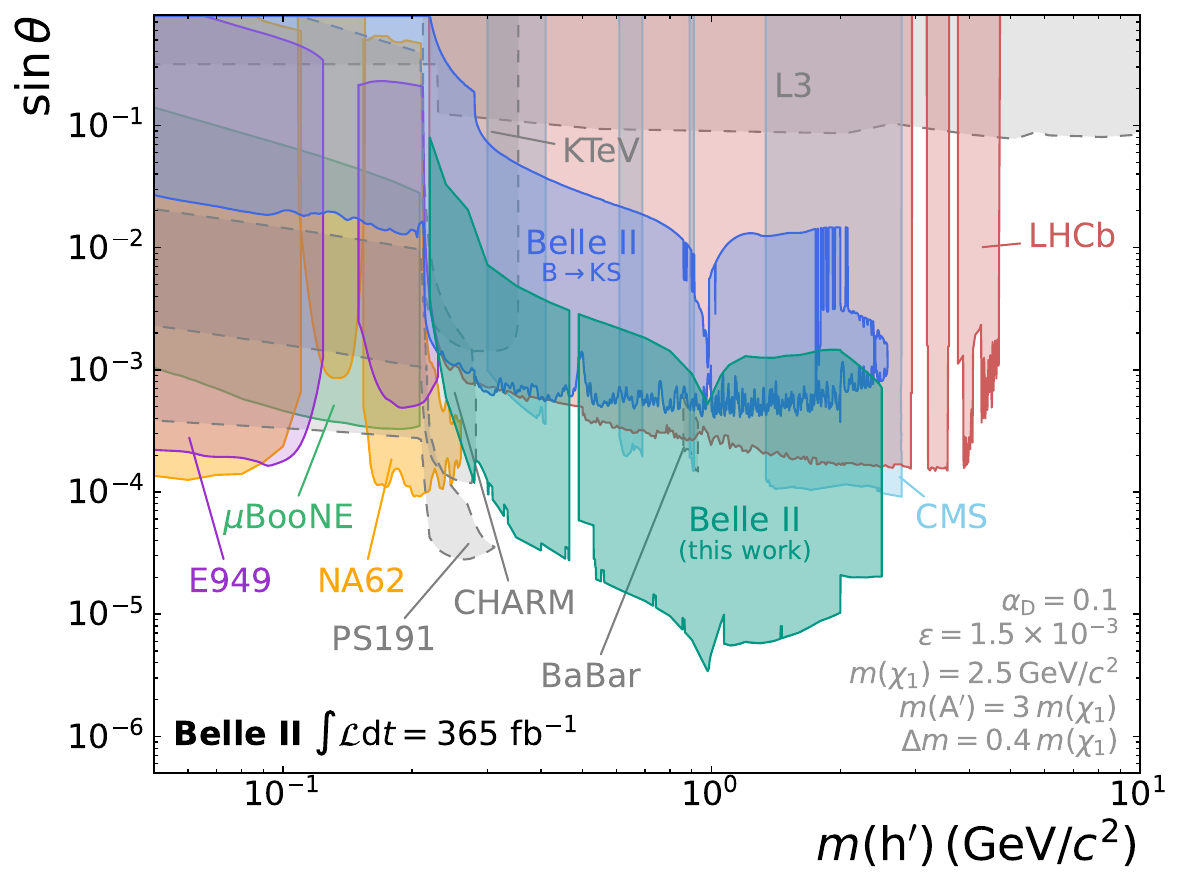}
 \caption{\label{fig:2} 
Exclusion regions at 95\,\% credibility level in the plane of the sine of the mixing angle $\theta$ and dark Higgs mass $\mdh$ from this work (teal) together with existing constraints at 90\,\% confidence level from 
PS191\,\cite{Gorbunov:2021ccu},
E949\,\cite{BNL-E949:2009dza},
NA62\,\cite{NA62:2020pwi,NA62:2021zjw},
KOTO\,\cite{KOTO:2020prk,Ferber:2023iso}, 
KTeV\,\cite{KTEV:2000ngj},
and BABAR\,\cite{BaBar:2015jvu,Winkler:2018qyg},
and at 95\,\% confidence level from
MicroBooNE\,\cite{MicroBooNE:2021usw,MicroBooNE:2022ctm,Ferber:2023iso},
L3\,\cite{L3:1996ome,Ferber:2023iso},
CHARM\,\cite{CHARM:1985anb,Winkler:2018qyg},
LHCb\,\cite{LHCb:2015nkv,LHCb:2016awg,Winkler:2018qyg},
\belletwo\,\cite{Belle-II:2023ueh},
and CMS\,\cite{CMS:2021sch} for $\alpha_D = 0.1$, $\map = 3\,\mchione$, $\Delta m=0.4\,\mchione$, $\epsilon = 1.5\times 10^{-3}$, and $\mchione = 2.5\,\gevcc$.
Constraints colored in gray with dashed outline are reinterpretations not performed by the experimental collaborations.
All constraints except for the one from this work do not require the presence of a dark photon or iDM.}
\end{figure}

\begin{figure}[ht]
\centering
\includegraphics[width=0.45\textwidth]{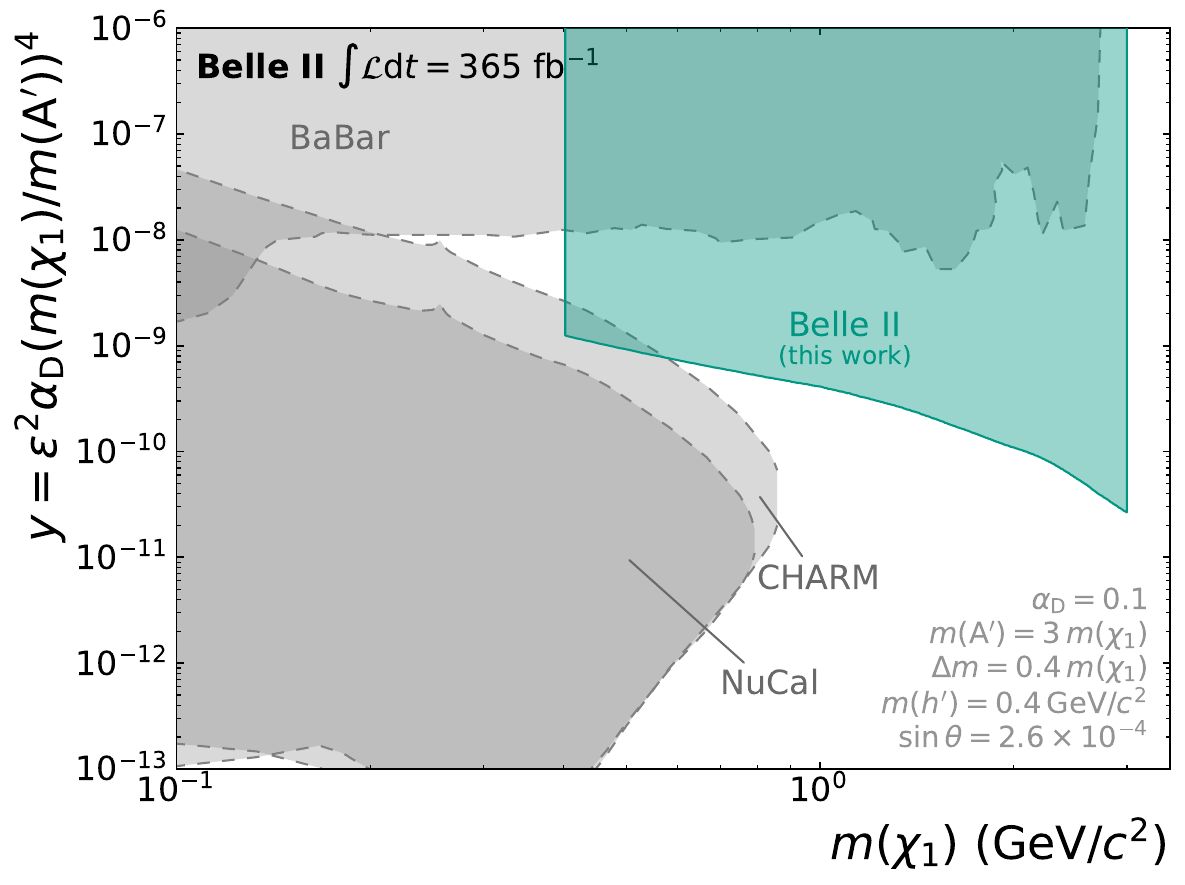}
 \caption{\label{fig:3}
Exclusion regions at 95\,\% credibility level in the plane of the dimensionless variable $y = \epsilon^2\alpha_D(m(\chi_1)/m(A'))^4$ and DM mass $\mchione$ from this work (teal) together with existing constraints at 90\,\% confidence level from
CHARM\,\cite{Gninenko:2012eq,Tsai:2019buq},
NuCal\,\cite{Blumlein:2011mv,Blumlein:2013cua,Tsai:2019buq}, and
BABAR\,\cite{BaBar:2017tiz,Duerr:2019dmv}) for $\alpha_D = 0.1$, $\map = 3\,\mchione$, $\Delta m=0.4\,\mchione$, $\sin \theta = 2.6\times 10^{-4}$, and $\mdh = 0.4\,\gevcc$. Constraints colored in gray with dashed outline are reinterpretations not performed by the experimental collaborations.
 All constraints except for the one from this work do not require the presence of a dark Higgs boson or iDM.
 }
 \end{figure}

For the calculation of the model-dependent upper limits on \mathprodbfmodeldep{} we multiply the individual likelihoods weighted by the theoretical \dh{} branching fractions from Ref.\,\cite{Winkler:2018qyg}.
For each \dh{} mass value, we determine the value of $\sin \theta$ such that the resulting predicted value of \mathprodbfmodeldep, equals the 95\% excluded \mathprodbfmodeldep.
To calculate the prediction, we fix $\sigma_\mathrm{prod}$ and the \chitwo{} branching fractions to the theoretical values from Ref.\,\cite{Duerr:2020muu} taking into account ISR.
\Cref{fig:2} shows the observed upper limit on $\sin\theta$ for one specific choice of model parameters.
Similarly, for each $m(\chione{})$, we determine the value of $y$ such that the resulting predicted value of \mathprodbfmodeldep, equals the 95\% excluded \mathprodbfmodeldep.
\Cref{fig:3} shows the observed upper limit on $y$ for a specific choice of model parameters.
In general, $\sigma_\mathrm{prod}$ increases with $\epsilon^2$, the lifetime of the \dh{} increases with $1/(\sin \theta)^2$, and the lifetime of the \chitwo{} increases with $1/\epsilon^2$.
Additional plots and detailed numerical results for many more parameter combinations can be found in the Supplemental Material~\cite{aux:2024}.

In conclusion, we report the first search for a dark Higgs in association with inelastic DM, using $\lumi~\invfb$ of \belletwo $\epem$ data.
We do not observe a significant excess above the background.
We set 95\%~credibility level upper limits on \mathprodbf.
Depending on the combination of model parameters, the limits improve over existing searches by up to 2 orders of magnitude.

%% file: acknowledgements.tex
\textit{Acknowledgments} -- We thank Felix Kahlhoefer for useful discussions during the preparation of this manuscript.
This work, based on data collected using the Belle II detector, which was built and commissioned prior to March 2019,
was supported by
Higher Education and Science Committee of the Republic of Armenia Grant No.~23LCG-1C011;
Australian Research Council and Research Grants
No.~DP200101792, 
No.~DP210101900, 
No.~DP210102831, 
No.~DE220100462, 
No.~LE210100098, 
and
No.~LE230100085; 
Austrian Federal Ministry of Education, Science and Research,
Austrian Science Fund (FWF) Grants
DOI:~10.55776/P34529,
DOI:~10.55776/J4731,
DOI:~10.55776/J4625,
DOI:~10.55776/M3153,
and
DOI:~10.55776/PAT1836324,
and
Horizon 2020 ERC Starting Grant No.~947006 ``InterLeptons'';
Natural Sciences and Engineering Research Council of Canada, Compute Canada and CANARIE;
National Key R\&D Program of China under Contracts No.~2024YFA1610503,
and
No.~2024YFA1610504
National Natural Science Foundation of China and Research Grants
No.~11575017,
No.~11761141009,
No.~11705209,
No.~11975076,
No.~12135005,
No.~12150004,
No.~12161141008,
No.~12475093,
and
No.~12175041,
and Shandong Provincial Natural Science Foundation Project~ZR2022JQ02;
the Czech Science Foundation Grant No.~22-18469S 
and
Charles University Grant Agency project No.~246122;
European Research Council, Seventh Framework PIEF-GA-2013-622527,
Horizon 2020 ERC-Advanced Grants No.~267104 and No.~884719,
Horizon 2020 ERC-Consolidator Grant No.~819127,
Horizon 2020 Marie Sklodowska-Curie Grant Agreement No.~700525 ``NIOBE''
and
No.~101026516,
and
Horizon 2020 Marie Sklodowska-Curie RISE project JENNIFER2 Grant Agreement No.~822070 (European grants);
L'Institut National de Physique Nucl\'{e}aire et de Physique des Particules (IN2P3) du CNRS
and
L'Agence Nationale de la Recherche (ANR) under Grant No.~ANR-21-CE31-0009 (France);
BMBF, DFG, HGF, MPG, AvH Foundation, and the state of Baden-Wuerttemberg through bwHPC (Germany);
Department of Atomic Energy under Project Identification No.~RTI 4002,
Department of Science and Technology,
and
UPES SEED funding programs
No.~UPES/R\&D-SEED-INFRA/17052023/01 and
No.~UPES/R\&D-SOE/20062022/06 (India);
Israel Science Foundation Grant No.~2476/17,
U.S.-Israel Binational Science Foundation Grant No.~2016113, and
Israel Ministry of Science Grant No.~3-16543;
Istituto Nazionale di Fisica Nucleare and the Research Grants BELLE2,
and
the ICSC – Centro Nazionale di Ricerca in High Performance Computing, Big Data and Quantum Computing, funded by European Union – NextGenerationEU;
Japan Society for the Promotion of Science, Grant-in-Aid for Scientific Research Grants
No.~16H03968,
No.~16H03993,
No.~16H06492,
No.~16K05323,
No.~17H01133,
No.~17H05405,
No.~18K03621,
No.~18H03710,
No.~18H05226,
No.~19H00682, 
No.~20H05850,
No.~20H05858,
No.~22H00144,
No.~22K14056,
No.~22K21347,
No.~23H05433,
No.~26220706,
and
No.~26400255,
and
the Ministry of Education, Culture, Sports, Science, and Technology (MEXT) of Japan;  
National Research Foundation (NRF) of Korea Grants
No.~2016R1-D1A1B-02012900,
No.~2018R1-A6A1A-06024970,
No.~2021R1-A6A1A-03043957,
No.~2021R1-F1A-1060423,
No.~2021R1-F1A-1064008,
No.~2022R1-A2C-1003993,
No.~2022R1-A2C-1092335,
No.~RS-2023-00208693,
No.~RS-2024-00354342
and
No.~RS-2022-00197659,
Radiation Science Research Institute,
Foreign Large-Size Research Facility Application Supporting project,
the Global Science Experimental Data Hub Center, the Korea Institute of
Science and Technology Information (K24L2M1C4)
and
KREONET/GLORIAD;
Universiti Malaya RU grant, Akademi Sains Malaysia, and Ministry of Education Malaysia;
Frontiers of Science Program Contracts
No.~FOINS-296,
No.~CB-221329,
No.~CB-236394,
No.~CB-254409,
and
No.~CB-180023, and SEP-CINVESTAV Research Grant No.~237 (Mexico);
the Polish Ministry of Science and Higher Education and the National Science Center;
the Ministry of Science and Higher Education of the Russian Federation
and
the HSE University Basic Research Program, Moscow;
University of Tabuk Research Grants
No.~S-0256-1438 and No.~S-0280-1439 (Saudi Arabia), and
Researchers Supporting Project number (RSPD2025R873), King Saud University, Riyadh,
Saudi Arabia;
Slovenian Research Agency and Research Grants
No.~J1-50010
and
No.~P1-0135;
Ikerbasque, Basque Foundation for Science,
State Agency for Research of the Spanish Ministry of Science and Innovation through Grant No. PID2022-136510NB-C33, Spain,
Agencia Estatal de Investigacion, Spain
Grant No.~RYC2020-029875-I
and
Generalitat Valenciana, Spain
Grant No.~CIDEGENT/2018/020;
The Knut and Alice Wallenberg Foundation (Sweden), Contracts No.~2021.0174 and No.~2021.0299;
National Science and Technology Council,
and
Ministry of Education (Taiwan);
Thailand Center of Excellence in Physics;
TUBITAK ULAKBIM (Turkey);
National Research Foundation of Ukraine, Project No.~2020.02/0257,
and
Ministry of Education and Science of Ukraine;
the U.S. National Science Foundation and Research Grants
No.~PHY-1913789 
and
No.~PHY-2111604, 
and the U.S. Department of Energy and Research Awards
No.~DE-AC06-76RLO1830, 
No.~DE-SC0007983, 
No.~DE-SC0009824, 
No.~DE-SC0009973, 
No.~DE-SC0010007, 
No.~DE-SC0010073, 
No.~DE-SC0010118, 
No.~DE-SC0010504, 
No.~DE-SC0011784, 
No.~DE-SC0012704, 
No.~DE-SC0019230, 
No.~DE-SC0021274, 
No.~DE-SC0021616, 
No.~DE-SC0022350, 
No.~DE-SC0023470; 
and
the Vietnam Academy of Science and Technology (VAST) under Grants
No.~NVCC.05.12/22-23
and
No.~DL0000.02/24-25.

These acknowledgments are not to be interpreted as an endorsement of any statement made
by any of our institutes, funding agencies, governments, or their representatives.

We thank the SuperKEKB team for delivering high-luminosity collisions;
the KEK cryogenics group for the efficient operation of the detector solenoid magnet and IBBelle on site;
the KEK Computer Research Center for on-site computing support; the NII for SINET6 network support;
and the raw-data centers hosted by BNL, DESY, GridKa, IN2P3, INFN, 
and the University of Victoria.

%% file: endmatter.tex
\title{End Matter}
\maketitle

\appendix
\makeatletter
\renewcommand{\c@secnumdepth}{0}
\makeatother

\section{Signal Extraction}
\label{sec:bkg_only}

In the hypothesis of uniform background, the expected number of background events in the signal window is $\mu/f$, where the nuisance parameter $\mu$ is the expected number of background events in the SB and $f$ the ratio between the widths of the SB and the signal window. 
The likelihood for observing $n$ events in the signal window with the background-only hypothesis is
\begin{align}
\label{eq:1}
    \mathcal{L}(n, \mu) = \frac{(\mu/f)^{n}}{n!} e^{-(\mu/f)} \times \frac{\mu^{N^\mathrm{SB}_\mathrm{obs}}}{N^\mathrm{SB}_\mathrm{obs}!} e^{-\mu}, 
\end{align} 
with $N_{\mathrm{obs}}^{\mathrm{SB}}$ being the number of observed events in the sideband.
Eq.~\ref{eq:1} incorporates all the relevant statistical fluctuations through Poissonian priors.
For the $\pi^+\pi^-$ final state, we include the additional systematic uncertainty of the background model by substituting $\mu \to \mu(1+x\delta)$, with the nuisance parameter $x$, that is constrained by an additional Gaussian prior.
The dependency on the nuisance parameters $\mu$ and $x$ is removed by marginalization of $\mathcal{L}$
\begin{align}
  \mathcal{L}(n) = \int_0^\infty \mathrm{d}\mu \int_{-\infty}^\infty \mathrm{d}x\mathcal{L}(n,\mu,x).  
\end{align}
We calculate the $p$-value, which describes the probability of observing at least $N_{\mathrm{obs}}$ events in the signal window given the background expectation, as
\begin{align}
    p_0 = \sum_{n = N_\mathrm{obs}}^\infty \mathcal{L}(n).
\end{align}
In the case of multiple final states, as in the background-only hypothesis the final states are independent of each other, for each scan point the combined $p$-value is given by the product of the individual $p$-values.
From the $p$-value the significance $Z$ is calculated via
\begin{align}
    Z = \Phi^{-1}(1-p_0)
\end{align}
where $\Phi^{-1}$ is the quantile of the standard Gaussian.

\section{Upper Limit Calculation}
\label{sec:upper_limit}

In the presence of signal, the likelihood of observing $N_\mathrm{obs}$ events in the signal window with a background expectation is given by
\begin{align}
    \mathcal{L}(n_{\mathrm{sig}}, \mu) = \frac{(n_{\mathrm{sig}}+\mu/f)^{N_{\mathrm{obs}}}}{N_{\mathrm{obs}}!} e^{-(n_{\mathrm{sig}}+\mu/f)} \times \frac{\mu^{N^\mathrm{SB}_\mathrm{obs}}}{N^\mathrm{SB}_\mathrm{obs}!} e^{-\mu}.
\end{align}
The number of signal events $n_\mathrm{sig}$ can be expressed in terms of the signal cross section $\sigma_\mathrm{sig}$ via
\begin{align}
    n_\mathrm{sig} = \sigma_\mathrm{sig} \epsilon_\mathrm{sig} L
\end{align}
with the signal efficiency $\epsilon_\mathrm{sig}$ and the integrated luminosity $L$.
Systematic uncertainties on $\epsilon_\mathrm{sig}$ and $L$ are incorporated into the likelihood by adding additional nuisance parameters $y$ and $z$, respectively.
These nuisance parameters are again constrained by Gaussian priors.
The dependence on all nuisance parameters $\mu$ and $\vec{\theta}=(x,y,z)$ is removed by marginalization
\begin{align}
      \mathcal{L}(\sigma_\mathrm{sig}) = \int_0^\infty \mathrm{d}\mu \int_{-\infty}^\infty \mathrm{d}\vec{\theta} \mathcal{L}(\sigma_\mathrm{sig},\mu,\vec{\theta}).
\end{align}
Using this marginalized likelihood we compute $\alpha=95\%$ CL upper limits on $\sigma_\mathrm{sig}$ via
\begin{align}
    \alpha = \int_{0}^{\sigma_\mathrm{sig}^\mathrm{up}} \mathrm{d}\sigma_\mathrm{sig} \mathcal{L}(\sigma_\mathrm{sig}) \Theta(\sigma_\mathrm{sig})
\end{align}
with a uniform prior $\Theta(\sigma_\mathrm{sig})$ on the cross section.


%% file: supplemental_content.tex
\title{Supplemental Material: Search for a Dark Higgs Boson Produced in Association with Inelastic Dark Matter at the \belletwo Experiment}
\maketitle

\renewcommand{\thefigure}{S\arabic{figure}}
\setcounter{figure}{0}

This material is submitted as supplemental material for the Electronic Physics Auxiliary Publication Service.

\section{Double-Sided Crystal Ball Function}

The double-sided Crystal Ball function used for the determination of the signal width $\sigma$ is defined as 
\begin{align}
    f\left(x; \vec{\Theta}\right) &= N \cdot \begin{cases}
      A_l \left( B_l - \frac{x-\mu}{\sigma} \right)^{-n_l} & \text{for } \frac{x-\mu}{\sigma} < - \alpha_l\\
      \exp\left( -\frac{(x-\mu)^2}{2{\sigma}^2} \right)  & \text{for } - \alpha_l \leq \frac{x-\mu}{\sigma} \leq \alpha_r\\
      A_r \left( B_r - \frac{x-\mu}{\sigma} \right)^{-n_r} & \text{for } \frac{x-\mu}{\sigma} > \alpha_r\\
    \end{cases},
\end{align}
with
\begin{align}
    \vec{\Theta} &= \left(\mu, \sigma, \alpha_l, \alpha_r, n_l, n_r\right),\\
    A_{l/r} &= \left( \frac{n_{l/r}}{|\alpha_{l/r}|} \right)^{n_{l/r}} \exp \left( -\frac{|\alpha_{l/r}|^2}{2} \right),\\
    B_{l/r} &=  \frac{n_{l/r}}{|\alpha_{l/r}|}  - |\alpha_{l/r}|.
\end{align}

\section{Observed Events}
We found 8 events in the $\pi^+\pi^-$ final state, and 1 event in the $K^+K^-$ final state passing all selection requirements.
The reconstructed \dh{} mass and vertex positions of the \dh{} and the \chitwo for these events are summarized in Tab.~\ref{tab:allevents}.

\begin{table}
    \centering
    \caption{
    Reconstructed \dh{} mass and vertex positions of the \dh{} and the \chitwo for all events passing the final event selection.
    $z$ and $\rho=\sqrt{x^2+y^2}$ are the longitudinal and transverse vertex positions.}
    \label{tab:allevents}
    \begin{tabular}{c|c|c|c|c|c}
         $M(x^+x^-)$ & $\rho(\dh)$ & $z(\dh)$ & $\rho(\chi_2)$ & $z(\chi_2)$ & Final State\\
         (in \gevcc) & (in cm) & (in cm) & (in cm) & (in cm) & \\
         \hline
        0.306 & 22.208 & 17.772 & 0.015 & 0.058  & $\dh \to \pi^+ \pi^-$\\
        0.332 & 35.565 & 1.304 & 0.013 & 0.006  & $\dh \to \pi^+ \pi^-$\\
         0.461 & 16.944 & -9.659 & 0.006 & -0.044 & $\dh \to \pi^+ \pi^-$\\
        0.532 & 78.67 & 57.687 & 0.009 & 0.096  & $\dh \to \pi^+ \pi^-$\\
        0.534 & 90.034 & -0.179 & 30.308 & 41.711  & $\dh \to \pi^+ \pi^-$\\
        0.558 & 41.612 & -19.201 & 0.002 & 0.058  & $\dh \to \pi^+ \pi^-$\\
        0.737 & 13.776 & 10.743 & 0.025 & -0.04  & $\dh \to \pi^+ \pi^-$\\
        0.861 & 0.08 & 0.16 & 0.795 & 0.341  & $\dh \to \pi^+ \pi^-$\\
        1.455 & 13.334 & 21.314 & 0.011 & -0.03  & $\dh \to K^+ K^-$\\
    \end{tabular}
\end{table}

\begin{figure}[htp!]
\centering
\includegraphics[width=1\linewidth]{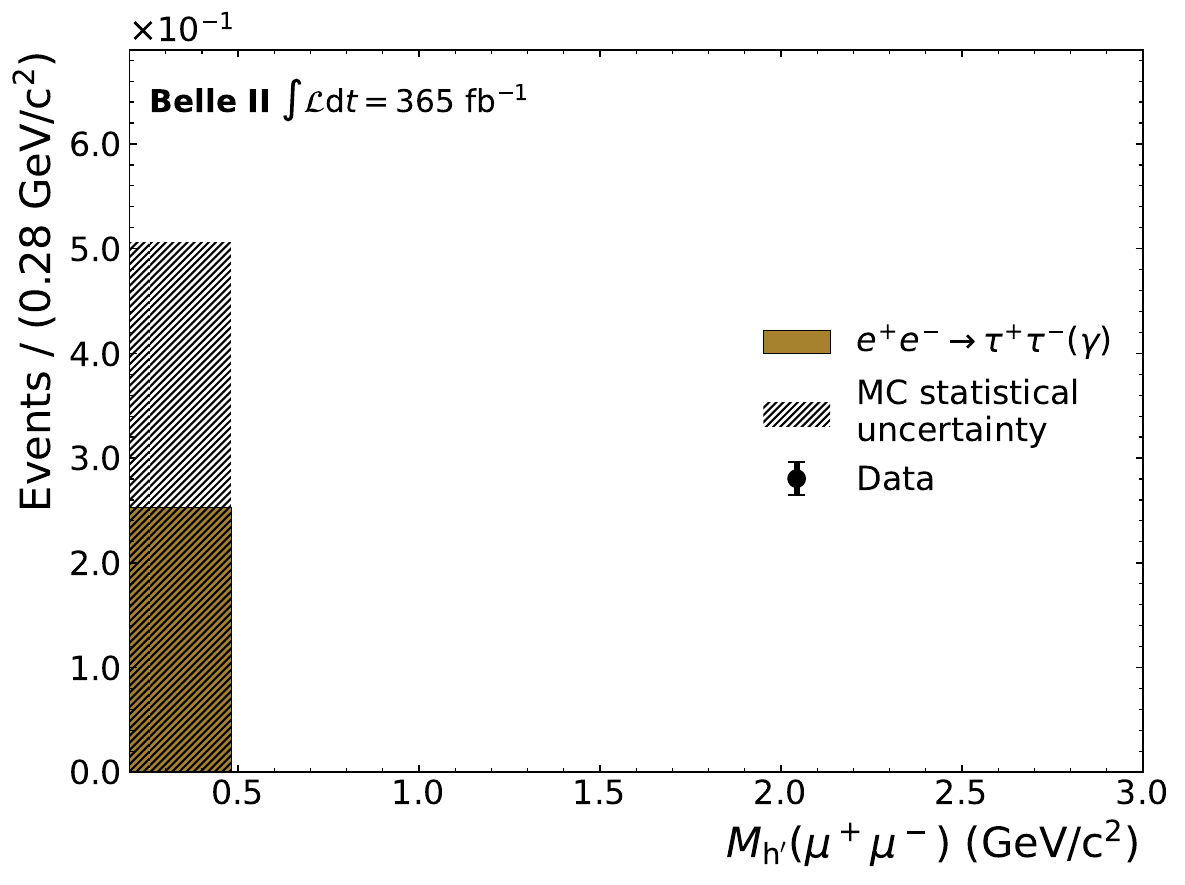}\\
\includegraphics[width=1\linewidth]{figures_supplemental/invM_selected_full_unblinded_final_selection_5.pdf}\\
\includegraphics[width=1\linewidth]{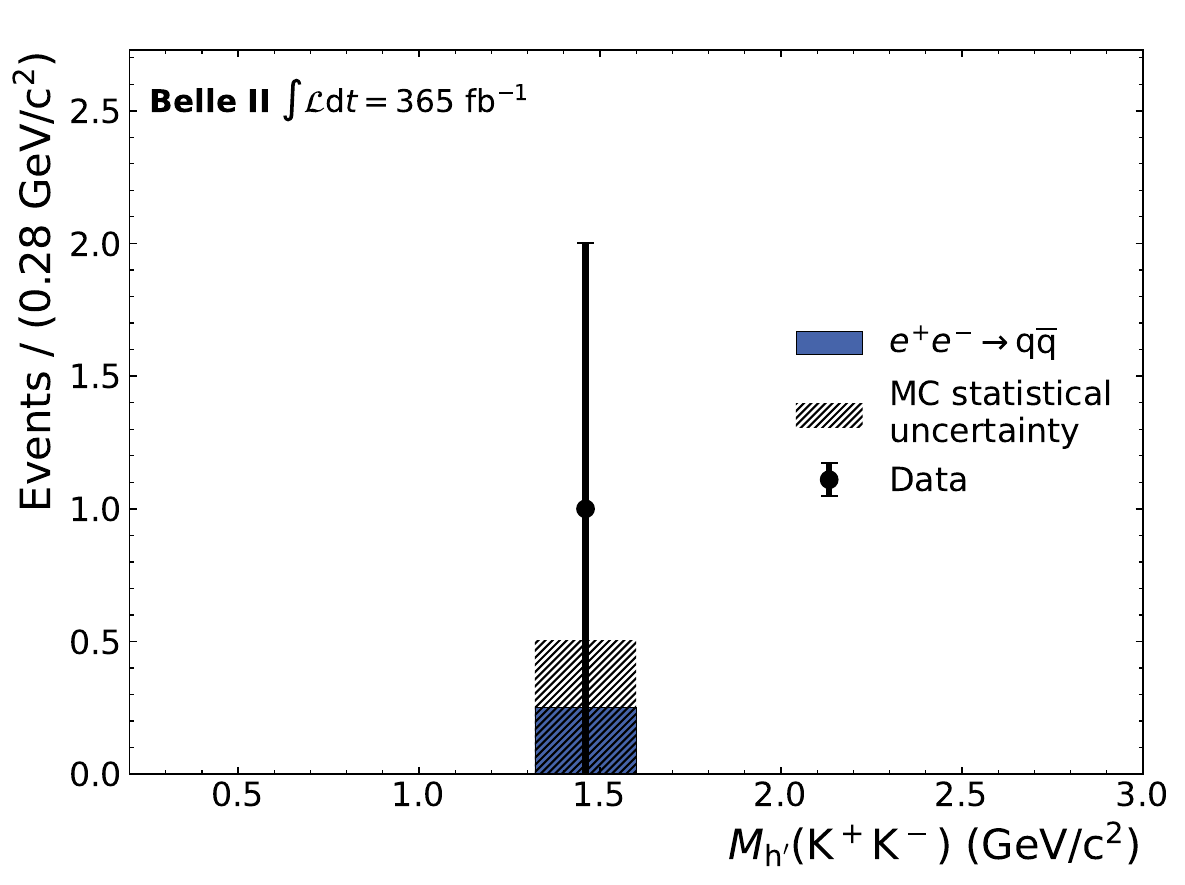}
 \caption{Distribution of $M(x^+x^-)$ together with the stacked contributions from the various simulated SM background samples for $\dh\to\mu^+\mu^-$ (top), $\dh\to\pi^+\pi^-$ (center), and $\dh\to K^+K^-$ (bottom) candidates. Simulation is normalized to a luminosity of \lumiinvfb. \label{fig:events_all}}
\end{figure}

\begin{figure}[htp!]
\centering
\includegraphics[width=1\linewidth]{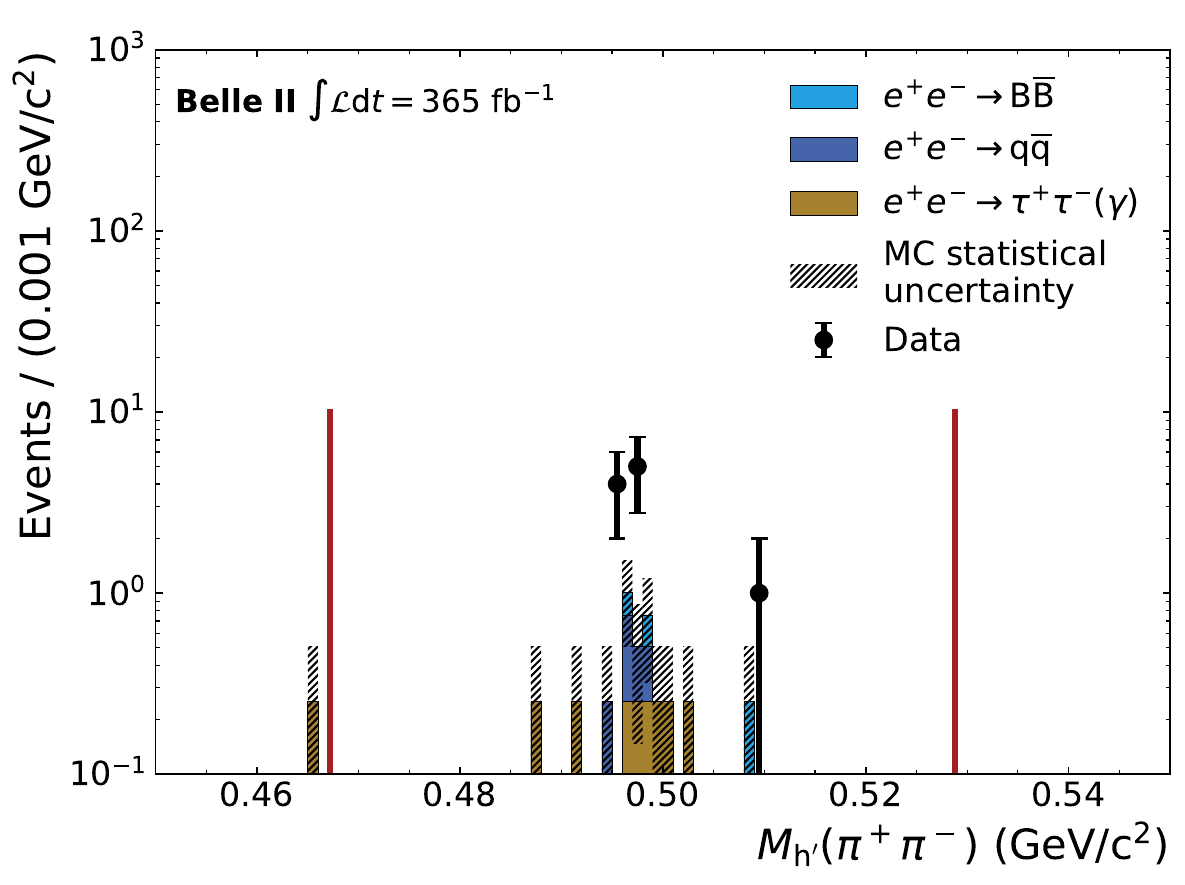}\\
\includegraphics[width=1\linewidth]{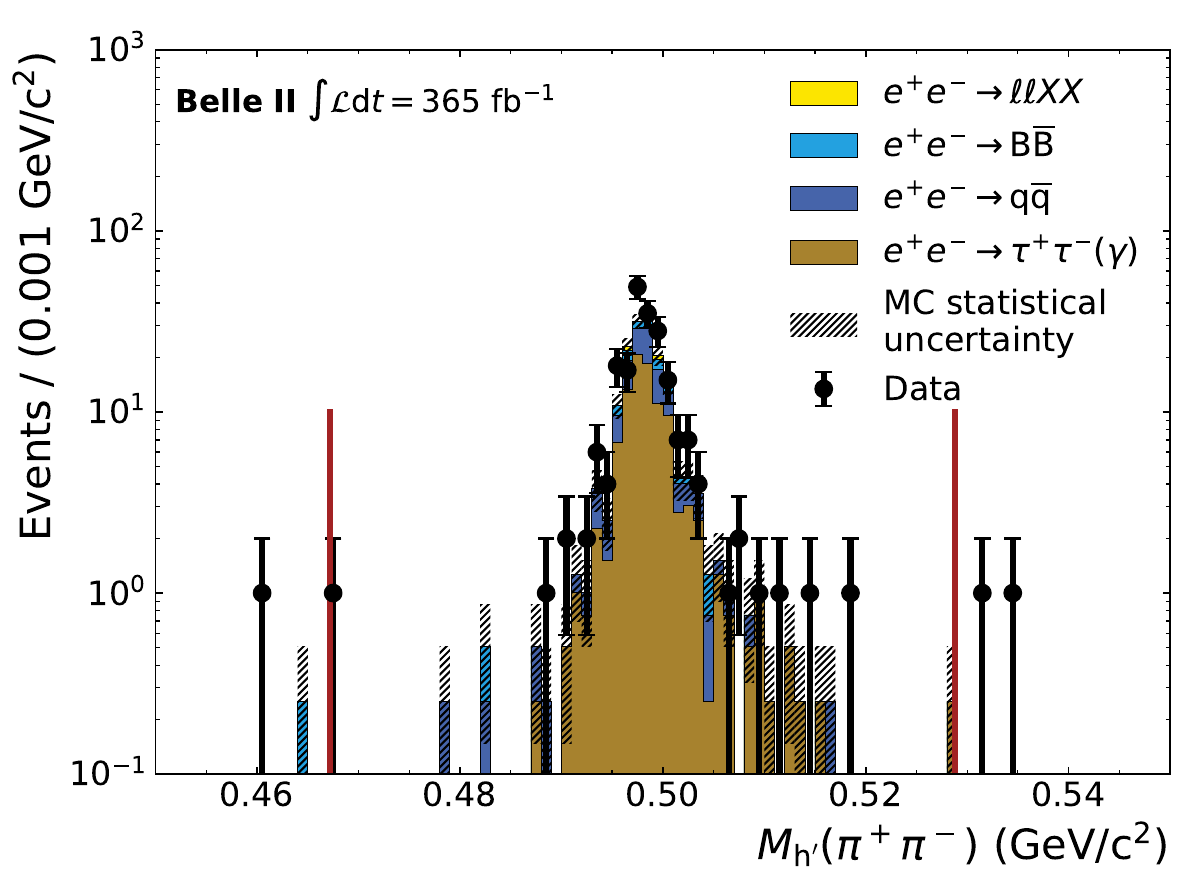}\\
\includegraphics[width=1\linewidth]{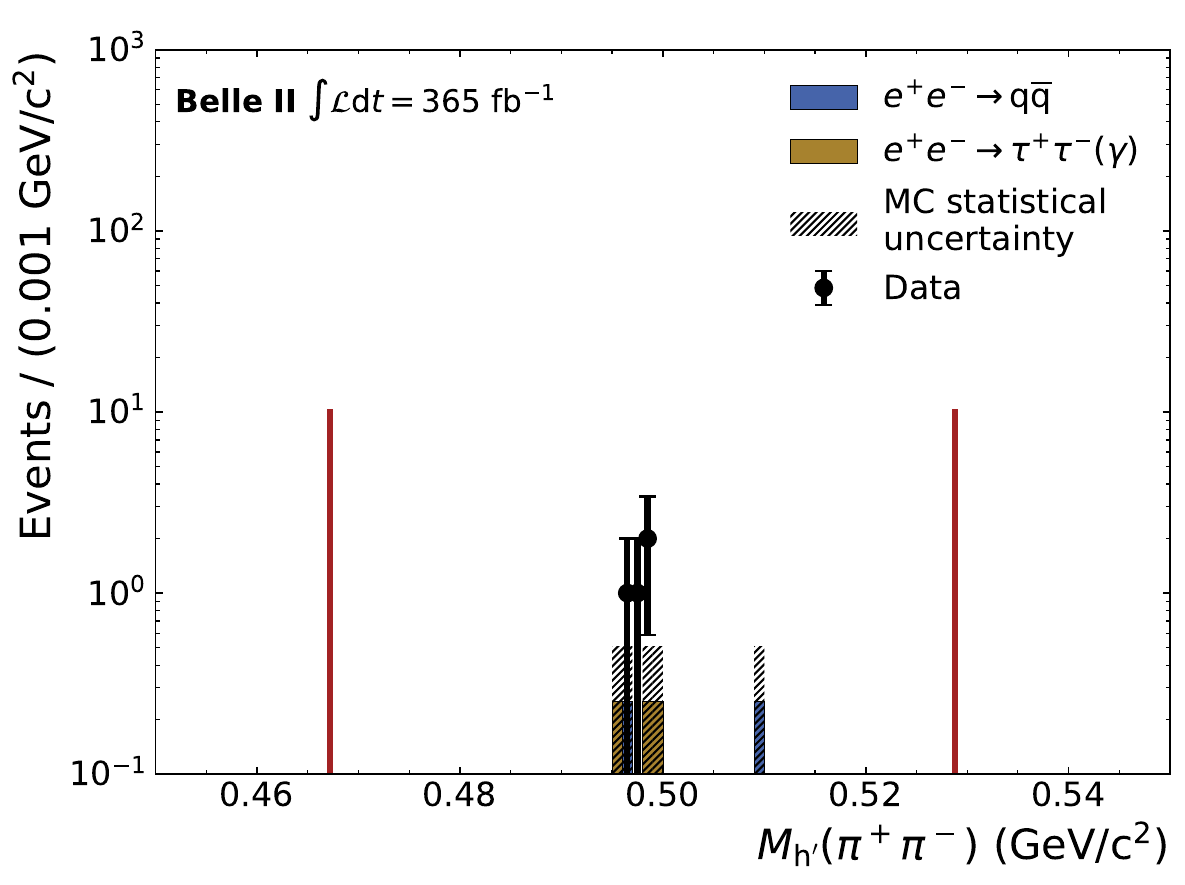}
 \caption{Distribution of $M(\pi^+\pi^-)$ together with the stacked contributions from the various simulated SM background samples for $\dh\to\mu^+\mu^-$ (top), $\dh\to\pi^+\pi^-$ (center), and $\dh\to K^+K^-$ (bottom) candidates. The red lines indicate the mass range {$0.467 < M(\pi^+\pi^-) < 0.529$\,\gevcc} that is rejected to reduce background from \KS decays. Simulation is normalized to a luminosity of \lumiinvfb. \label{fig:events_k0s} }
\end{figure}

Reconstructed \dh{} mass distributions for $\dh\to\mu^+\mu^-$, $\dh\to\pi^+\pi^-$, and $\dh\to K^+K^-$ are shown in Fig.~\ref{fig:events_all}.
The same distributions but with a limited mass range around the \KS veto region without the \KS veto applied are shown in Fig.~\ref{fig:events_k0s}.
Due to mis-reconstruction, contributions from \KS{} are also visible in the non-pion final states.

\clearpage

\section{Upper Limits}
This section of the supplemental material contains the model-independent and model-dependent upper limits (95\% credibility level) on the \prodbf\ as a function of \dh\ mass for several variations of $\mchione$, $\Delta m$, $c\tau(\chi_2)$, and $\map$.
All studied combinations are summarized in Tab.~\ref{tab:chosen_parameters_mphi_theta} and the corresponding figures are referenced in the table.
\begin{table}
    \centering
    \caption{
        Model parameter variations used for the scans in the \mdh{}-$\theta$-plane.
        Identical parameter values in each column share the same colour for better visualization.
        The last column references the corresponding figure.
    }
    \label{tab:chosen_parameters_mphi_theta}
    \renewcommand{\arraystretch}{1.5}
    \begin{tabular}{c|c|c|c|c|c}
        $\alpha_D$ & \mchione{} & $\Delta m$ & $c\tau(\chitwo)$ & \map & Figures\\
        \hline
        0.1 & 2.5\,\gevcc  & 0.4\,\mchione & 0.1\,\cm & 3\,\mchione & \ref{fig:model_independent3},\ref{fig:model_dependent1}\\
        0.1 & 2.5\,\gevcc  & 0.4\,\mchione & \textcolor{purple}{0.01\,\cm} & 3\,\mchione & \ref{fig:model_independent3},\ref{fig:model_dependent1}\\
        0.1 & 2.5\,\gevcc  & 0.4\,\mchione & \textcolor{cyan}{1.0\,\cm} & 3\,\mchione & \ref{fig:model_independent4},\ref{fig:model_dependent2}\\
        0.1 & 2.5\,\gevcc  & 0.4\,\mchione & \textcolor{green}{100.0\,\cm} & 3\,\mchione & \ref{fig:model_independent4},\ref{fig:model_dependent2}\\
        0.1 & 2.5\,\gevcc  & 0.4\,\mchione & 0.1\,\cm & \textcolor{purple}{4\,\mchione} & \ref{fig:model_independent8},\ref{fig:model_dependent3}\\
        0.1 & 2.5\,\gevcc  & 0.4\,\mchione & \textcolor{cyan}{1.0\,\cm} & \textcolor{purple}{4\,\mchione} & \ref{fig:model_independent8},\ref{fig:model_dependent3}\\
        0.1 & \textcolor{purple}{1.25\,\gevcc}  & 0.4\,\mchione & 0.1\,\cm & 3\,\mchione & \ref{fig:model_independent10},\ref{fig:model_dependent5}\\
        0.1 & 2.5\,\gevcc  & \textcolor{purple}{0.2\,\mchione} & 0.1\,\cm & 3\,\mchione & \ref{fig:model_independent1},\ref{fig:model_dependent1}\\
        0.1 & 2.5\,\gevcc  & \textcolor{purple}{0.2\,\mchione} & \textcolor{purple}{0.01\,\cm} & 3\,\mchione & \ref{fig:model_independent1},\ref{fig:model_dependent1}\\
        0.1 & 2.5\,\gevcc  & \textcolor{purple}{0.2\,\mchione} & \textcolor{cyan}{1.0\,\cm} & 3\,\mchione & \ref{fig:model_independent2},\ref{fig:model_dependent2}\\
        0.1 & 2.5\,\gevcc  & \textcolor{purple}{0.2\,\mchione} & \textcolor{green}{100.0\,\cm} & 3\,\mchione & \ref{fig:model_independent2},\ref{fig:model_dependent2}\\
        0.1 & 2.5\,\gevcc  & \textcolor{purple}{0.2\,\mchione} & 0.1\,\cm & \textcolor{purple}{4\,\mchione} & \ref{fig:model_independent7},\ref{fig:model_dependent3}\\
        0.1 & 2.5\,\gevcc  & \textcolor{purple}{0.2\,\mchione} & \textcolor{cyan}{1.0\,\cm} & \textcolor{purple}{4\,\mchione} & \ref{fig:model_independent7},\ref{fig:model_dependent3}\\
        0.1 & 2.5\,\gevcc  & \textcolor{cyan}{1.0\,\mchione} & 0.1\,\cm & 3\,\mchione & \ref{fig:model_independent5},\ref{fig:model_dependent1}\\
        0.1 & 2.5\,\gevcc  & \textcolor{cyan}{1.0\,\mchione} & \textcolor{purple}{0.01\,\cm} & 3\,\mchione & \ref{fig:model_independent5},\ref{fig:model_dependent1}\\
        0.1 & 2.5\,\gevcc  & \textcolor{cyan}{1.0\,\mchione} & \textcolor{cyan}{1.0\,\cm} & 3\,\mchione & \ref{fig:model_independent6},\ref{fig:model_dependent2}\\
        0.1 & 2.5\,\gevcc  & \textcolor{cyan}{1.0\,\mchione} & \textcolor{green}{100.0\,\cm} & 3\,\mchione & \ref{fig:model_independent6},\ref{fig:model_dependent2}\\
        0.1 & 2.5\,\gevcc  & \textcolor{cyan}{1.0\,\mchione} & 0.1\,\cm & \textcolor{purple}{4\,\mchione} & \ref{fig:model_independent9},\ref{fig:model_dependent3}\\
        0.1 & 2.5\,\gevcc  & \textcolor{cyan}{1.0\,\mchione} & \textcolor{cyan}{1.0\,\cm} & \textcolor{purple}{4\,\mchione} & \ref{fig:model_independent9},\ref{fig:model_dependent3}\\
        0.1 & \textcolor{purple}{1.25\,\gevcc}  & \textcolor{cyan}{1.0\,\mchione} & 0.1\,\cm & 3\,\mchione & \ref{fig:model_independent10},\ref{fig:model_dependent5}\\
        \textcolor{purple}{0.5} & 2.5\,\gevcc  & 0.4\,\mchione & \textcolor{purple}{0.01\,\cm} & 3\,\mchione & \ref{fig:model_dependent4}\\
        \textcolor{purple}{0.5} & 2.5\,\gevcc  & 0.4\,\mchione & \textcolor{cyan}{1.0\,\cm} & 3\,\mchione & \ref{fig:model_dependent4}\\
        \textcolor{purple}{0.5} & 2.5\,\gevcc  & \textcolor{purple}{0.2\,\mchione} & \textcolor{purple}{0.01\,\cm} & 3\,\mchione & \ref{fig:model_dependent4}\\
        \textcolor{purple}{0.5} & 2.5\,\gevcc  & \textcolor{purple}{0.2\,\mchione} & \textcolor{cyan}{1.0\,\cm} & 3\,\mchione & \ref{fig:model_dependent4}\\
        \textcolor{purple}{0.5} & 2.5\,\gevcc  & \textcolor{cyan}{1.0\,\mchione} & \textcolor{purple}{0.01\,\cm} & 3\,\mchione & \ref{fig:model_dependent4}\\
        \textcolor{purple}{0.5} & 2.5\,\gevcc  & \textcolor{cyan}{1.0\,\mchione} & \textcolor{cyan}{1.0\,\cm} & 3\,\mchione & \ref{fig:model_dependent4}\\
    \end{tabular}
\end{table}
We also show the the model-independent and model-dependent upper limits (95\% credibility level) on the \prodbf\ as a function of \chione\ mass for several variations of $\mdh$, $\Delta m$, $c\tau(\dh)$, and $\map$.
All studied combinations are summarized in Tab.~\ref{tab:chosen_parameters_mdh} and the corresponding figures are referenced in the table.
\begin{table}
    \centering
    \caption{
        Model parameter variations used for the scans in the \mchione{}-$\epsilon$-plane.
        Identical parameter values in each column share the same colour for better visualization.
        The last column references the corresponding figure.
    }
    \label{tab:chosen_parameters_mdh}
    \renewcommand{\arraystretch}{1.5}
    \begin{tabular}{c|c|c|c|c|c}
        $\alpha_D$ & \mdh{} & $c\tau(\dh)$ & $\Delta m$ &  \map & Figures\\
        \hline
        0.1 & 0.4\,\gevcc & 21.54\,\cm & 0.4\,\mchione & 3\,\mchione & \ref{fig:model_independent14},\ref{fig:model_dependent8}\\
        0.1 & 0.4\,\gevcc & 21.54\,\cm & 0.4\,\mchione & \textcolor{purple}{4\,\mchione} & \ref{fig:model_independent14},\ref{fig:model_dependent8}\\
        0.1 & \textcolor{purple}{0.6\,\gevcc} & \textcolor{cyan}{1.0\,\cm} & 0.4\,\mchione & 3\,\mchione & \ref{fig:model_independent12},\ref{fig:model_dependent6}\\
        0.1 & \textcolor{purple}{0.6\gevcc} & 21.54\,\cm & 0.4\,\mchione & 3\,\mchione & \ref{fig:model_independent12},\ref{fig:model_dependent7}\\
        0.1 & \textcolor{cyan}{1.2\,\gevcc} & 21.54\,\cm & 0.4\,\mchione & 3\,\mchione & \ref{fig:model_independent13},\ref{fig:model_dependent7}\\
        0.1 & \textcolor{purple}{0.6\,\gevcc} & 21.54\,\cm & \textcolor{purple}{0.2\,\mchione} & 3\,\mchione & \ref{fig:model_independent11},\ref{fig:model_dependent6}\\
        0.1 & \textcolor{purple}{0.6\,\gevcc} & \textcolor{cyan}{1.0\,\cm} & \textcolor{purple}{0.2\,\mchione} & 3\,\mchione & \ref{fig:model_independent11},\ref{fig:model_dependent6}\\
        0.1 & \textcolor{cyan}{1.2\,\gevcc} & 21.54\,\cm & \textcolor{purple}{0.2\,\mchione} & 3\,\mchione & \ref{fig:model_independent13},\ref{fig:model_dependent7}\\
    \end{tabular}
\end{table}

\FloatBarrier

\begin{figure*}[htp!]
    \centering
    \includegraphics[width=0.45\textwidth]{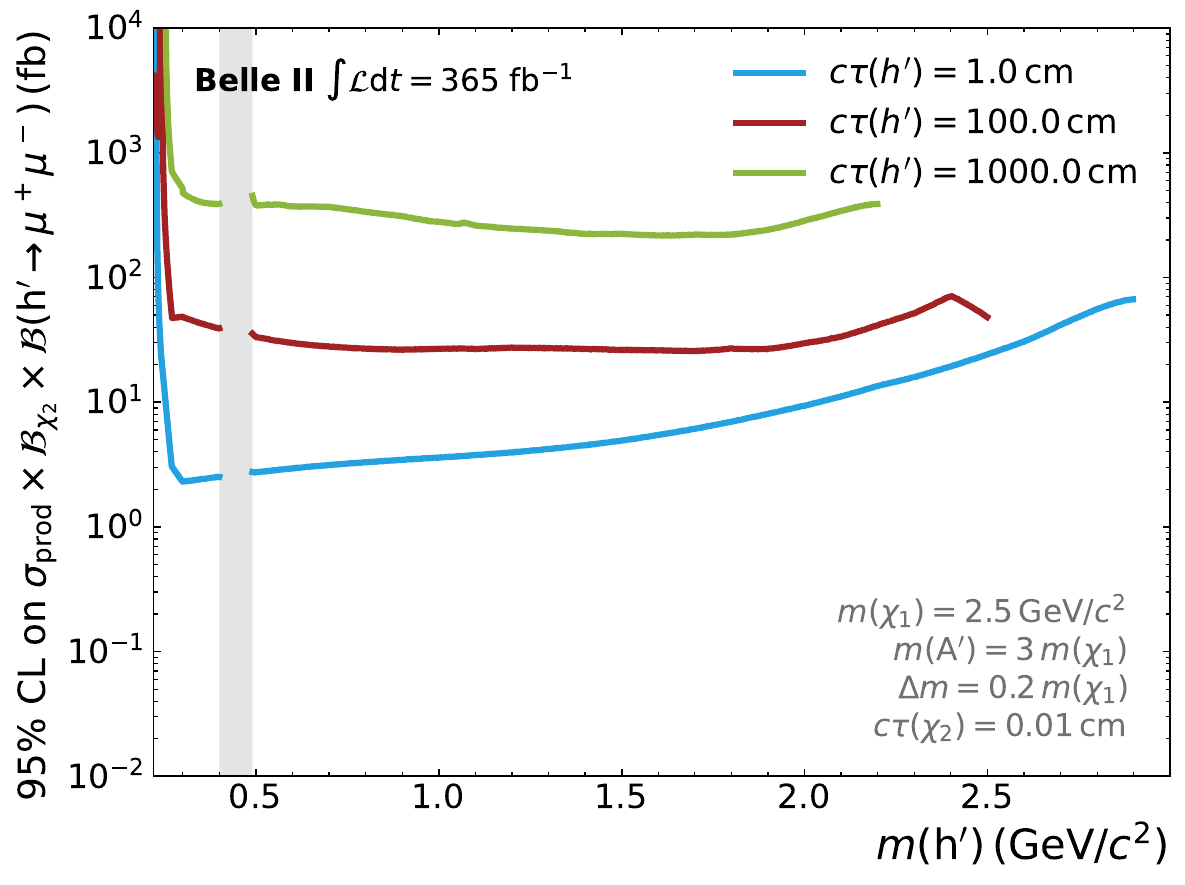}
    \includegraphics[width=0.45\textwidth]{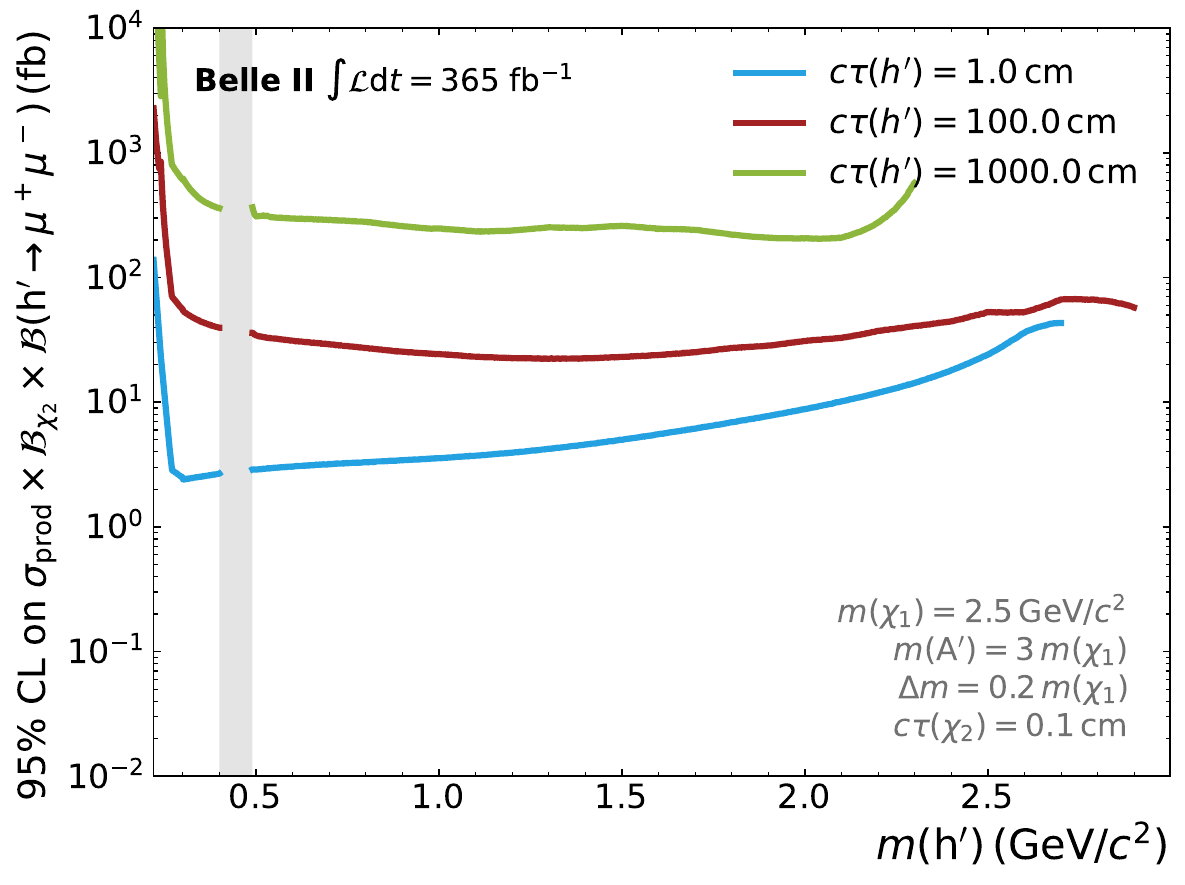}
    \includegraphics[width=0.45\textwidth]{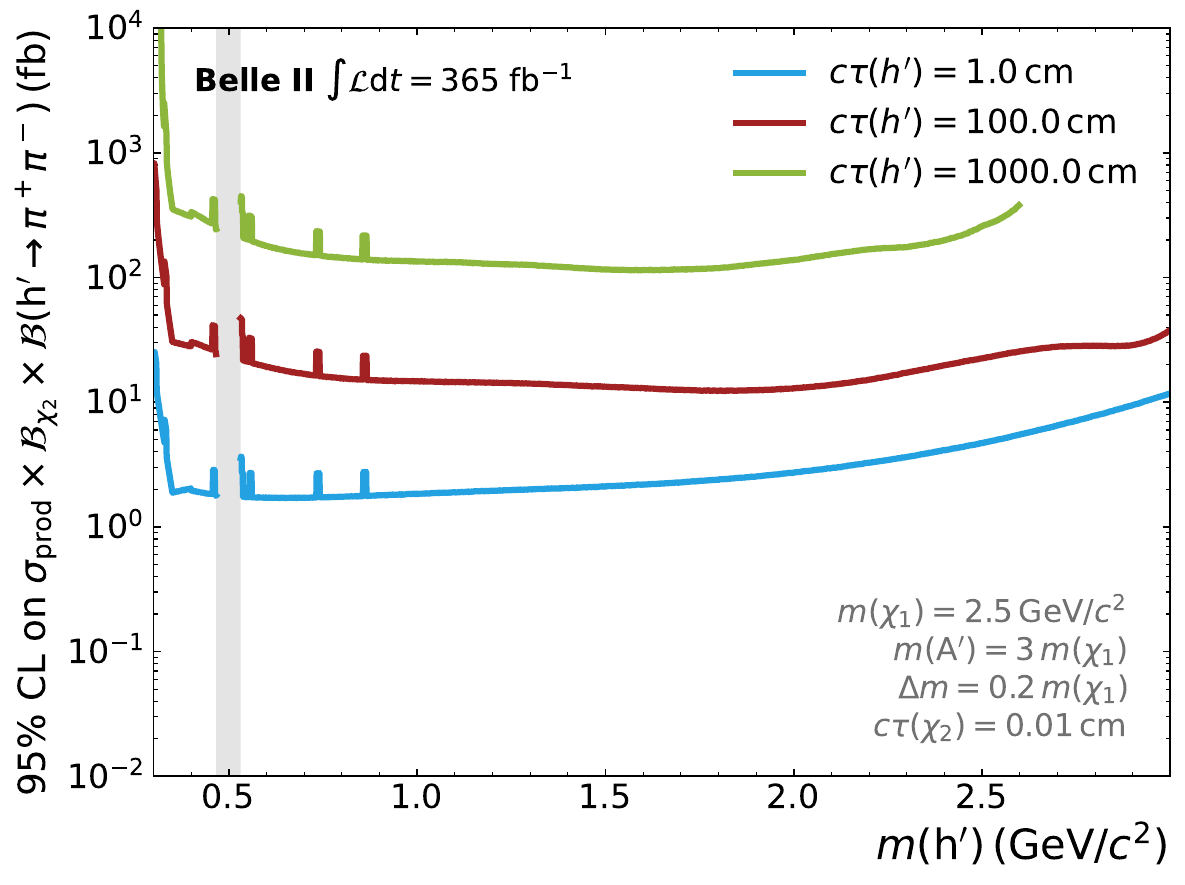}
    \includegraphics[width=0.45\textwidth]{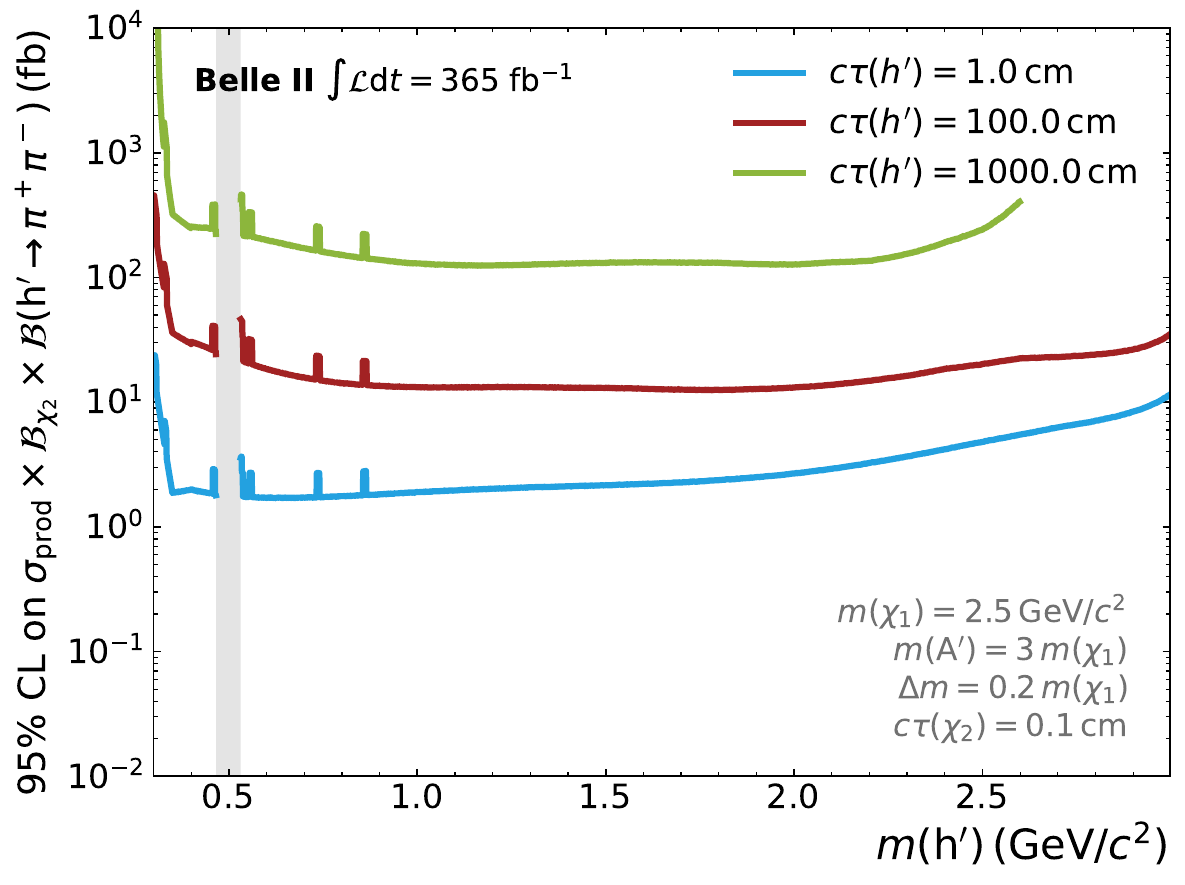}
    \includegraphics[width=0.45\textwidth]{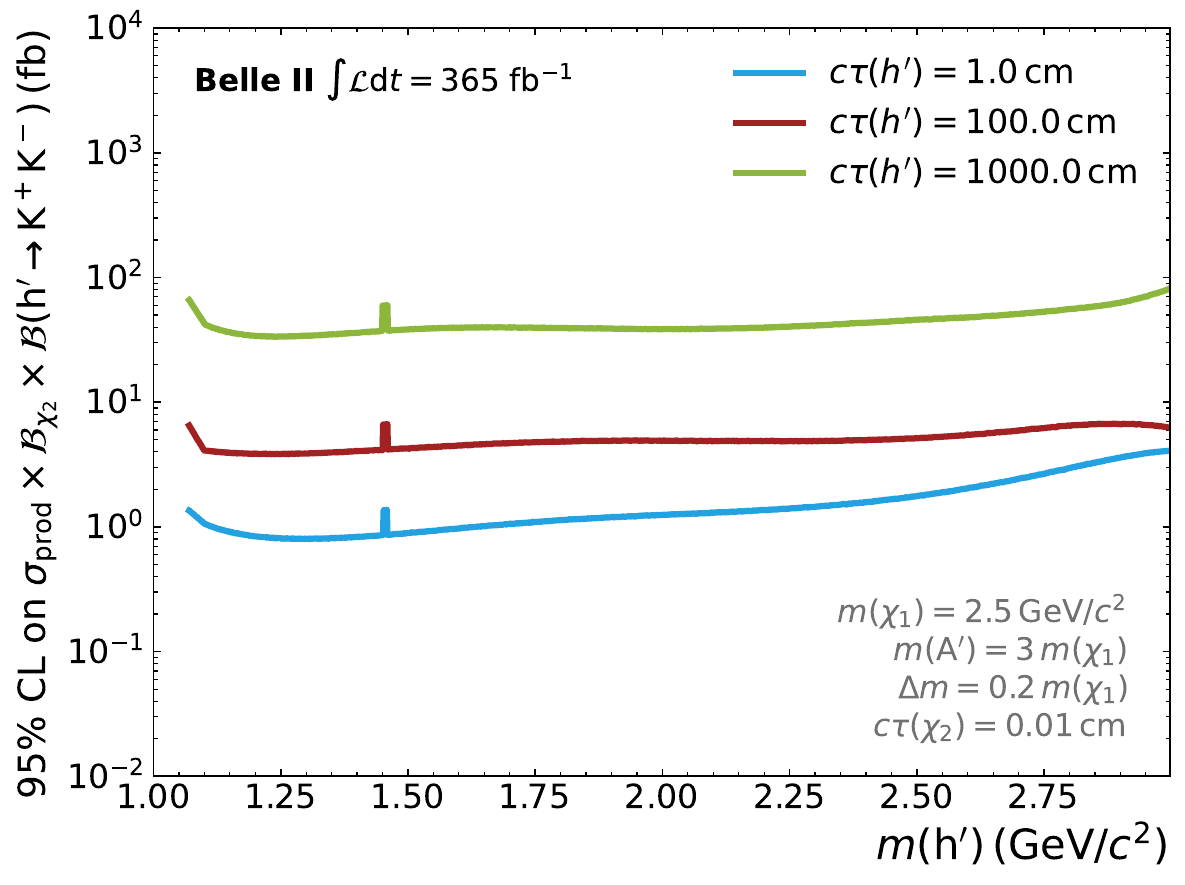}
    \includegraphics[width=0.45\textwidth]{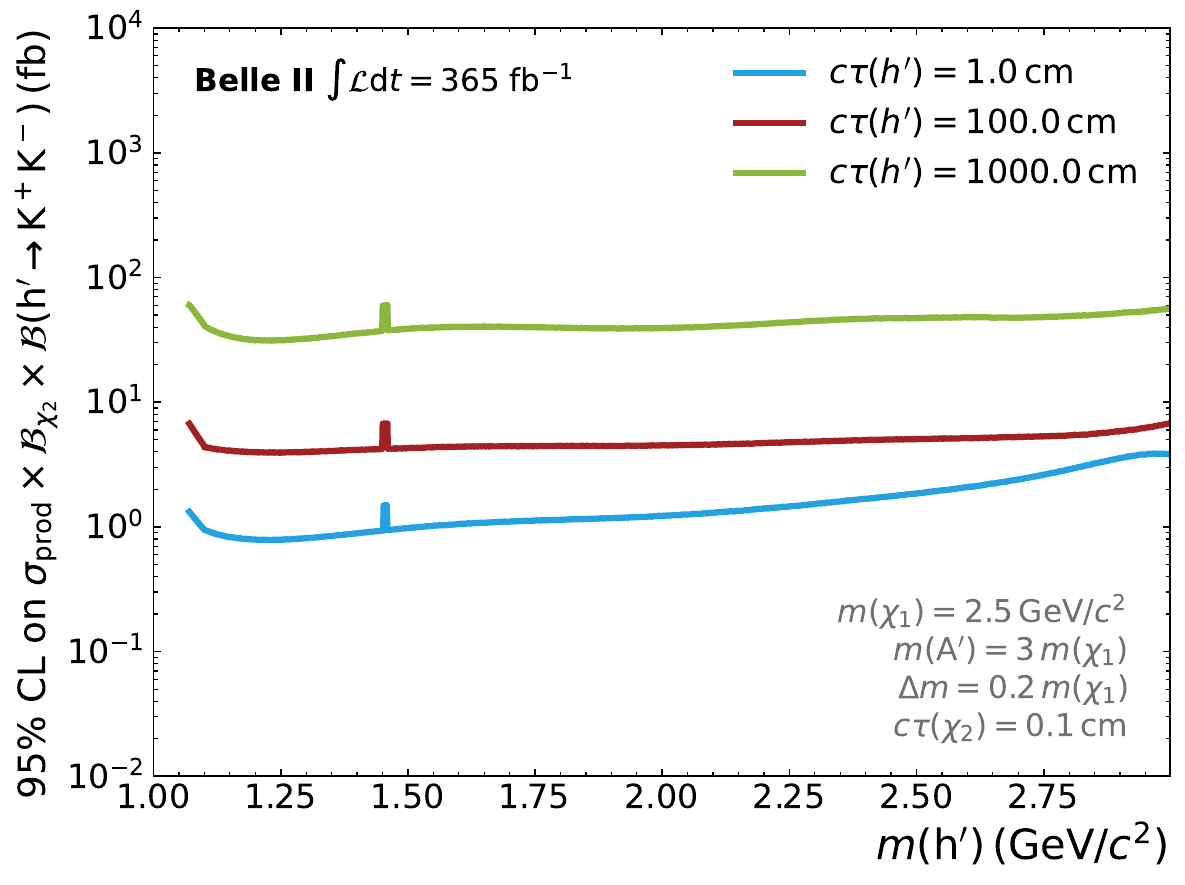}
    \caption{
Upper limits (95\% credibility level) on the \prodbf\ as function of dark Higgs mass $\mdh$ for \hbox{$c\tau(\dh) =1\,\text{cm}$}~(blue), \hbox{$c\tau(\dh) =100\,\text{cm}$}~(red), and \hbox{$c\tau(\dh) =1000\,\text{cm}$}~(green) for $\dh\to\mu^+\mu^-$\,(top), $\dh\to\pi^+\pi^-$\,(center) and $\dh\to K^+ K^-$\,(bottom).
The \chitwo{} lifetime is chosen as $c\tau(\chi_2) = 0.01\,\cm$ (left) and $c\tau(\chi_2) = 0.1\,\cm$ (right).
The remaining model parameters are chosen as $\mchione = 2.5\,\gevcc$, $\map = 3\,\mchione$, and $\Delta m = 0.2\,\mchione$.
The region corresponding to the fully-vetoed \KS mass region is marked in gray.
}
    \label{fig:model_independent1}
\end{figure*}

\begin{figure*}[htp!]
    \centering
    \includegraphics[width=0.45\textwidth]{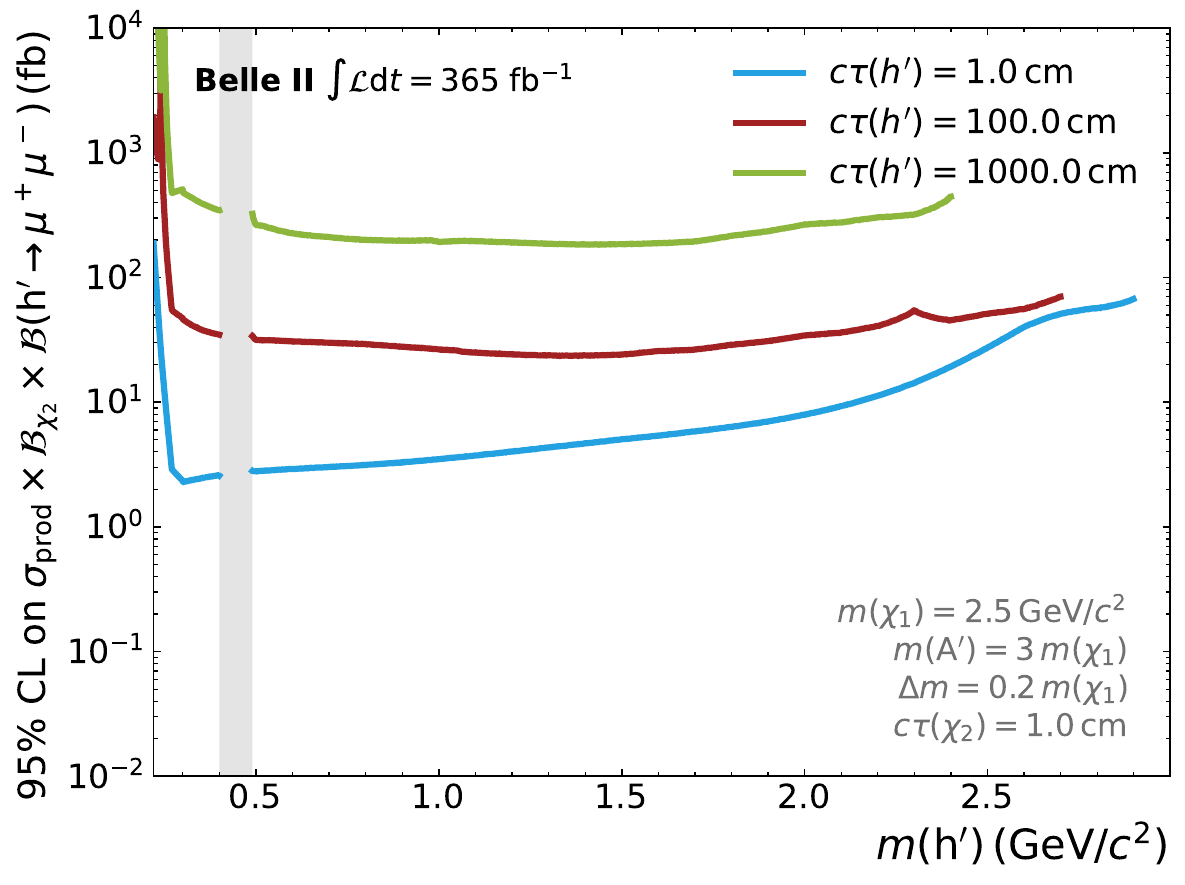}
    \includegraphics[width=0.45\textwidth]{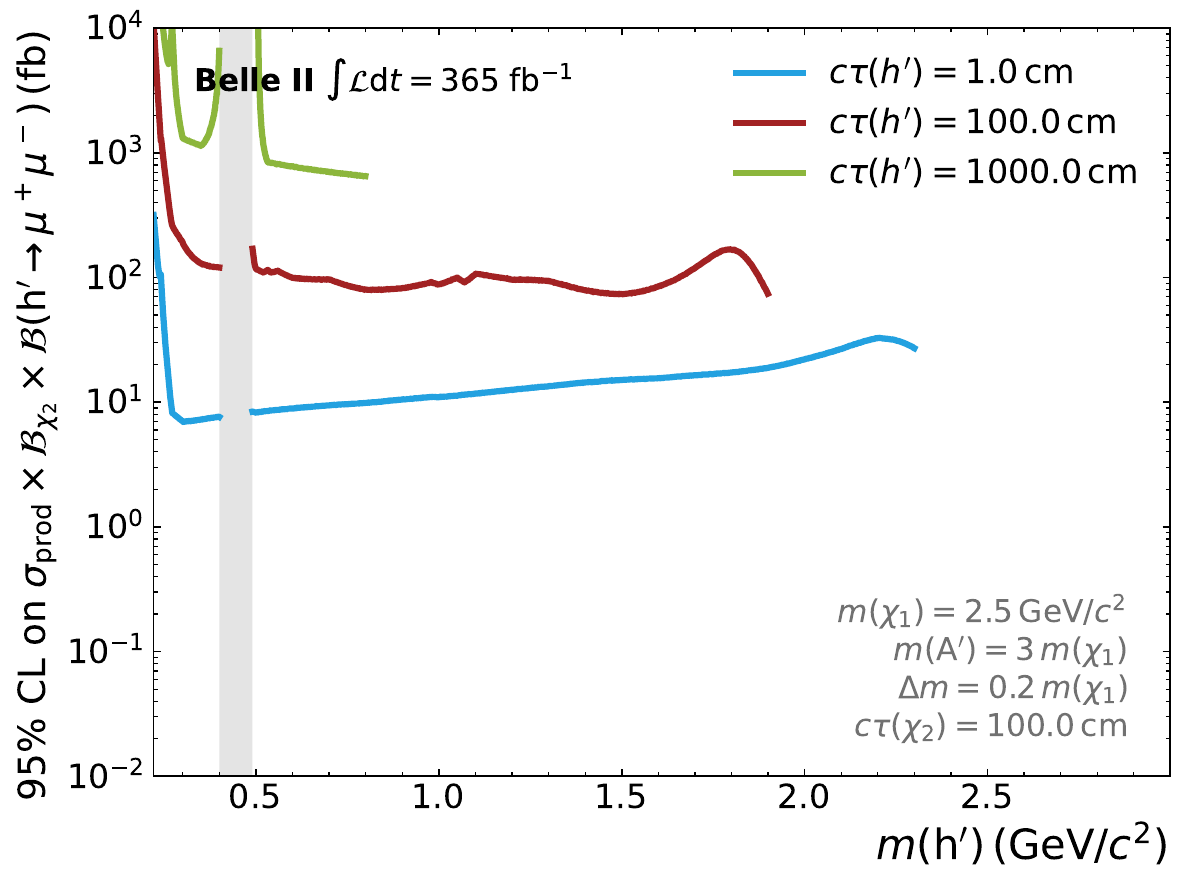}
    \includegraphics[width=0.45\textwidth]{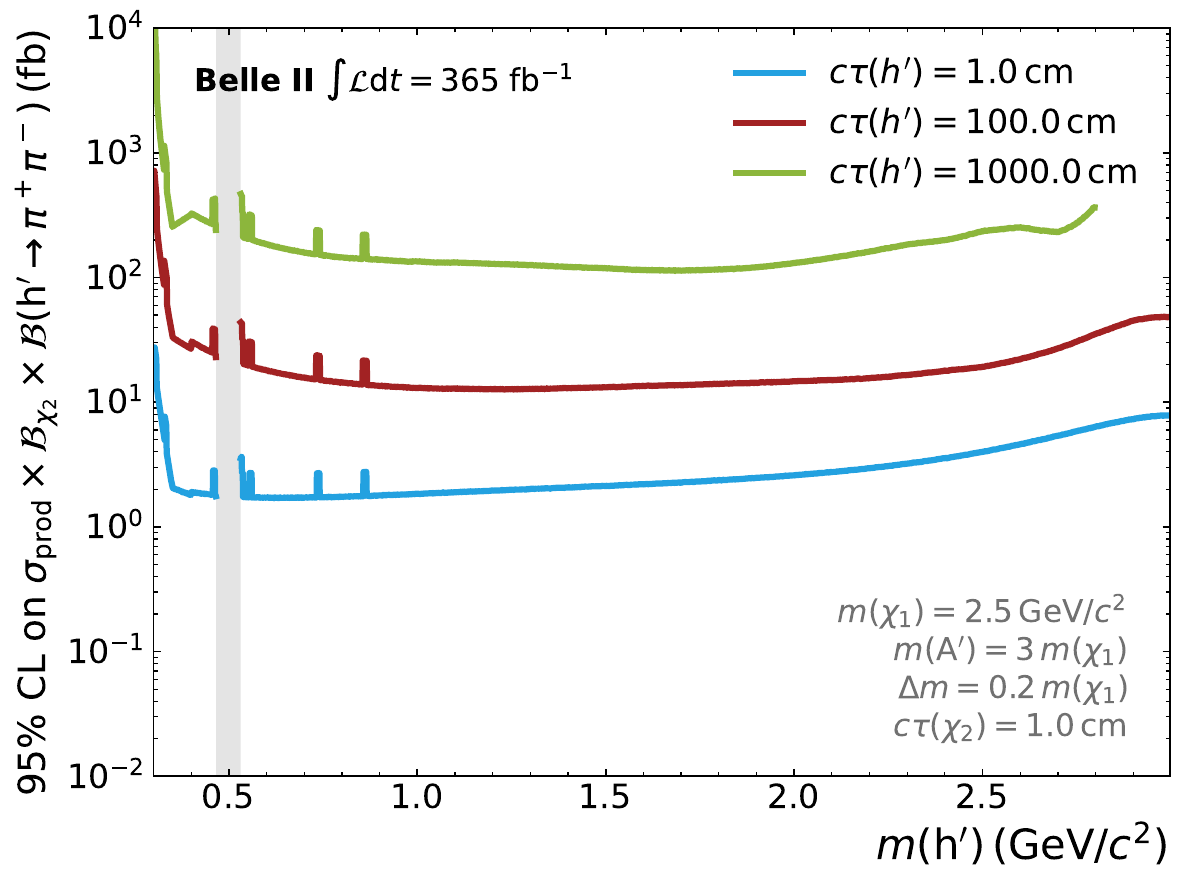}
    \includegraphics[width=0.45\textwidth]{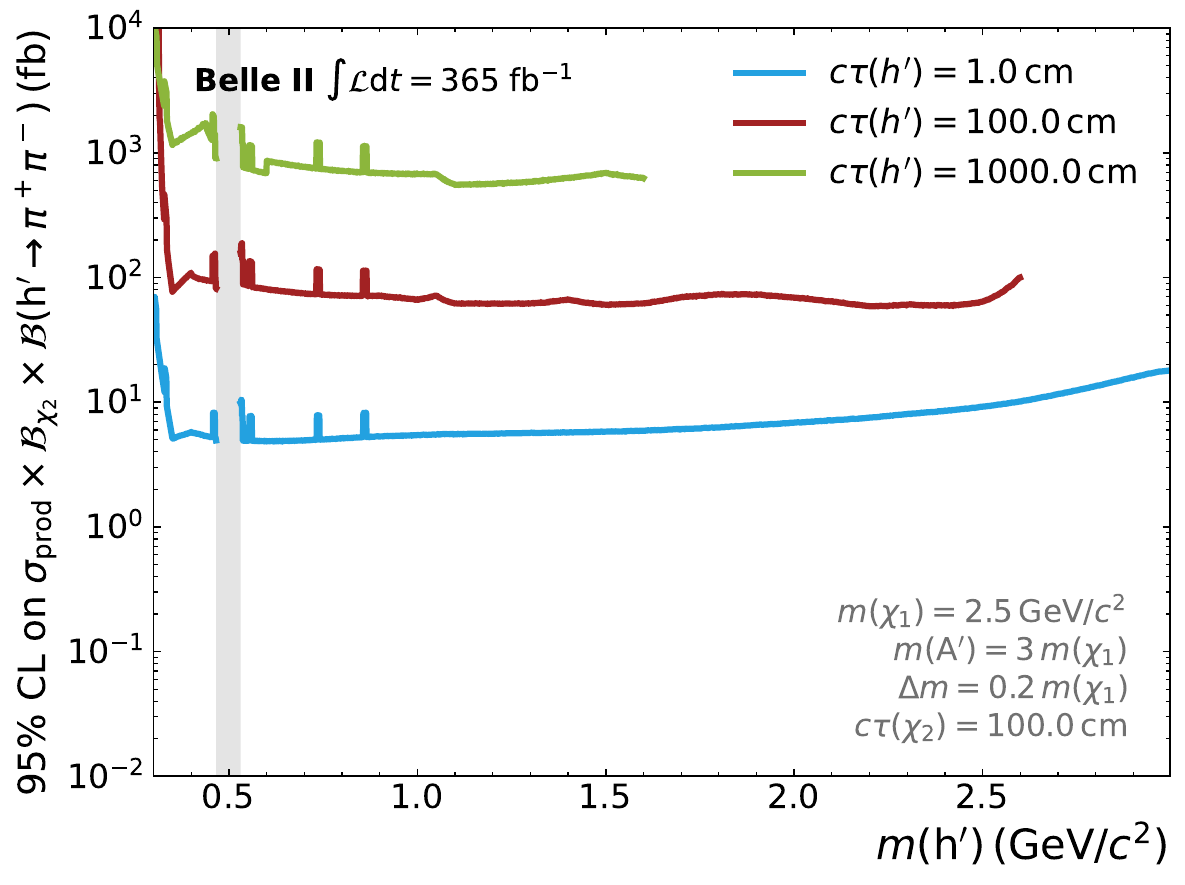}
    \includegraphics[width=0.45\textwidth]{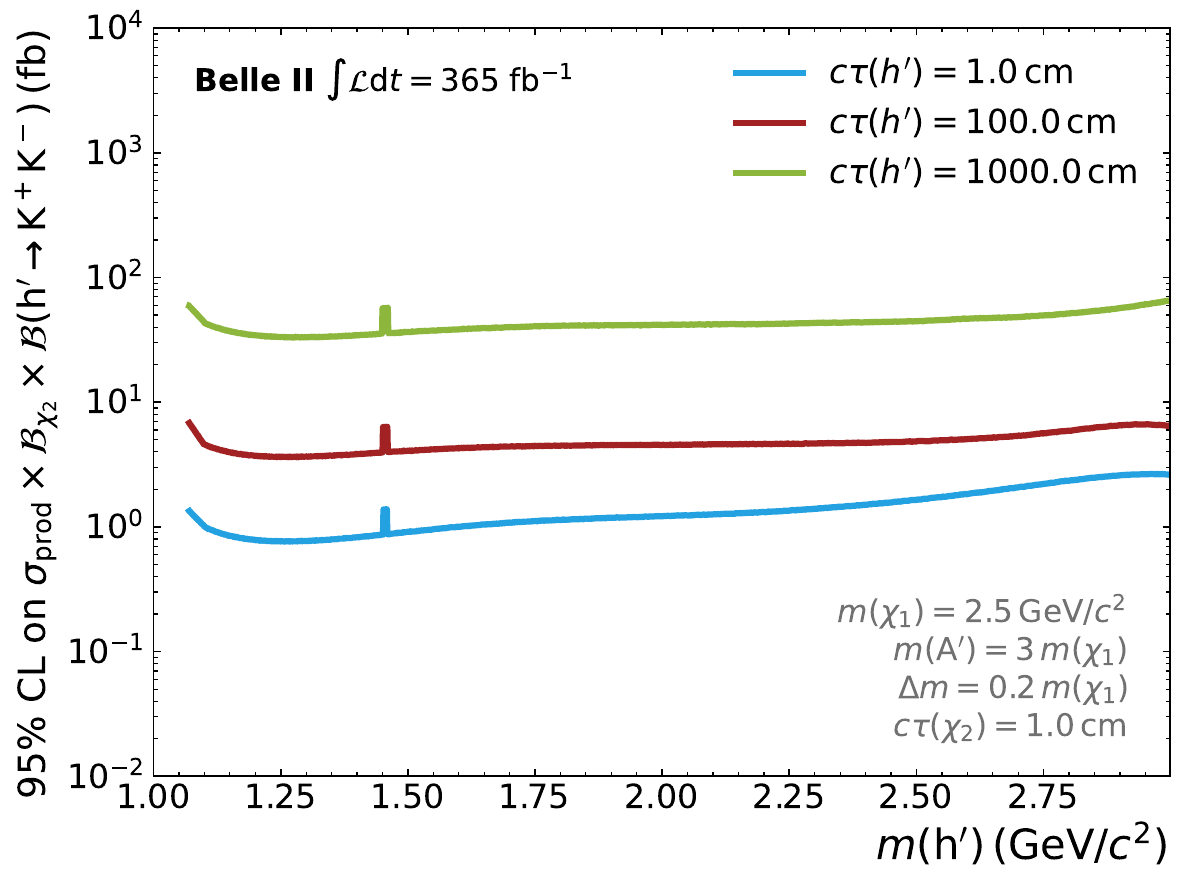}
    \includegraphics[width=0.45\textwidth]{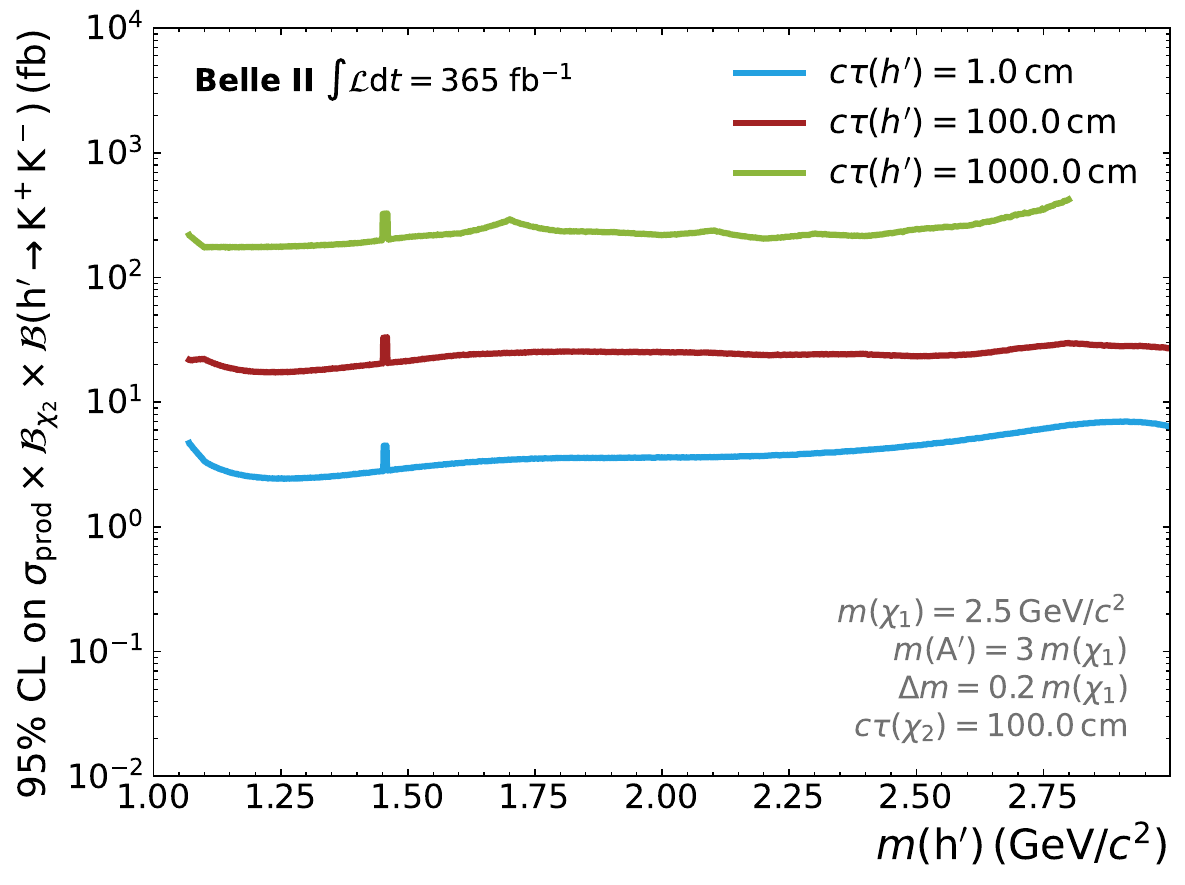}
    \caption{
Upper limits (95\% credibility level) on the \prodbf\ as function of dark Higgs mass $\mdh$ for \hbox{$c\tau(\dh) =1\,\text{cm}$}~(blue), \hbox{$c\tau(\dh) =100\,\text{cm}$}~(red), and \hbox{$c\tau(\dh) =1000\,\text{cm}$}~(green) for $\dh\to\mu^+\mu^-$\,(top), $\dh\to\pi^+\pi^-$\,(center) and $\dh\to K^+ K^-$\,(bottom).
The \chitwo{} lifetime is chosen as $c\tau(\chi_2) = 1.0\,\cm$ (left) and $c\tau(\chi_2) = 100.0\,\cm$ (right).
The remaining model parameters are chosen as $\mchione = 2.5\,\gevcc$, $\map = 3\,\mchione$, and $\Delta m = 0.2\,\mchione$.
The region corresponding to the fully-vetoed \KS mass region is marked in gray.
}
    \label{fig:model_independent2}
\end{figure*}

\begin{figure*}[htp!]
    \centering
    \includegraphics[width=0.45\textwidth]{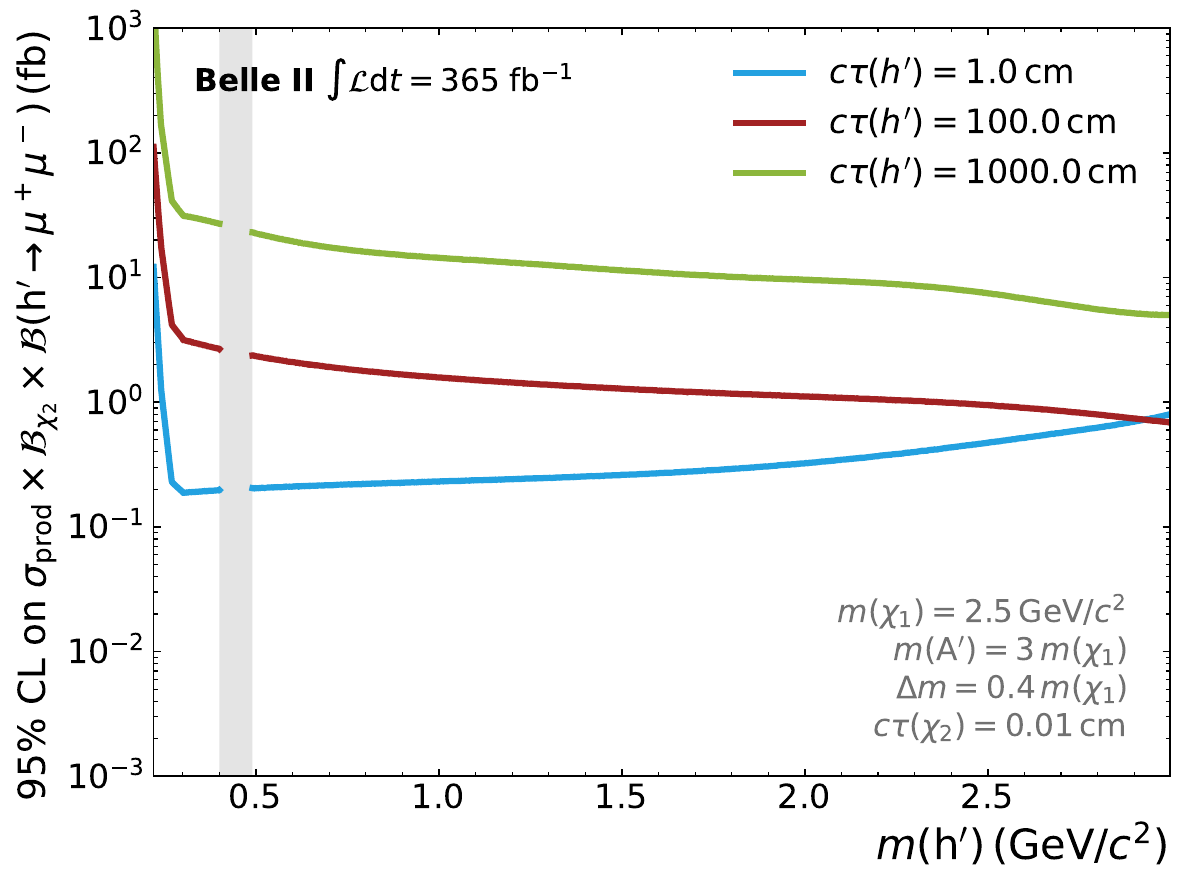}
    \includegraphics[width=0.45\textwidth]{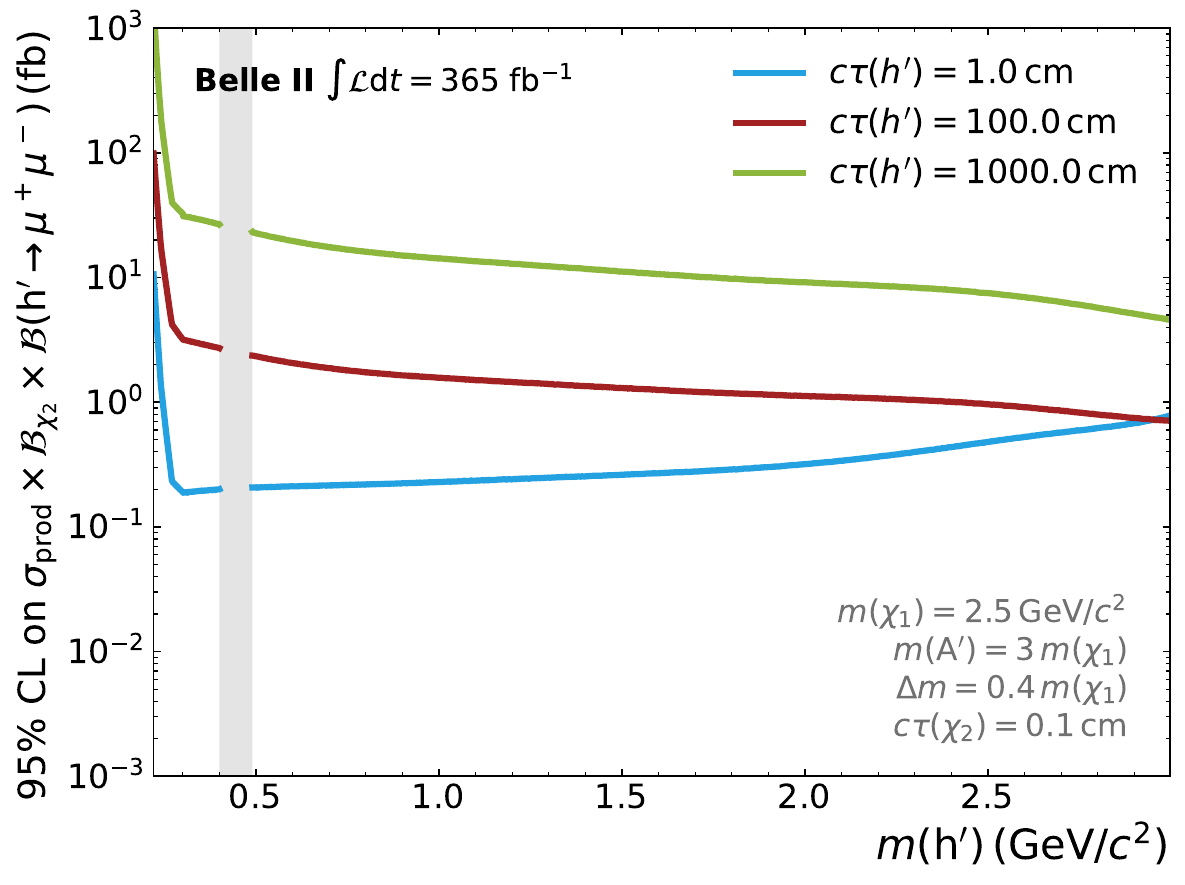}
    \includegraphics[width=0.45\textwidth]{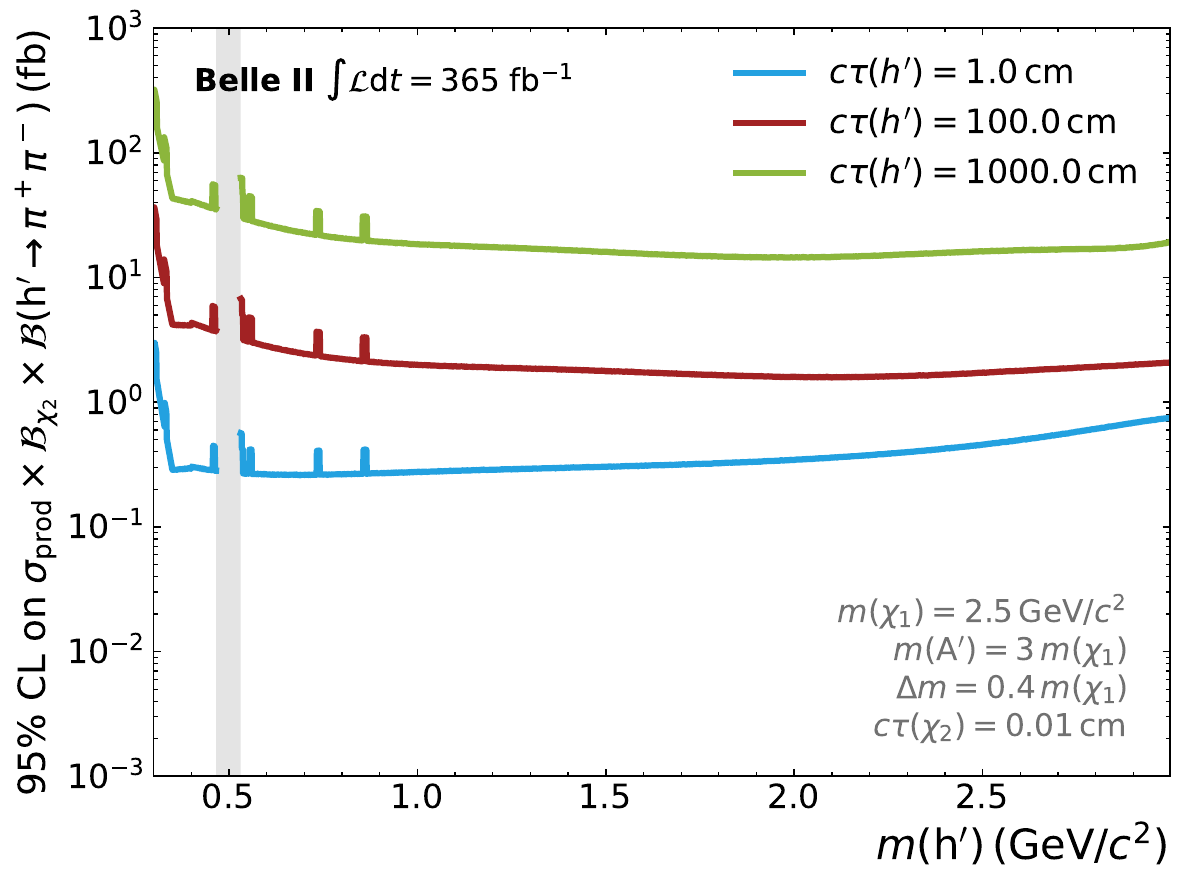}
    \includegraphics[width=0.45\textwidth]{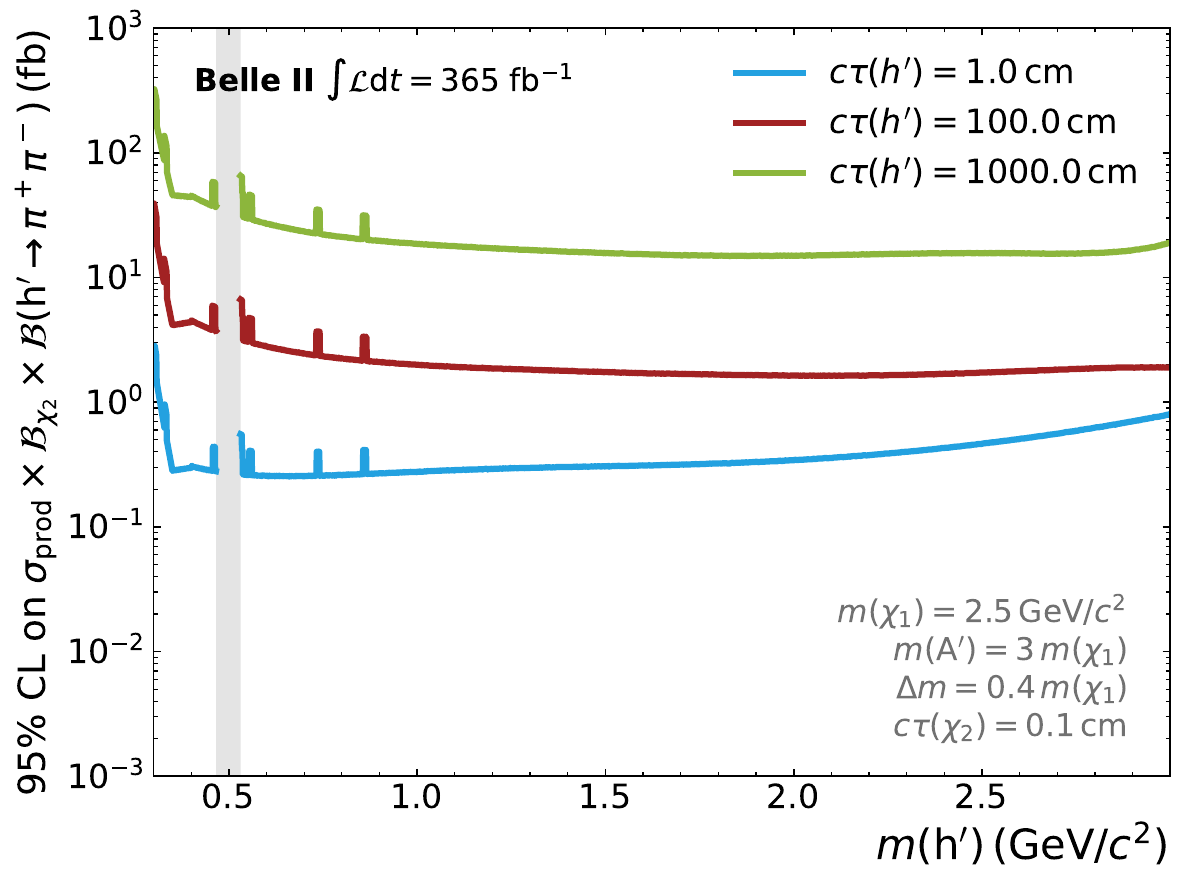}
    \includegraphics[width=0.45\textwidth]{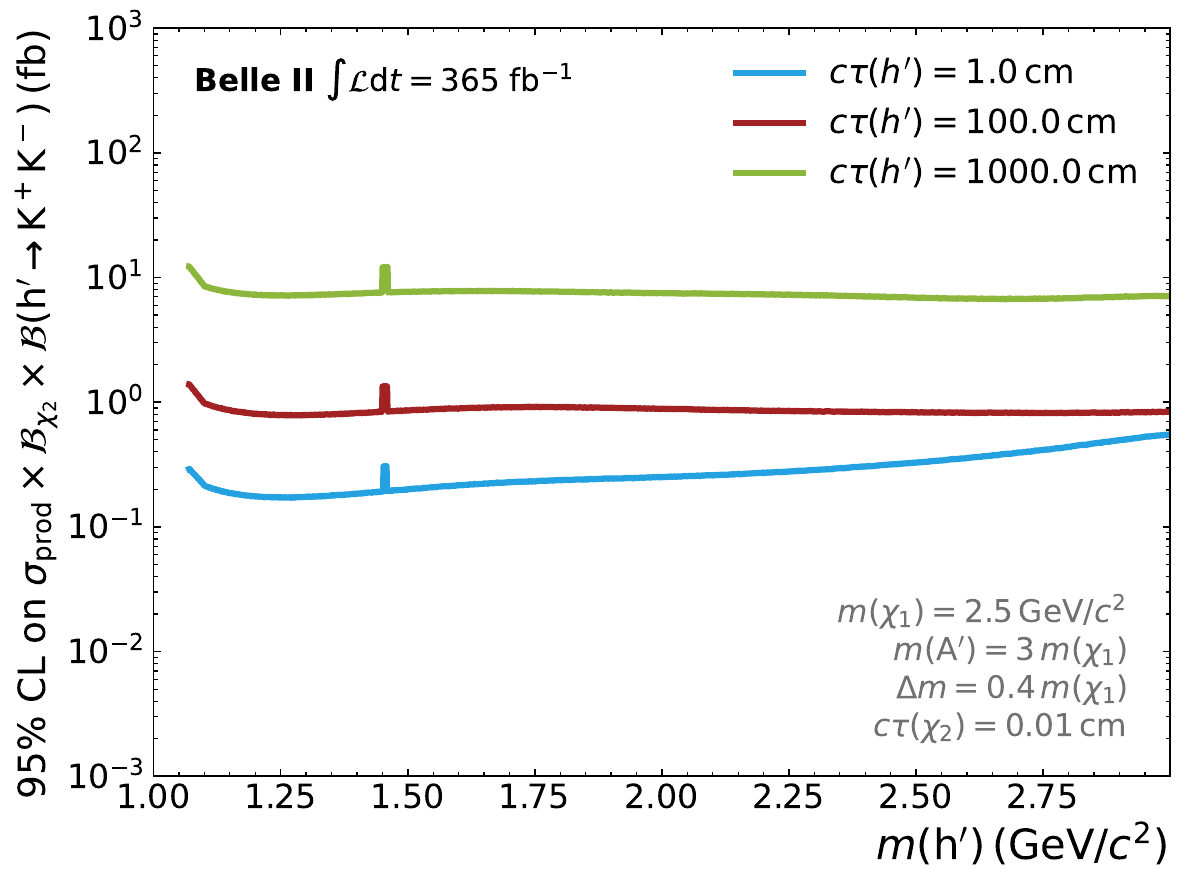}
    \includegraphics[width=0.45\textwidth]{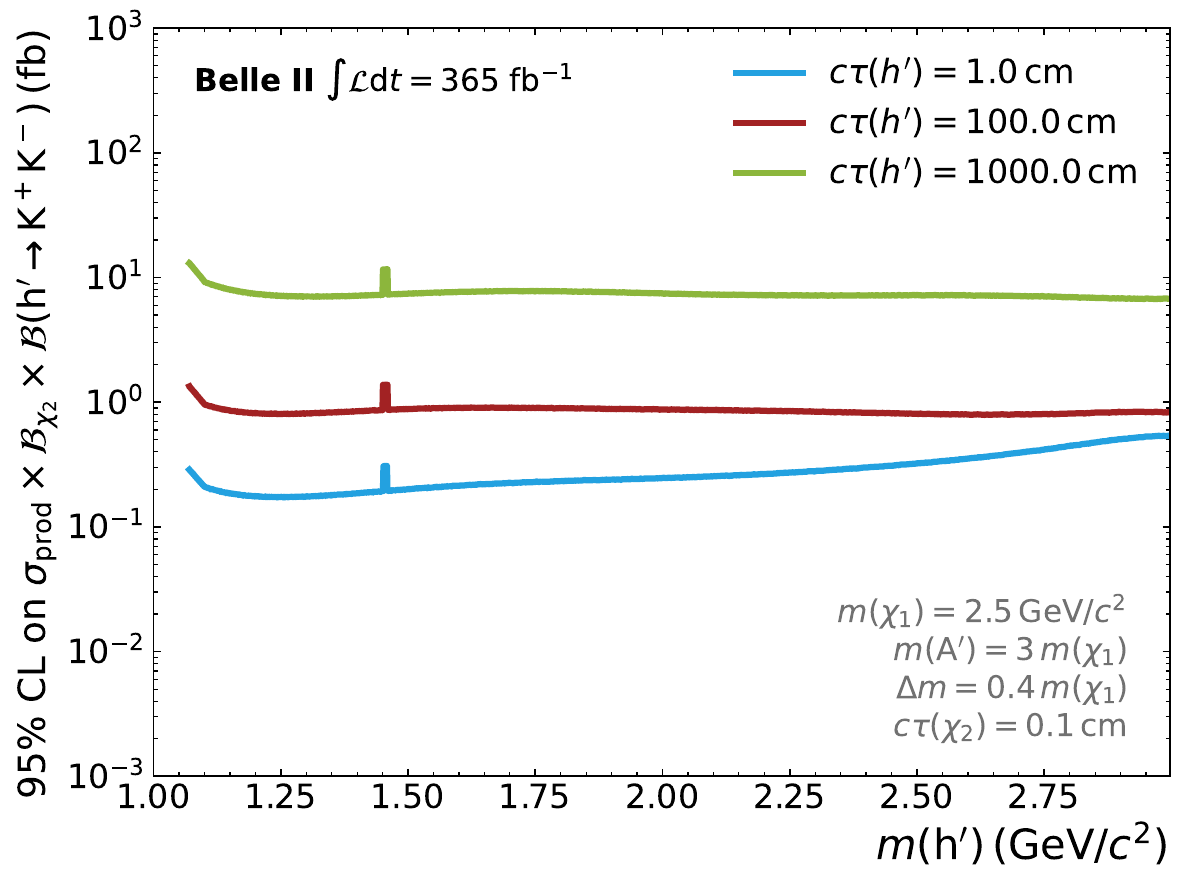}
    \caption{
Upper limits (95\% credibility level) on the \prodbf\ as function of dark Higgs mass $\mdh$ for \hbox{$c\tau(\dh) =1\,\text{cm}$}~(blue), \hbox{$c\tau(\dh) =100\,\text{cm}$}~(red), and \hbox{$c\tau(\dh) =1000\,\text{cm}$}~(green) for $\dh\to\mu^+\mu^-$\,(top), $\dh\to\pi^+\pi^-$\,(center) and $\dh\to K^+ K^-$\,(bottom).
The \chitwo{} lifetime is chosen as $c\tau(\chi_2) = 0.01\,\cm$ (left) and $c\tau(\chi_2) = 0.1\,\cm$ (right).
The remaining model parameters are chosen as $\mchione = 2.5\,\gevcc$, $\map = 3\,\mchione$, and $\Delta m = 0.4\,\mchione$.
The region corresponding to the fully-vetoed \KS mass region is marked in gray.
}
    \label{fig:model_independent3}
\end{figure*}

\begin{figure*}[htp!]
    \centering
    \includegraphics[width=0.45\textwidth]{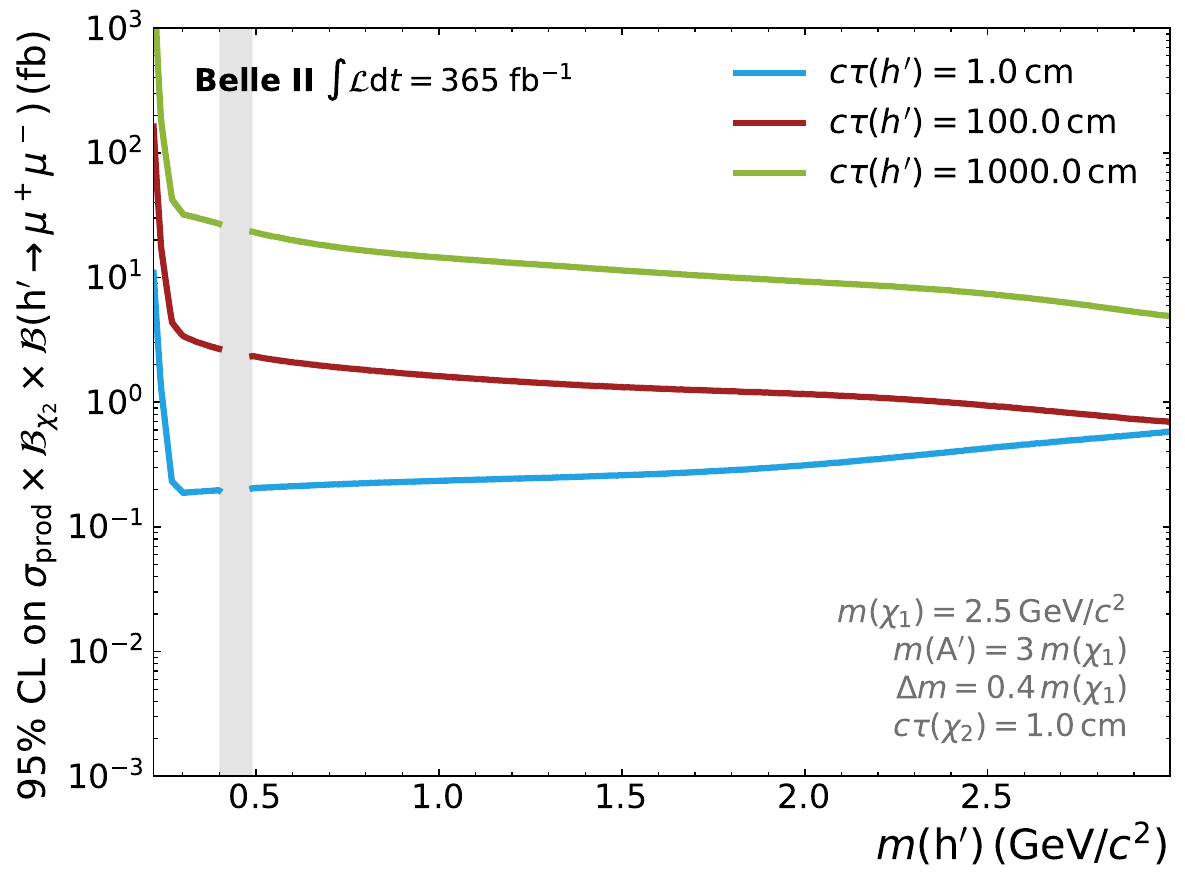}
    \includegraphics[width=0.45\textwidth]{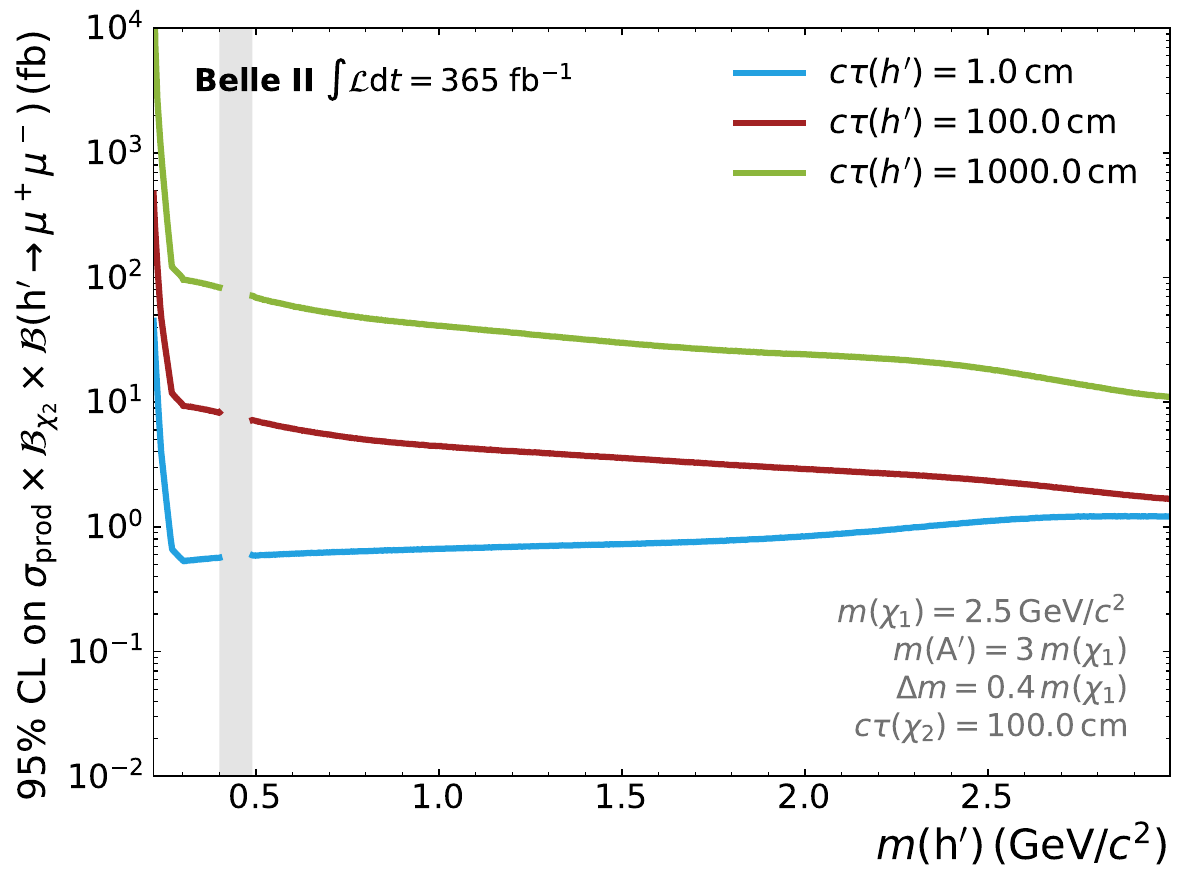}
    \includegraphics[width=0.45\textwidth]{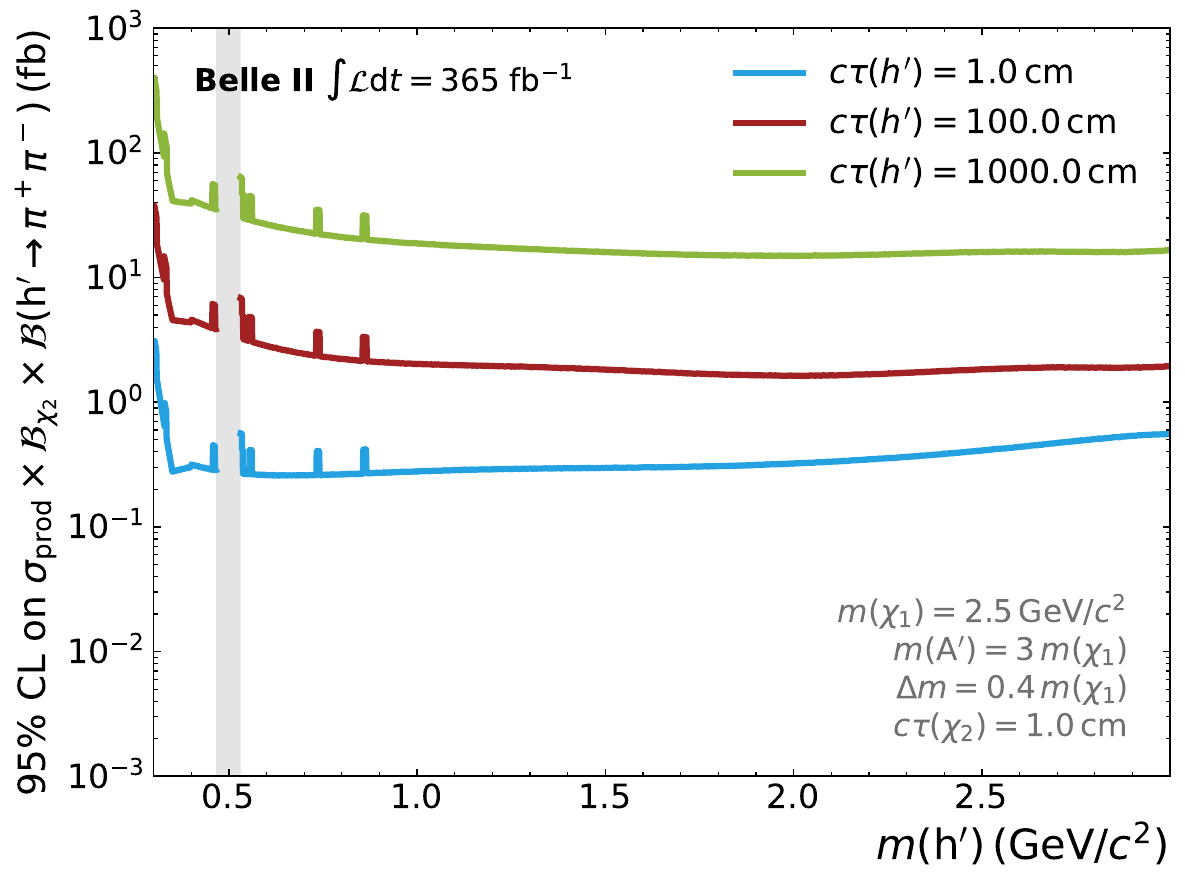}
    \includegraphics[width=0.45\textwidth]{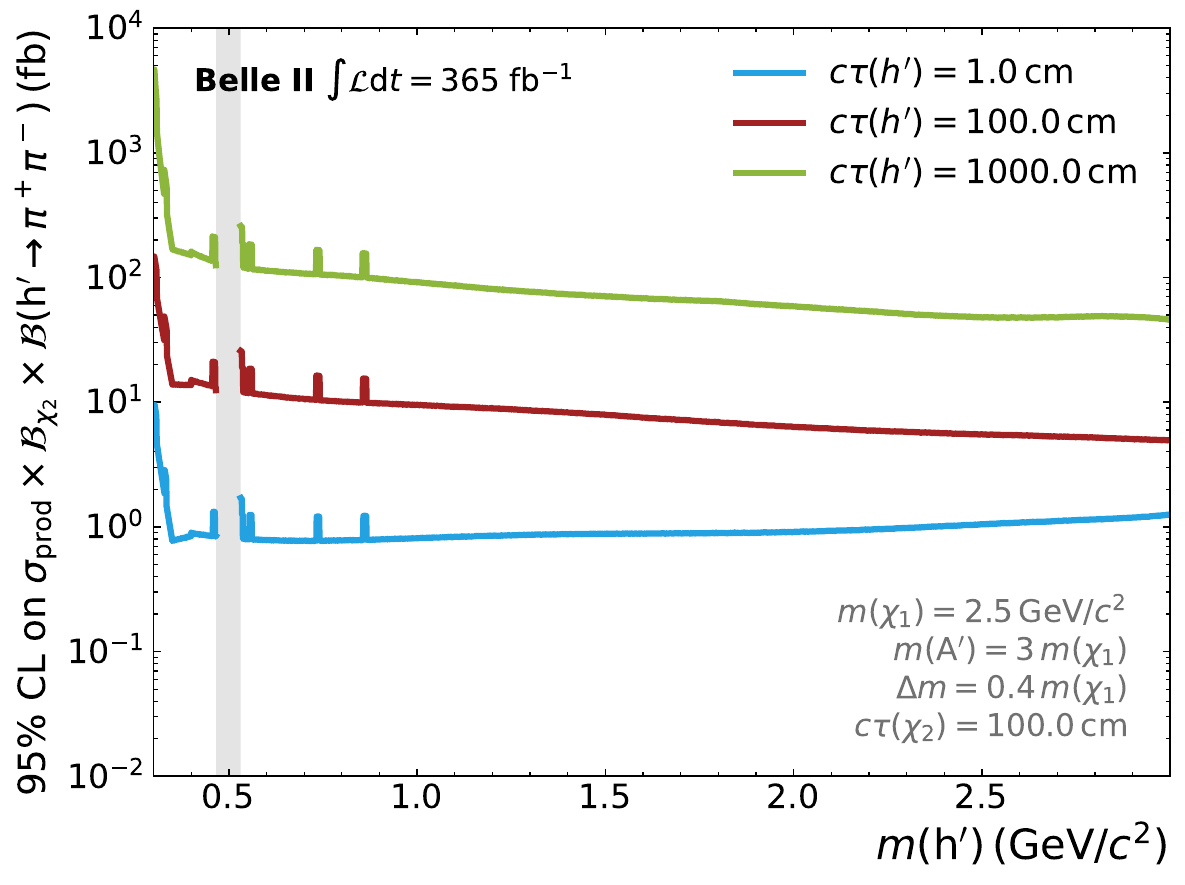}
    \includegraphics[width=0.45\textwidth]{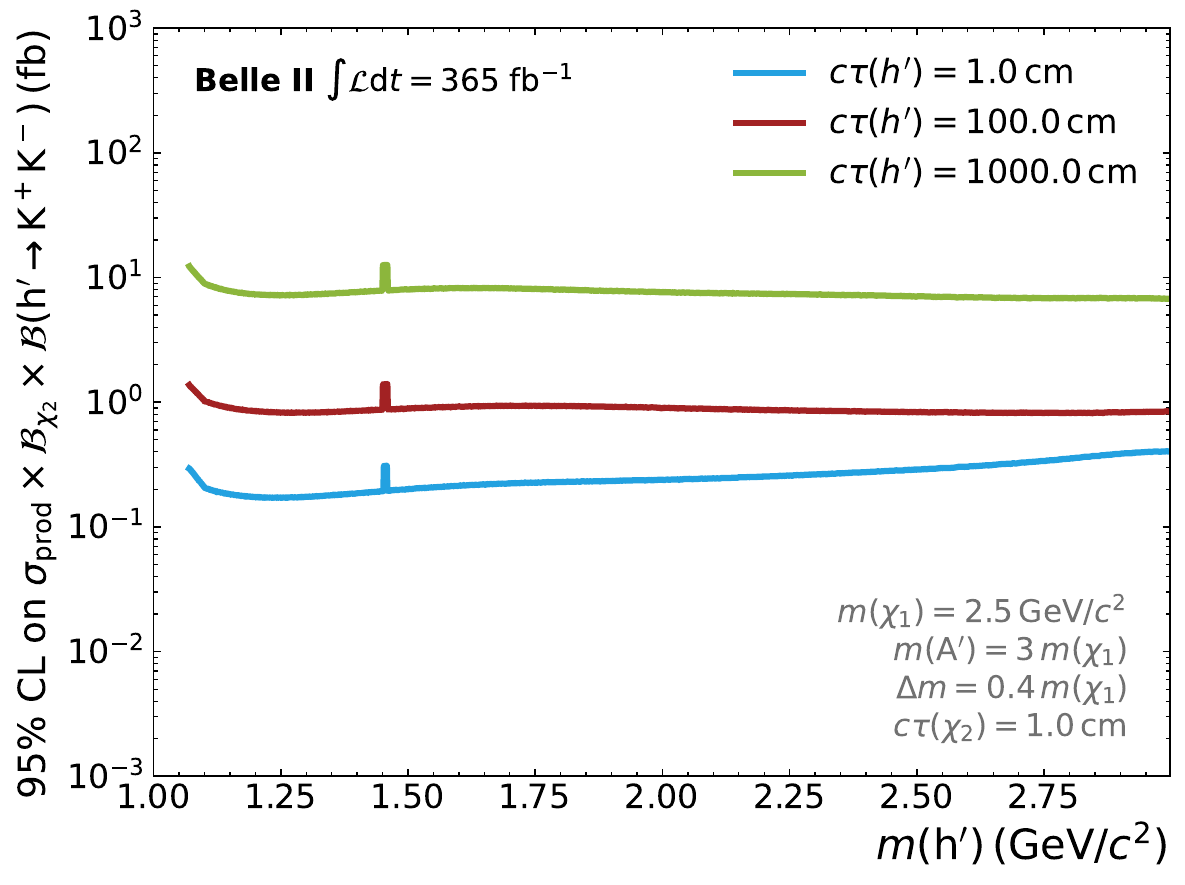}
    \includegraphics[width=0.45\textwidth]{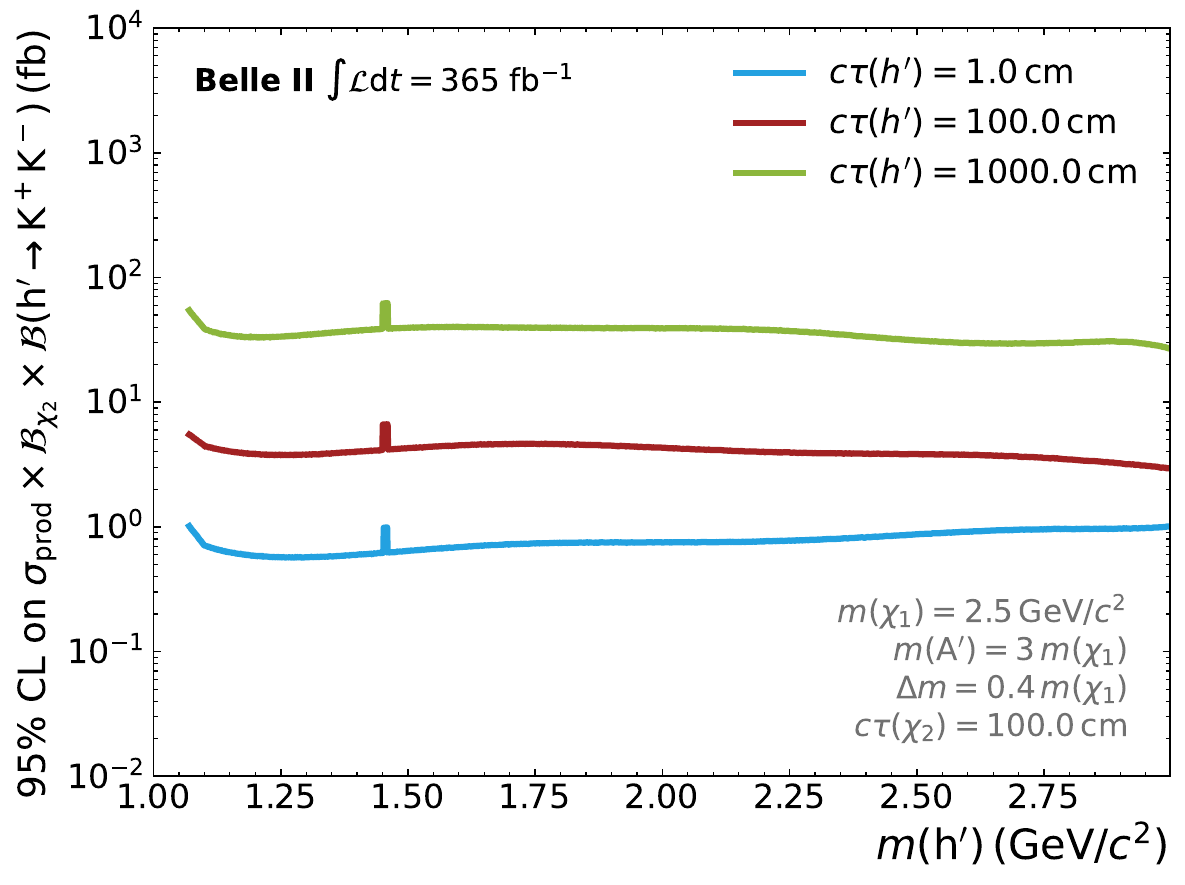}
    \caption{
Upper limits (95\% credibility level) on the \prodbf\ as function of dark Higgs mass $\mdh$ for \hbox{$c\tau(\dh) =1\,\text{cm}$}~(blue), \hbox{$c\tau(\dh) =100\,\text{cm}$}~(red), and \hbox{$c\tau(\dh) =1000\,\text{cm}$}~(green) for $\dh\to\mu^+\mu^-$\,(top), $\dh\to\pi^+\pi^-$\,(center) and $\dh\to K^+ K^-$\,(bottom).
The \chitwo{} lifetime is chosen as $c\tau(\chi_2) = 1.0\,\cm$ (left) and $c\tau(\chi_2) = 100.0\,\cm$ (right).
The remaining model parameters are chosen as $\mchione = 2.5\,\gevcc$, $\map = 3\,\mchione$, and $\Delta m = 0.4\,\mchione$.
The region corresponding to the fully-vetoed \KS mass region is marked in gray.
}
    \label{fig:model_independent4}
\end{figure*}

\begin{figure*}[htp!]
    \centering
    \includegraphics[width=0.45\textwidth]{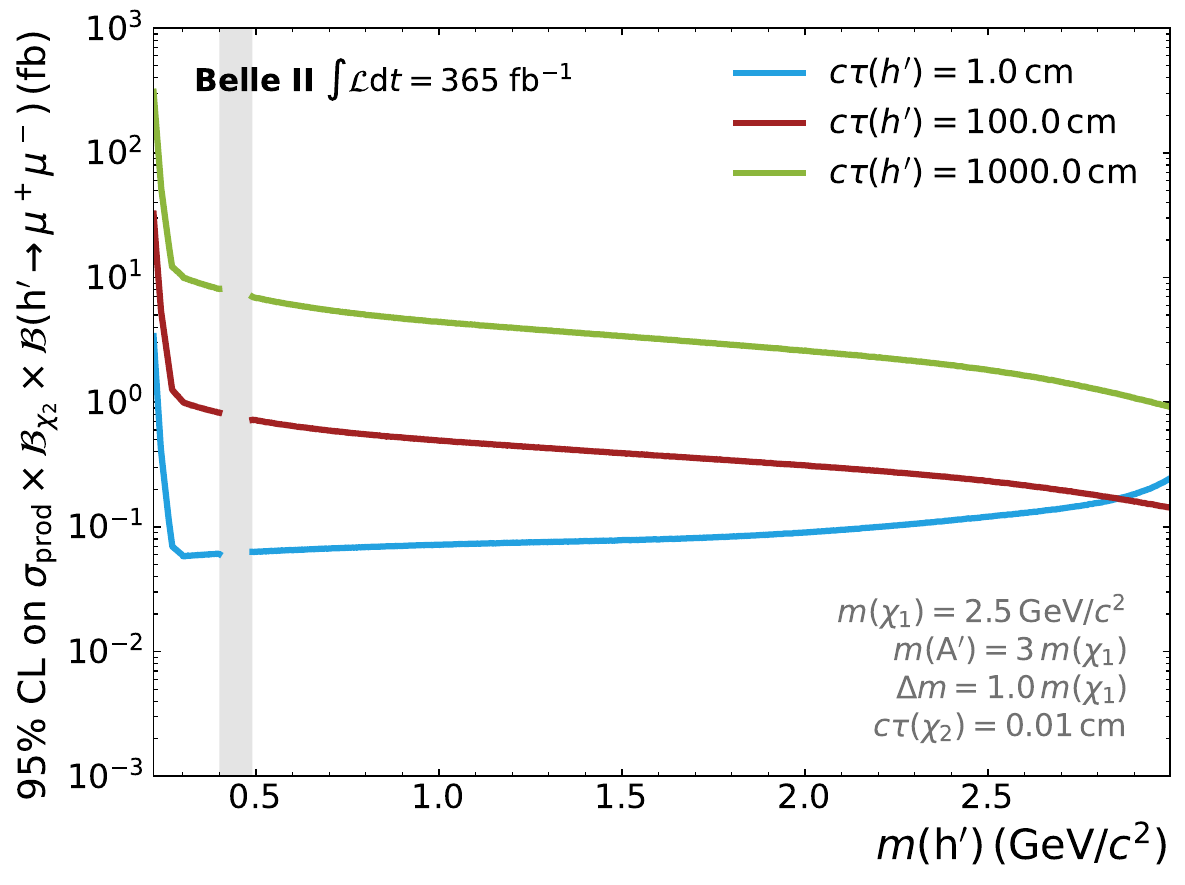}
    \includegraphics[width=0.45\textwidth]{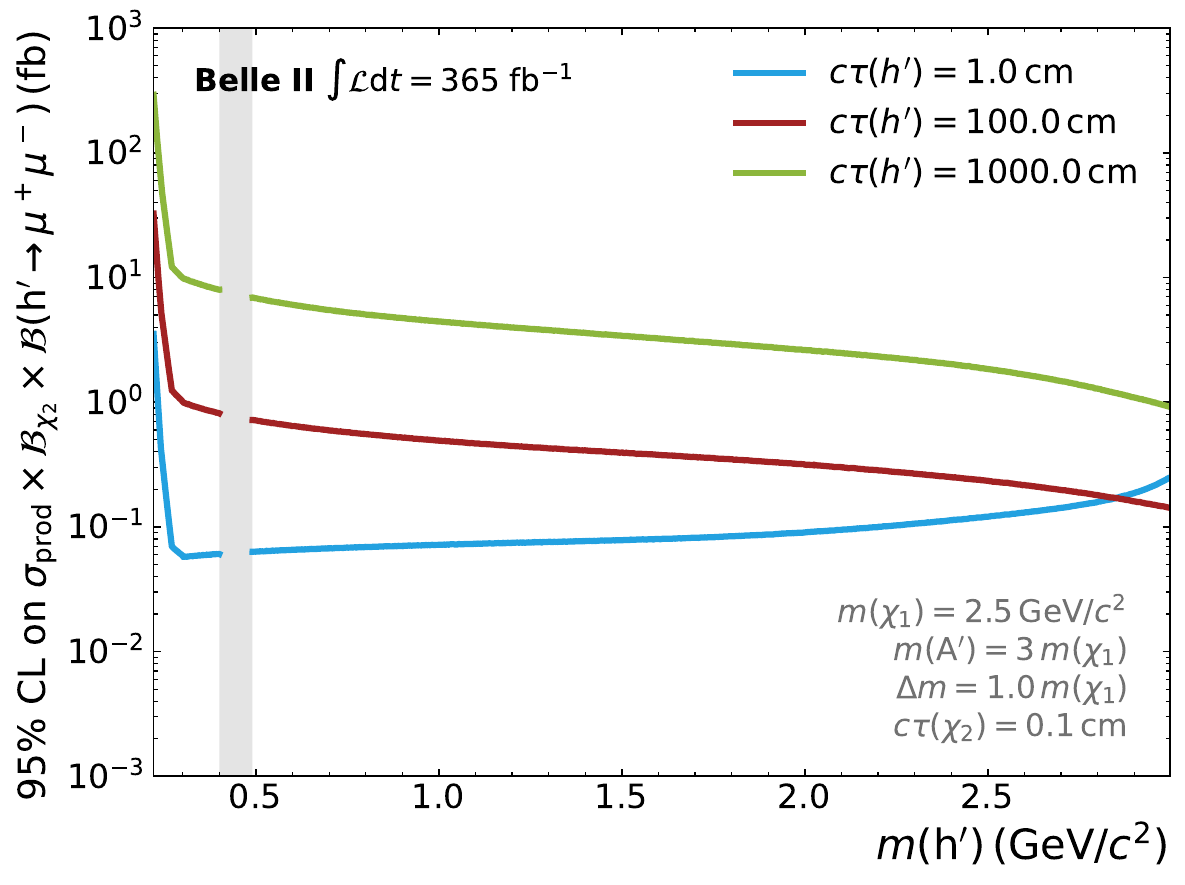}
    \includegraphics[width=0.45\textwidth]{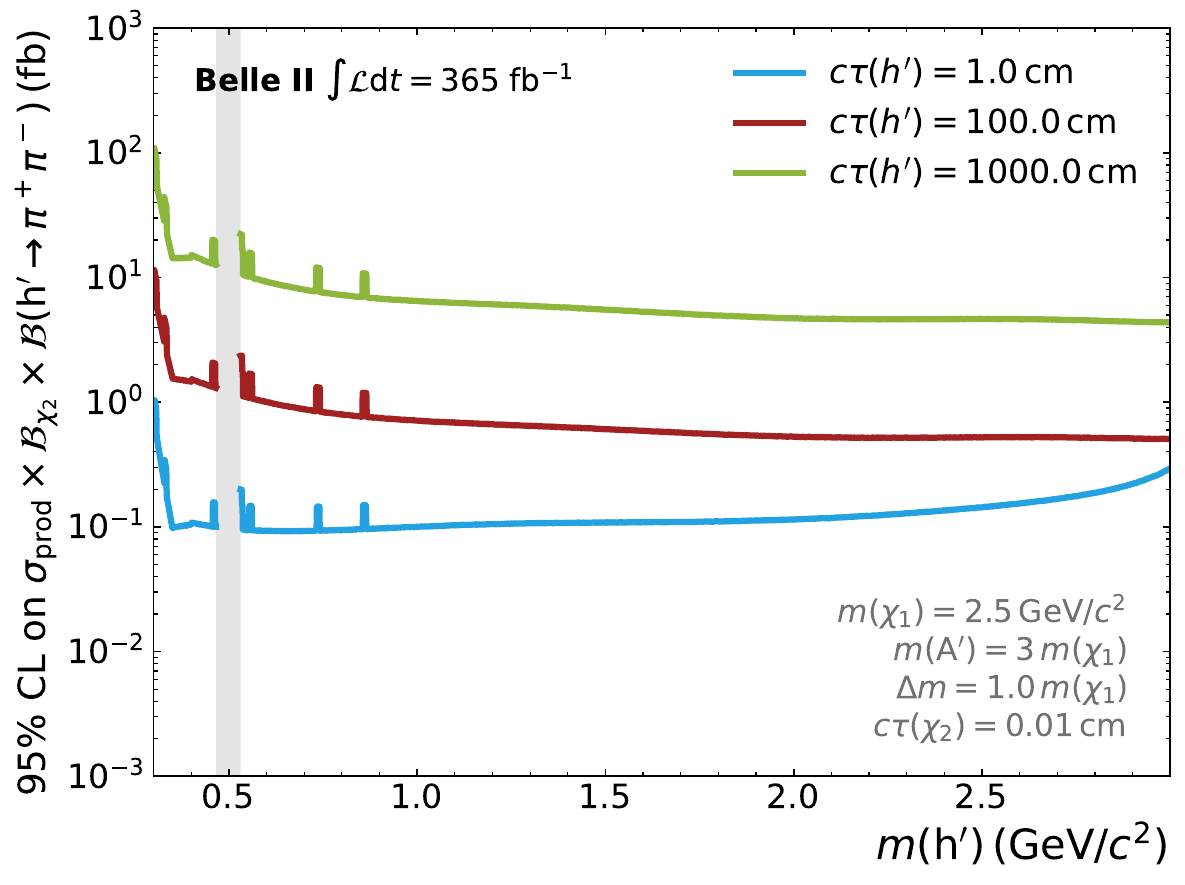}
    \includegraphics[width=0.45\textwidth]{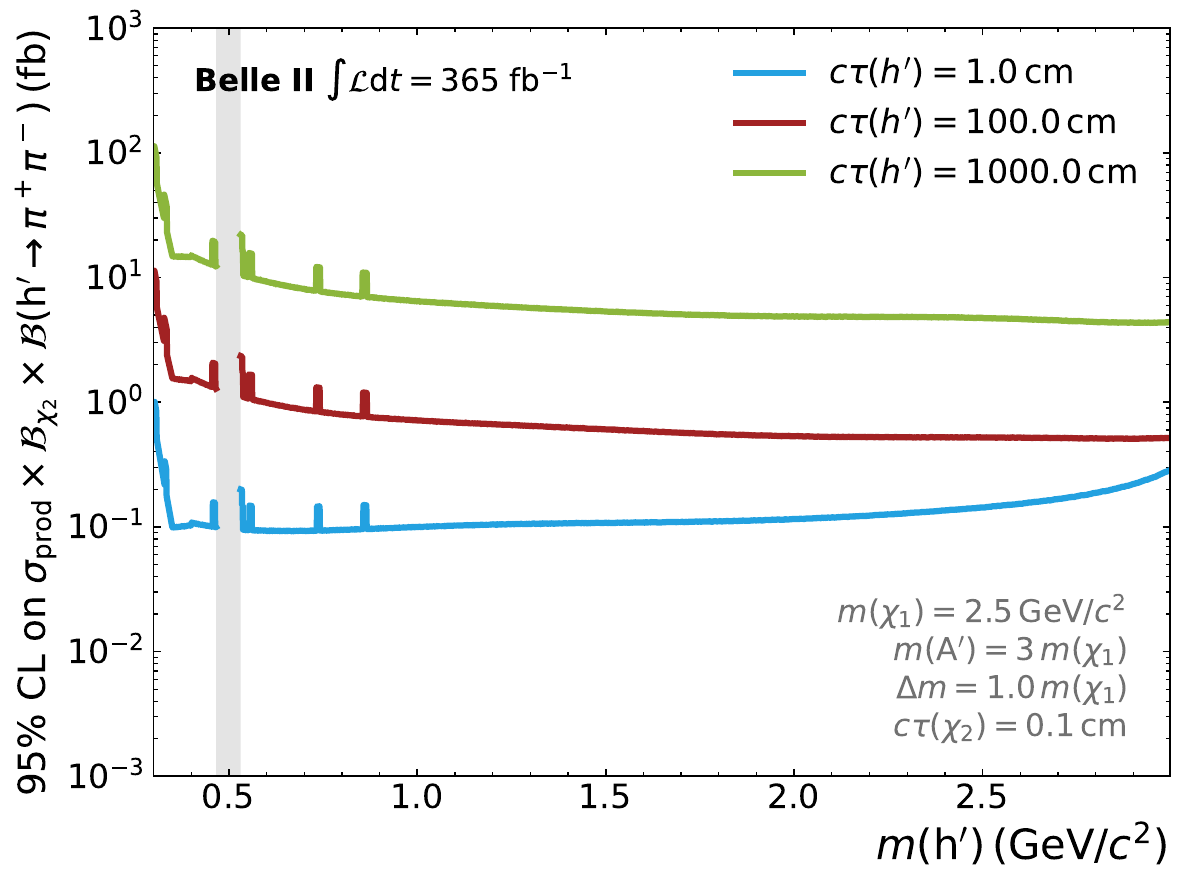}
    \includegraphics[width=0.45\textwidth]{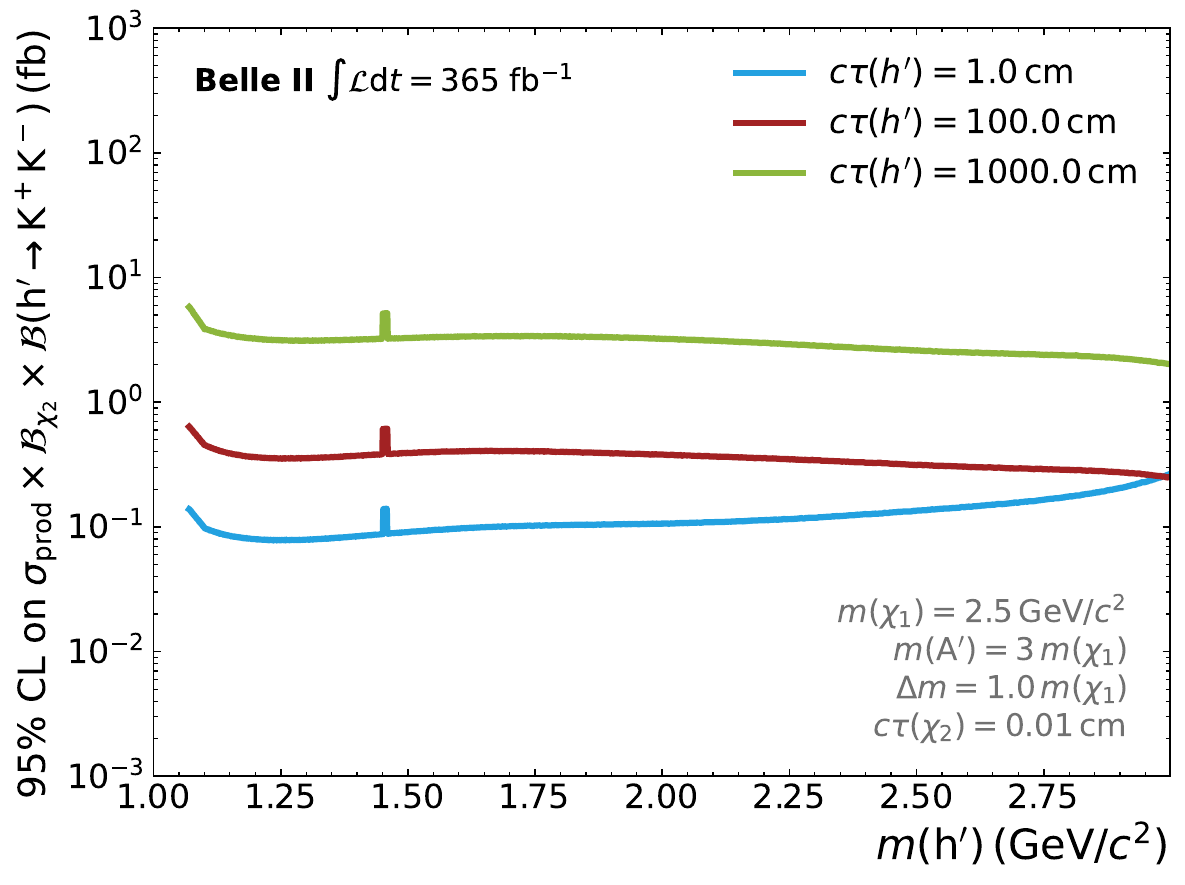}
    \includegraphics[width=0.45\textwidth]{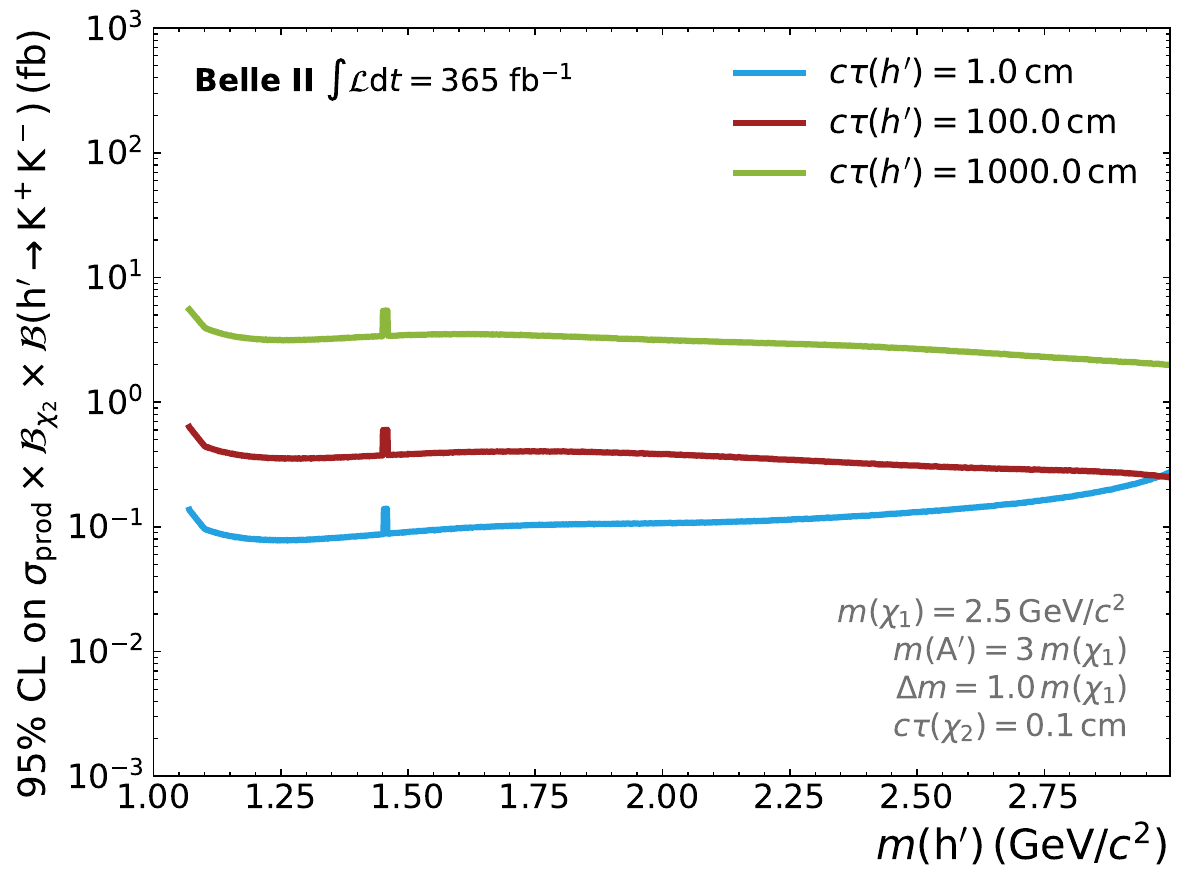}
    \caption{
Upper limits (95\% credibility level) on the \prodbf\ as function of dark Higgs mass $\mdh$ for \hbox{$c\tau(\dh) =1\,\text{cm}$}~(blue), \hbox{$c\tau(\dh) =100\,\text{cm}$}~(red), and \hbox{$c\tau(\dh) =1000\,\text{cm}$}~(green) for $\dh\to\mu^+\mu^-$\,(top), $\dh\to\pi^+\pi^-$\,(center) and $\dh\to K^+ K^-$\,(bottom).
The \chitwo{} lifetime is chosen as $c\tau(\chi_2) = 0.01\,\cm$ (left) and $c\tau(\chi_2) = 0.1\,\cm$ (right).
The remaining model parameters are chosen as $\mchione = 2.5\,\gevcc$, $\map = 3\,\mchione$, and $\Delta m = 1.0\,\mchione$.
The region corresponding to the fully-vetoed \KS mass region is marked in gray.
}
    \label{fig:model_independent5}
\end{figure*}

\begin{figure*}[htp!]
    \centering
    \includegraphics[width=0.45\textwidth]{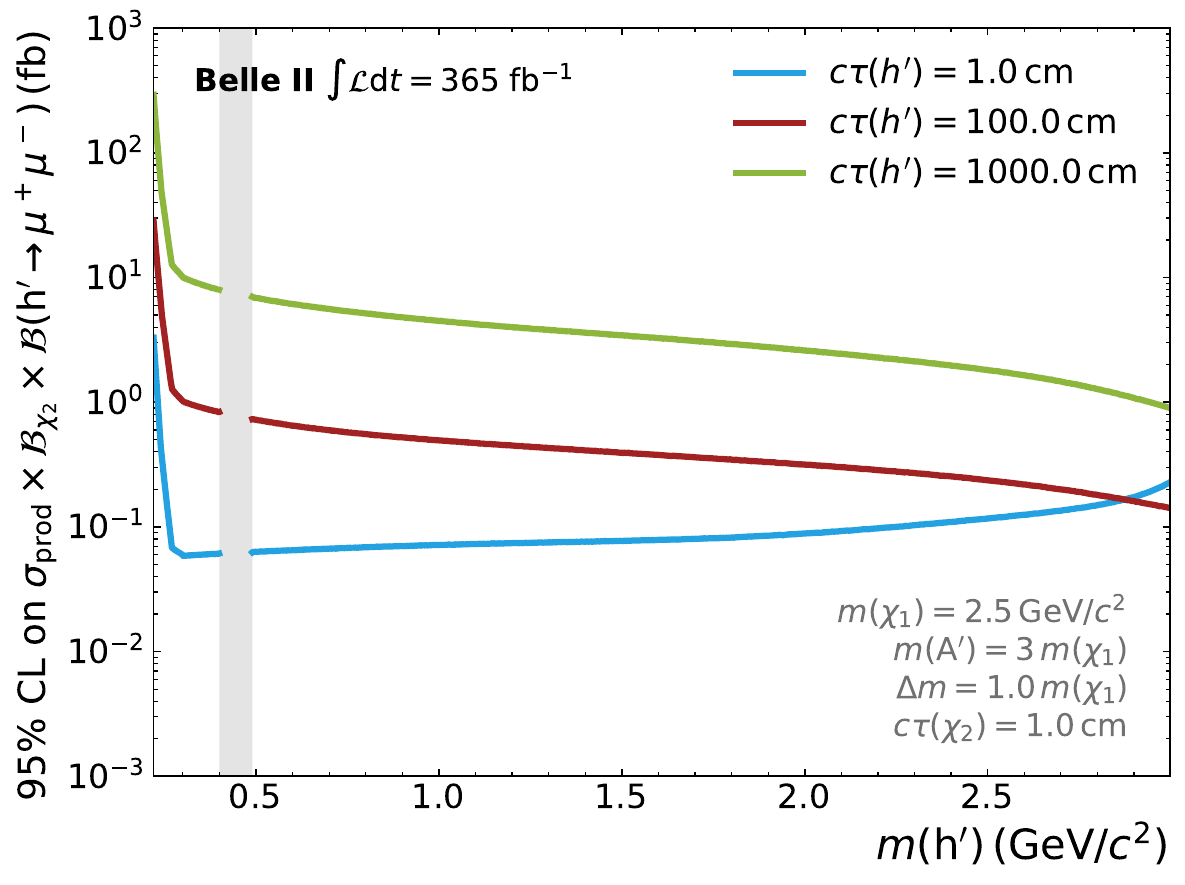}
    \includegraphics[width=0.45\textwidth]{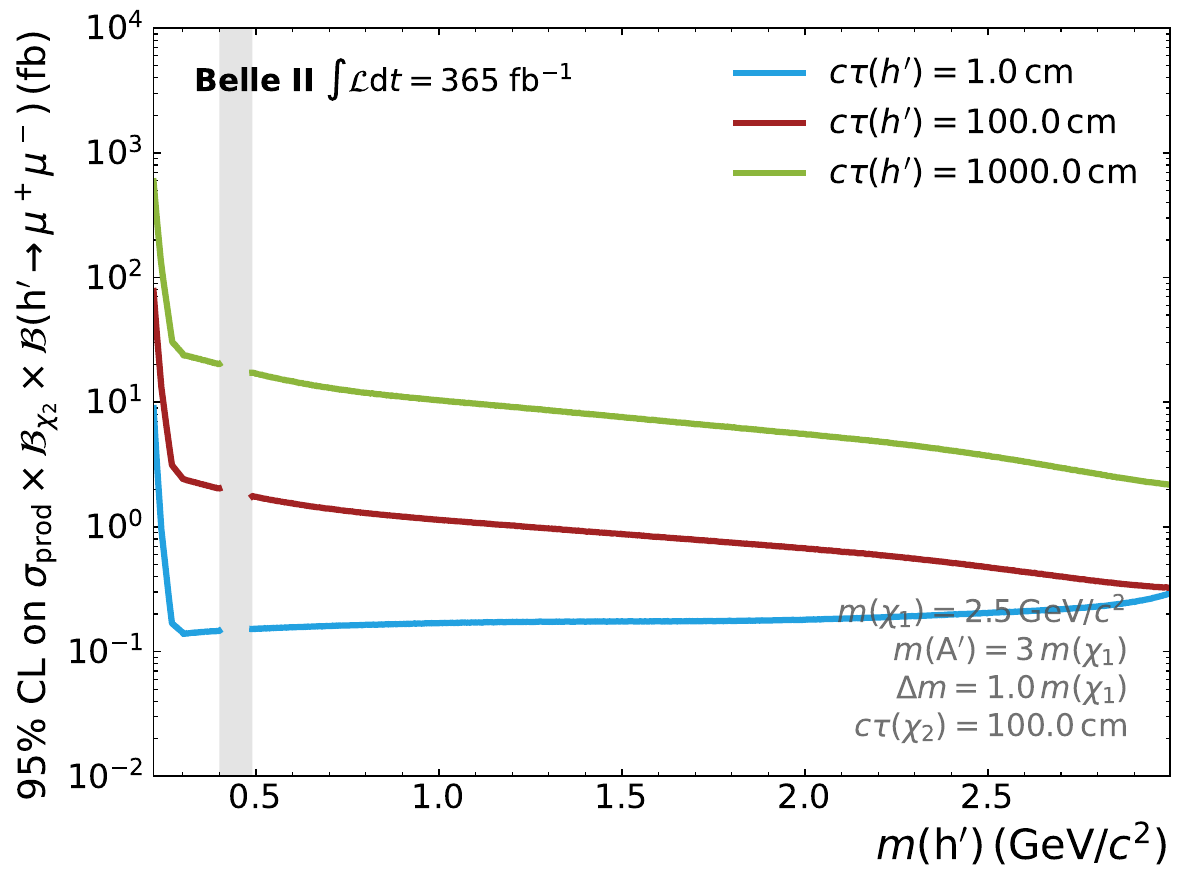}
    \includegraphics[width=0.45\textwidth]{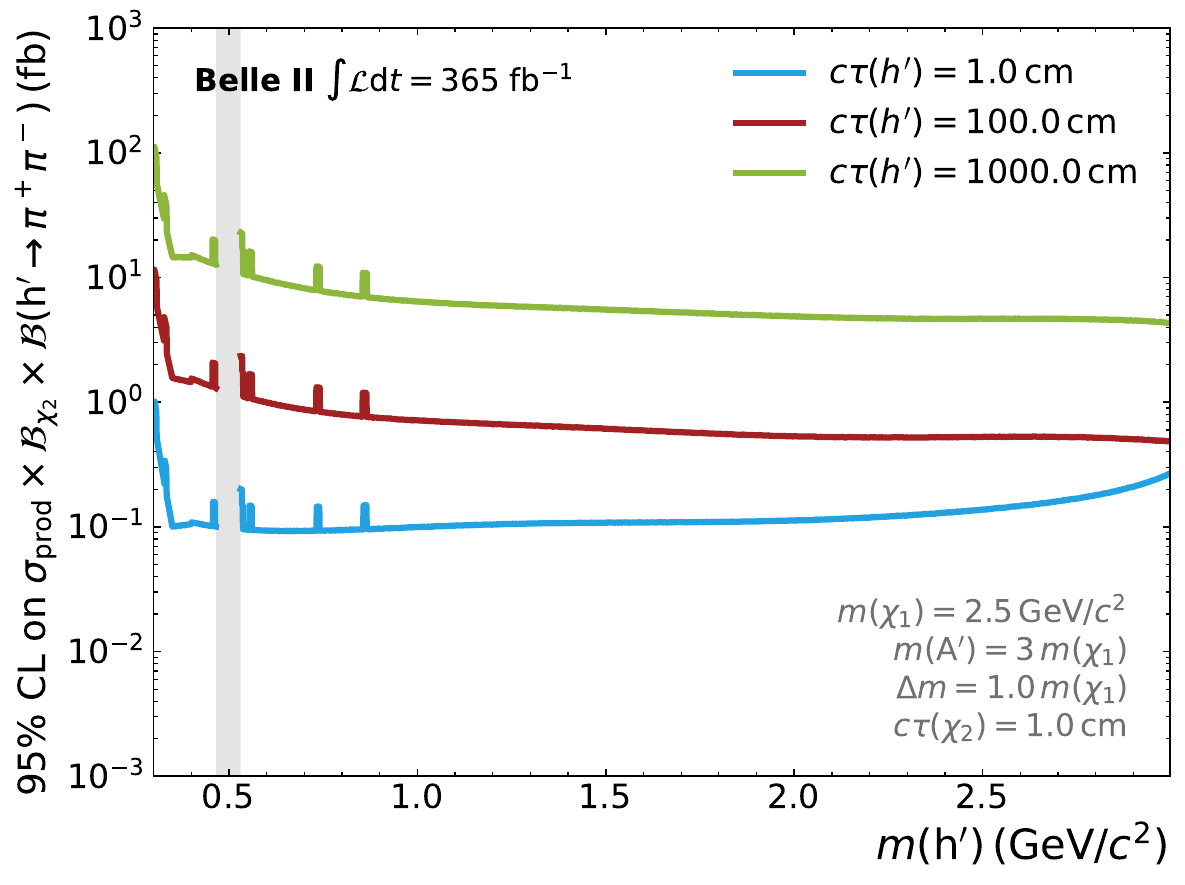}
    \includegraphics[width=0.45\textwidth]{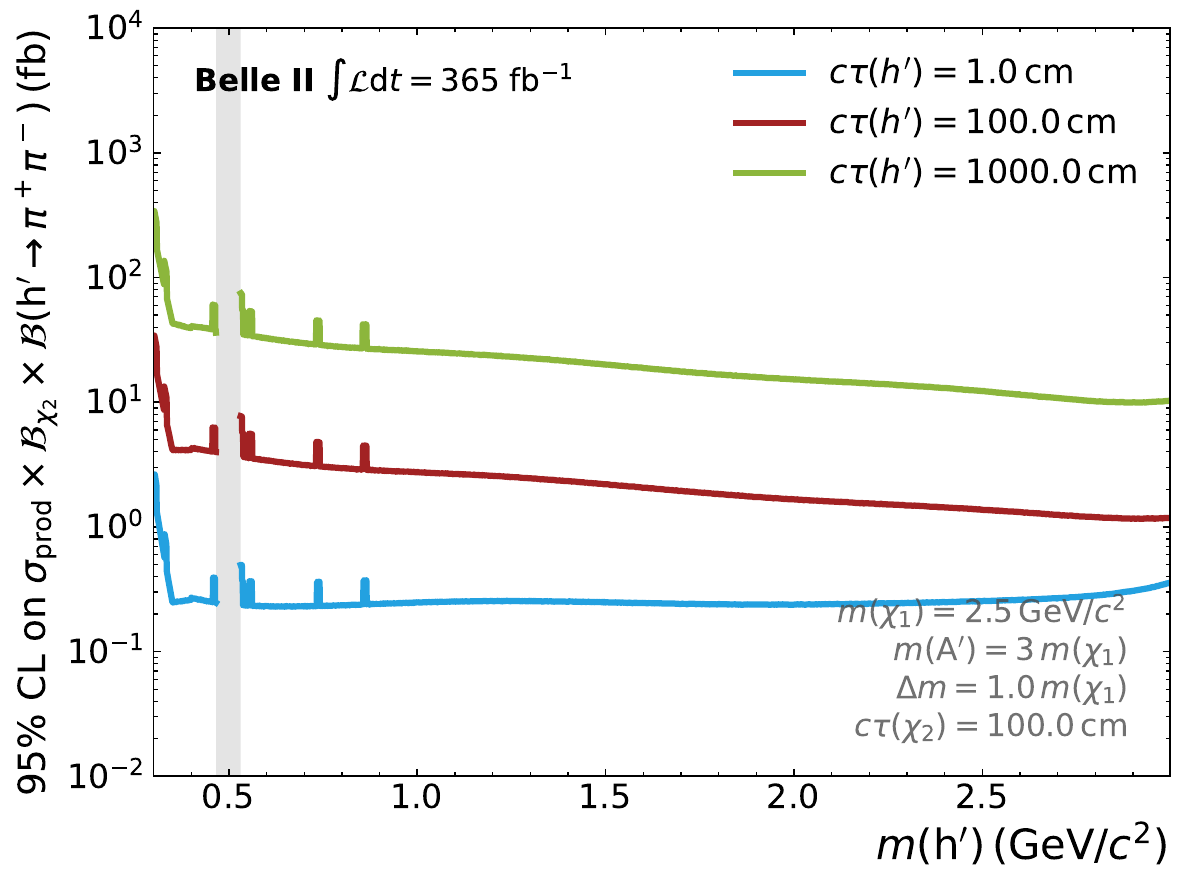}
    \includegraphics[width=0.45\textwidth]{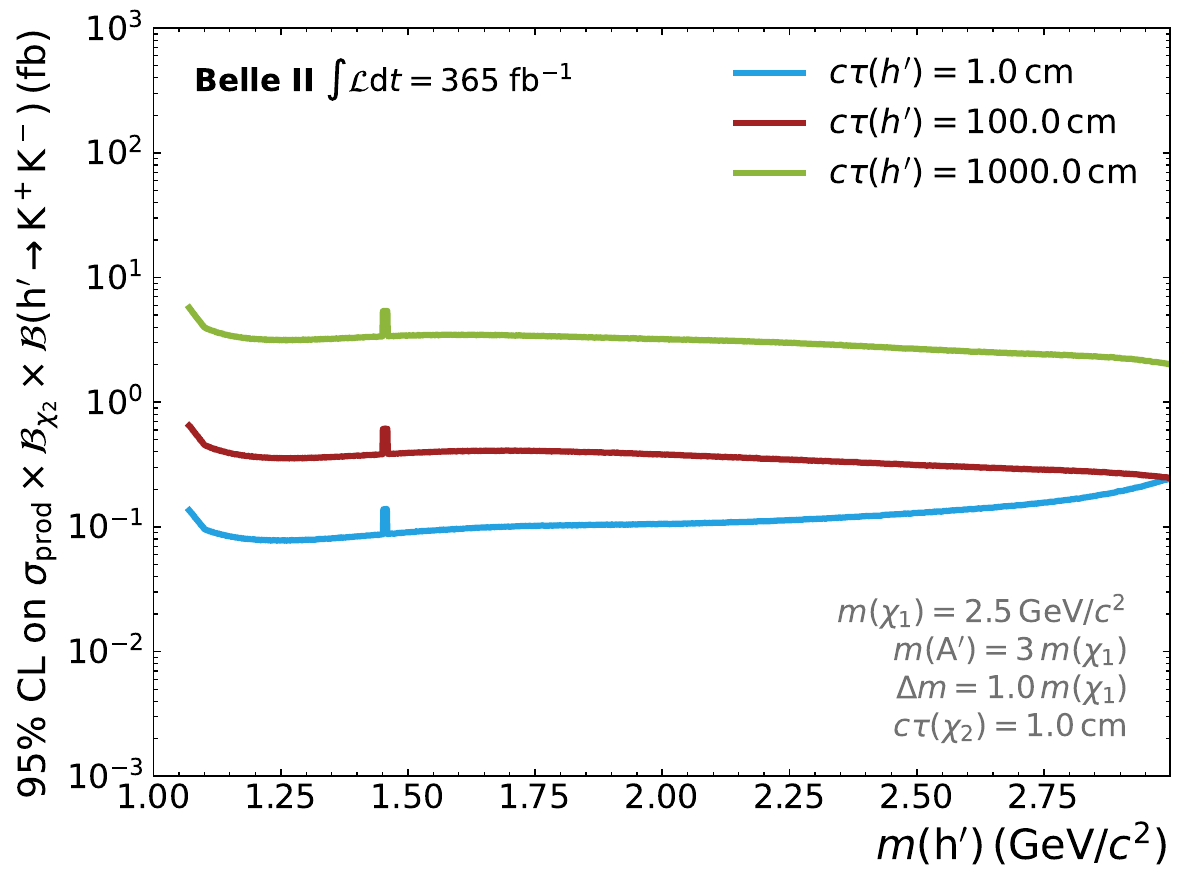}
    \includegraphics[width=0.45\textwidth]{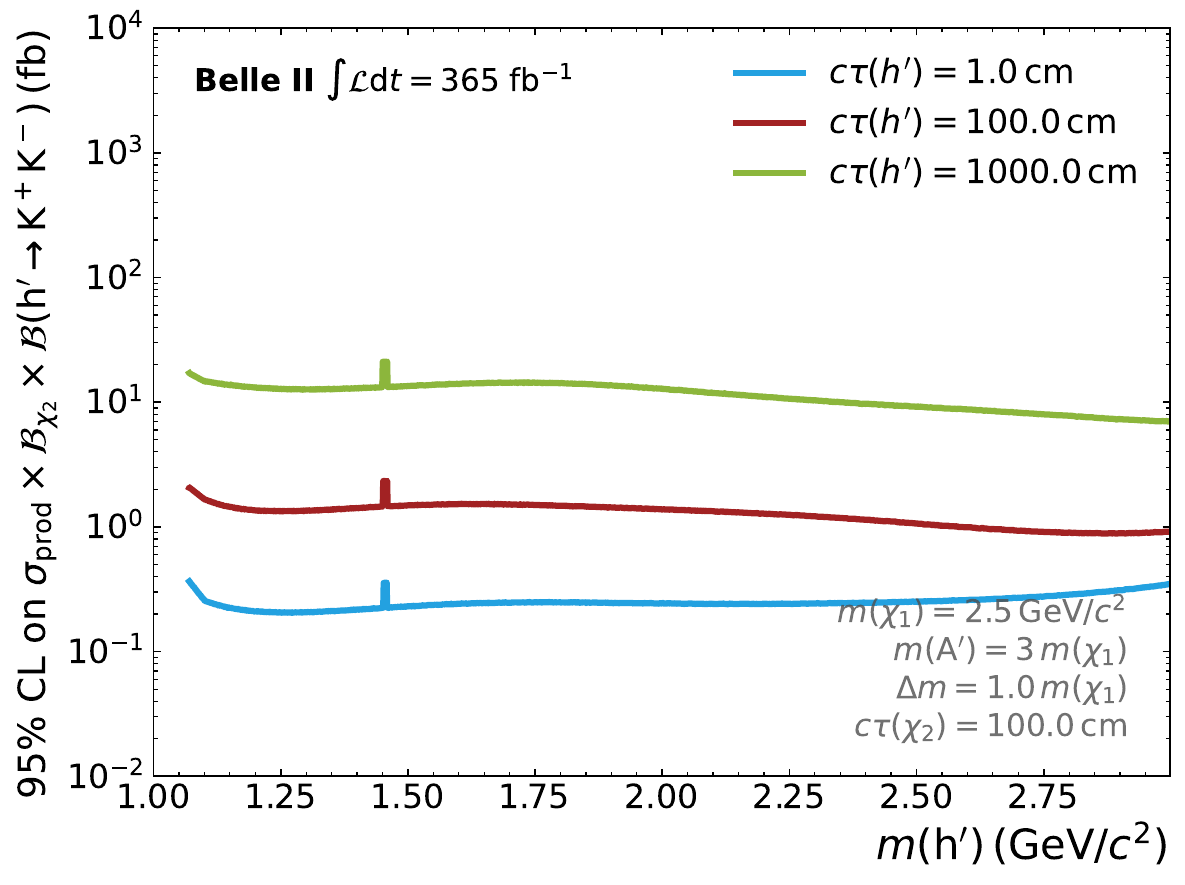}
    \caption{
Upper limits (95\% credibility level) on the \prodbf\ as function of dark Higgs mass $\mdh$ for \hbox{$c\tau(\dh) =1\,\text{cm}$}~(blue), \hbox{$c\tau(\dh) =100\,\text{cm}$}~(red), and \hbox{$c\tau(\dh) =1000\,\text{cm}$}~(green) for $\dh\to\mu^+\mu^-$\,(top), $\dh\to\pi^+\pi^-$\,(center) and $\dh\to K^+ K^-$\,(bottom).
The \chitwo{} lifetime is chosen as $c\tau(\chi_2) = 1.0\,\cm$ (left) and $c\tau(\chi_2) = 100.0\,\cm$ (right).
The remaining model parameters are chosen as $\mchione = 2.5\,\gevcc$, $\map = 3\,\mchione$, and $\Delta m = 1.0\,\mchione$.
The region corresponding to the fully-vetoed \KS mass region is marked in gray.
}
    \label{fig:model_independent6}
\end{figure*}

\begin{figure*}[htp!]
    \centering
    \includegraphics[width=0.45\textwidth]{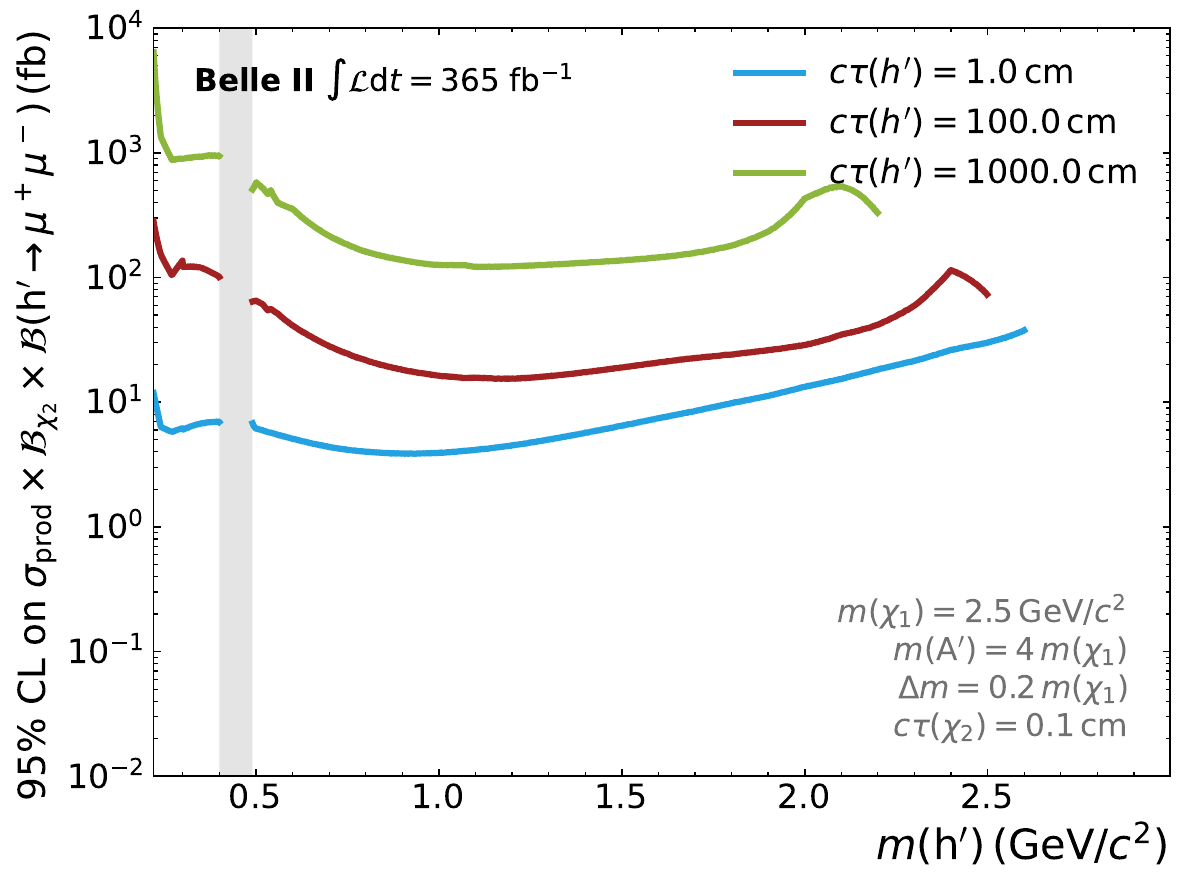}
    \includegraphics[width=0.45\textwidth]{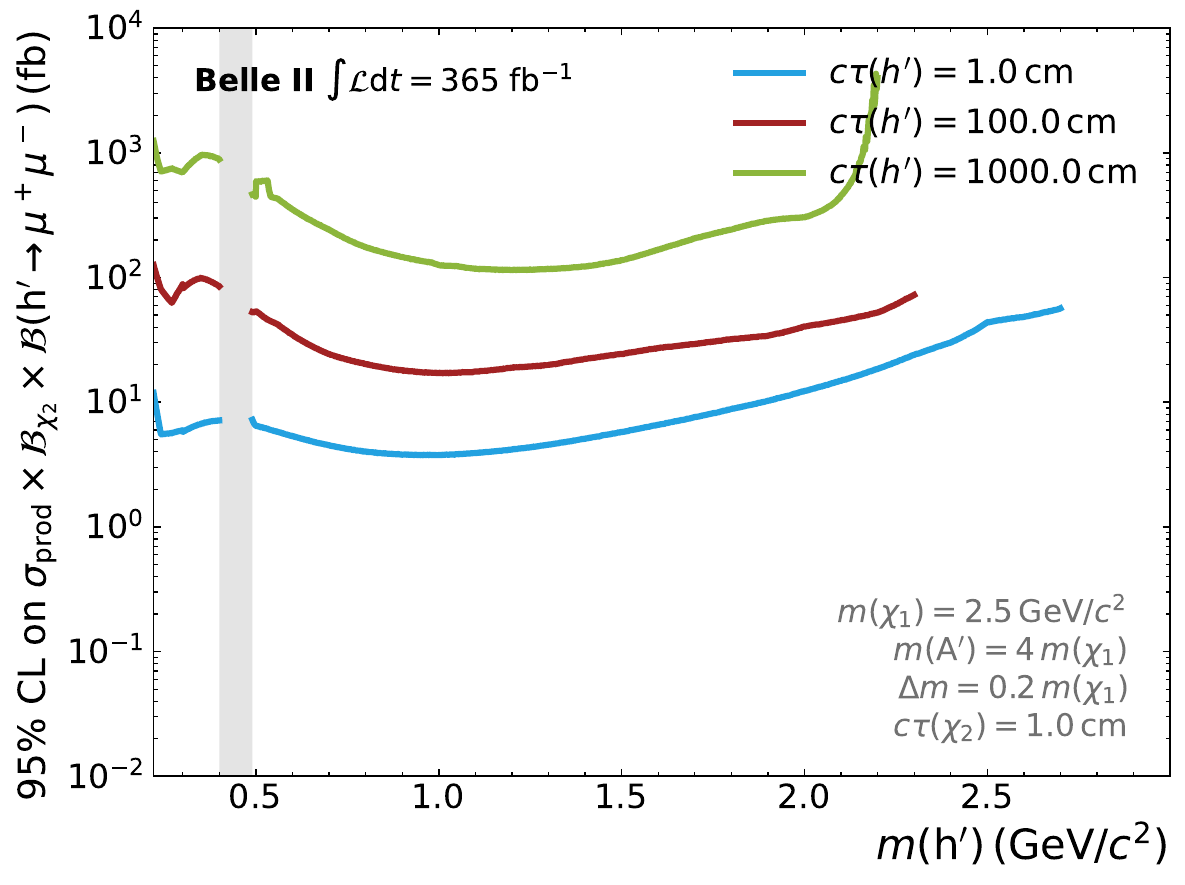}
    \includegraphics[width=0.45\textwidth]{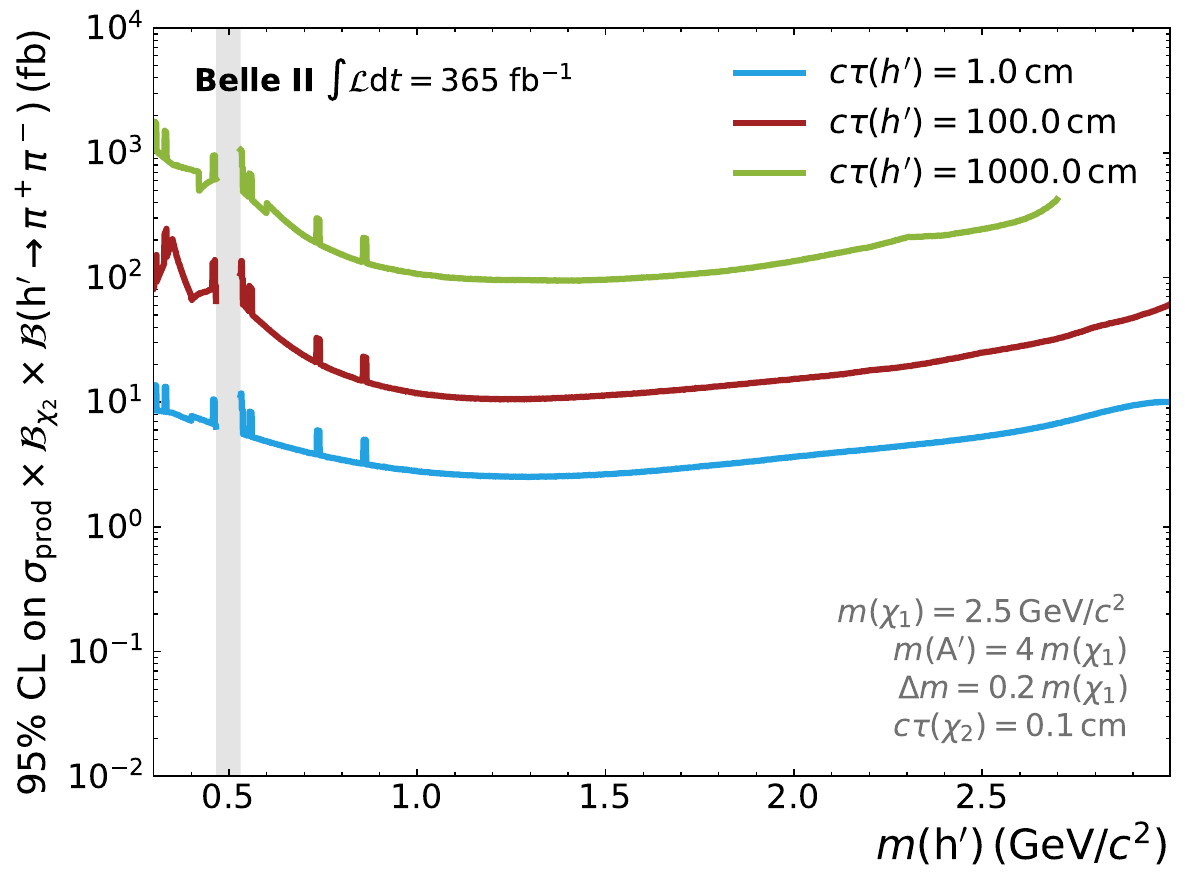}
    \includegraphics[width=0.45\textwidth]{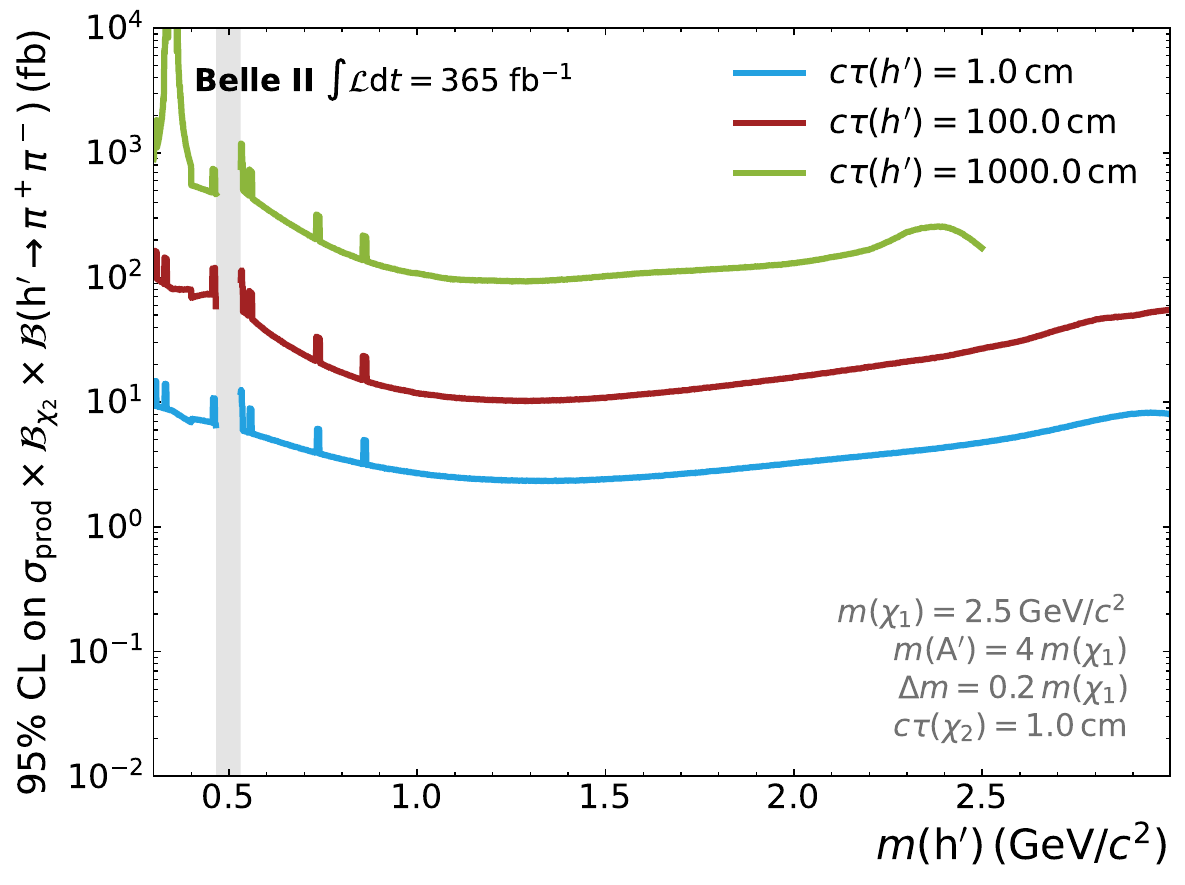}
    \includegraphics[width=0.45\textwidth]{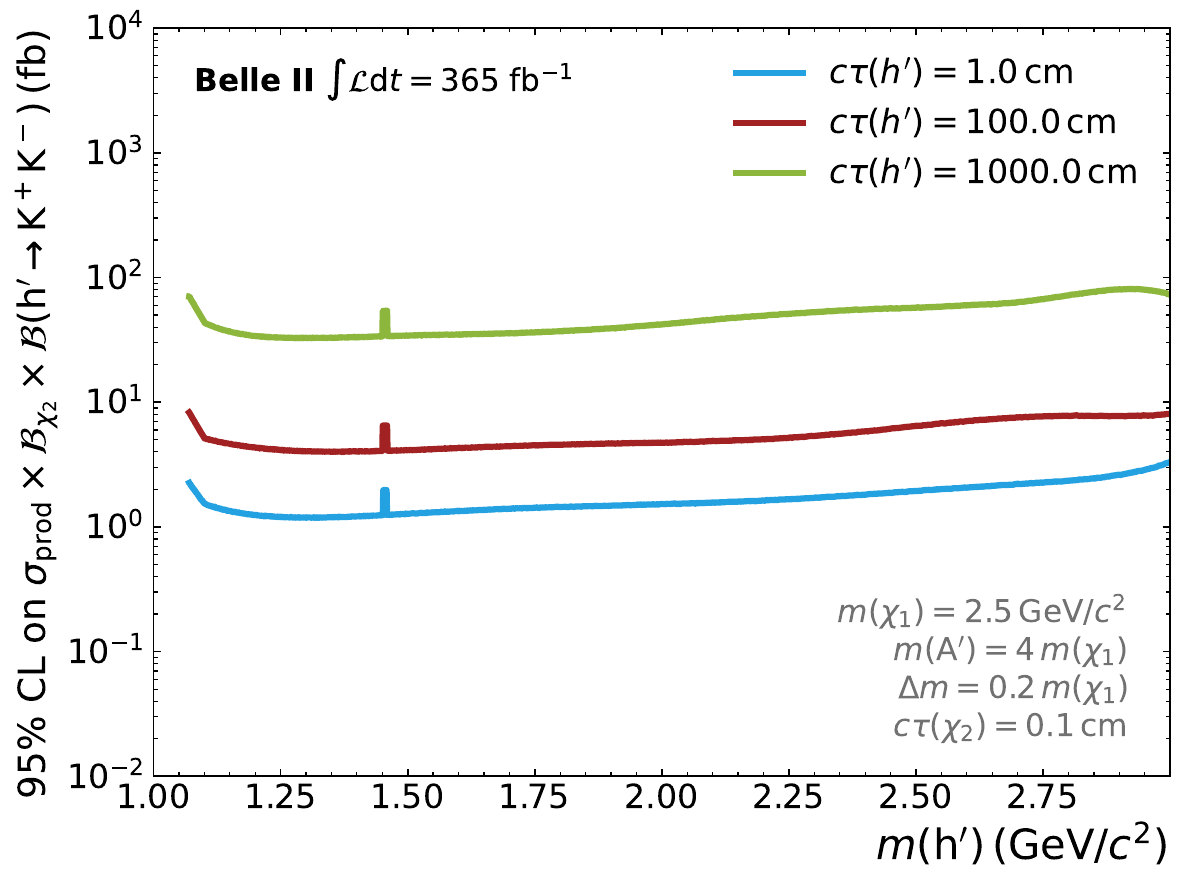}
    \includegraphics[width=0.45\textwidth]{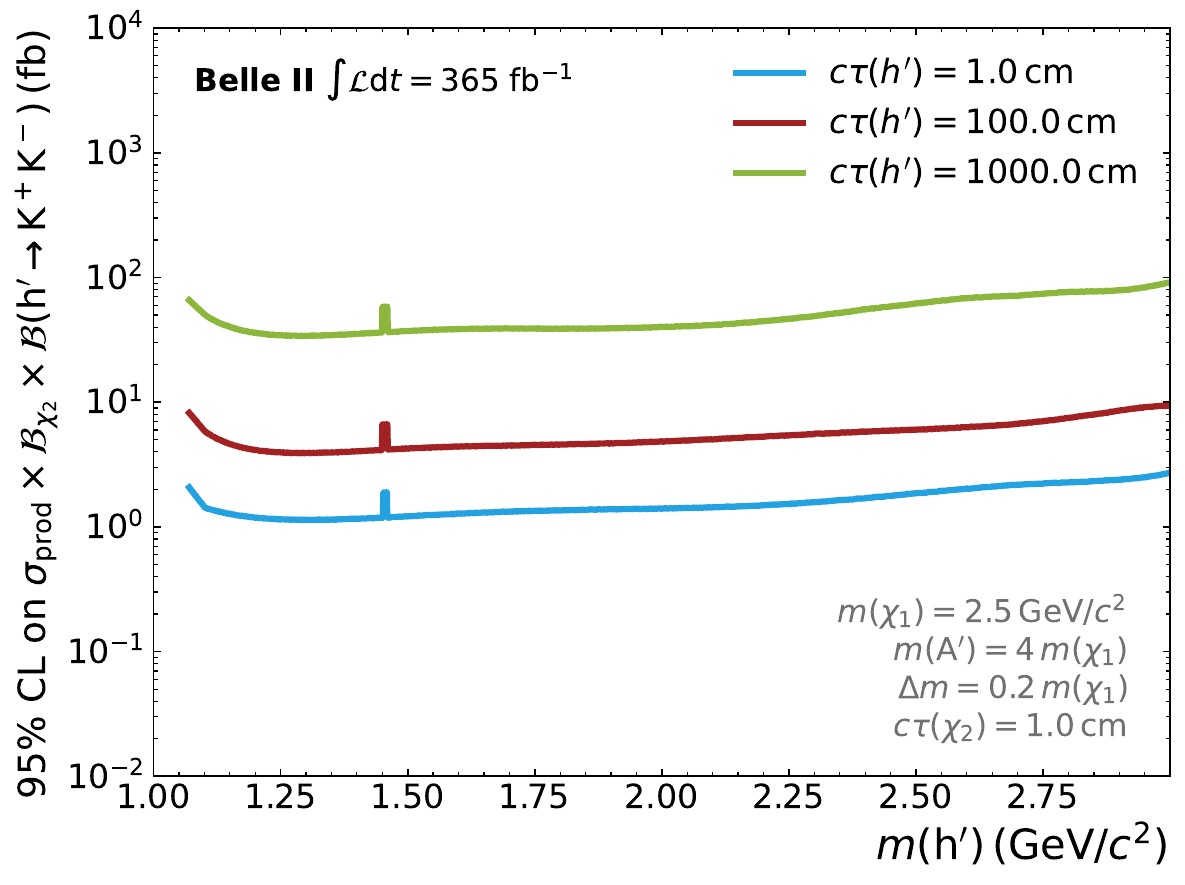}
    \caption{
Upper limits (95\% credibility level) on the \prodbf\ as function of dark Higgs mass $\mdh$ for \hbox{$c\tau(\dh) =1\,\text{cm}$}~(blue), \hbox{$c\tau(\dh) =100\,\text{cm}$}~(red), and \hbox{$c\tau(\dh) =1000\,\text{cm}$}~(green) for $\dh\to\mu^+\mu^-$\,(top), $\dh\to\pi^+\pi^-$\,(center) and $\dh\to K^+ K^-$\,(bottom).
The \chitwo{} lifetime is chosen as $c\tau(\chi_2) = 0.1\,\cm$ (left) and $c\tau(\chi_2) = 1.0\,\cm$ (right).
The remaining model parameters are chosen as $\mchione = 2.5\,\gevcc$, $\map = 4\,\mchione$, and $\Delta m = 0.2\,\mchione$.
The region corresponding to the fully-vetoed \KS mass region is marked in gray.
}
    \label{fig:model_independent7}
\end{figure*}

\begin{figure*}[htp!]
    \centering
    \includegraphics[width=0.45\textwidth]{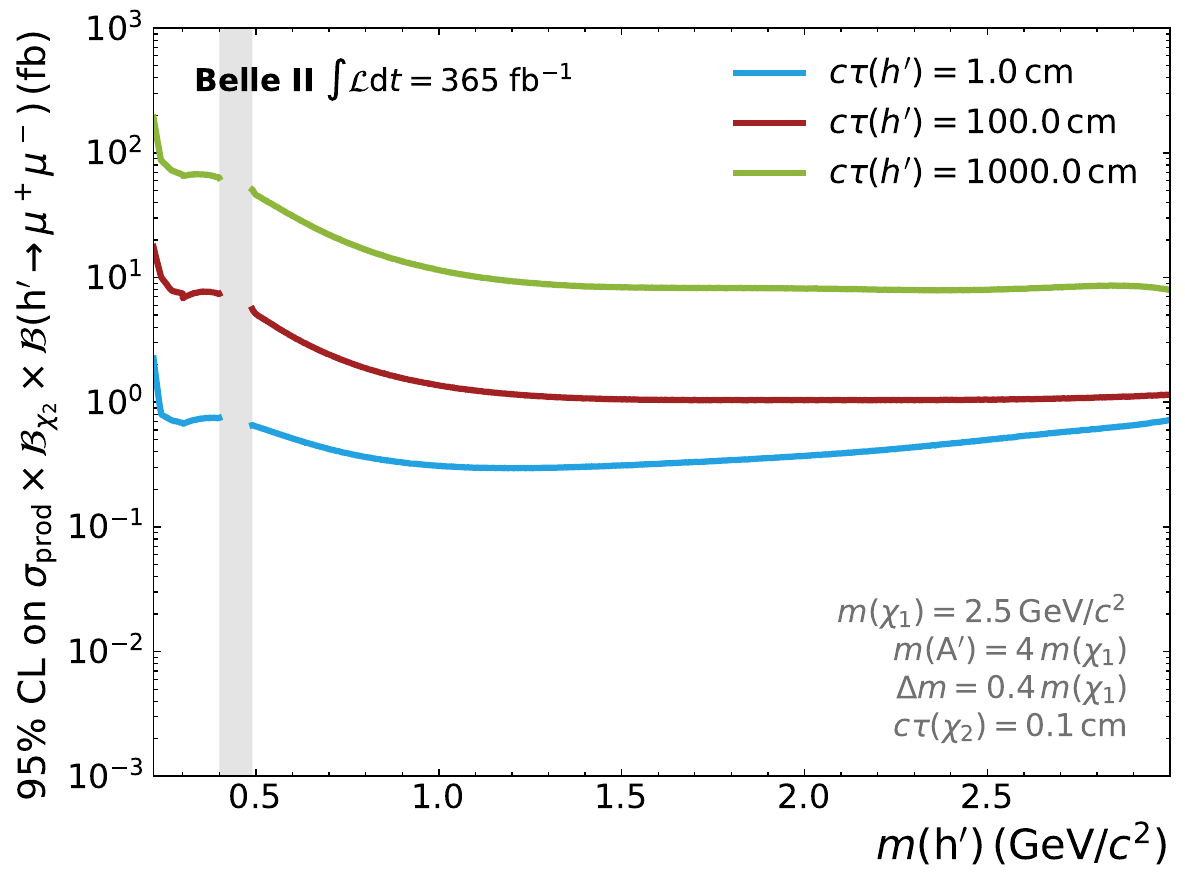}
    \includegraphics[width=0.45\textwidth]{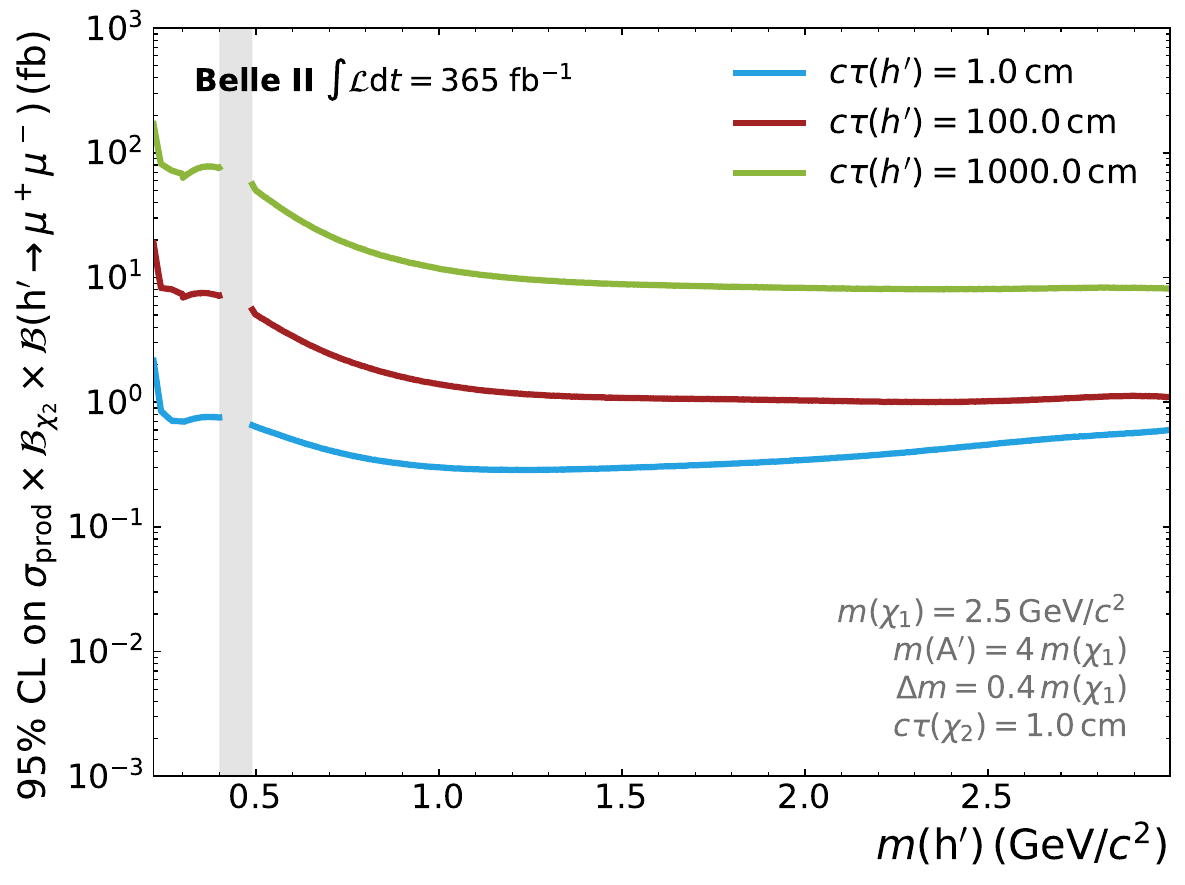}
    \includegraphics[width=0.45\textwidth]{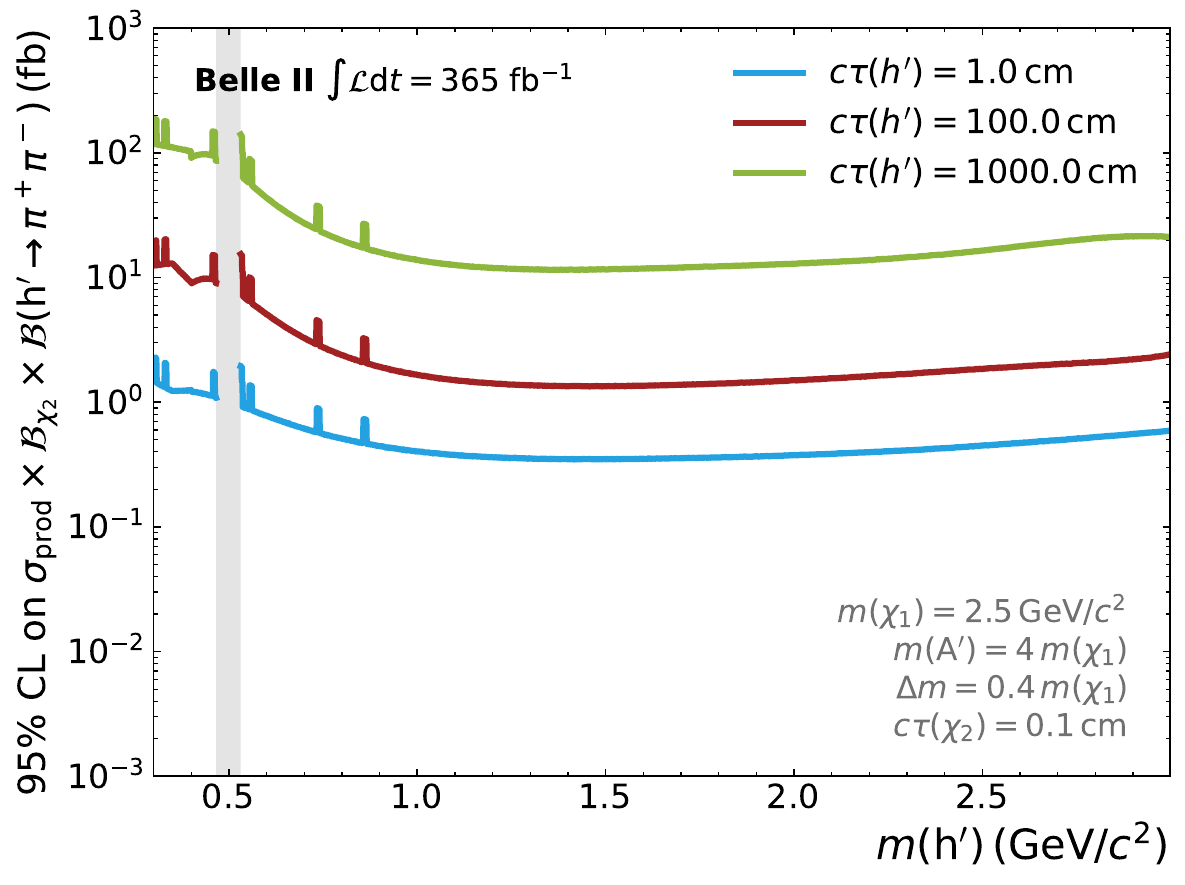}
    \includegraphics[width=0.45\textwidth]{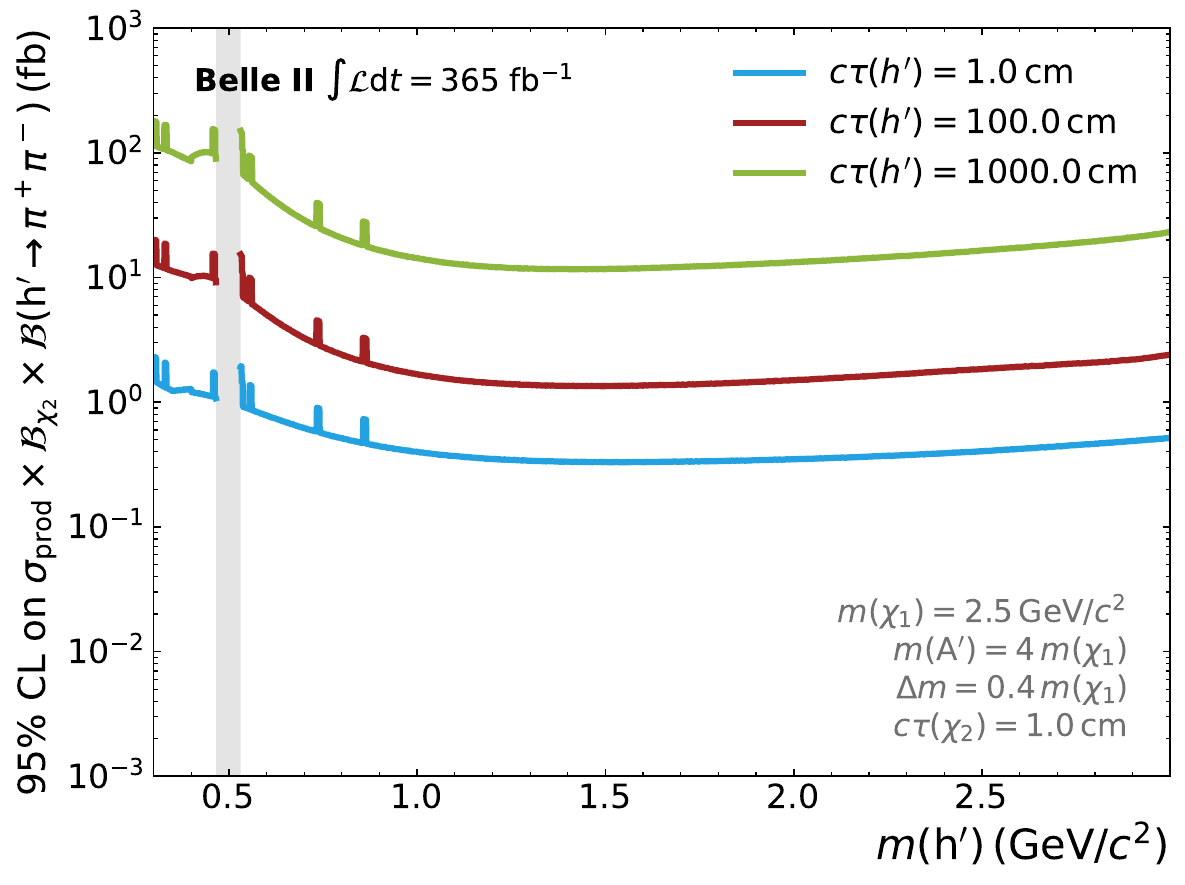}
    \includegraphics[width=0.45\textwidth]{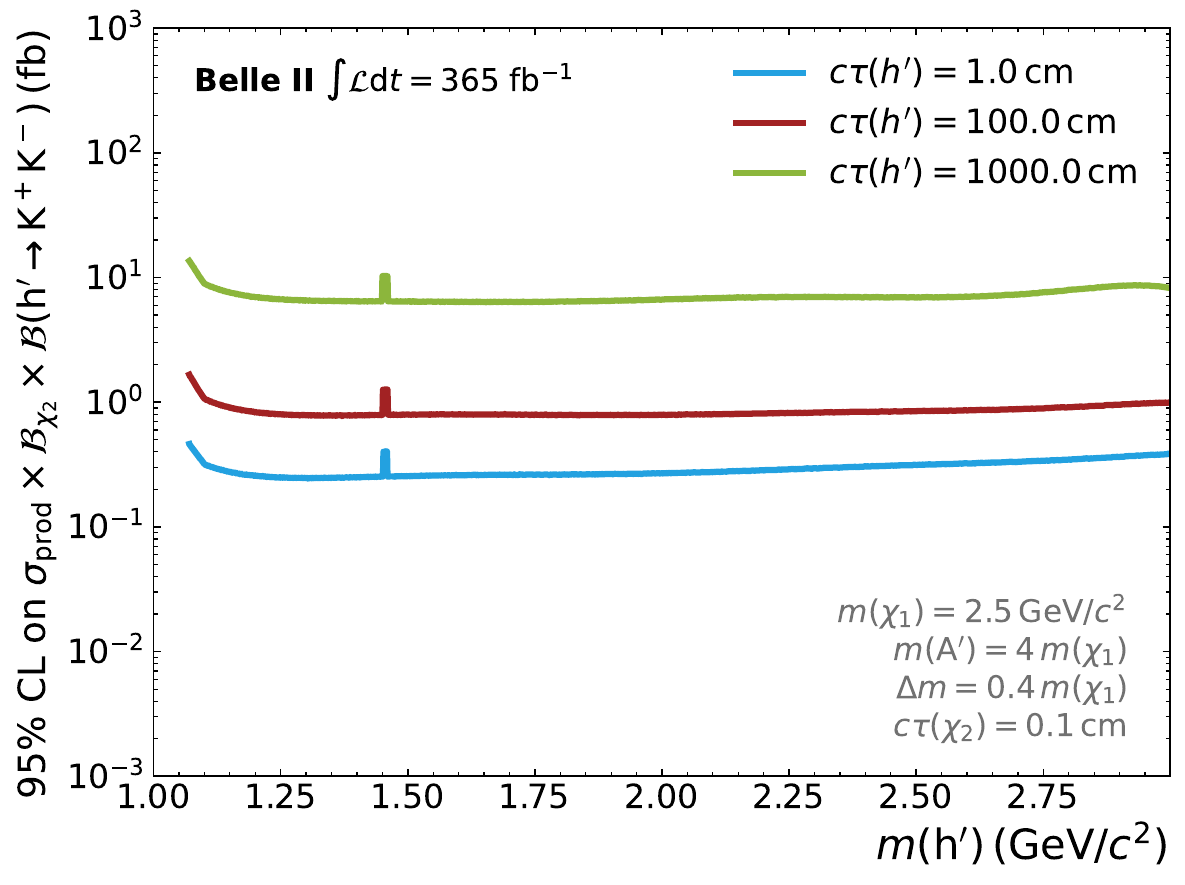}
    \includegraphics[width=0.45\textwidth]{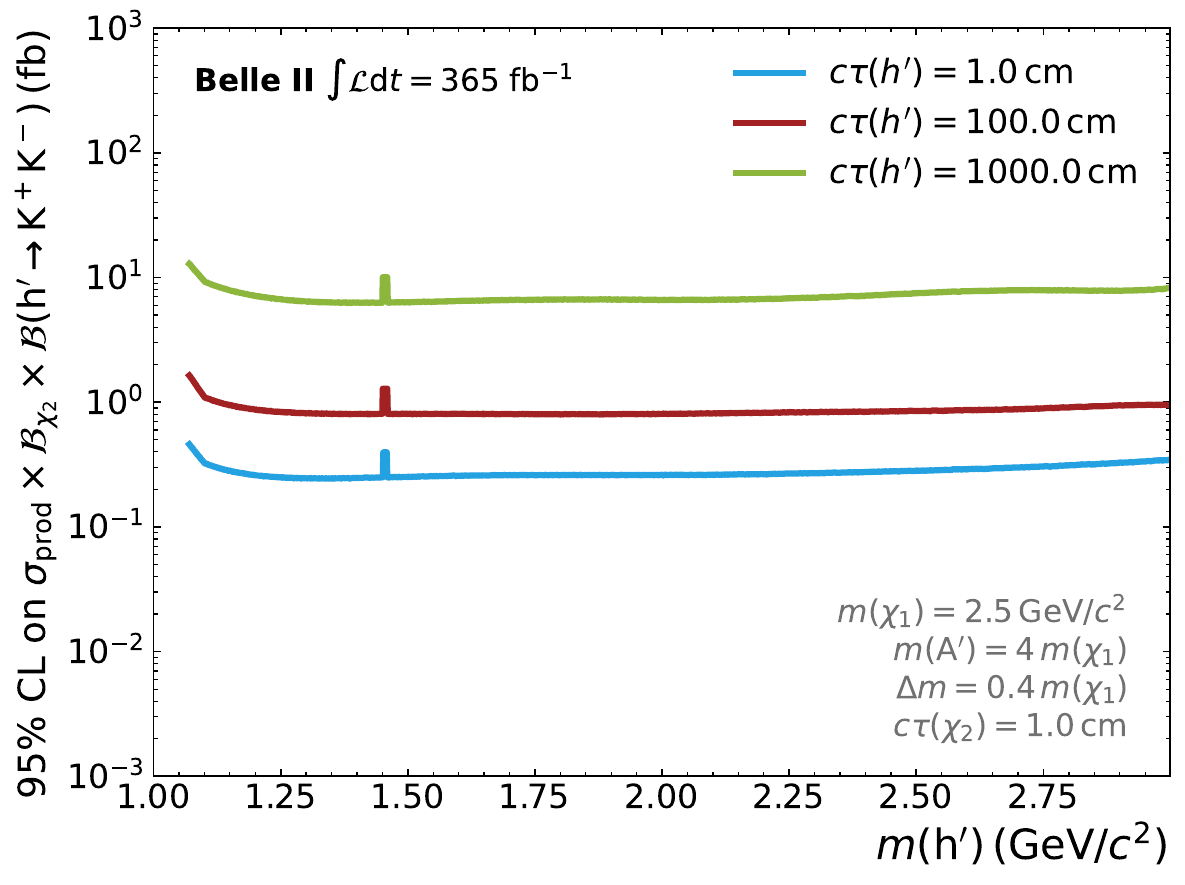}
    \caption{
Upper limits (95\% credibility level) on the \prodbf\ as function of dark Higgs mass $\mdh$ for \hbox{$c\tau(\dh) =1\,\text{cm}$}~(blue), \hbox{$c\tau(\dh) =100\,\text{cm}$}~(red), and \hbox{$c\tau(\dh) =1000\,\text{cm}$}~(green) for $\dh\to\mu^+\mu^-$\,(top), $\dh\to\pi^+\pi^-$\,(center) and $\dh\to K^+ K^-$\,(bottom).
The \chitwo{} lifetime is chosen as $c\tau(\chi_2) = 0.1\,\cm$ (left) and $c\tau(\chi_2) = 1.0\,\cm$ (right).
The remaining model parameters are chosen as $\mchione = 2.5\,\gevcc$, $\map = 4\,\mchione$, and $\Delta m = 0.4\,\mchione$.
The region corresponding to the fully-vetoed \KS mass region is marked in gray.
}
    \label{fig:model_independent8}
\end{figure*}

\begin{figure*}[htp!]
    \centering
    \includegraphics[width=0.45\textwidth]{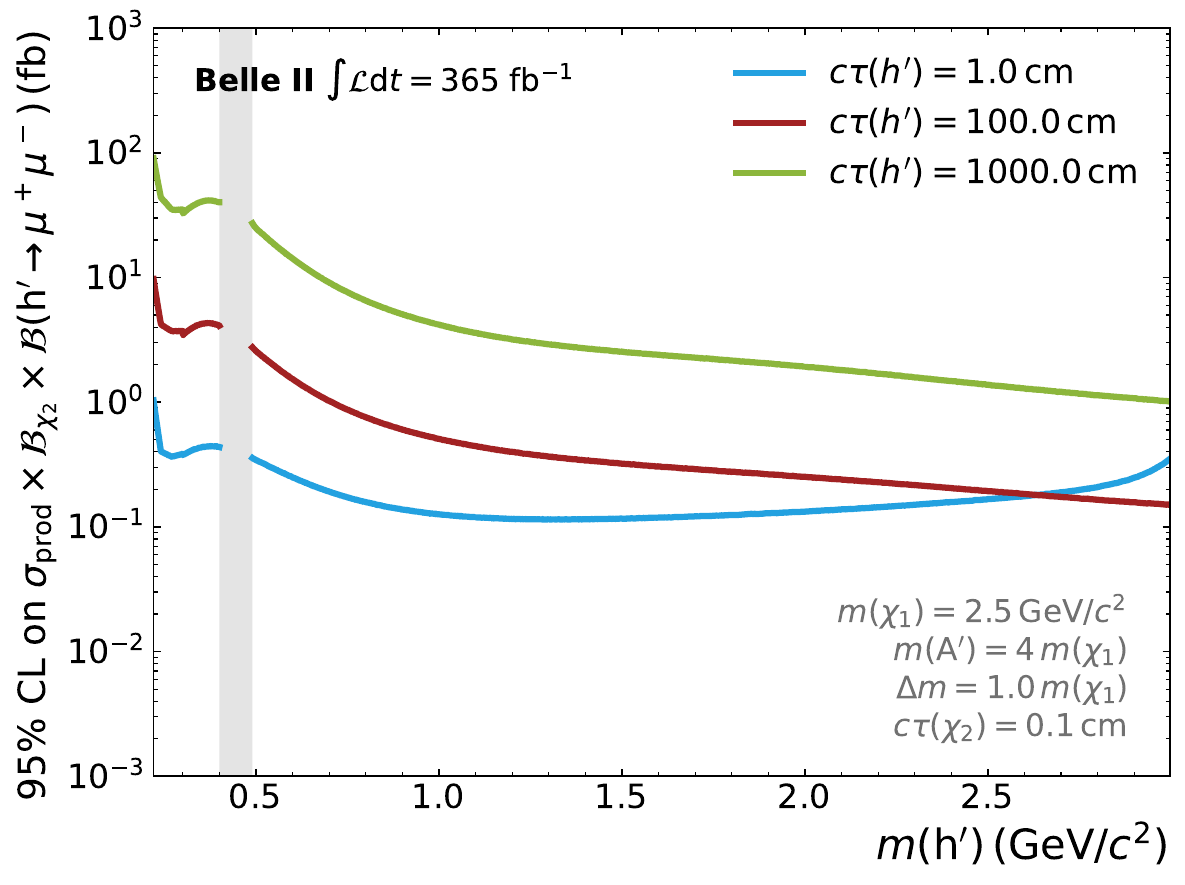}
    \includegraphics[width=0.45\textwidth]{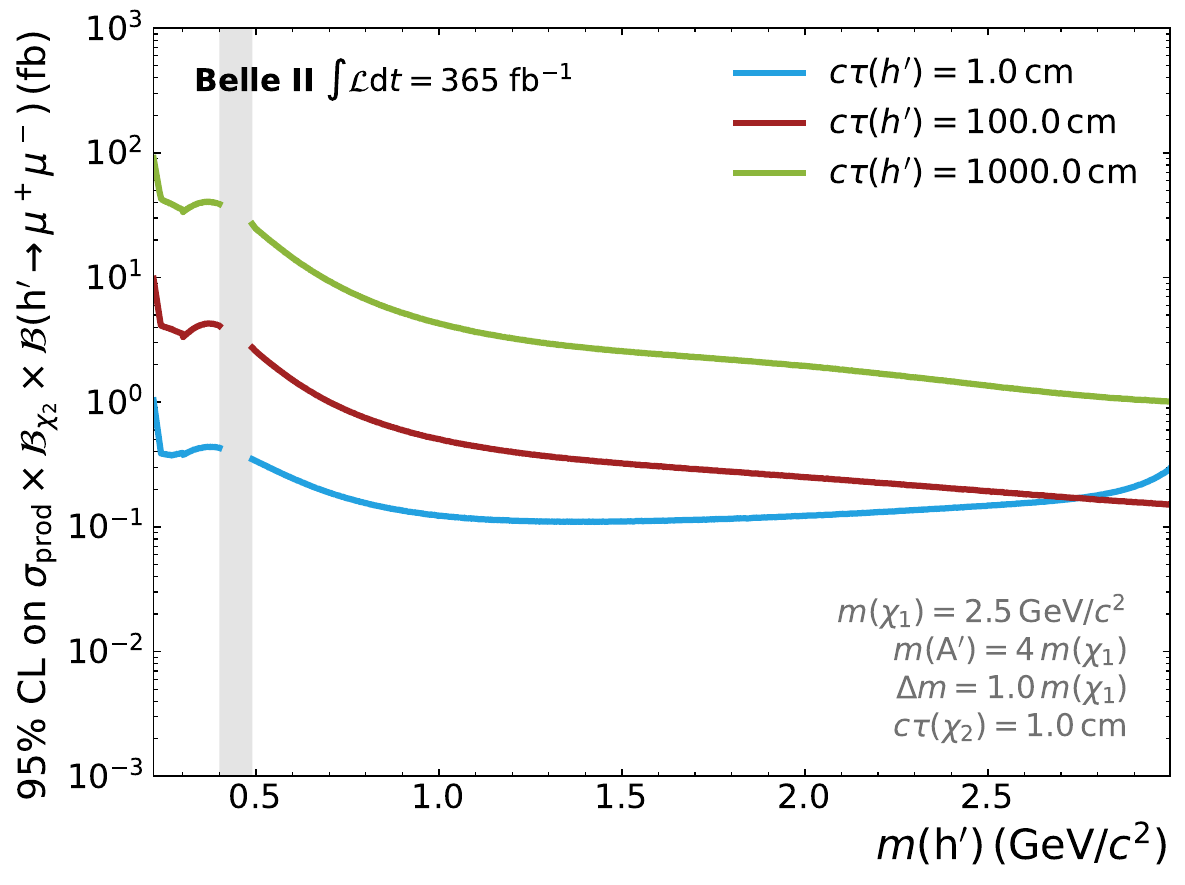}
    \includegraphics[width=0.45\textwidth]{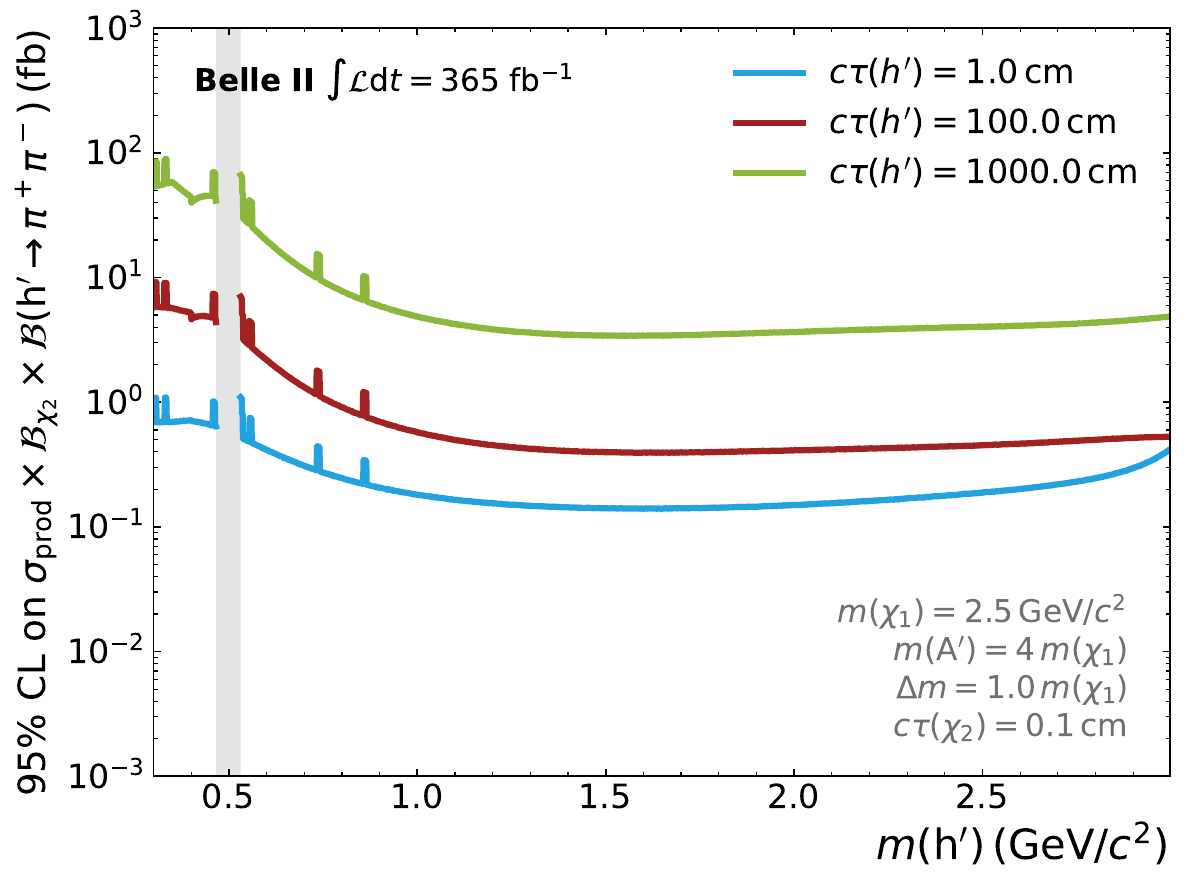}
    \includegraphics[width=0.45\textwidth]{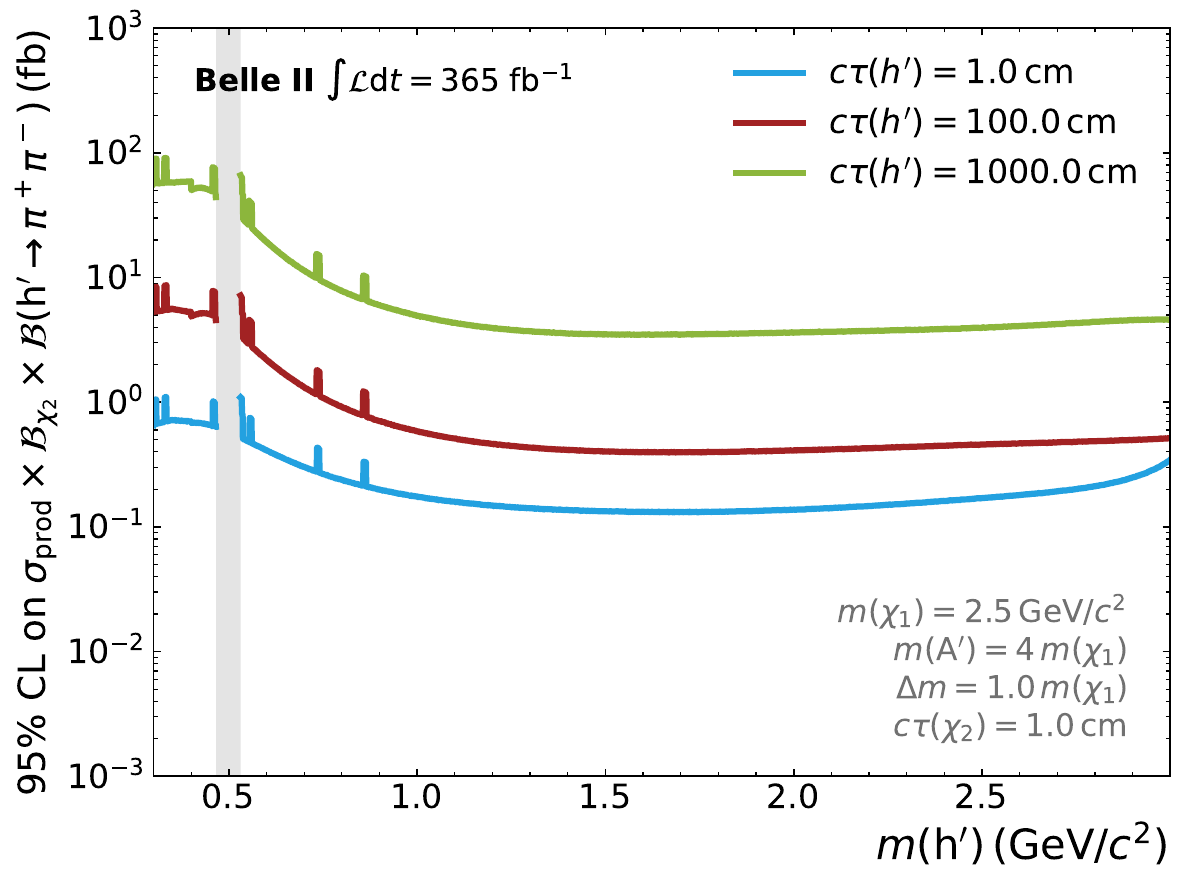}
    \includegraphics[width=0.45\textwidth]{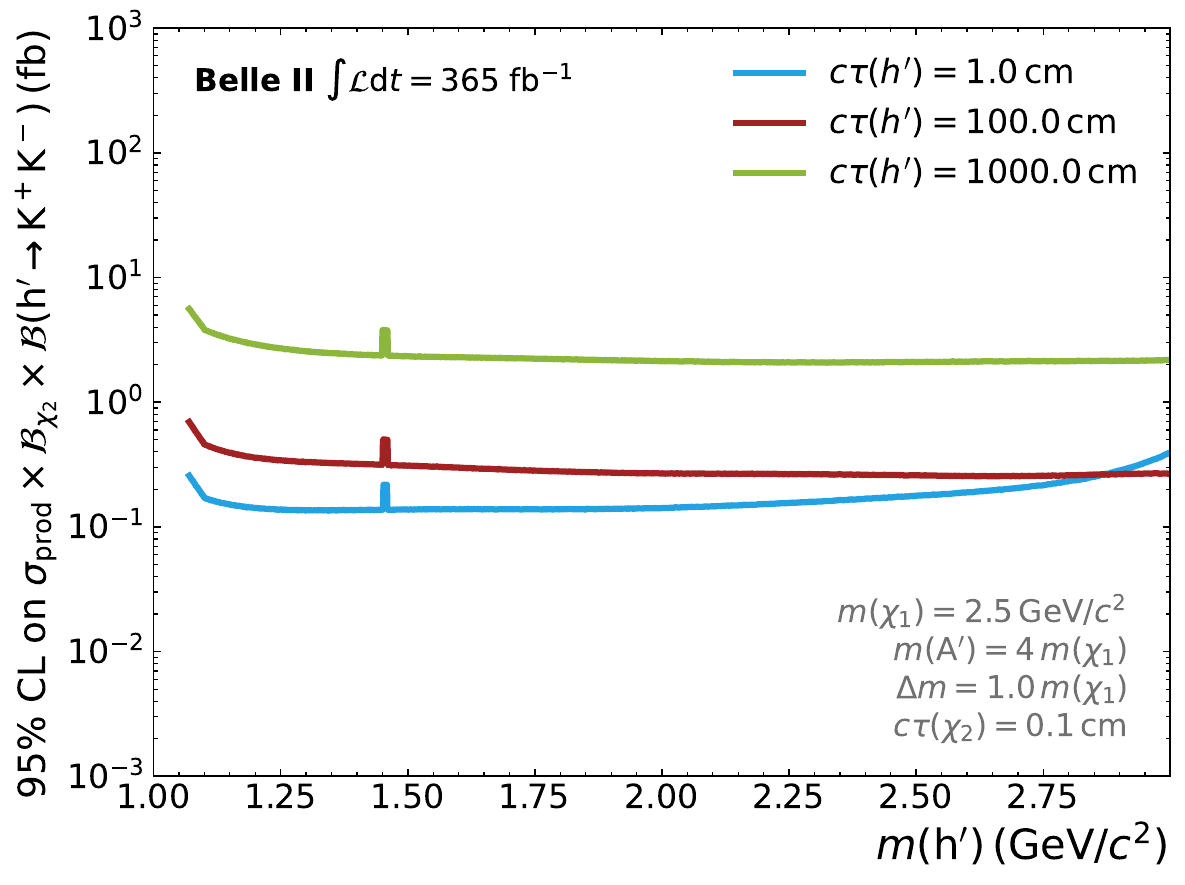}
    \includegraphics[width=0.45\textwidth]{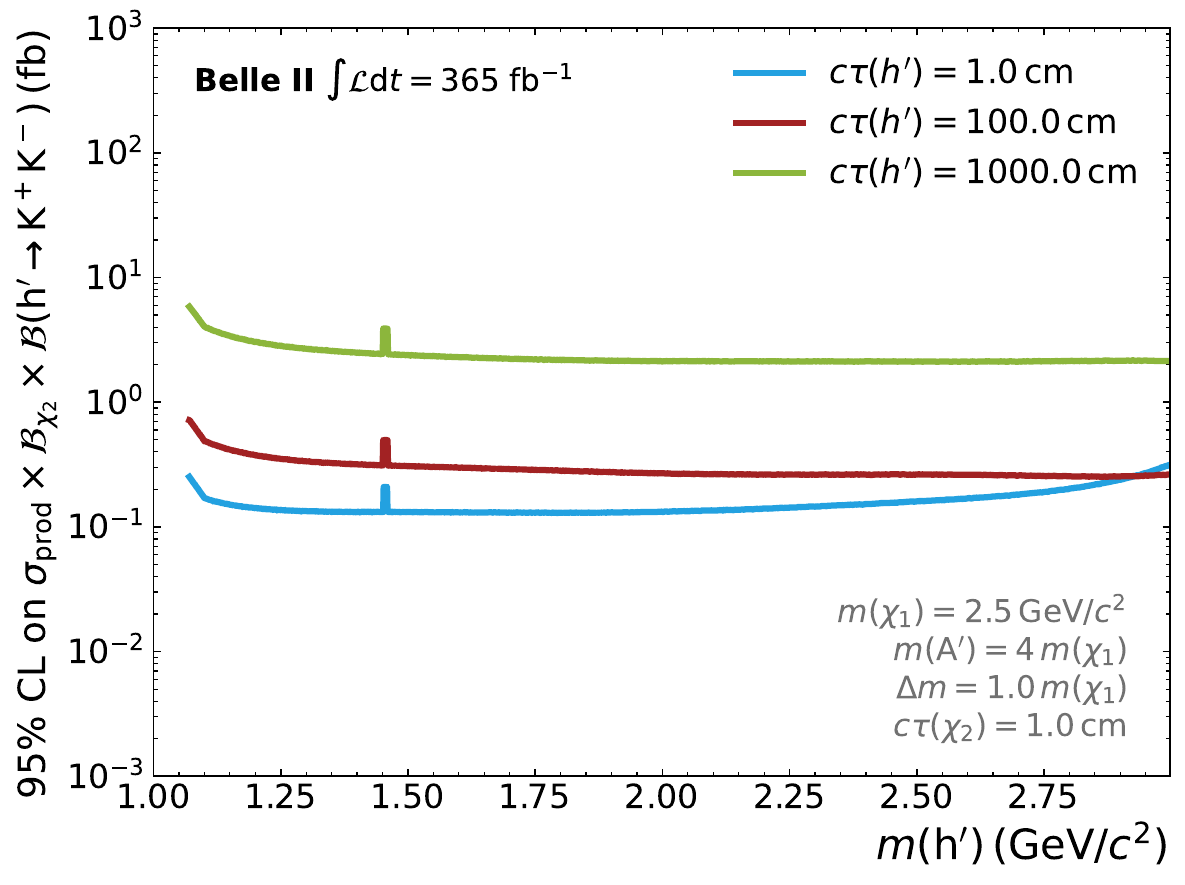}
    \caption{
Upper limits (95\% credibility level) on the \prodbf\ as function of dark Higgs mass $\mdh$ for \hbox{$c\tau(\dh) =1\,\text{cm}$}~(blue), \hbox{$c\tau(\dh) =100\,\text{cm}$}~(red), and \hbox{$c\tau(\dh) =1000\,\text{cm}$}~(green) for $\dh\to\mu^+\mu^-$\,(top), $\dh\to\pi^+\pi^-$\,(center) and $\dh\to K^+ K^-$\,(bottom).
The \chitwo{} lifetime is chosen as $c\tau(\chi_2) = 0.1\,\cm$ (left) and $c\tau(\chi_2) = 1.0\,\cm$ (right).
The remaining model parameters are chosen as $\mchione = 2.5\,\gevcc$, $\map = 4\,\mchione$, and $\Delta m = 1.0\,\mchione$.
The region corresponding to the fully-vetoed \KS mass region is marked in gray.
}
    \label{fig:model_independent9}
\end{figure*}

\begin{figure*}[htp!]
    \centering
    \includegraphics[width=0.45\textwidth]{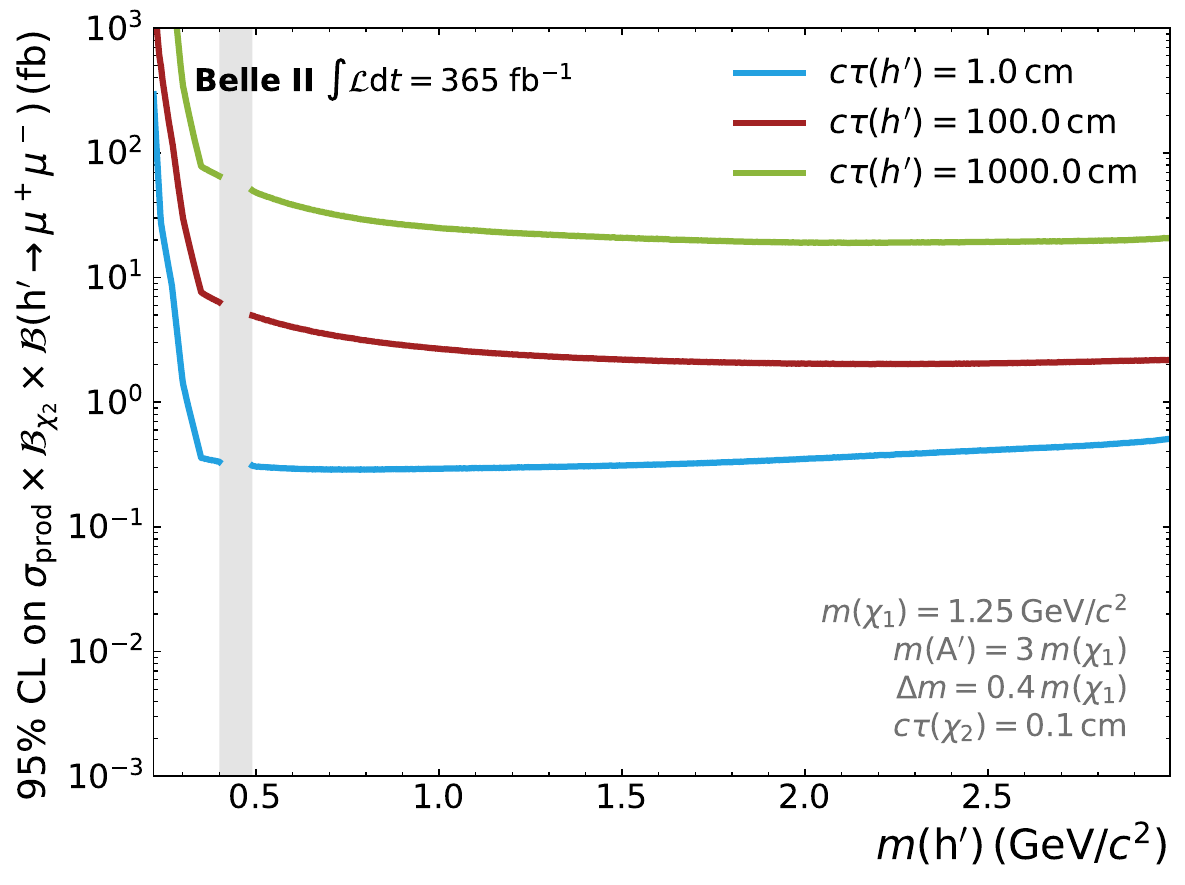}
    \includegraphics[width=0.45\textwidth]{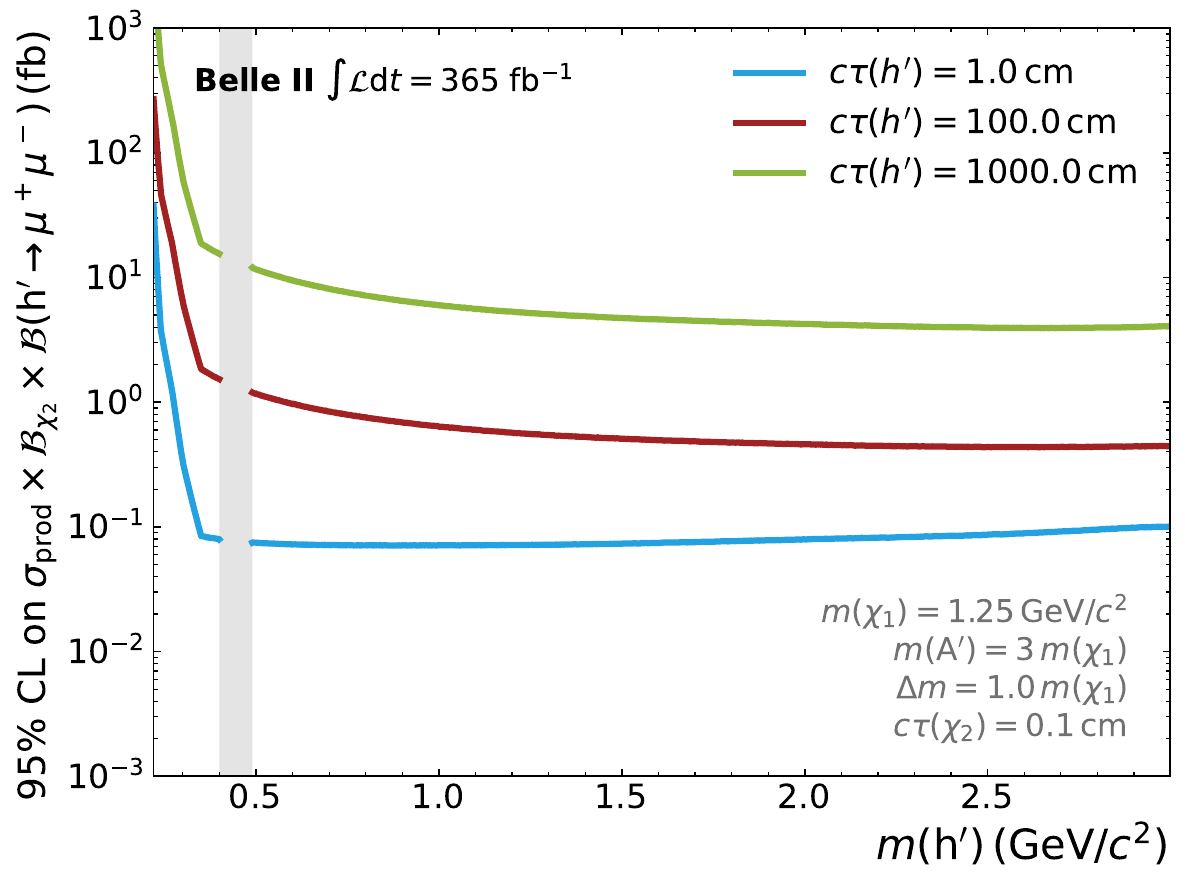}
    \includegraphics[width=0.45\textwidth]{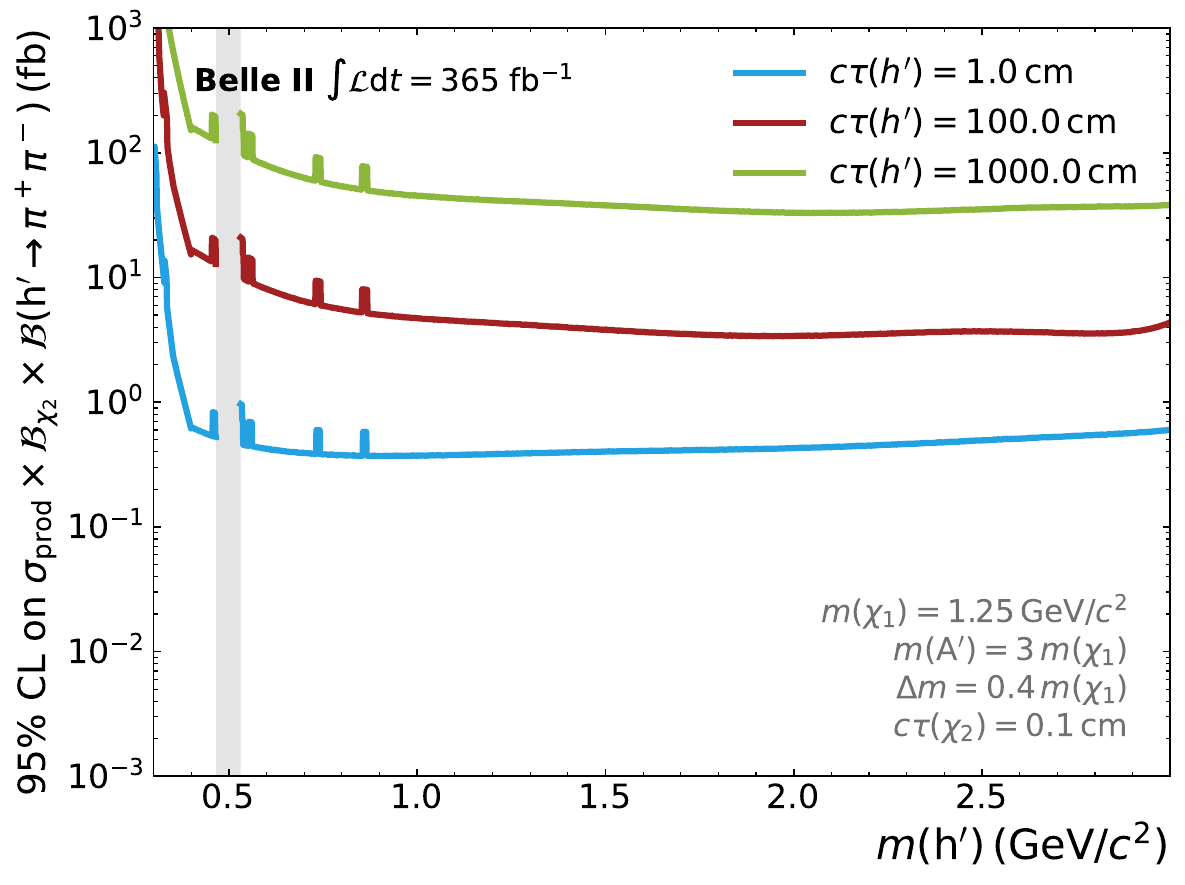}
    \includegraphics[width=0.45\textwidth]{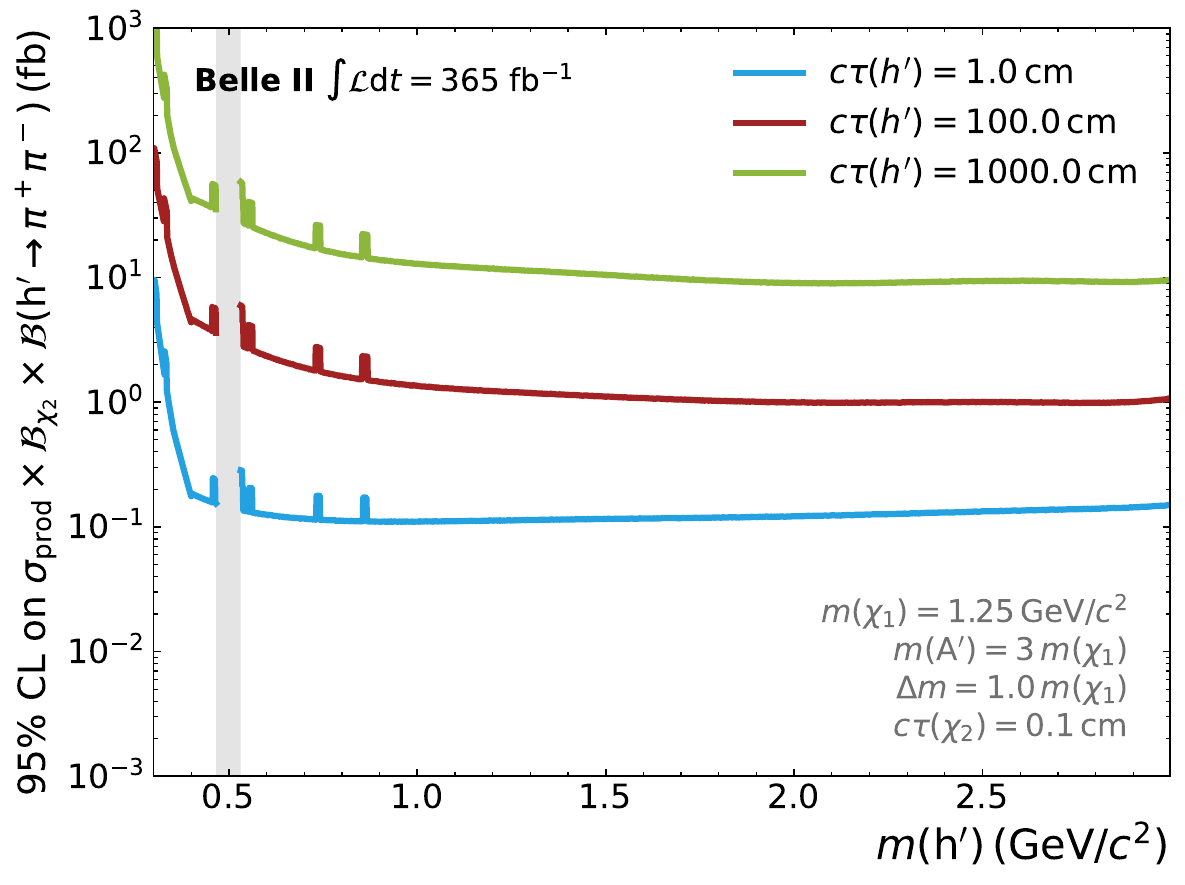}
    \includegraphics[width=0.45\textwidth]{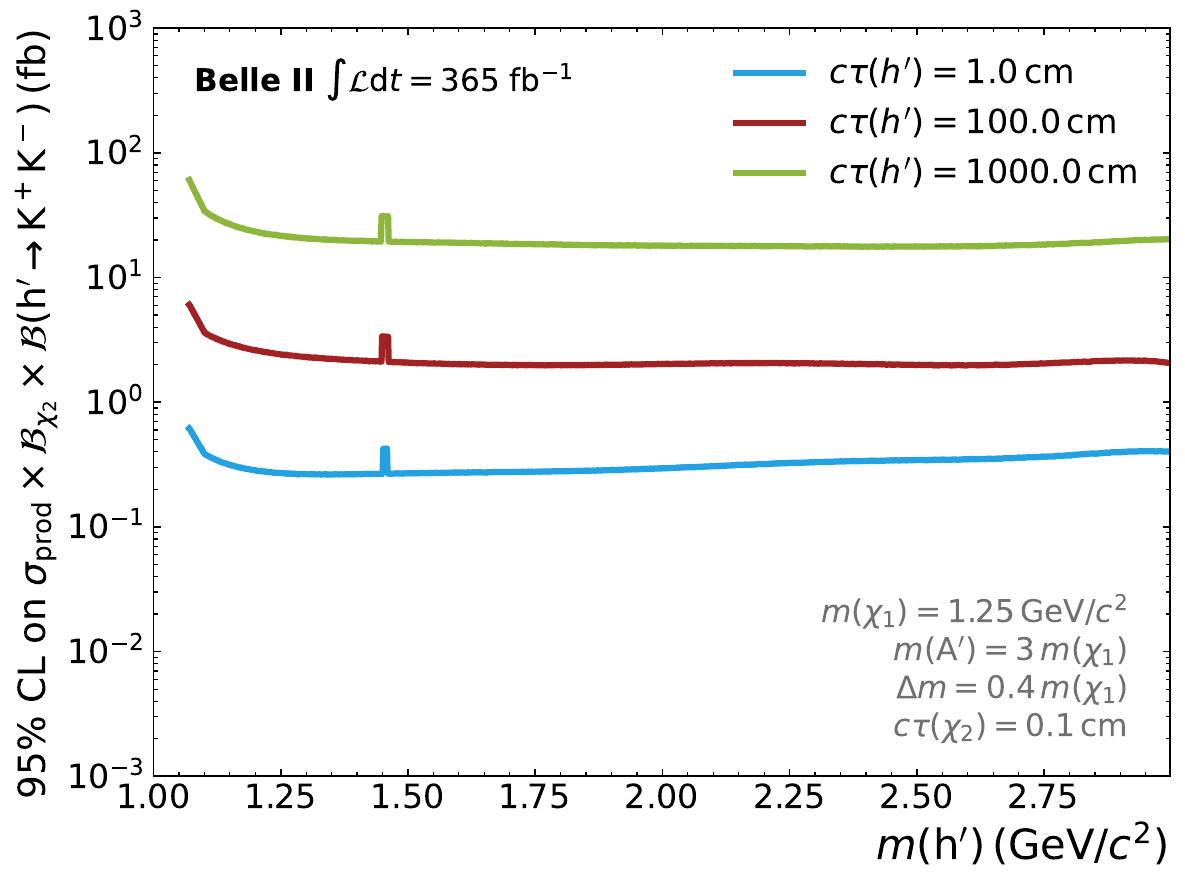}
    \includegraphics[width=0.45\textwidth]{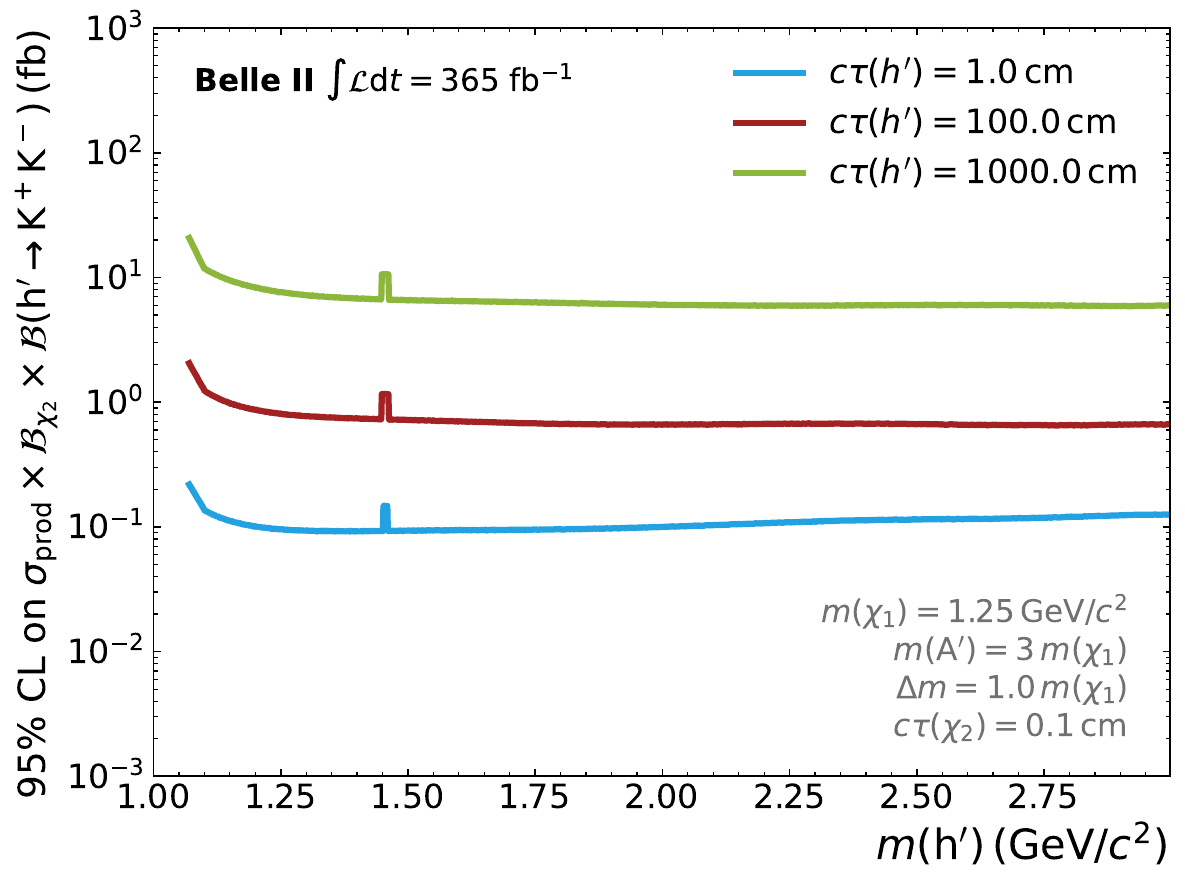}
    \caption{
Upper limits (95\% credibility level) on the \prodbf\ as function of dark Higgs mass $\mdh$ for \hbox{$c\tau(\dh) =1\,\text{cm}$}~(blue), \hbox{$c\tau(\dh) =100\,\text{cm}$}~(red), and \hbox{$c\tau(\dh) =1000\,\text{cm}$}~(green) for $\dh\to\mu^+\mu^-$\,(top), $\dh\to\pi^+\pi^-$\,(center) and $\dh\to K^+ K^-$\,(bottom).
The mass splitting is chosen as $\Delta m = 0.4\,\mchione$ (left) and $\Delta m = 1.0\,\mchione$ (right).
The remaining model parameters are chosen as $\mchione = 1.25\,\gevcc$, $\map = 3\,\mchione$, and $c\tau(\chi_2) = 0.1\,\cm$.
The region corresponding to the fully-vetoed \KS mass region is marked in gray.
}
    \label{fig:model_independent10}
\end{figure*}

\begin{figure*}[htp!]
    \centering
    \includegraphics[width=0.45\textwidth]{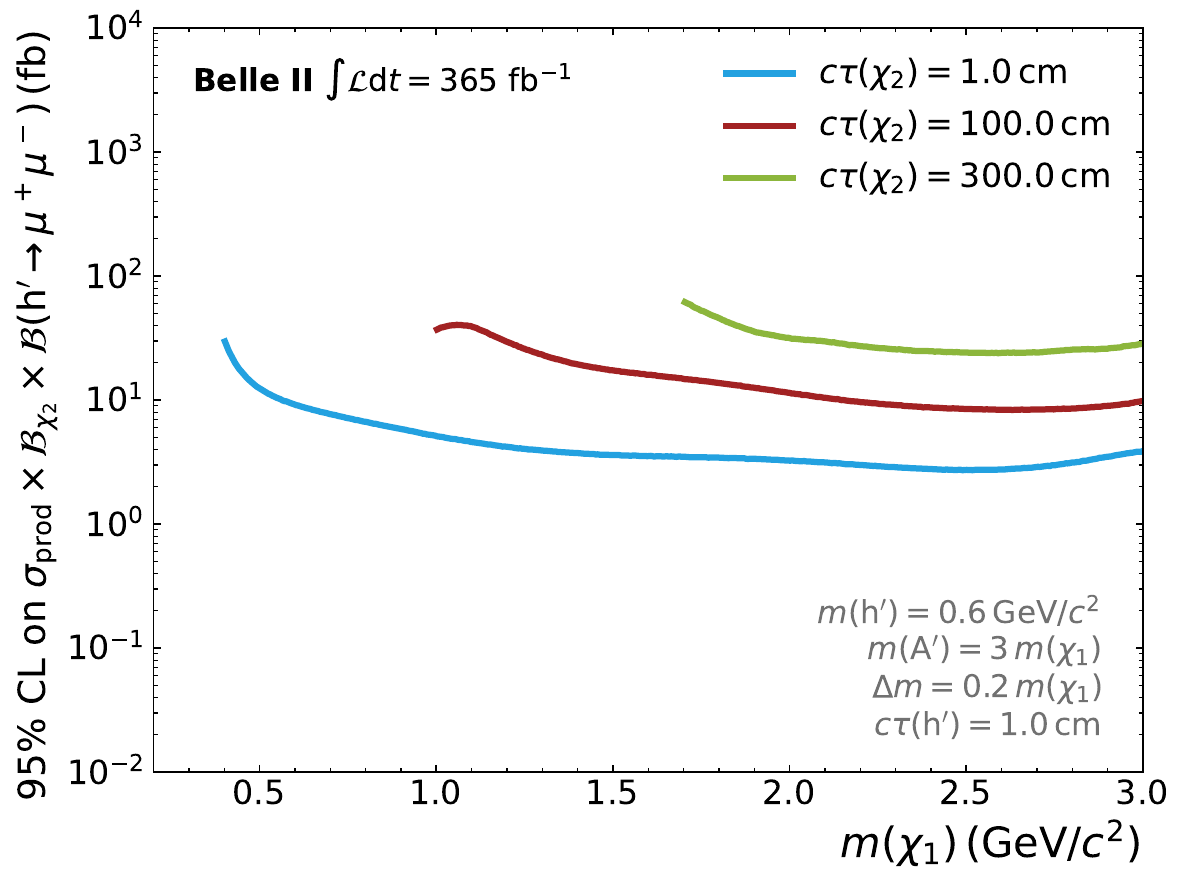}
    \includegraphics[width=0.45\textwidth]{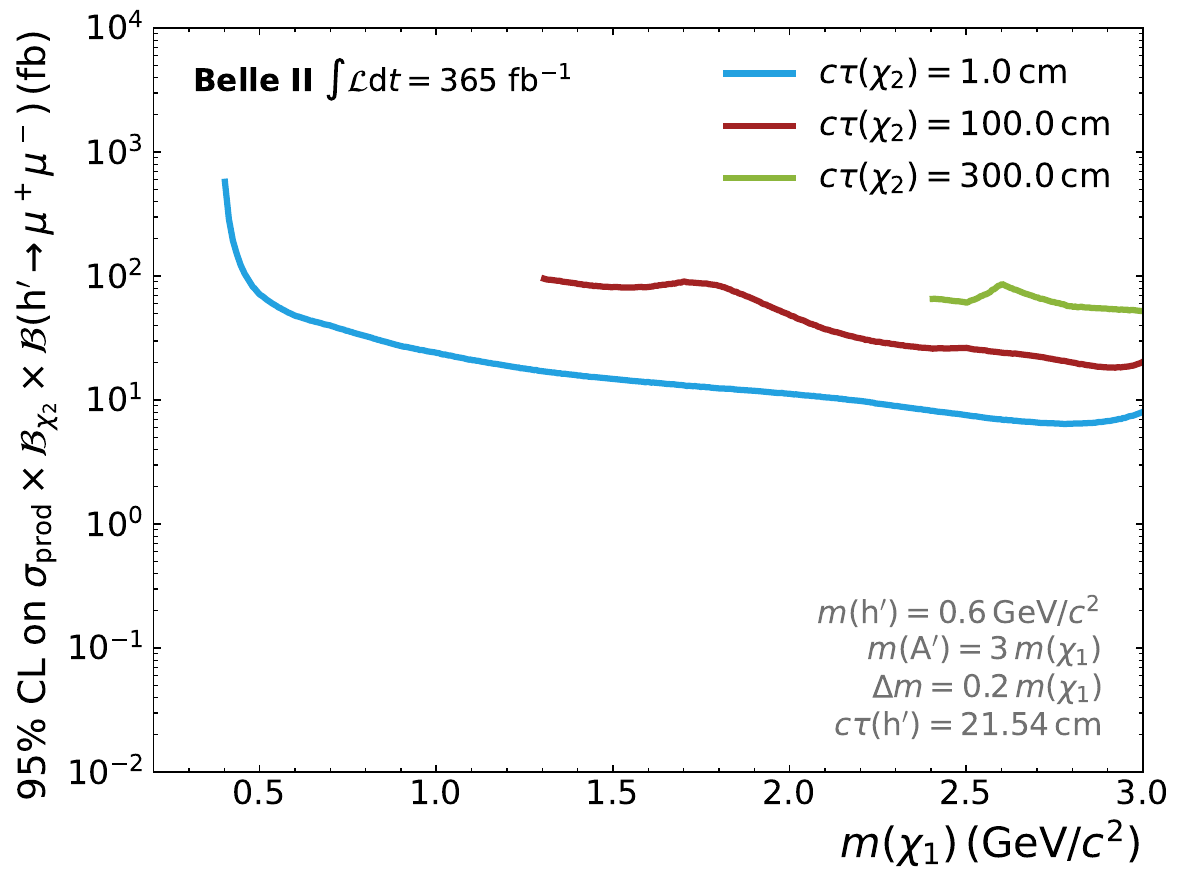}
    \includegraphics[width=0.45\textwidth]{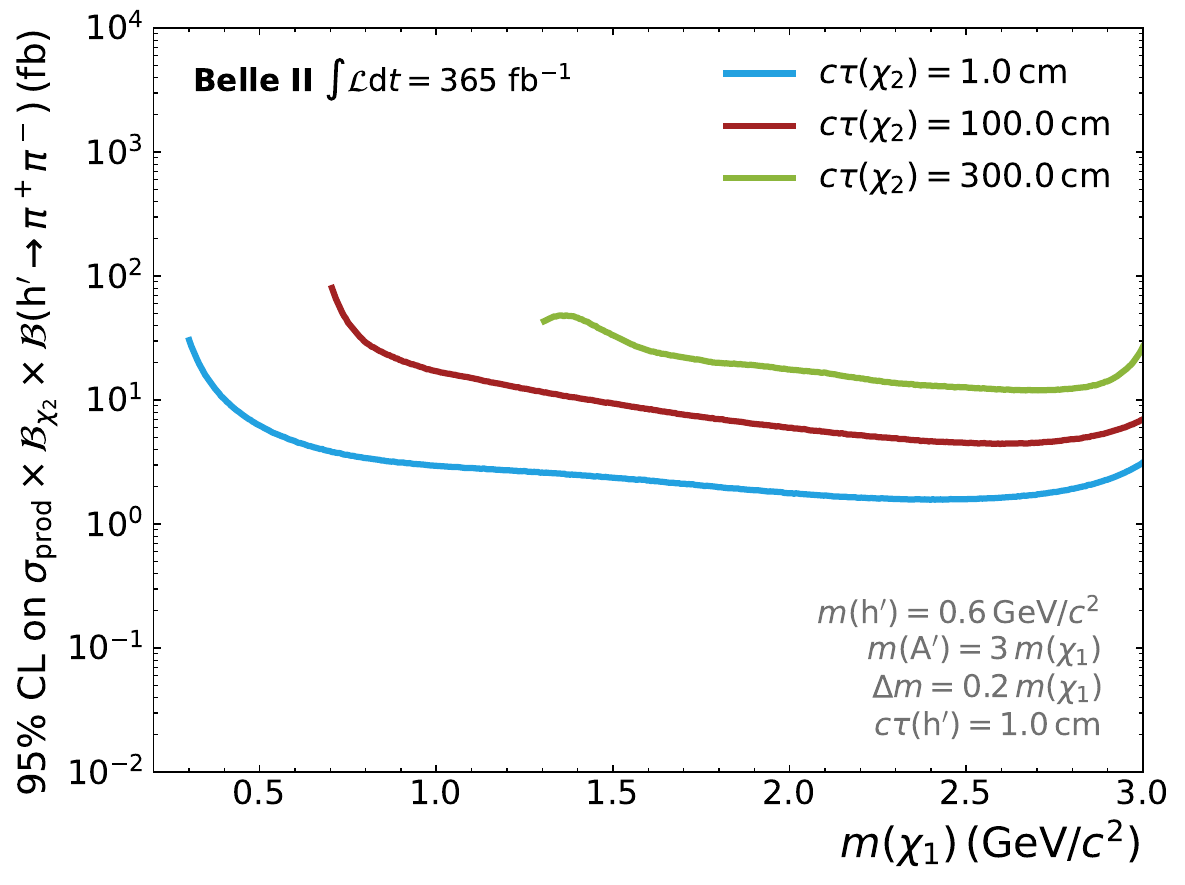}
    \includegraphics[width=0.45\textwidth]{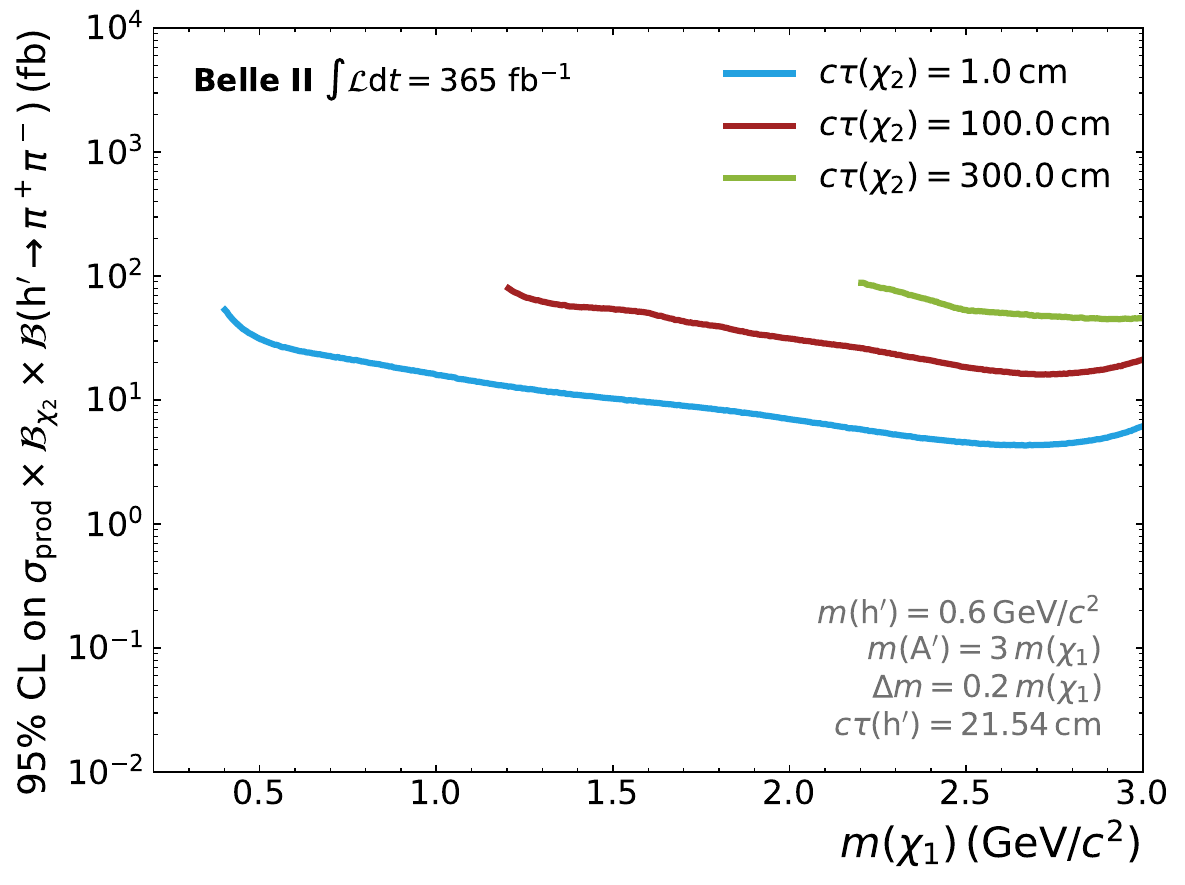}
    \caption{
Upper limits (95\% credibility level) on the \prodbf\ as function of dark Higgs mass $\mdh$ for \hbox{$c\tau(\chitwo) =1\,\text{cm}$}~(blue), \hbox{$c\tau(\chitwo) =100\,\text{cm}$}~(red), and \hbox{$c\tau(\chitwo) =300\,\text{cm}$}~(green) for $\dh\to\mu^+\mu^-$\,(top) and $\dh\to\pi^+\pi^-$\,(bottom).
The dark Higgs lifetime is chosen as $c\tau(\dh) =1.0\,\text{cm}$ (left) and $c\tau(\dh) =21.54\,\text{cm}$ (right).
 The remaining model parameters are chosen as $\mdh = 0.6\,\gevcc$, $\map = 3\,\mchione$, and $\Delta m = 0.2\,\mchione$.
    }
    \label{fig:model_independent11}
\end{figure*}

\begin{figure*}[htp!]
    \centering
    \includegraphics[width=0.45\textwidth]{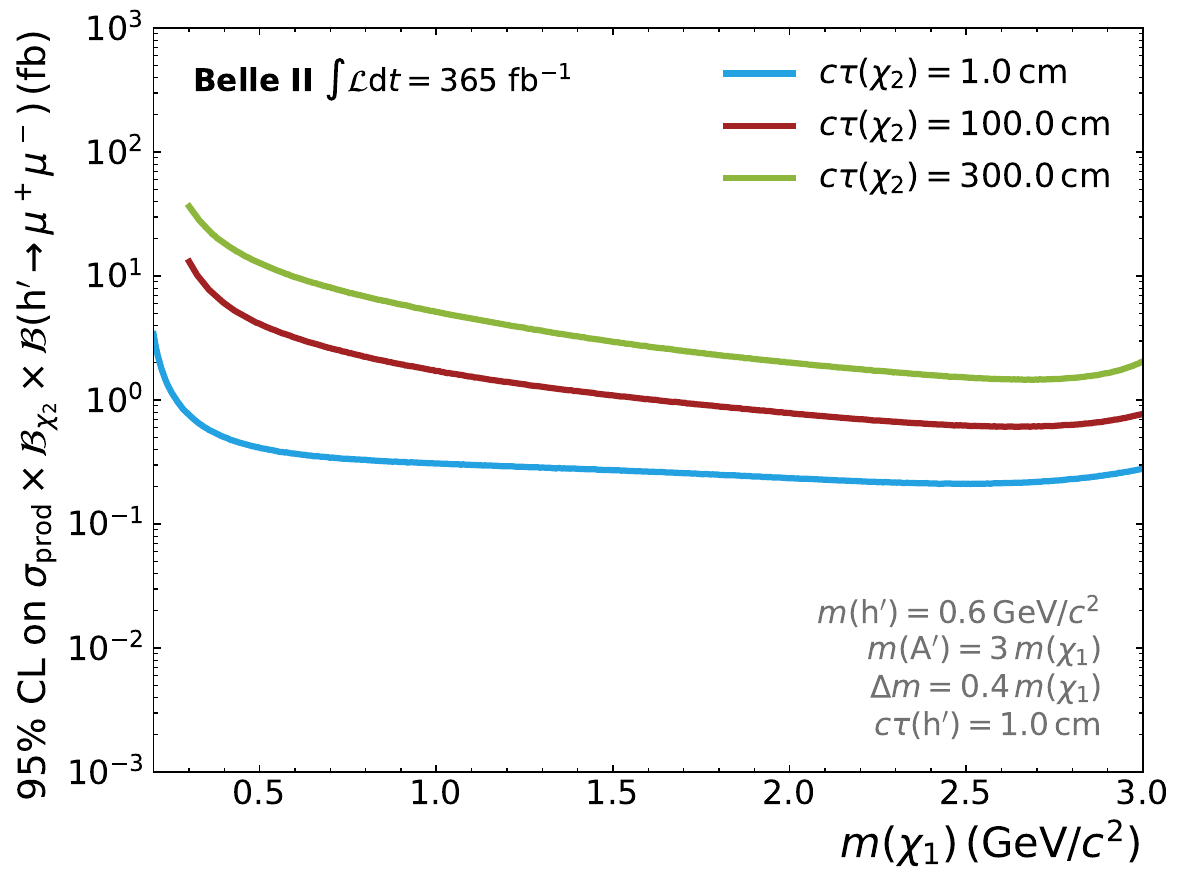}
    \includegraphics[width=0.45\textwidth]{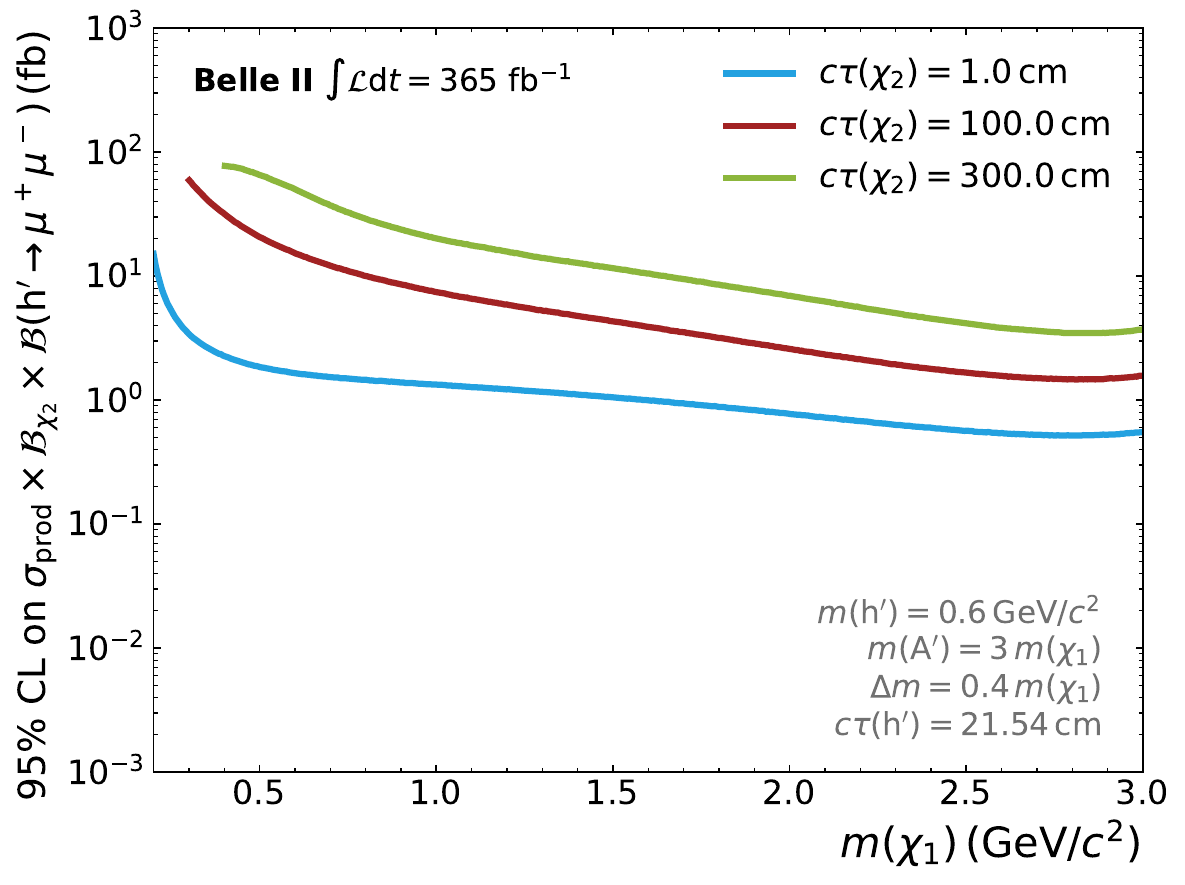}
    \includegraphics[width=0.45\textwidth]{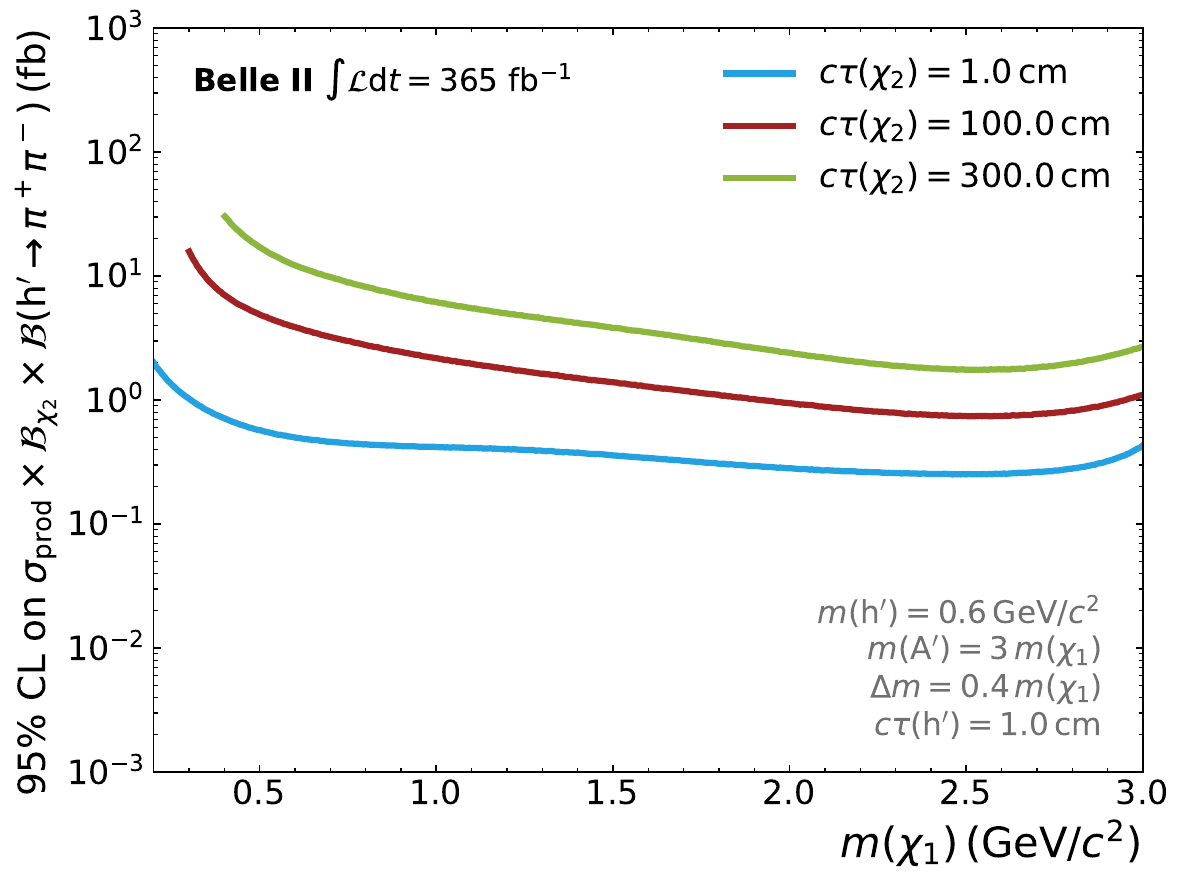}
    \includegraphics[width=0.45\textwidth]{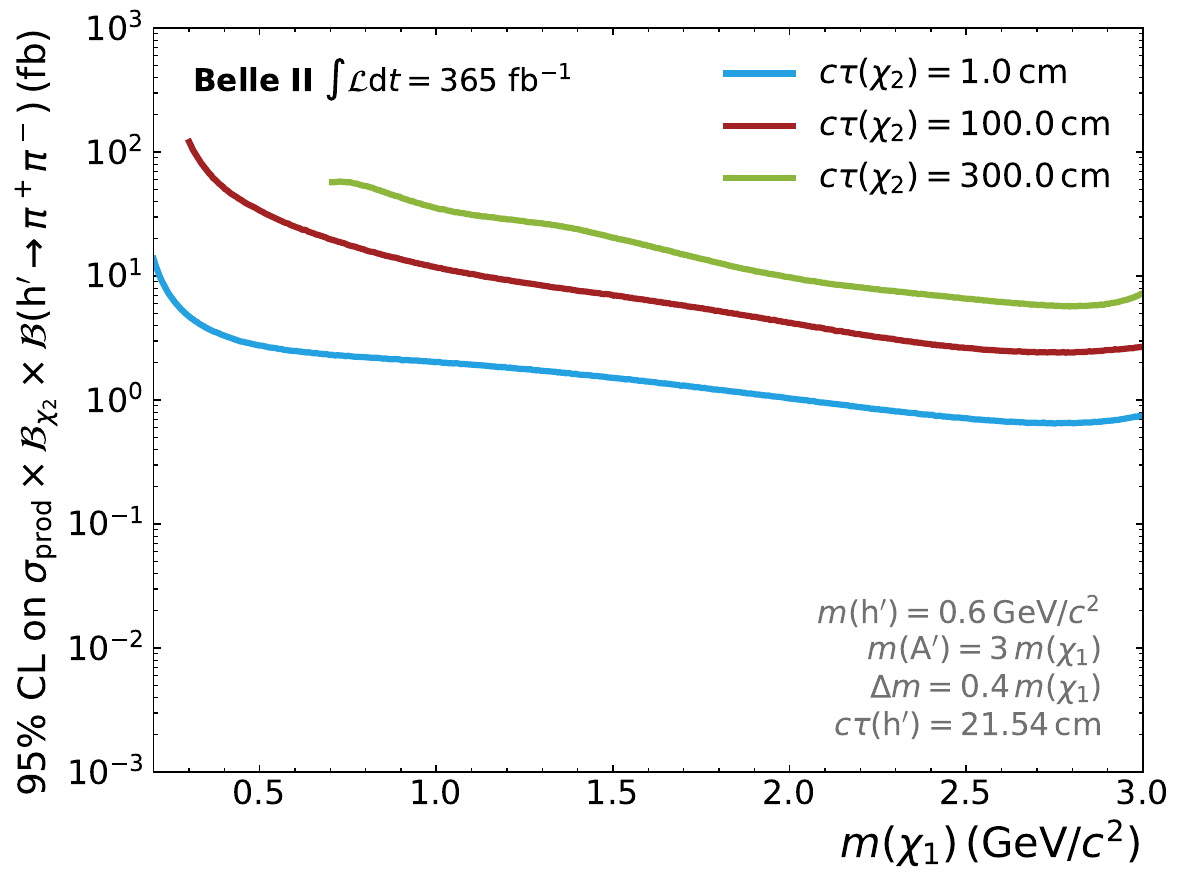}
    \caption{
Upper limits (95\% credibility level) on the \prodbf\ as function of dark Higgs mass $\mdh$ for \hbox{$c\tau(\chitwo) =1\,\text{cm}$}~(blue), \hbox{$c\tau(\chitwo) =100\,\text{cm}$}~(red), and \hbox{$c\tau(\chitwo) =300\,\text{cm}$}~(green) for $\dh\to\mu^+\mu^-$\,(top) and $\dh\to\pi^+\pi^-$\,(bottom).
The dark Higgs lifetime is chosen as $c\tau(\dh) =1.0\,\text{cm}$ (left) and $c\tau(\dh) =21.54\,\text{cm}$ (right).
The remaining model parameters are chosen as $\mdh = 0.6\,\gevcc$, $\map = 3\,\mchione$, and $\Delta m = 0.4\,\mchione$.
}
    \label{fig:model_independent12}
\end{figure*}

\begin{figure*}[htp!]
    \centering
    \includegraphics[width=0.45\textwidth]{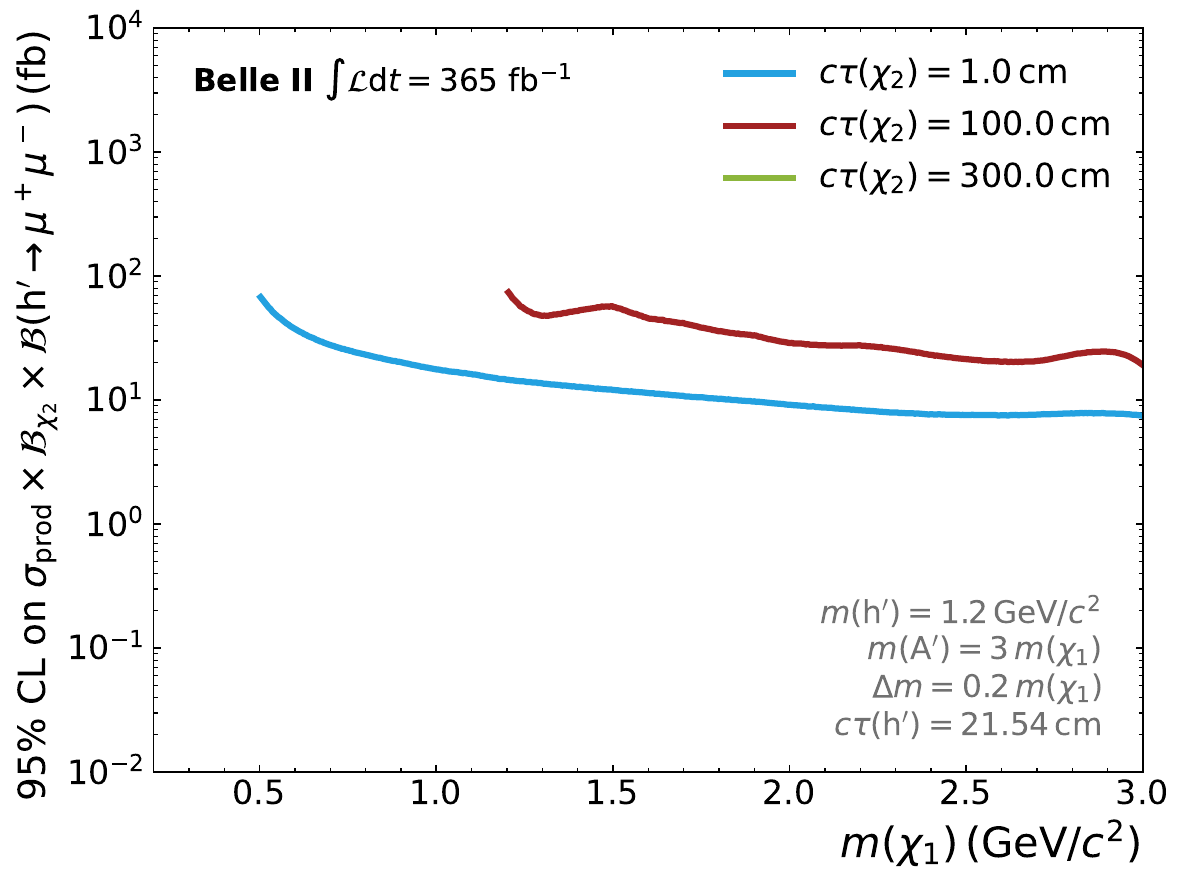}
    \includegraphics[width=0.45\textwidth]{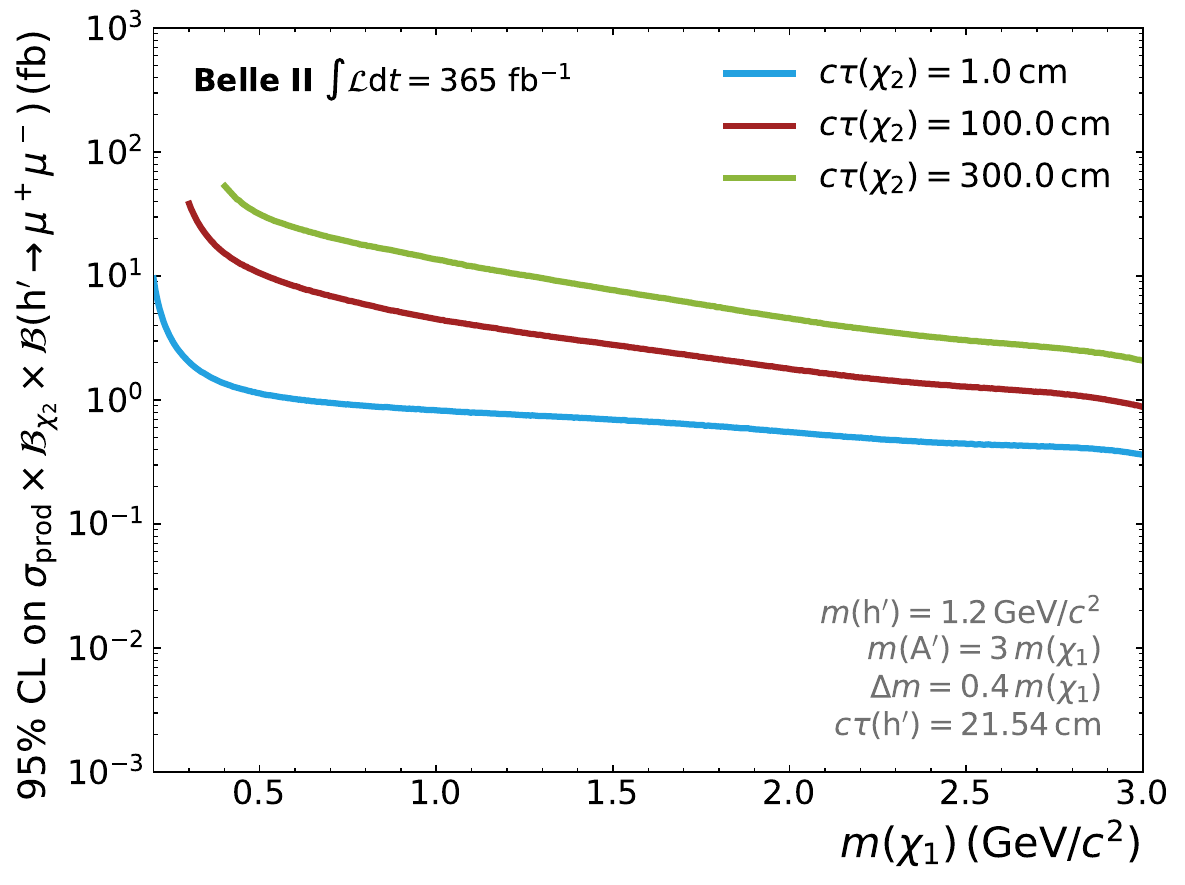}
    \includegraphics[width=0.45\textwidth]{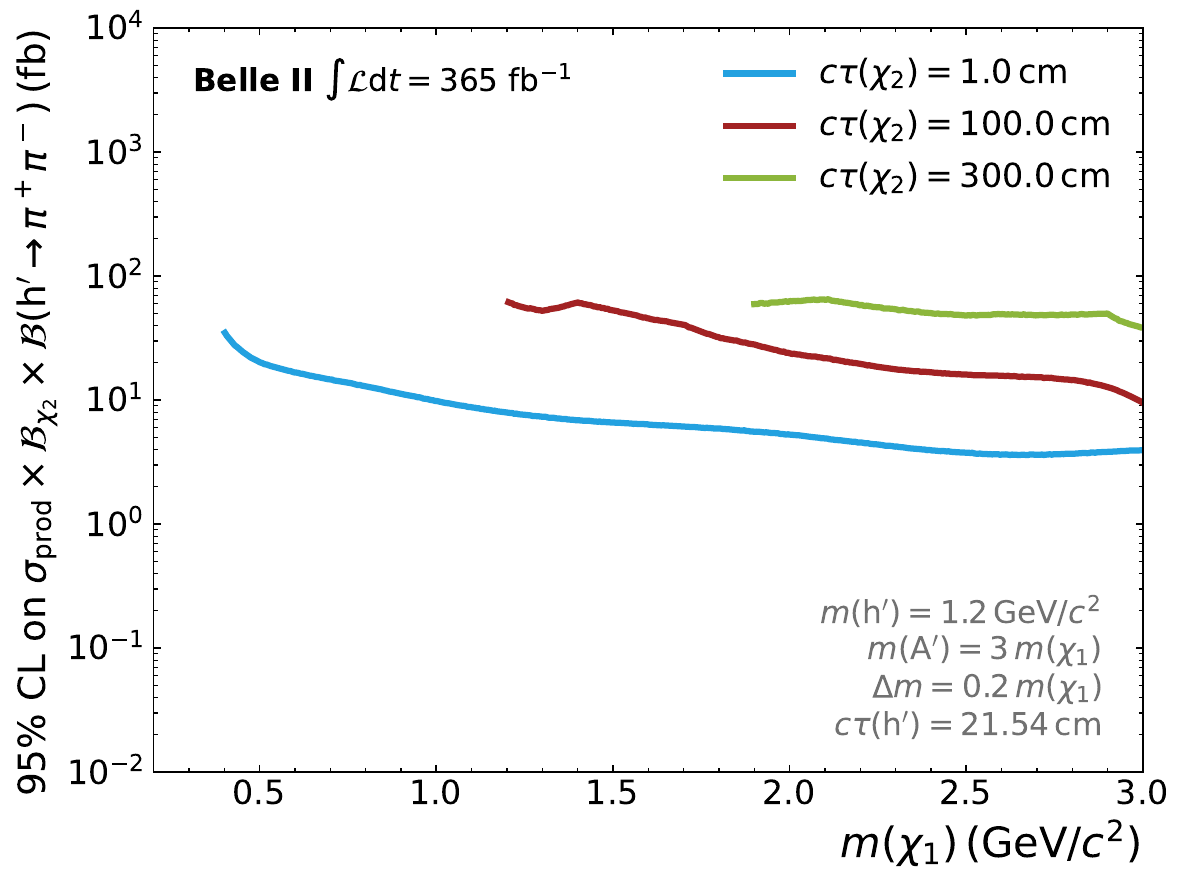}
    \includegraphics[width=0.45\textwidth]{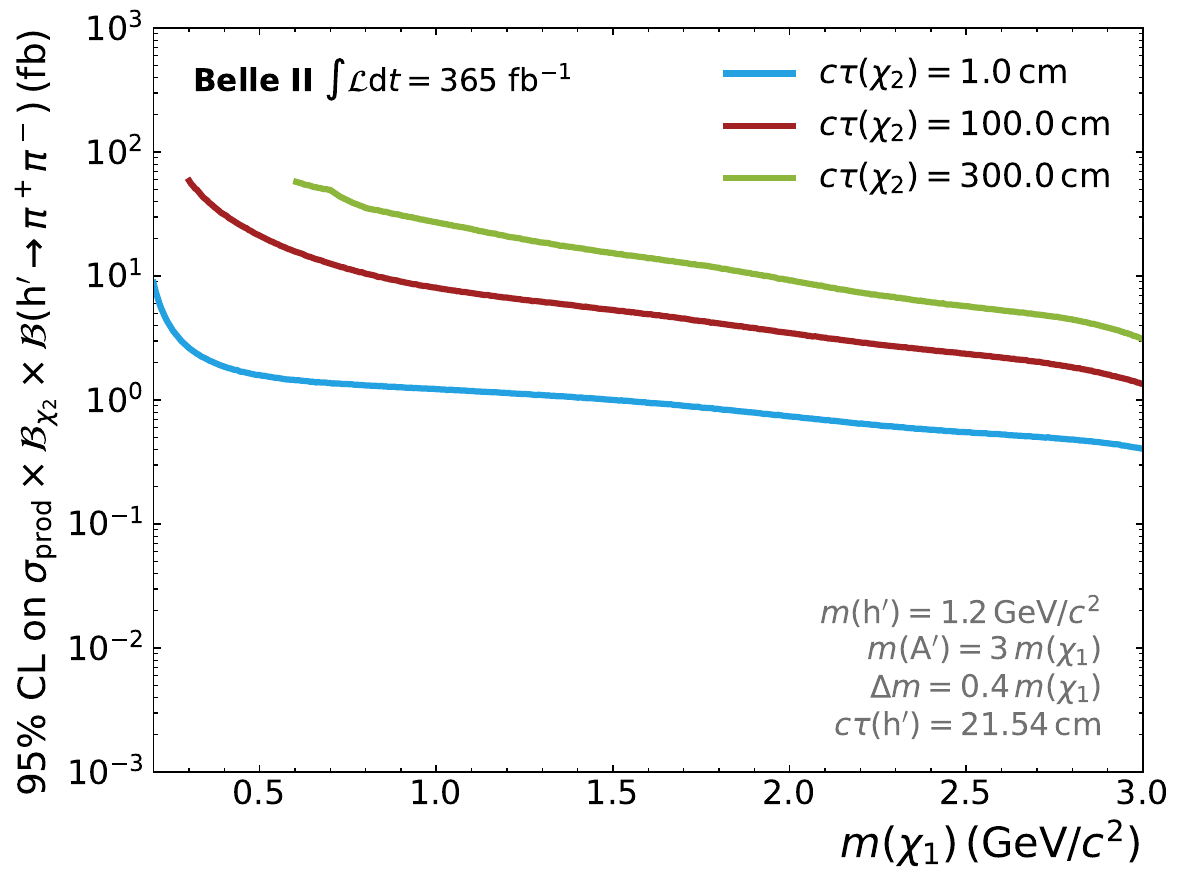}
    \includegraphics[width=0.45\textwidth]{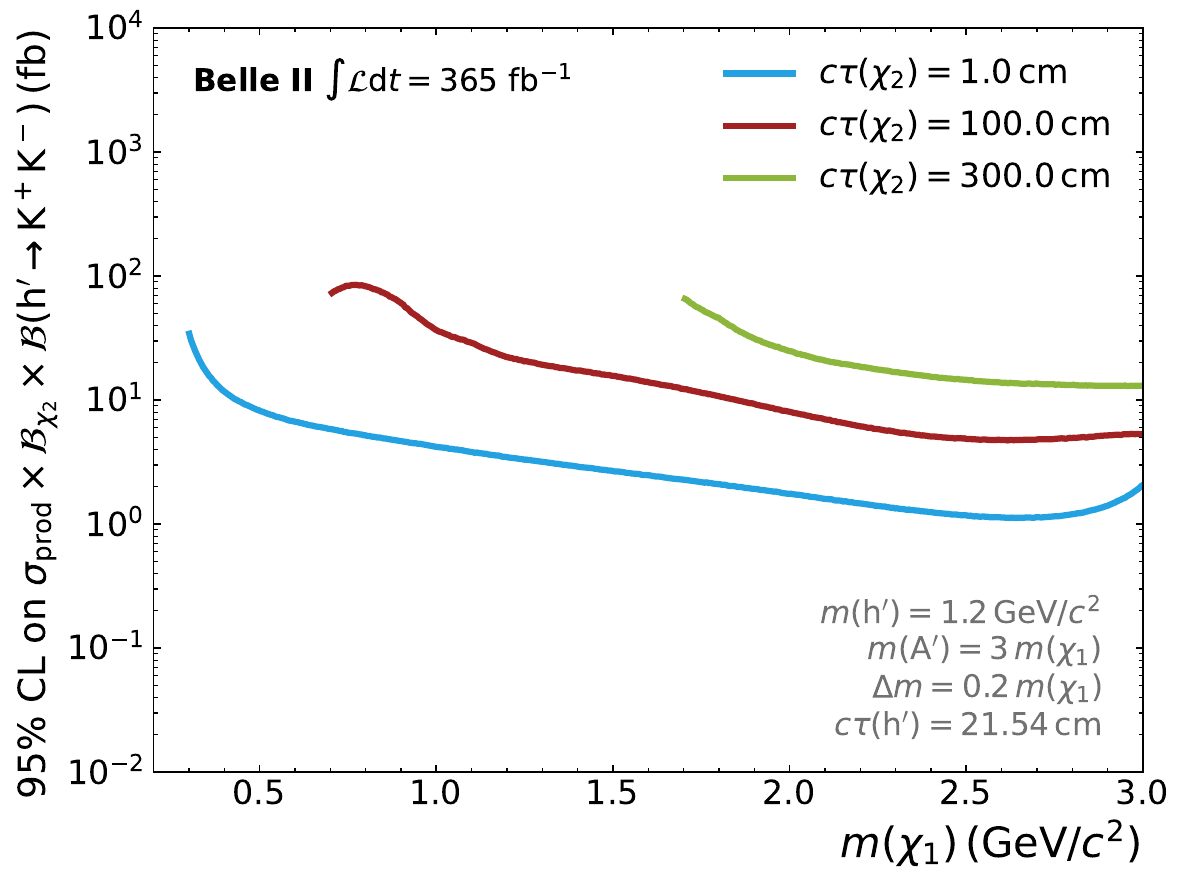}
    \includegraphics[width=0.45\textwidth]{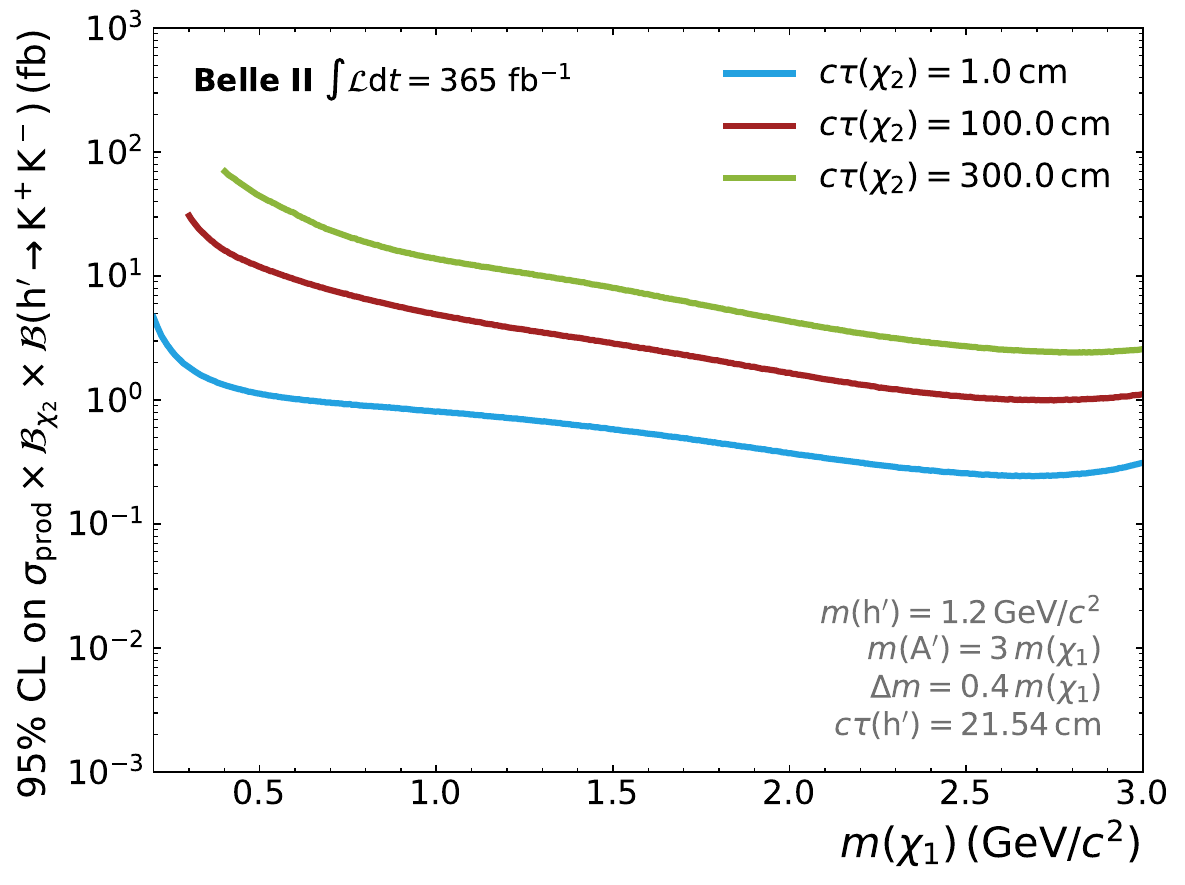}
    \caption{
Upper limits (95\% credibility level) on the \prodbf\ as function of dark Higgs mass $\mdh$ for \hbox{$c\tau(\chitwo) =1\,\text{cm}$}~(blue), \hbox{$c\tau(\chitwo) =100\,\text{cm}$}~(red), and \hbox{$c\tau(\chitwo) =300\,\text{cm}$}~(green) for $\dh\to\mu^+\mu^-$\,(top), $\dh\to\pi^+\pi^-$\,(center) and $\dh\to K^+ K^-$\,(bottom).
The mass splitting is chosen as $\Delta m = 0.2\,\mchione$ (left) and $\Delta m = 0.4\,\mchione$ (right).
The remaining model parameters are chosen as $\mdh = 0.6\,\gevcc$, $c\tau(\dh) =21.54\,\text{cm}$, and $\map = 3\,\mchione$.
}
    \label{fig:model_independent13}
\end{figure*}

\begin{figure*}[htp!]
    \centering
    \includegraphics[width=0.45\textwidth]{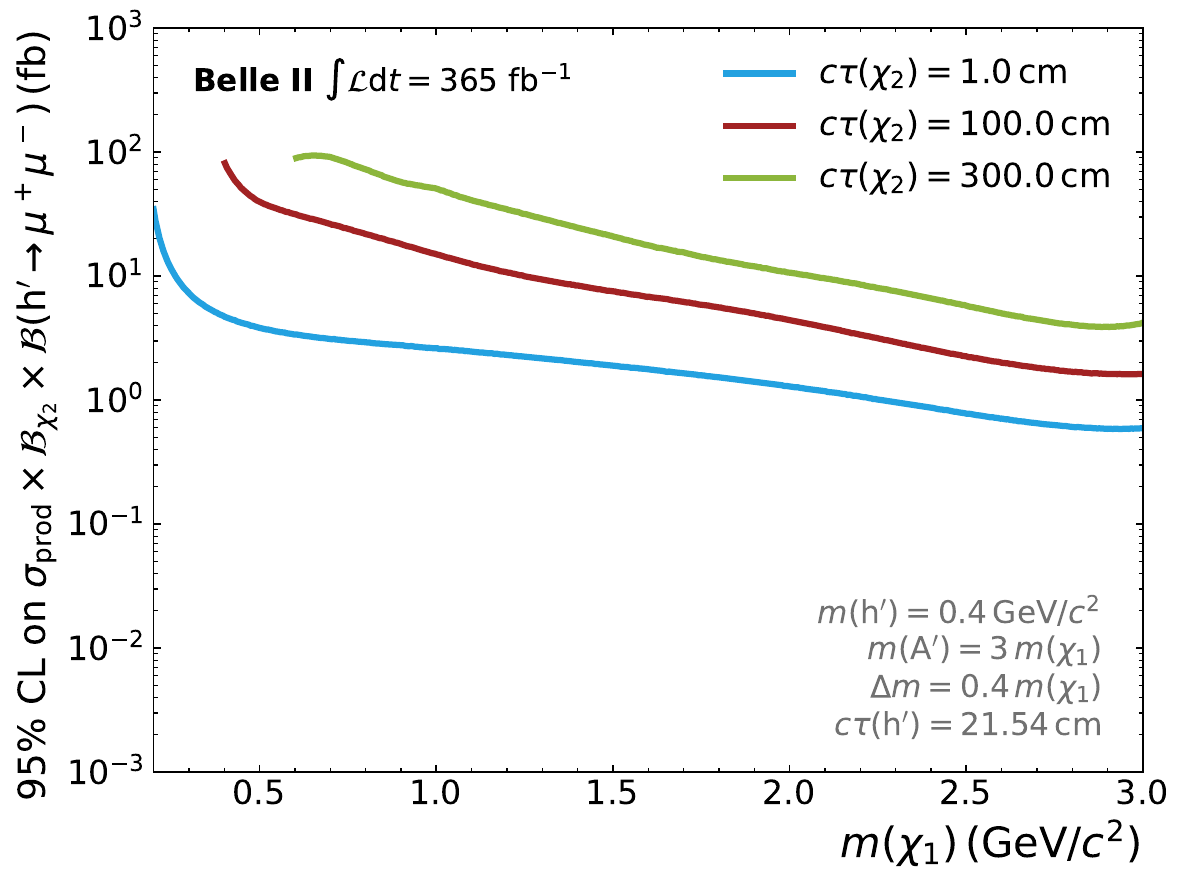}
    \includegraphics[width=0.45\textwidth]{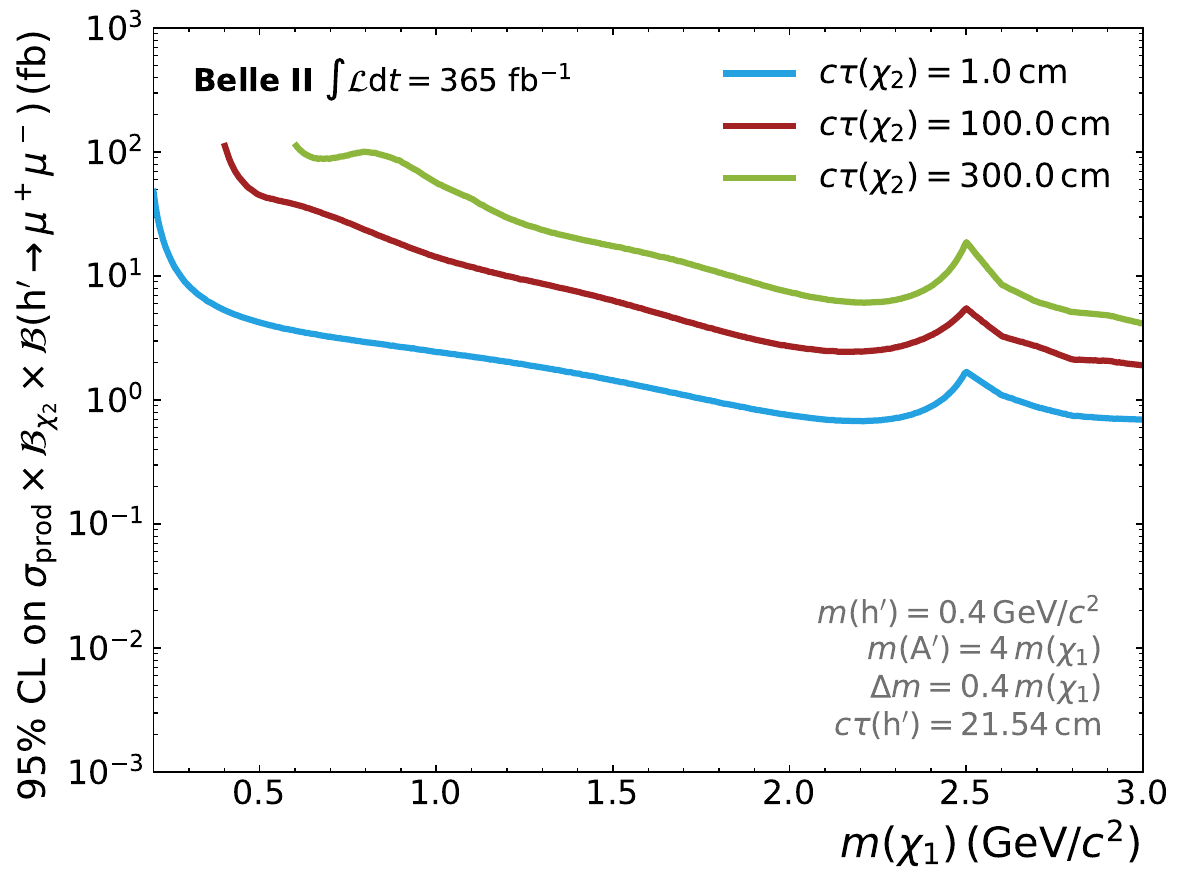}
    \includegraphics[width=0.45\textwidth]{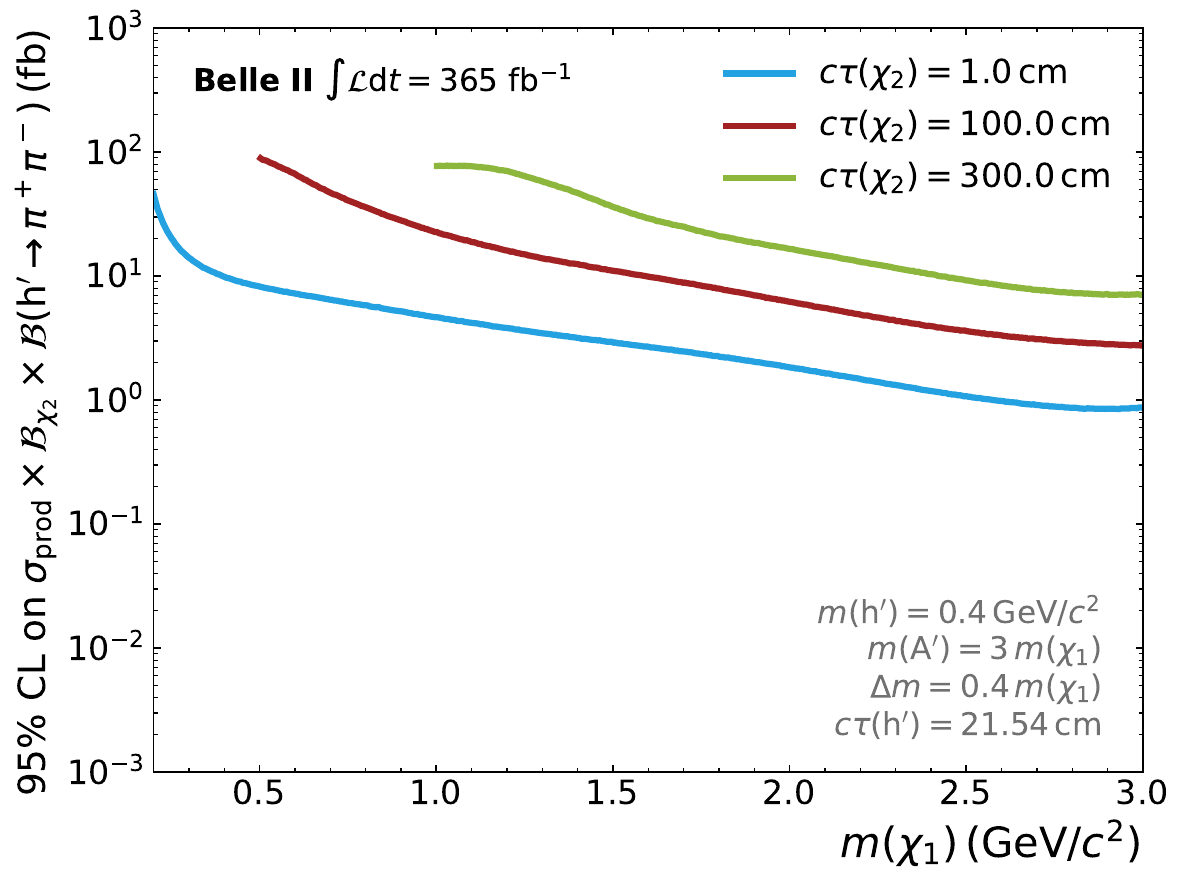}
    \includegraphics[width=0.45\textwidth]{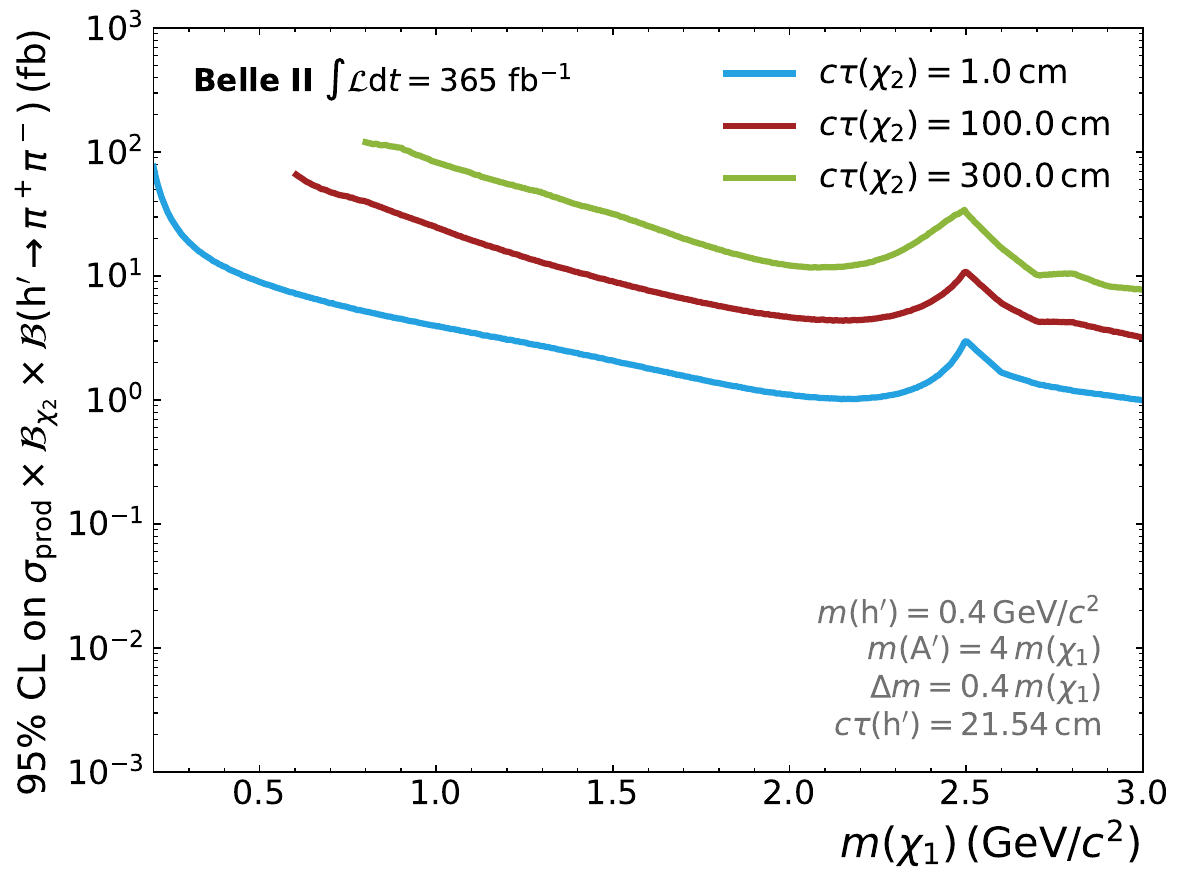}
    \caption{
Upper limits (95\% credibility level) on the \prodbf\ as function of dark Higgs mass $\mdh$ for \hbox{$c\tau(\chitwo) =1\,\text{cm}$}~(blue), \hbox{$c\tau(\chitwo) =100\,\text{cm}$}~(red), and \hbox{$c\tau(\chitwo) =300\,\text{cm}$}~(green) for $\dh\to\mu^+\mu^-$\,(top) and $\dh\to\pi^+\pi^-$\,(bottom).
The dark photon mass is chosen as $\map = 3\,\mchione$ (left) and $\map = 4\,\mchione$ (right).
The remaining model parameters are chosen as $\mdh = 0.4\,\gevcc$, $c\tau(\dh) =21.54\,\text{cm}$, and $\Delta m = 0.4\,\mchione$.
}
    \label{fig:model_independent14}
\end{figure*}


\begin{figure*}[htp!]
    \centering
    \includegraphics[width=0.45\textwidth]{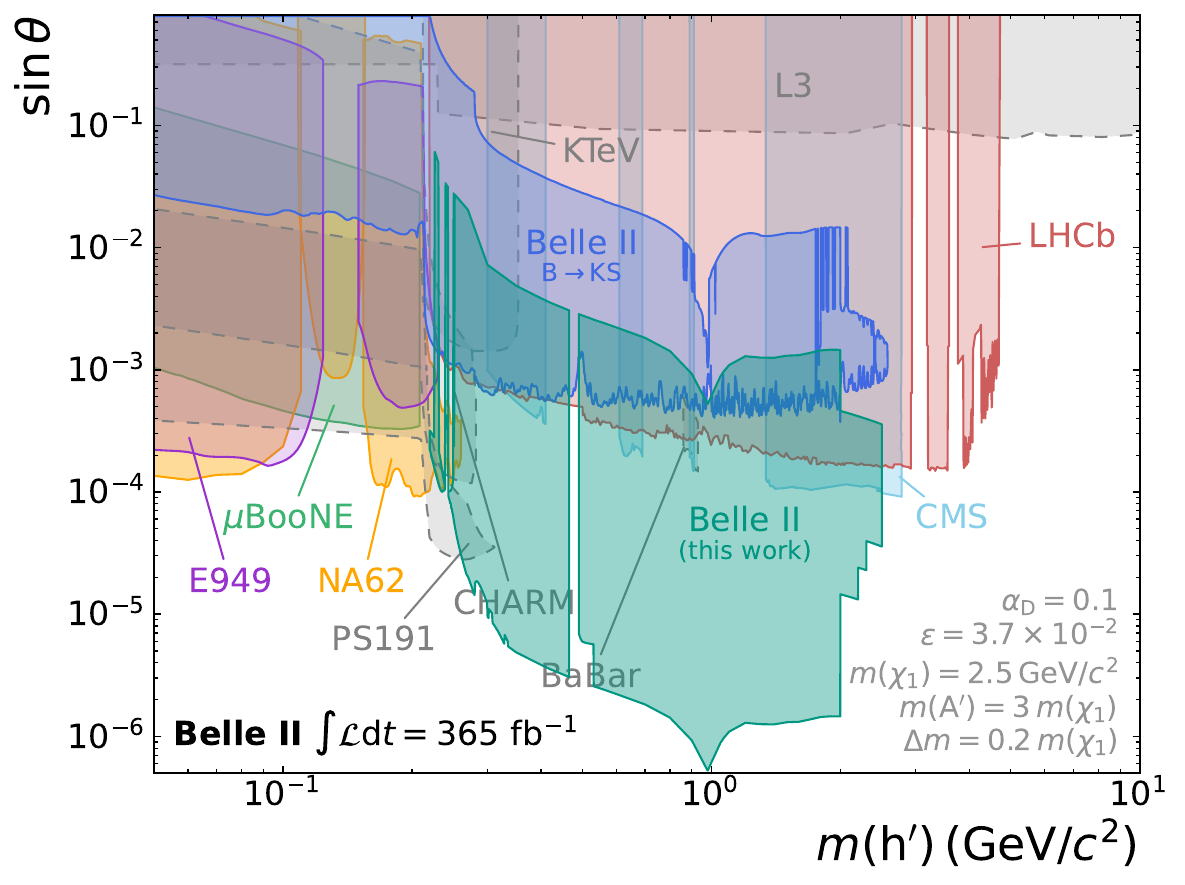}
    \includegraphics[width=0.45\textwidth]{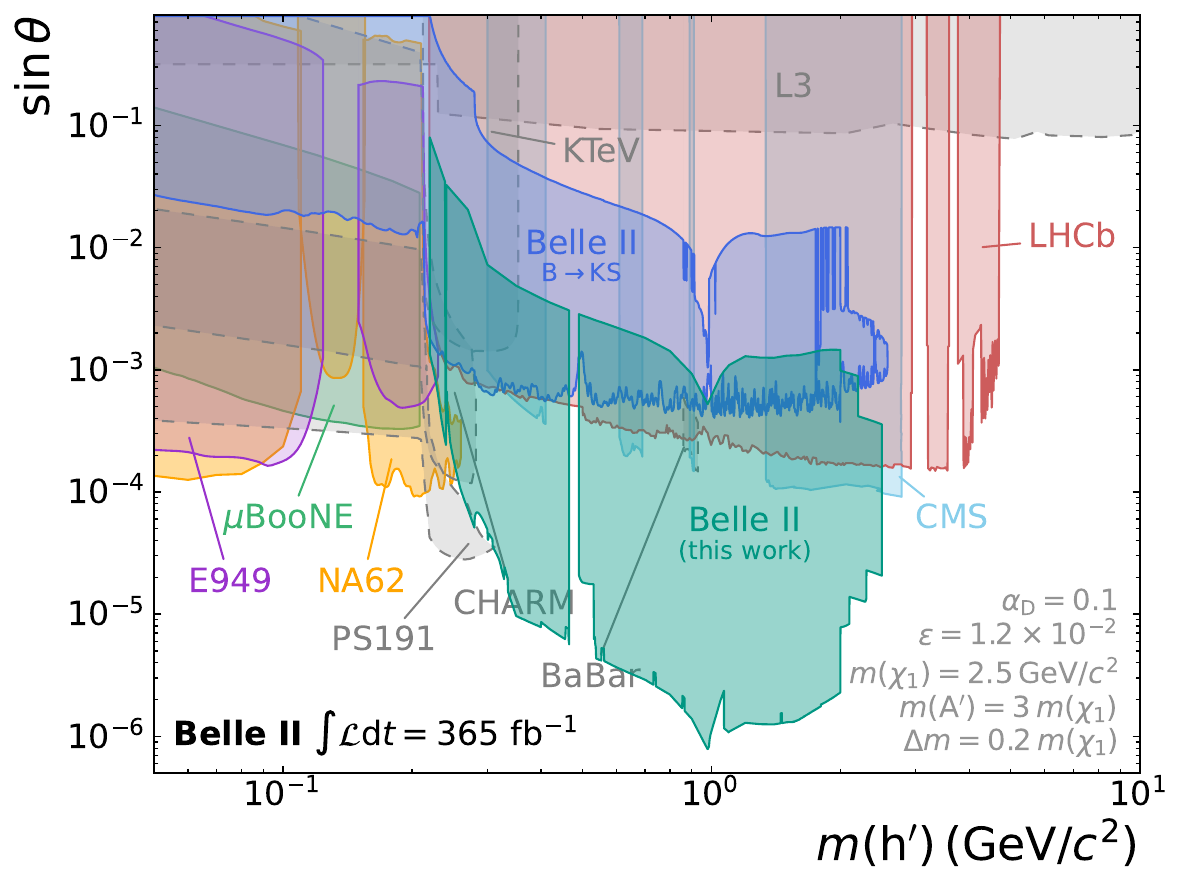}
    \includegraphics[width=0.45\textwidth]{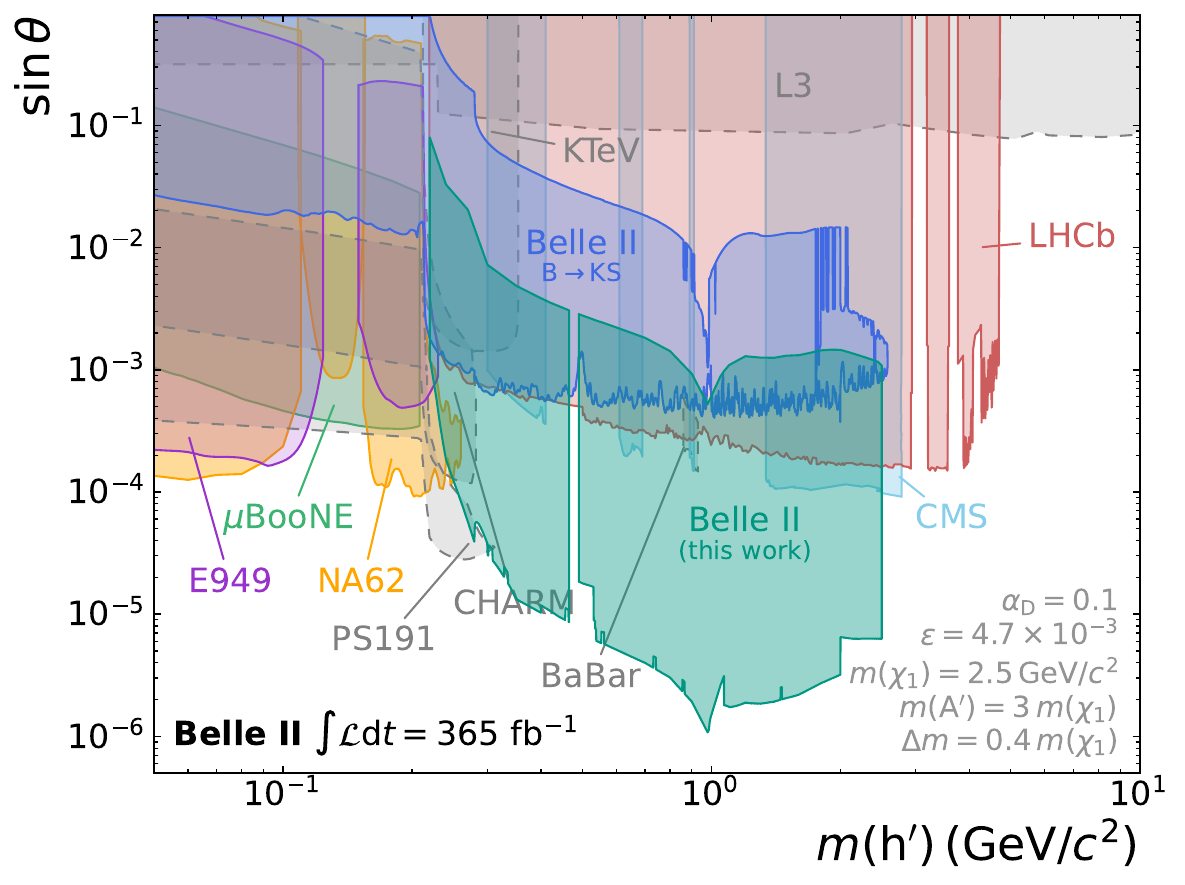}
    \includegraphics[width=0.45\textwidth]{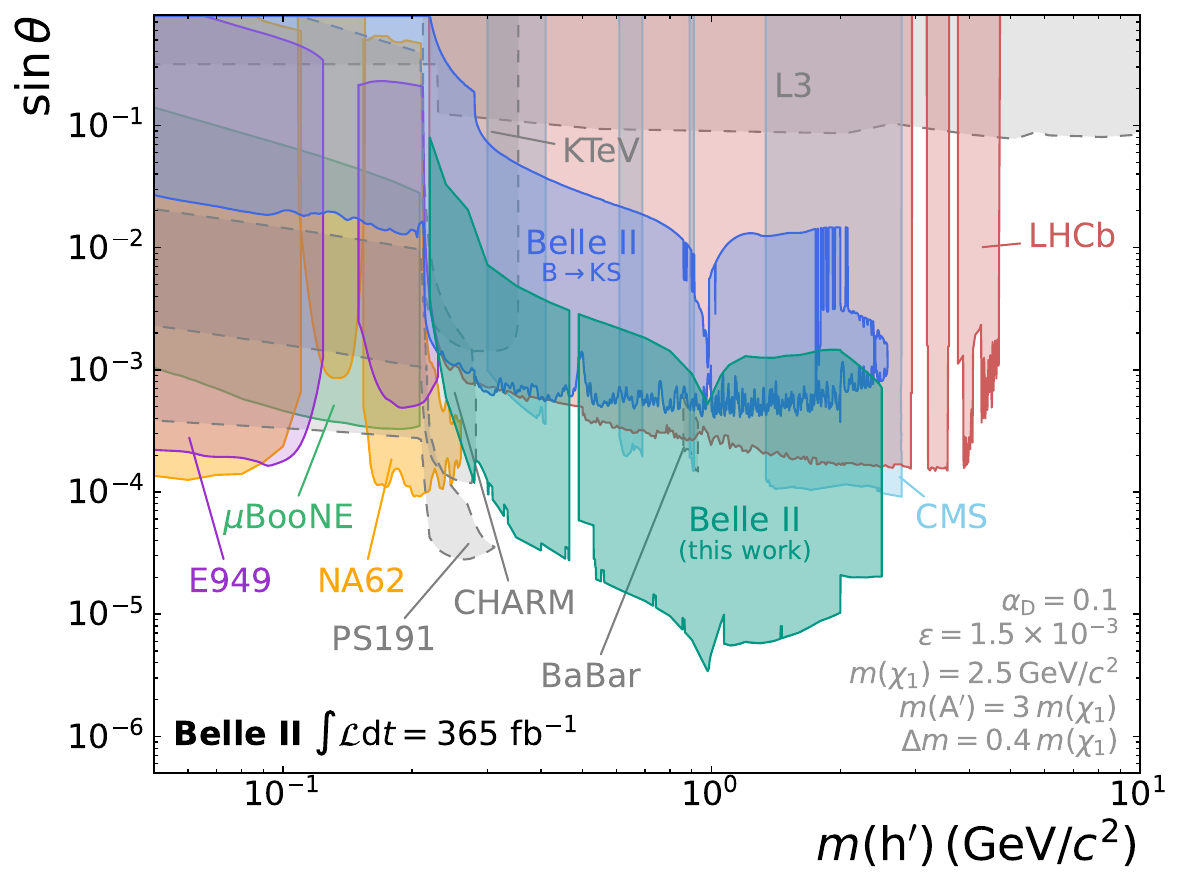}
    \includegraphics[width=0.45\textwidth]{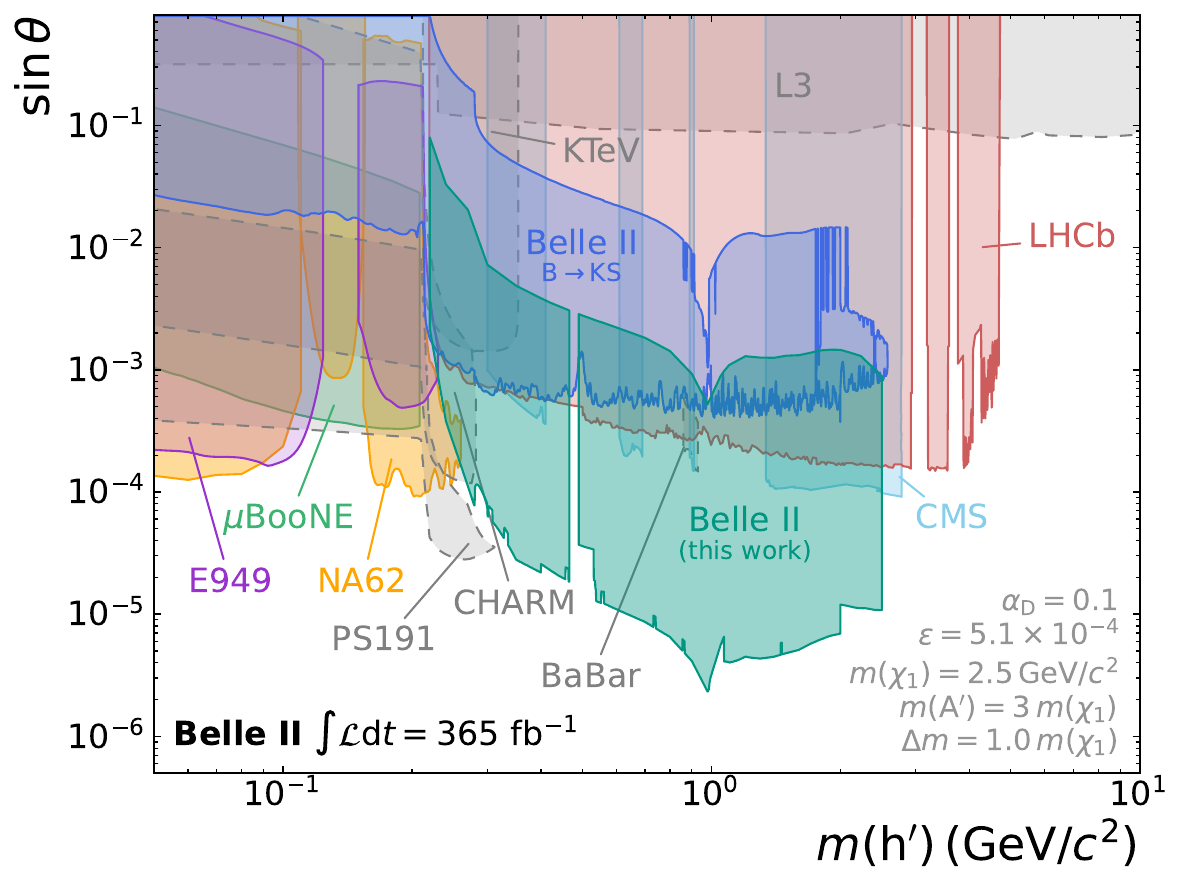}
    \includegraphics[width=0.45\textwidth]{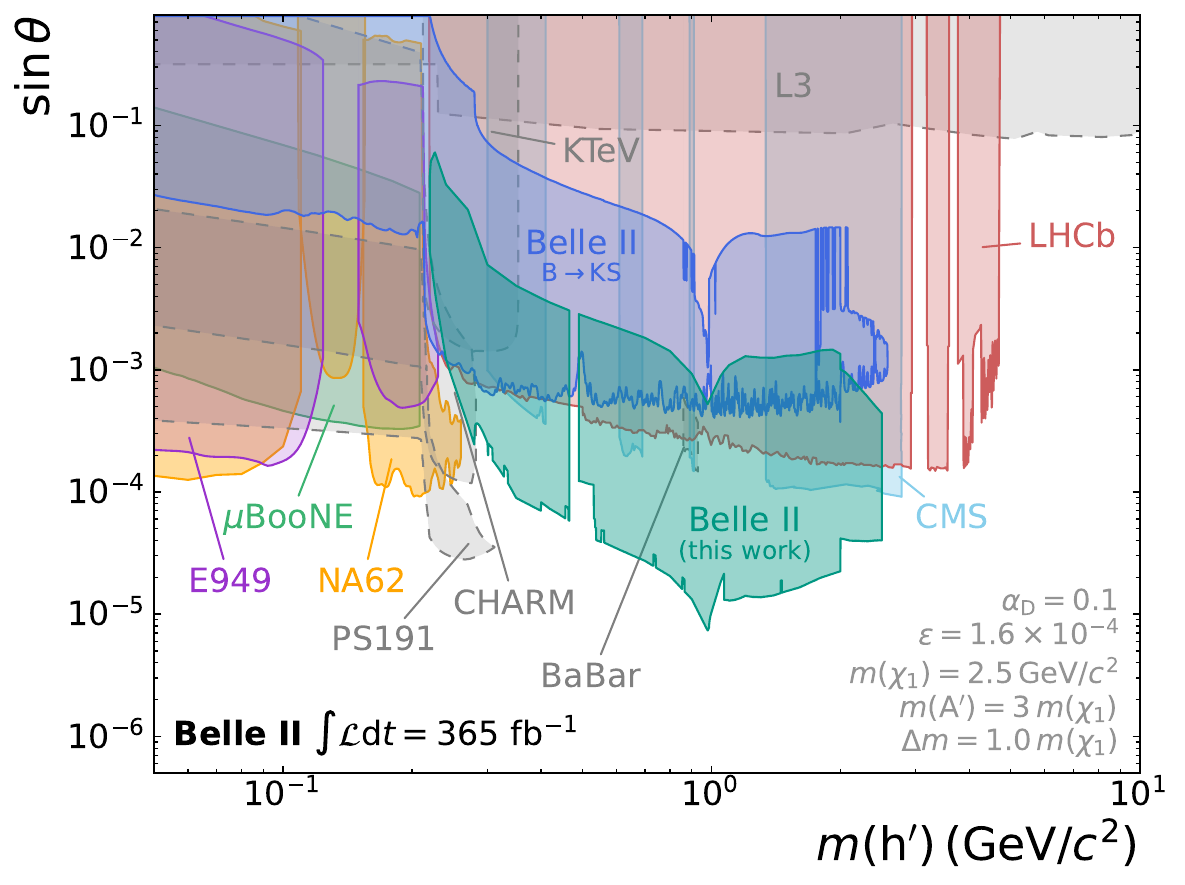}
    \caption{
Exclusion regions at 95\,\% credibility level in the plane of the sine of the mixing angle $\theta$ and dark Higgs mass $\mdh$ from this work (teal) together with existing constraints at 90\,\% credibility level from 
PS191\,\cite{Gorbunov:2021ccu},
E949\,\cite{BNL-E949:2009dza},
NA62\,\cite{NA62:2020pwi,NA62:2021zjw},
KOTO\,\cite{KOTO:2020prk,Ferber:2023iso}, 
KTeV\,\cite{KTEV:2000ngj},
and BABAR\,\cite{BaBar:2015jvu,Winkler:2018qyg},
and at 95\,\% credibility level from
MicroBooNE\,\cite{MicroBooNE:2021usw,MicroBooNE:2022ctm,Ferber:2023iso},
L3\,\cite{L3:1996ome,Ferber:2023iso},
CHARM\,\cite{CHARM:1985anb,Winkler:2018qyg},
LHCb\,\cite{LHCb:2015nkv,LHCb:2016awg,Winkler:2018qyg},
\belletwo\,\cite{Belle-II:2023ueh},
and CMS\,\cite{CMS:2023bay}  for $\alpha_D = 0.1$, $\mchione = 2.5\,\gevcc$, and $\map = 3\,\mchione$.
The mass splitting is set to $\Delta m = 0.2\,\mchione$ (top), $\Delta m = 0.4\,\mchione$ (center), and $\Delta m = 1.0\,\mchione$ (bottom).
Plots on the left assume a \chitwo{} lifetime of $c\tau(\chi_2) = 0.01\,\cm$ and on the right $c\tau(\chi_2) = 0.1\,\cm$.
This results in different mixing parameters $\epsilon$, which are reported in each plot.
All constraints but the one from this work do not depend on the presence of a dark photon or iDM.
}
    \label{fig:model_dependent1}
\end{figure*}

\begin{figure*}[htp!]
    \centering
    \includegraphics[width=0.45\textwidth]{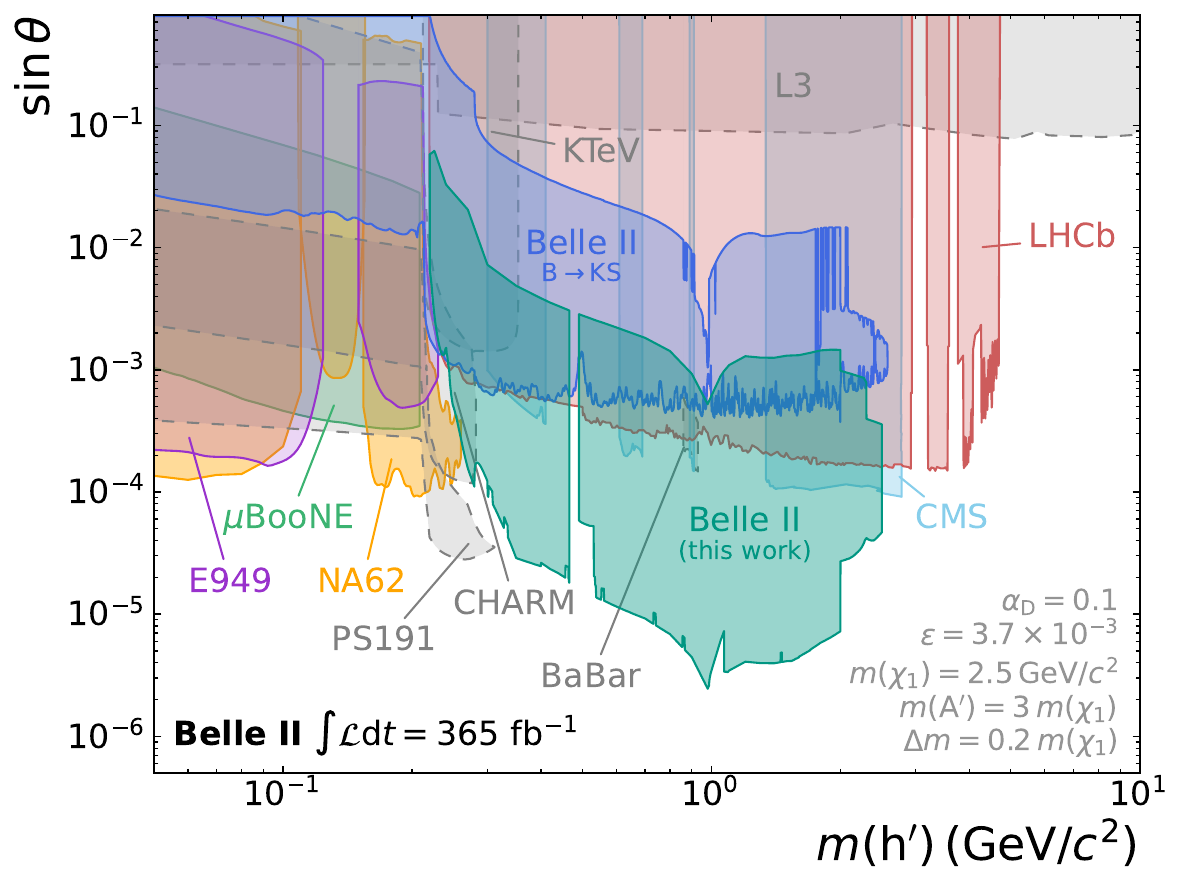}
    \includegraphics[width=0.45\textwidth]{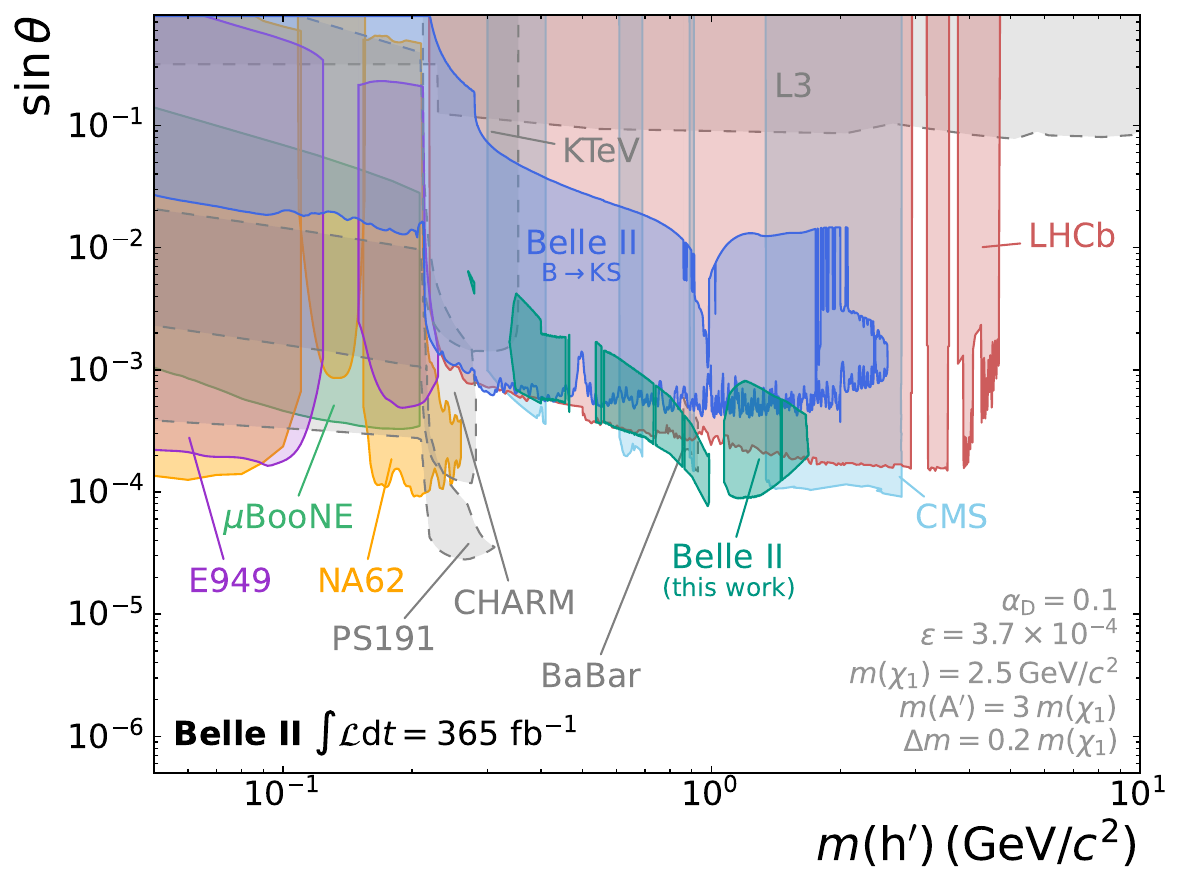}
    \includegraphics[width=0.45\textwidth]{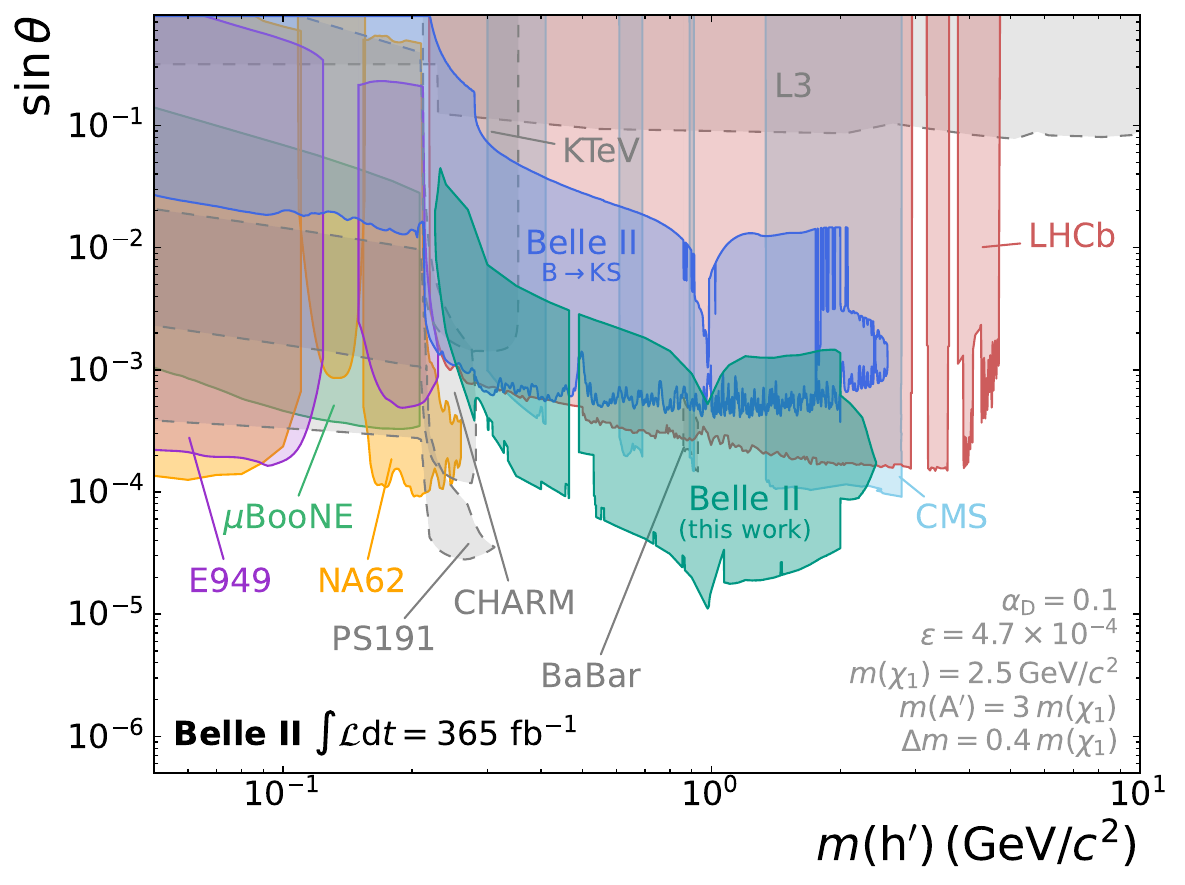}
    \includegraphics[width=0.45\textwidth]{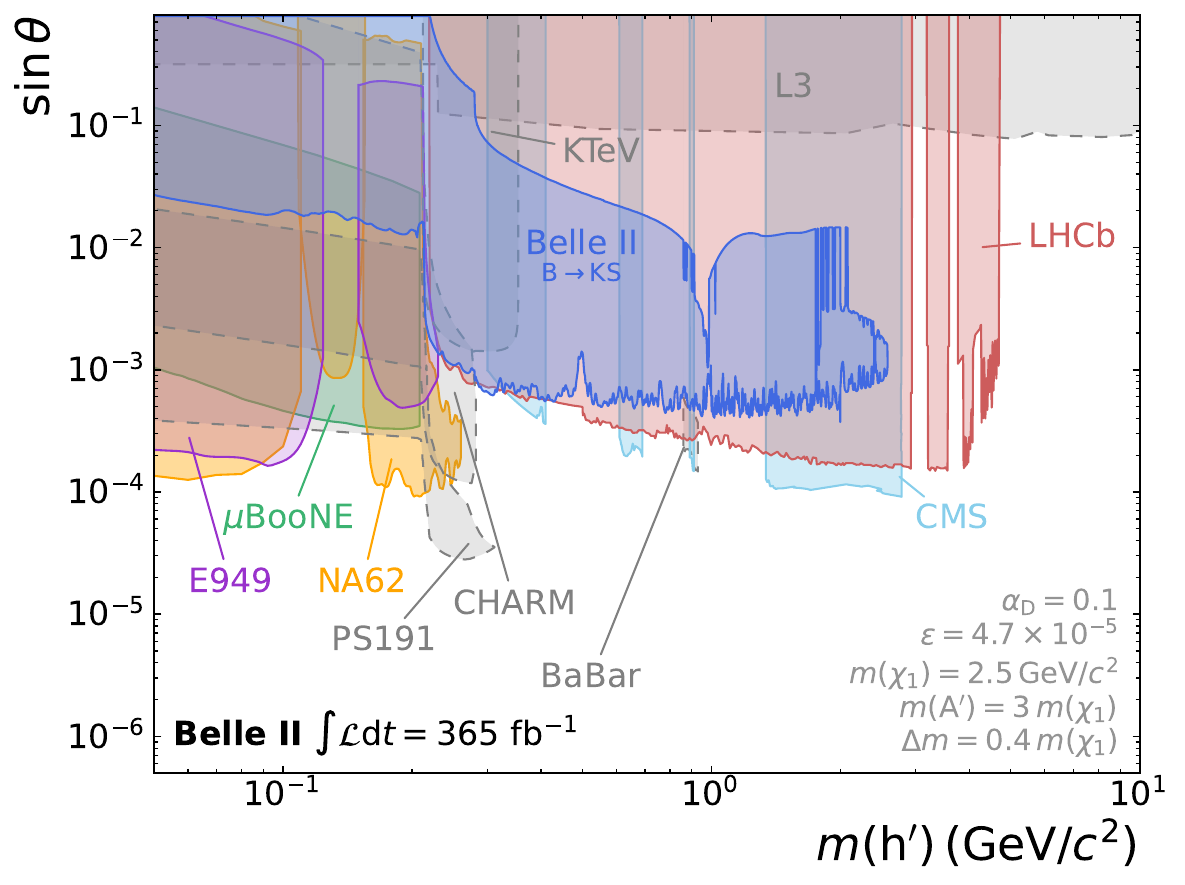}
    \includegraphics[width=0.45\textwidth]{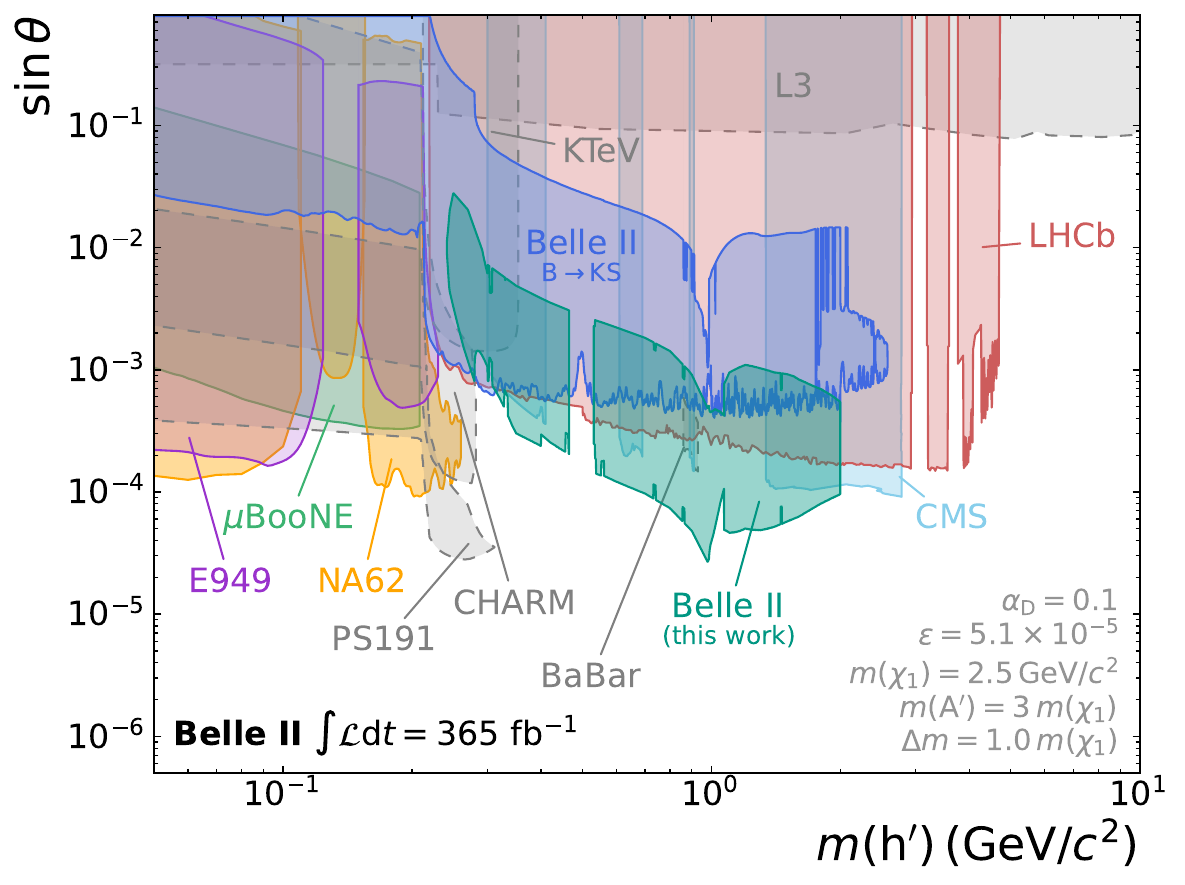}
    \includegraphics[width=0.45\textwidth]{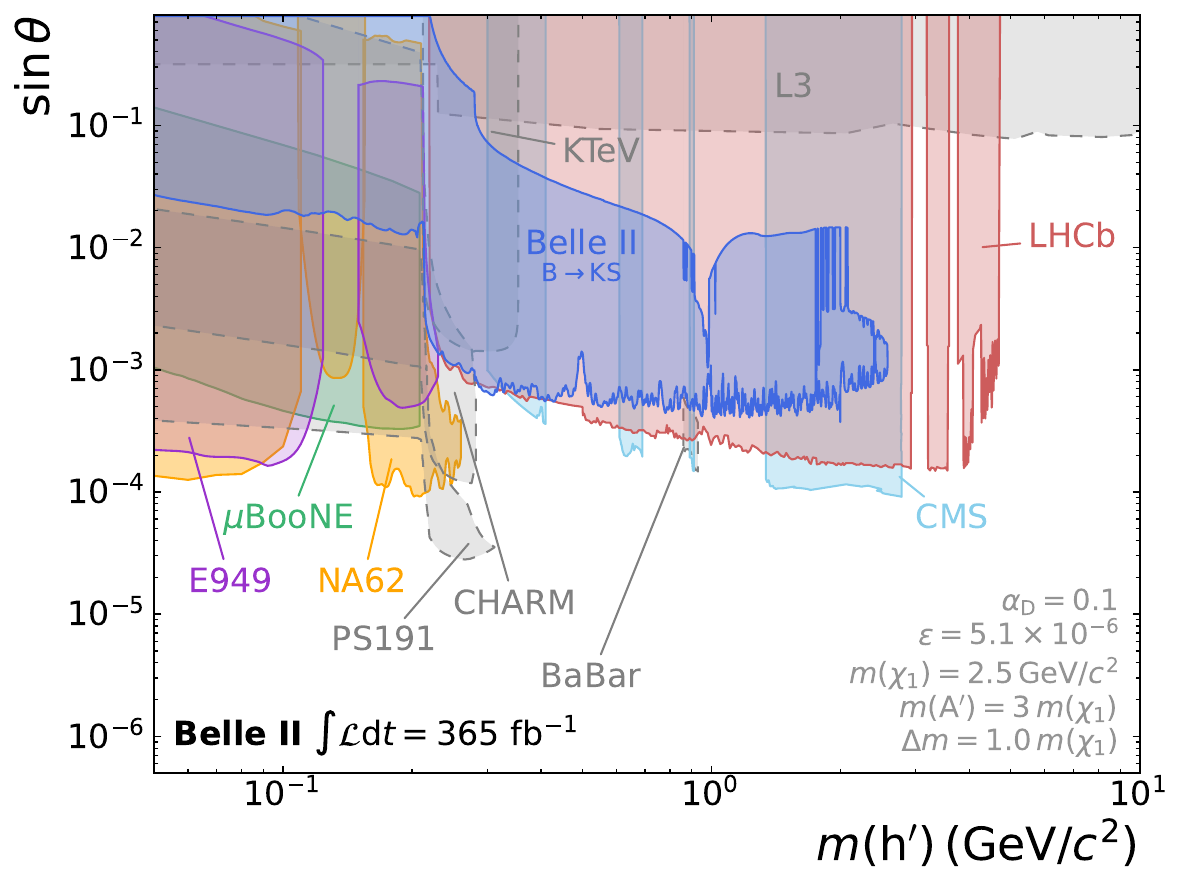}
    \caption{
Exclusion regions at 95\,\% credibility level in the plane of the sine of the mixing angle $\theta$ and dark Higgs mass $\mdh$ from this work (teal) together with existing constraints at 90\,\% credibility level from 
PS191\,\cite{Gorbunov:2021ccu},
E949\,\cite{BNL-E949:2009dza},
NA62\,\cite{NA62:2020pwi,NA62:2021zjw},
KOTO\,\cite{KOTO:2020prk,Ferber:2023iso}, 
KTeV\,\cite{KTEV:2000ngj},
and BABAR\,\cite{BaBar:2015jvu,Winkler:2018qyg},
and at 95\,\% credibility level from
MicroBooNE\,\cite{MicroBooNE:2021usw,MicroBooNE:2022ctm,Ferber:2023iso},
L3\,\cite{L3:1996ome,Ferber:2023iso},
CHARM\,\cite{CHARM:1985anb,Winkler:2018qyg},
LHCb\,\cite{LHCb:2015nkv,LHCb:2016awg,Winkler:2018qyg},
\belletwo\,\cite{Belle-II:2023ueh},
and CMS\,\cite{CMS:2023bay}  for $\alpha_D = 0.1$, $\mchione = 2.5\,\gevcc$, and $\map = 3\,\mchione$.
The mass splitting is set to $\Delta m = 0.2\,\mchione$ (top), $\Delta m = 0.4\,\mchione$ (center), and $\Delta m = 1.0\,\mchione$ (bottom).
Plots on the left assume a \chitwo{} lifetime of $c\tau(\chi_2) = 1.0\,\cm$ and on the right $c\tau(\chi_2) = 100.0\,\cm$.
This results in different mixing parameters $\epsilon$, which are reported in each plot.
For the model parameter configurations in the center right and lower right plot Belle~II is not sensitive.
All constraints but the one from this work do not depend on the presence of a dark photon or iDM.
}
    \label{fig:model_dependent2}
\end{figure*}

\begin{figure*}[htp!]
    \centering
    \includegraphics[width=0.45\textwidth]{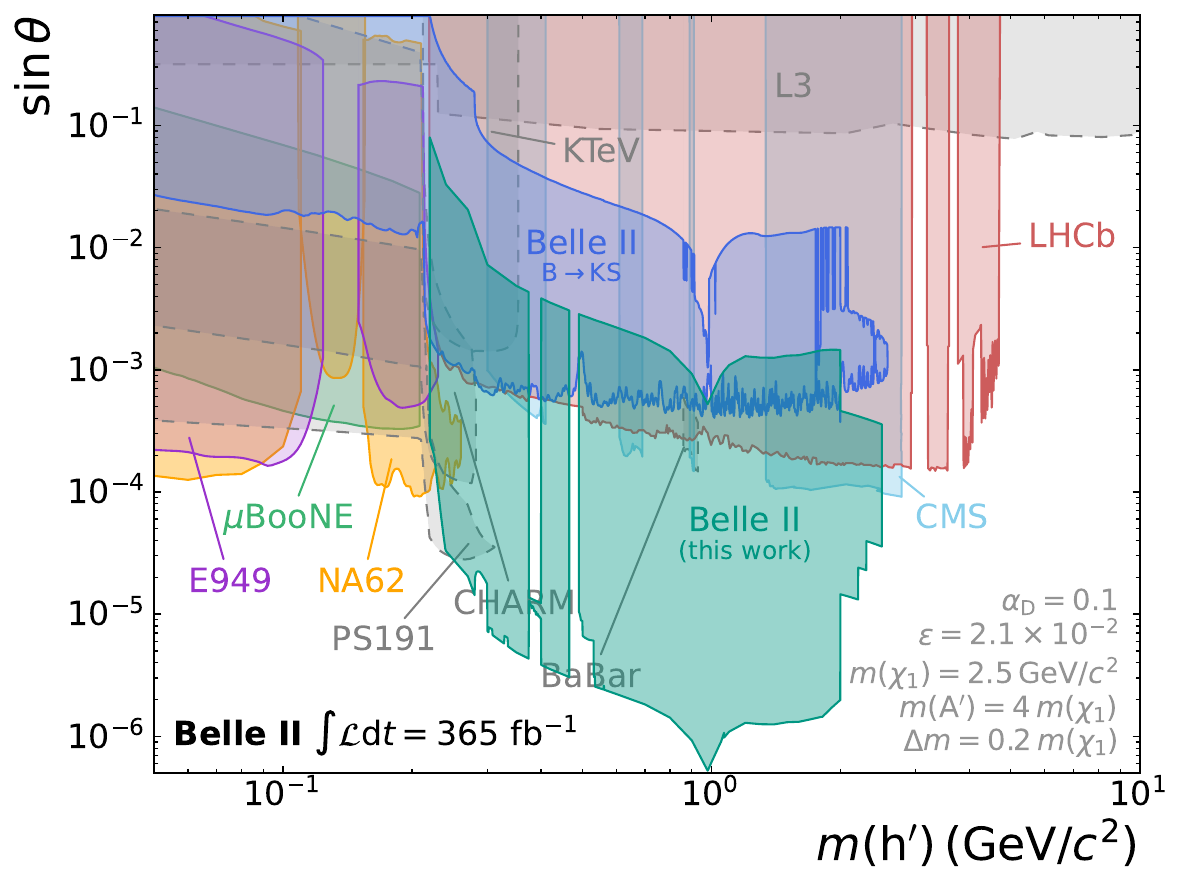}
    \includegraphics[width=0.45\textwidth]{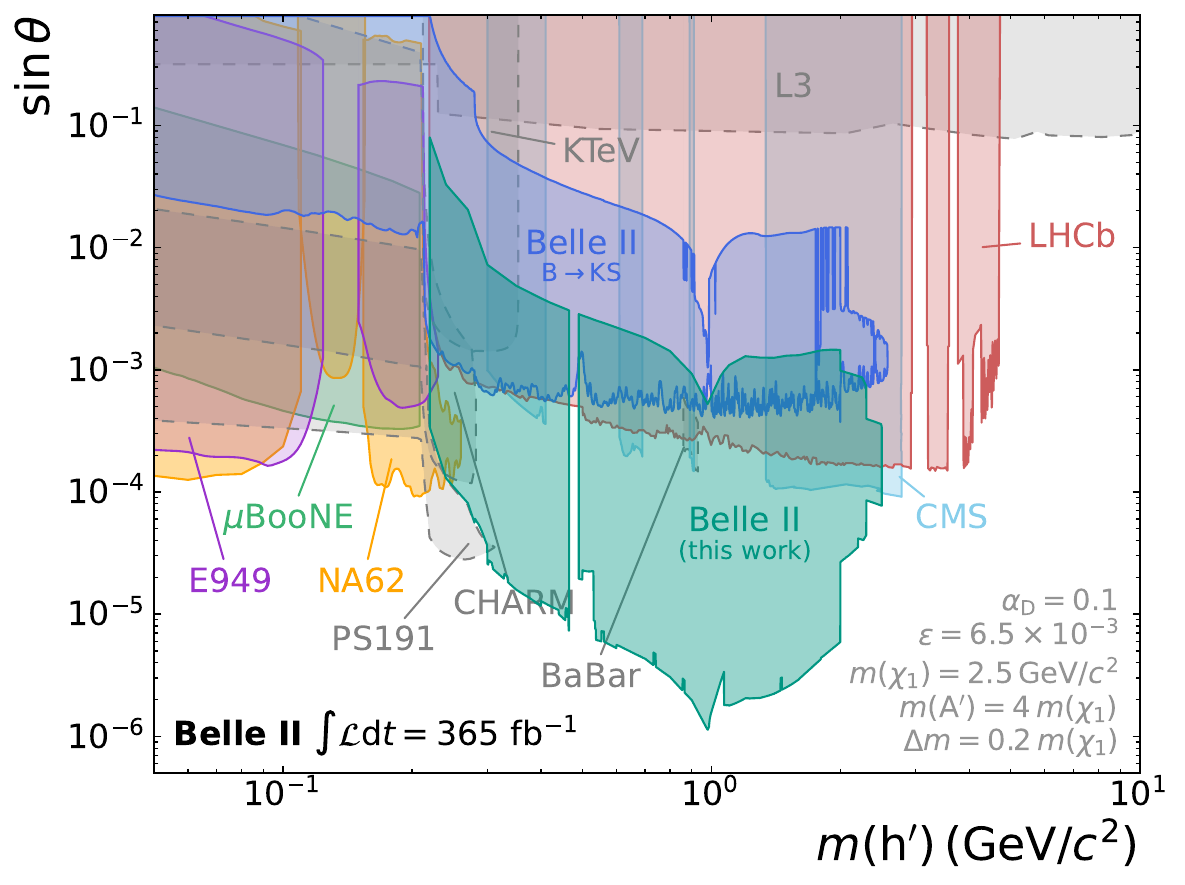}
    \includegraphics[width=0.45\textwidth]{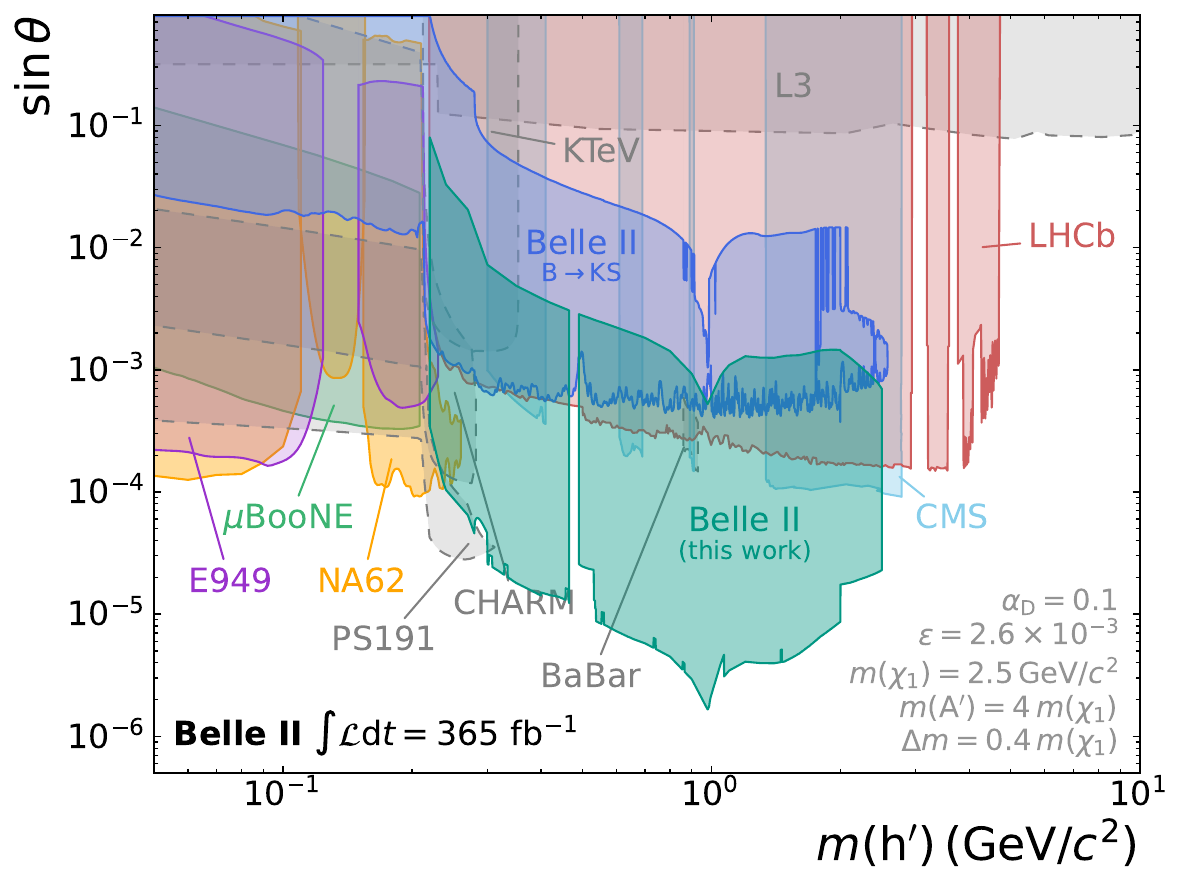}
    \includegraphics[width=0.45\textwidth]{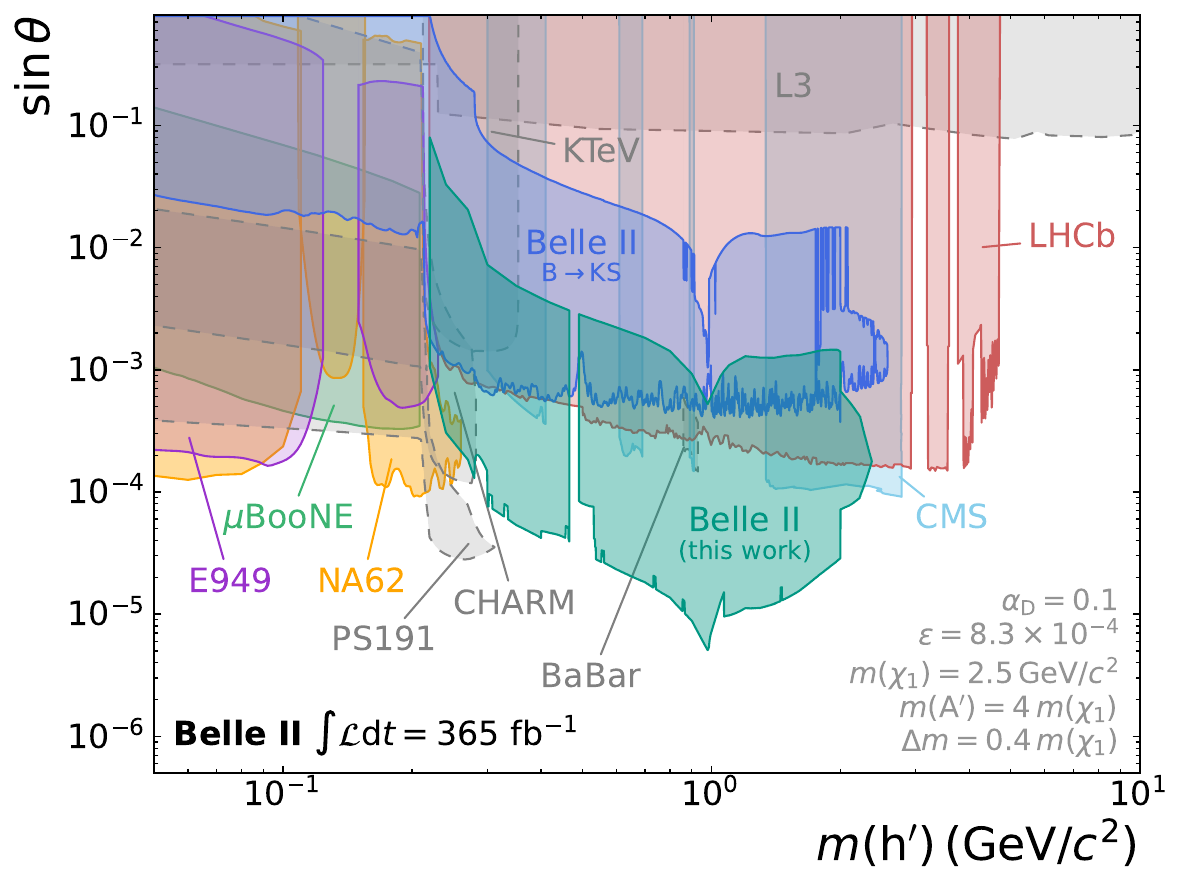}
    \includegraphics[width=0.45\textwidth]{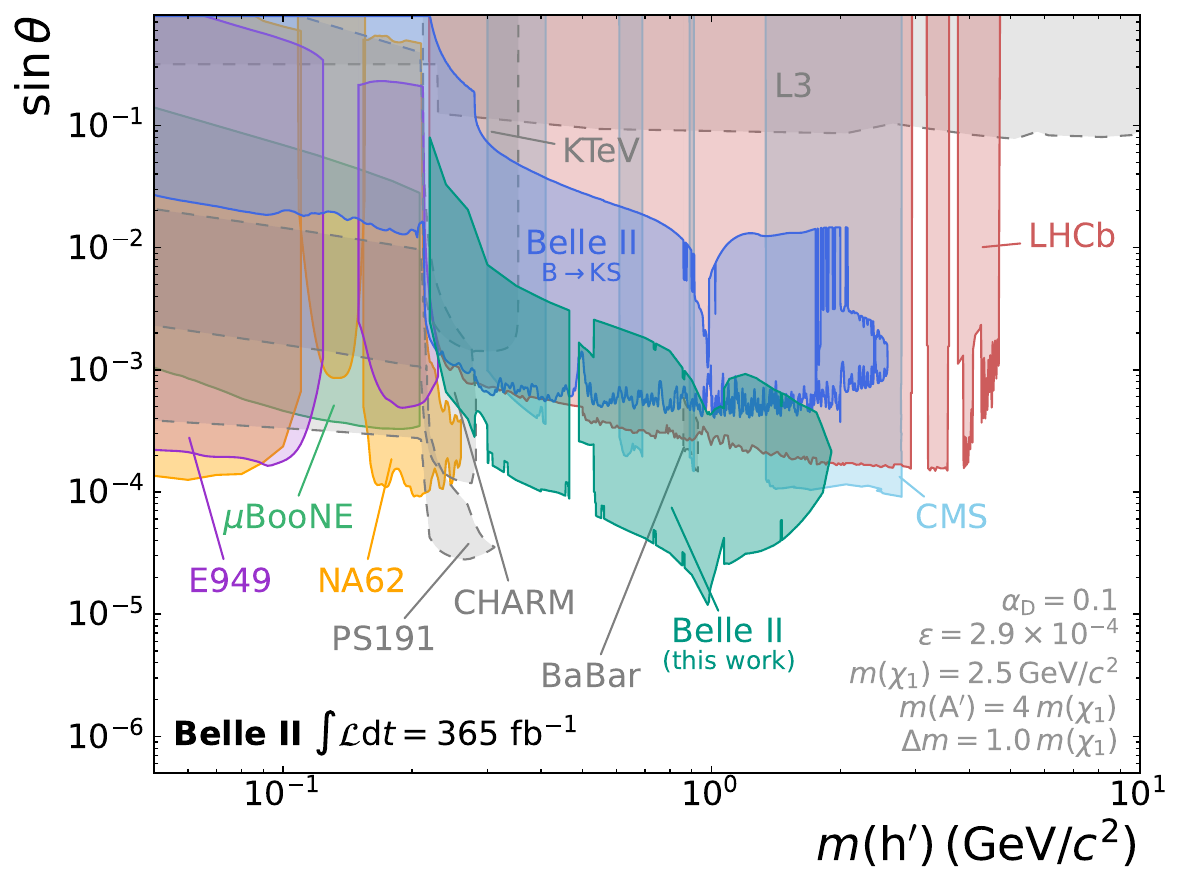}
    \includegraphics[width=0.45\textwidth]{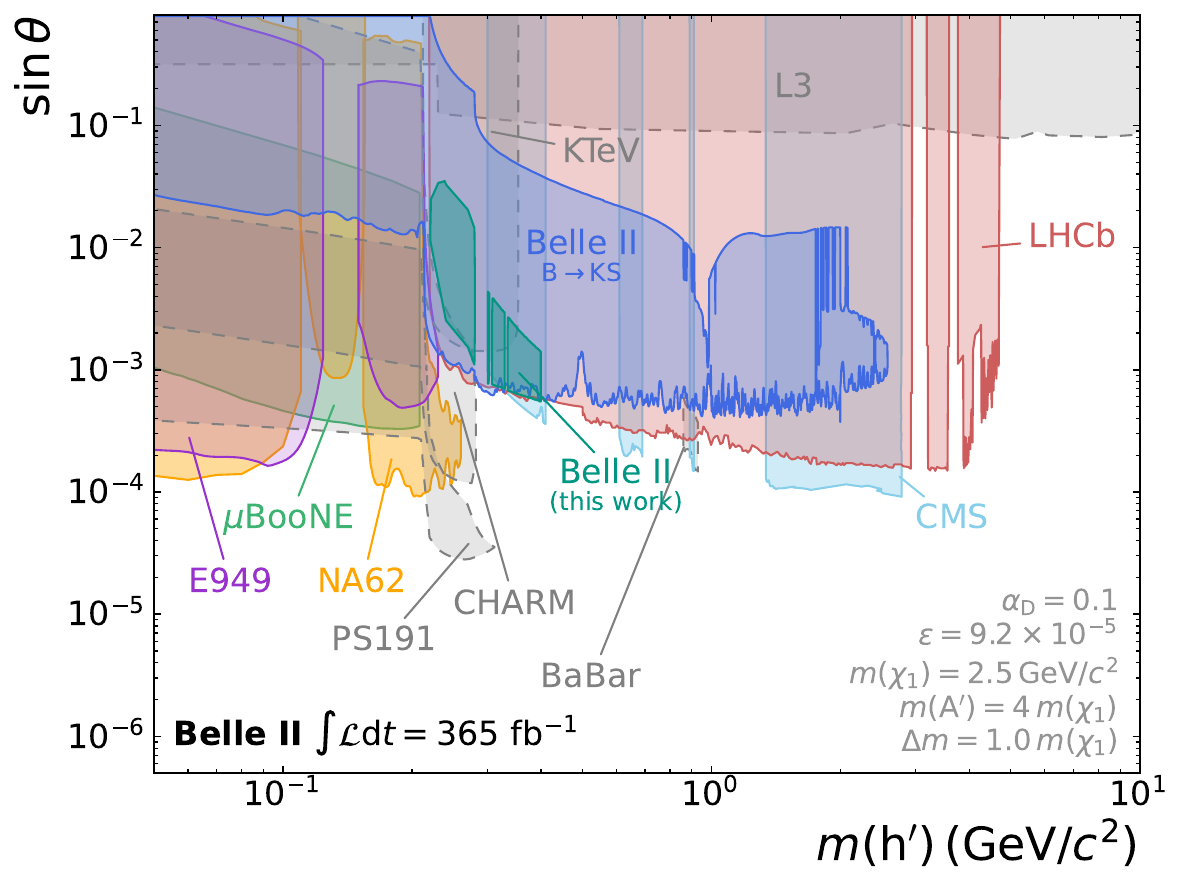}
    \caption{
Exclusion regions at 95\,\% credibility level in the plane of the sine of the mixing angle $\theta$ and dark Higgs mass $\mdh$ from this work (teal) together with existing constraints at 90\,\% credibility level from 
PS191\,\cite{Gorbunov:2021ccu},
E949\,\cite{BNL-E949:2009dza},
NA62\,\cite{NA62:2020pwi,NA62:2021zjw},
KOTO\,\cite{KOTO:2020prk,Ferber:2023iso}, 
KTeV\,\cite{KTEV:2000ngj},
and BABAR\,\cite{BaBar:2015jvu,Winkler:2018qyg},
and at 95\,\% credibility level from
MicroBooNE\,\cite{MicroBooNE:2021usw,MicroBooNE:2022ctm,Ferber:2023iso},
L3\,\cite{L3:1996ome,Ferber:2023iso},
CHARM\,\cite{CHARM:1985anb,Winkler:2018qyg},
LHCb\,\cite{LHCb:2015nkv,LHCb:2016awg,Winkler:2018qyg},
\belletwo\,\cite{Belle-II:2023ueh},
and CMS\,\cite{CMS:2023bay})  for $\alpha_D = 0.1$, $\mchione = 2.5\,\gevcc$, and $\map = 4\,\mchione$.
The mass splitting is set to $\Delta m = 0.2\,\mchione$ (top), $\Delta m = 0.4\,\mchione$ (center), and $\Delta m = 1.0\,\mchione$ (bottom).
Plots on the left assume a \chitwo{} lifetime of $c\tau(\chi_2) = 0.1\,\cm$ and on the right $c\tau(\chi_2) = 1.0\,\cm$.
This results in different mixing parameters $\epsilon$, which are reported in each plot.
All constraints but the one from this work do not depend on the presence of a dark photon or iDM.
}
    \label{fig:model_dependent3}
\end{figure*}

\begin{figure*}[htp!]
    \centering
    \includegraphics[width=0.45\textwidth]{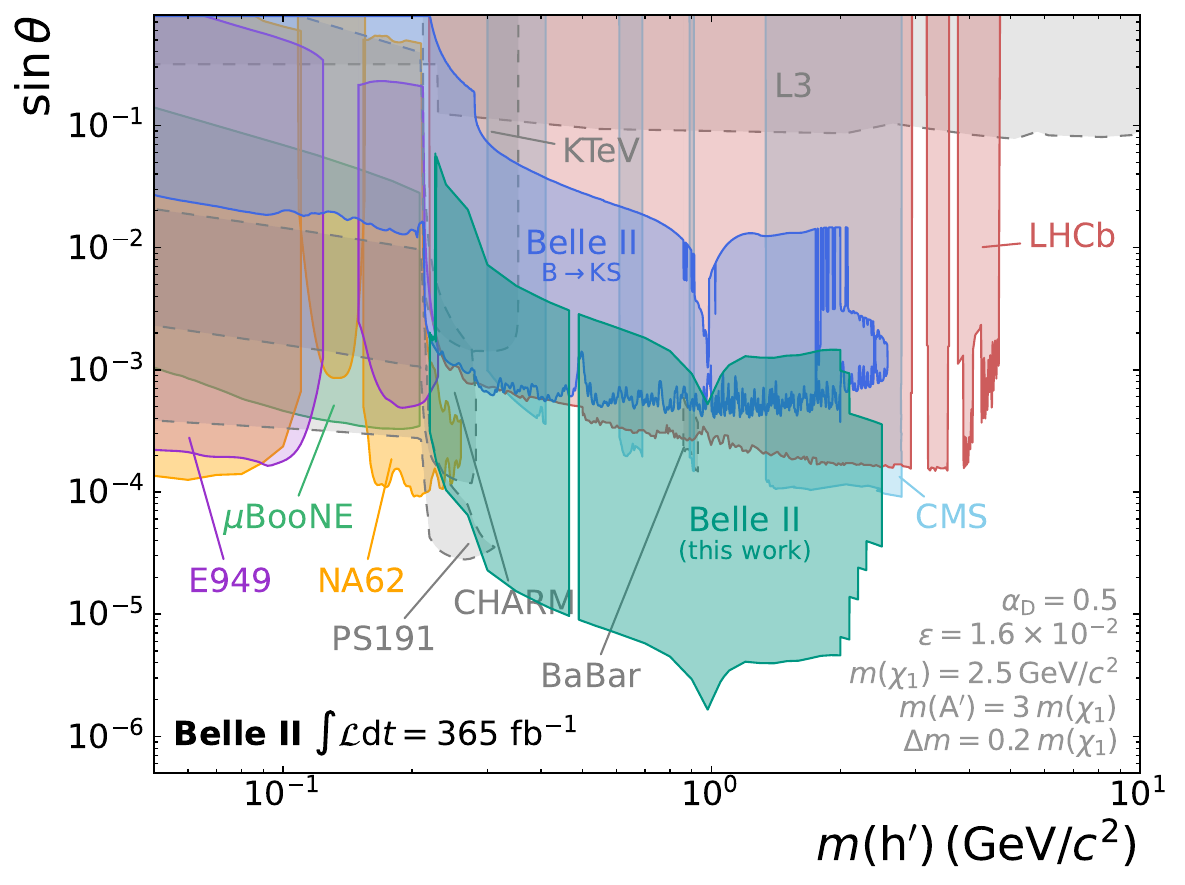}
    \includegraphics[width=0.45\textwidth]{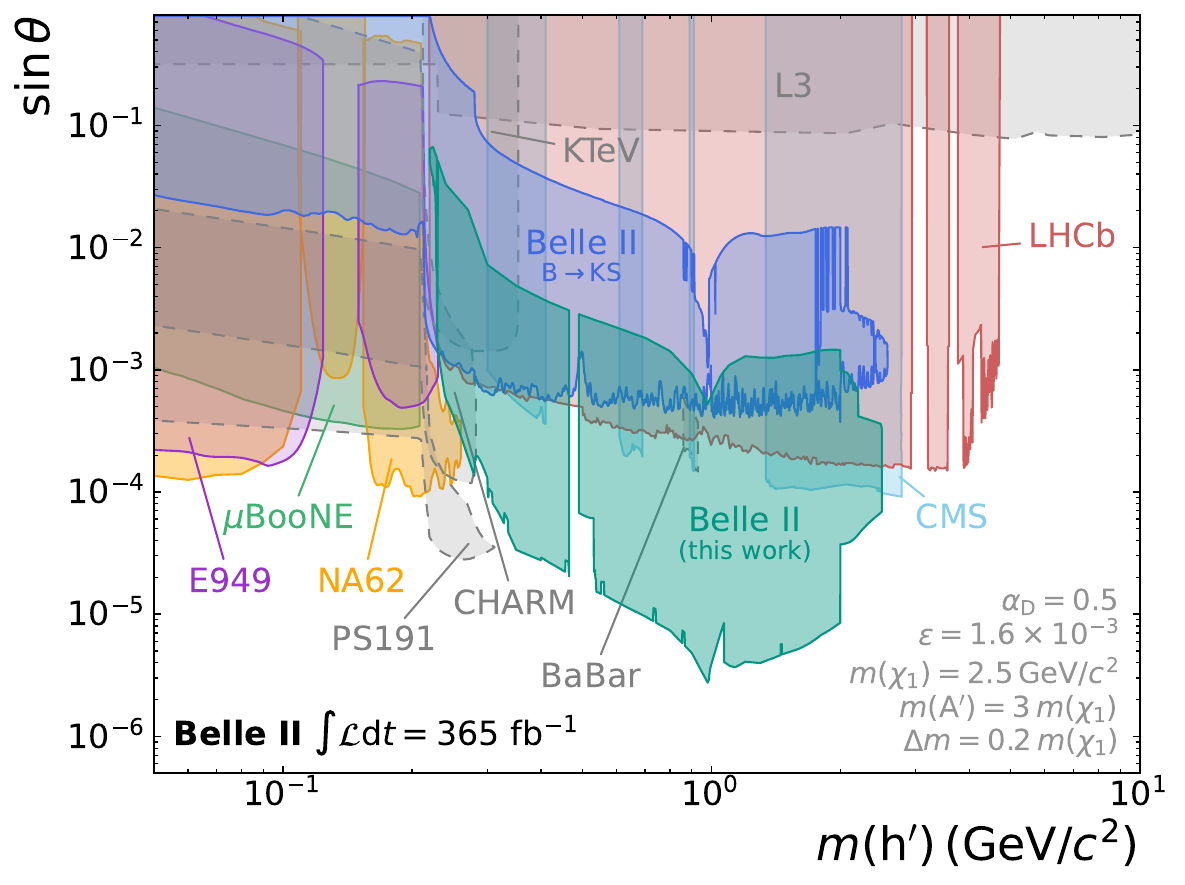}
    \includegraphics[width=0.45\textwidth]{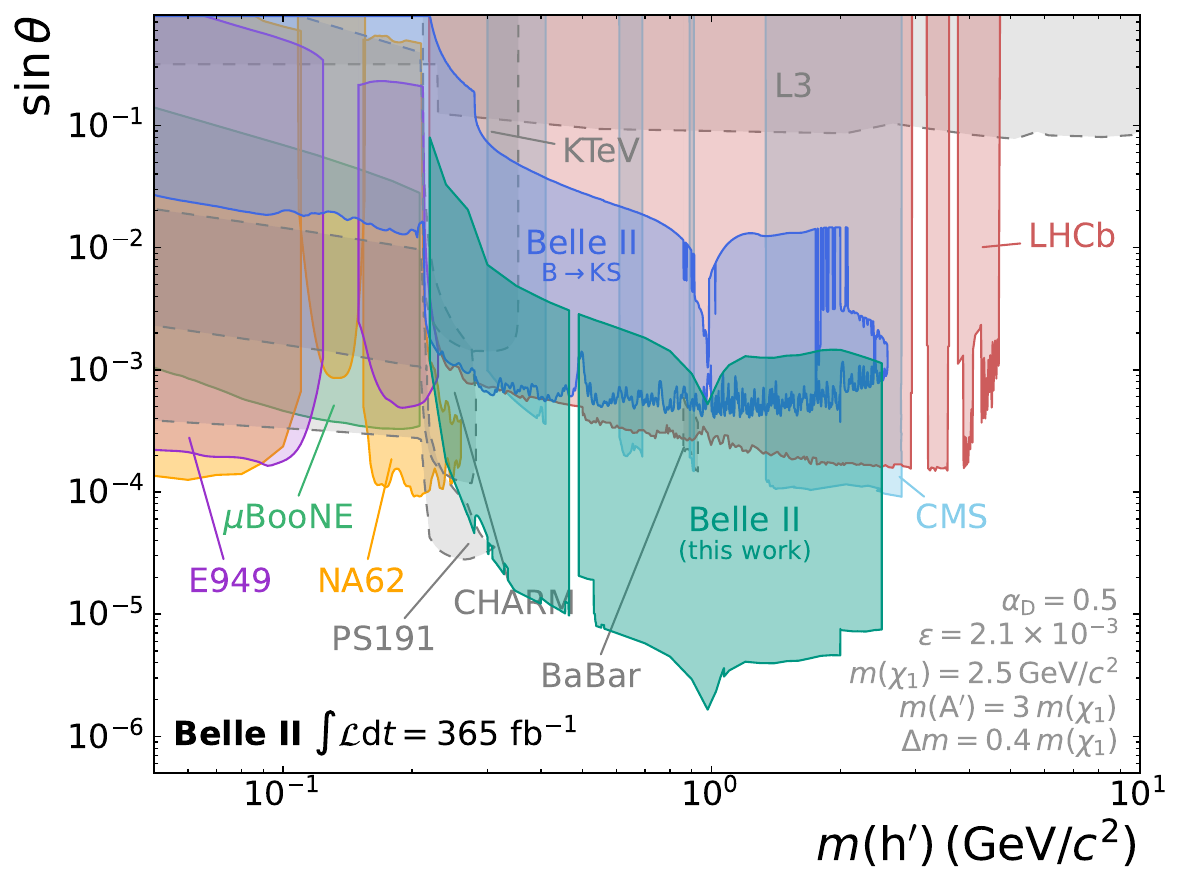}
    \includegraphics[width=0.45\textwidth]{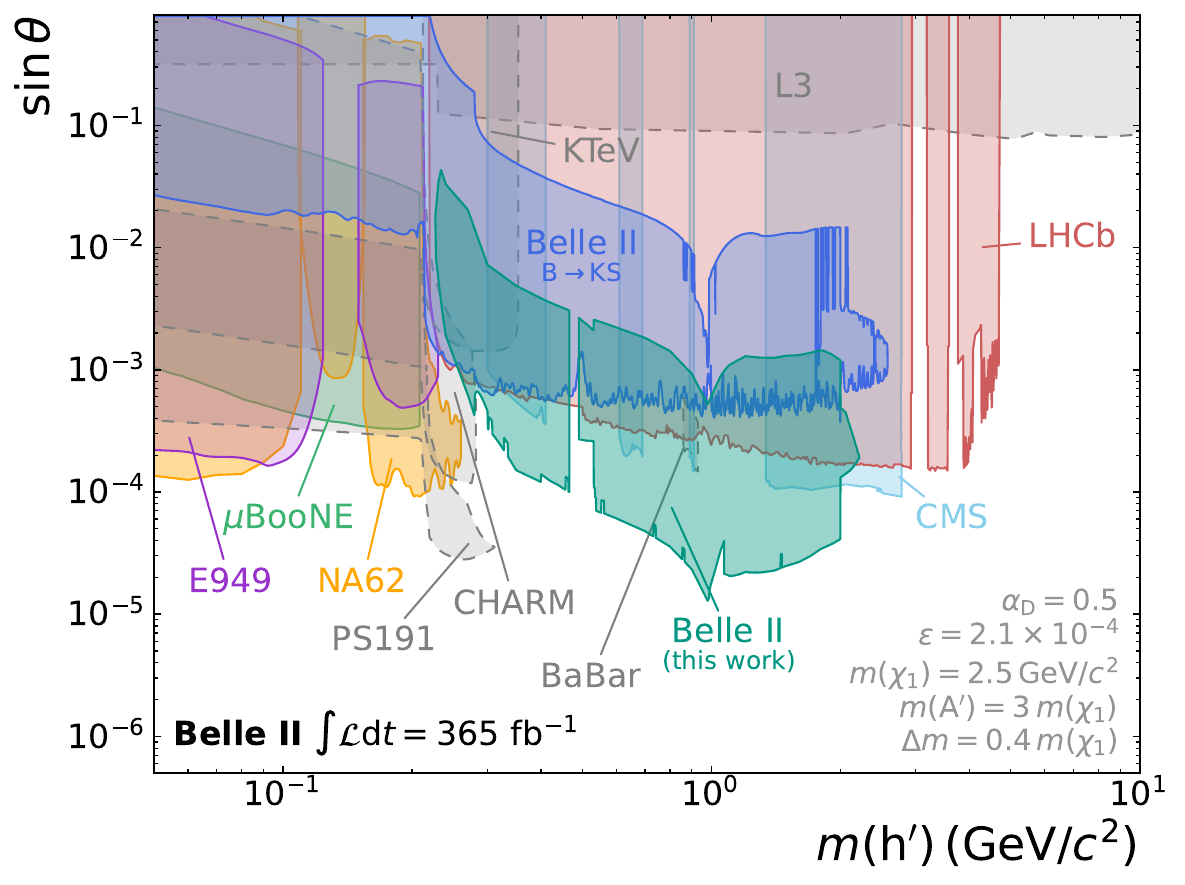}
    \includegraphics[width=0.45\textwidth]{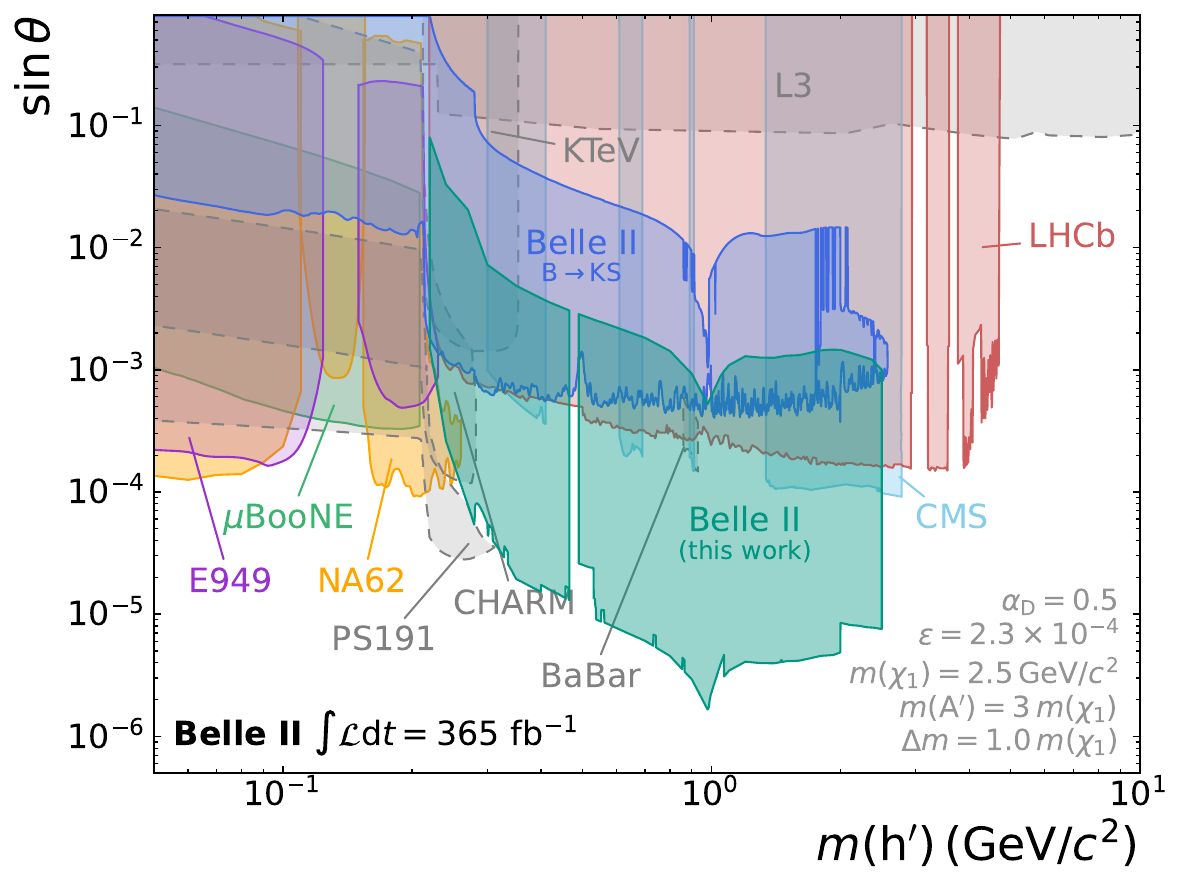}
    \includegraphics[width=0.45\textwidth]{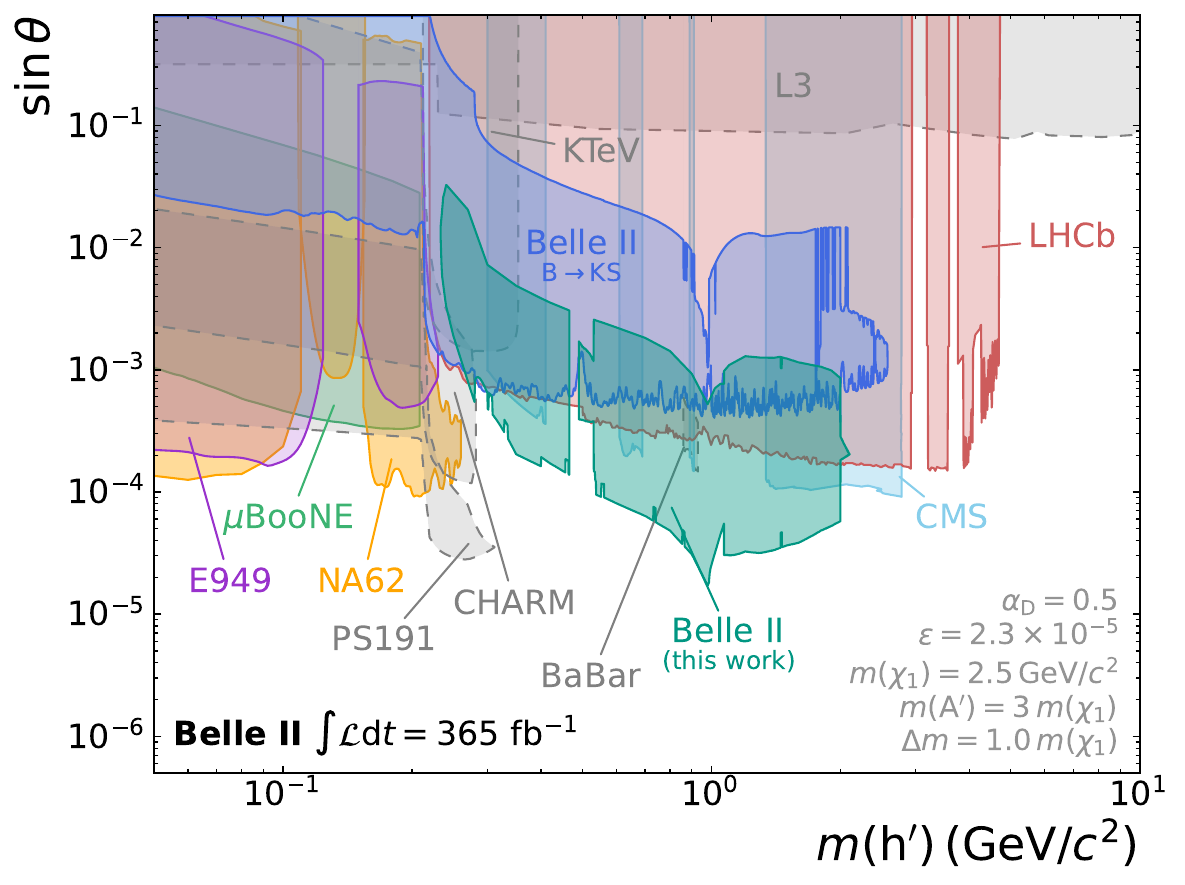}
    \caption{
Exclusion regions at 95\,\% credibility level in the plane of the sine of the mixing angle $\theta$ and dark Higgs mass $\mdh$ from this work (teal) together with existing constraints at 90\,\% credibility level from 
PS191\,\cite{Gorbunov:2021ccu},
E949\,\cite{BNL-E949:2009dza},
NA62\,\cite{NA62:2020pwi,NA62:2021zjw},
KOTO\,\cite{KOTO:2020prk,Ferber:2023iso}, 
KTeV\,\cite{KTEV:2000ngj},
and BABAR\,\cite{BaBar:2015jvu,Winkler:2018qyg},
and at 95\,\% credibility level from
MicroBooNE\,\cite{MicroBooNE:2021usw,MicroBooNE:2022ctm,Ferber:2023iso},
L3\,\cite{L3:1996ome,Ferber:2023iso},
CHARM\,\cite{CHARM:1985anb,Winkler:2018qyg},
LHCb\,\cite{LHCb:2015nkv,LHCb:2016awg,Winkler:2018qyg},
\belletwo\,\cite{Belle-II:2023ueh},
and CMS\,\cite{CMS:2023bay}  for $\alpha_D = 0.5$, $\mchione = 2.5\,\gevcc$, and $\map = 3\,\mchione$.
The mass splitting is set to $\Delta m = 0.2\,\mchione$ (top), $\Delta m = 0.4\,\mchione$ (center), and $\Delta m = 1.0\,\mchione$ (bottom).
Plots on the left assume a \chitwo{} lifetime of $c\tau(\chi_2) = 0.01\,\cm$ and on the right $c\tau(\chi_2) = 1.0\,\cm$.
This results in different mixing parameters $\epsilon$, which are reported in each plot.
All constraints but the one from this work do not depend on the presence of a dark photon or iDM.
}
    \label{fig:model_dependent4}
\end{figure*}

\begin{figure*}[htp!]
    \centering
    \includegraphics[width=0.45\textwidth]{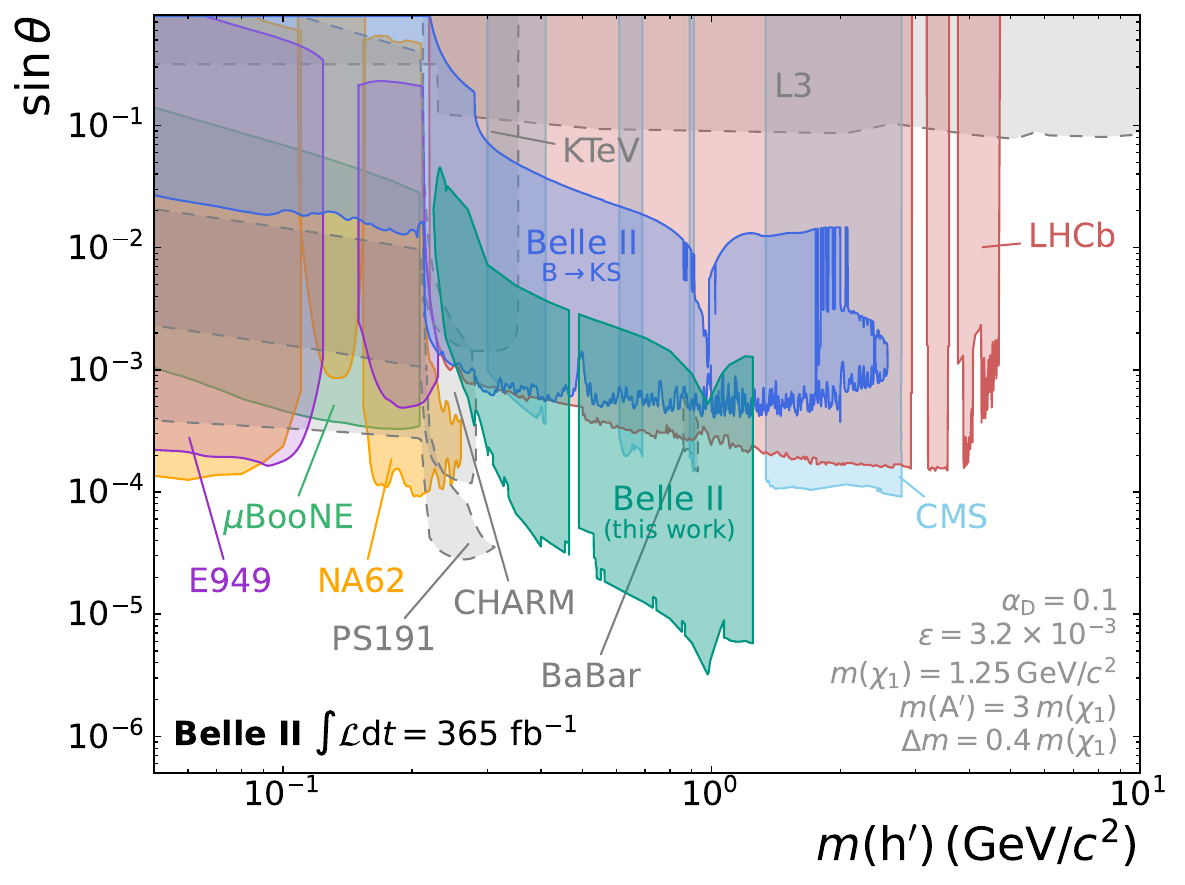}
    \includegraphics[width=0.45\textwidth]{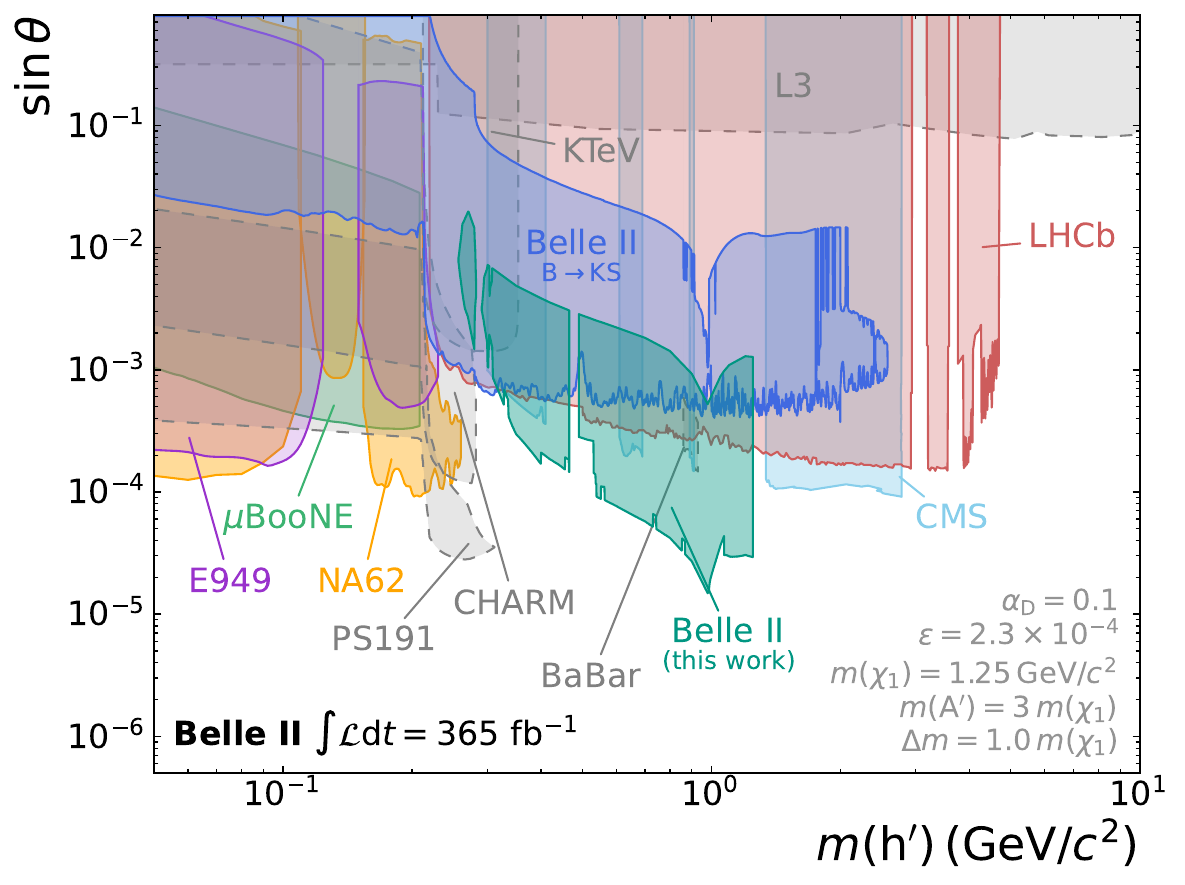}
    \caption{
Exclusion regions at 95\,\% credibility level in the plane of the sine of the mixing angle $\theta$ and dark Higgs mass $\mdh$ from this work (teal) together with existing constraints at 90\,\% credibility level from 
PS191\,\cite{Gorbunov:2021ccu},
E949\,\cite{BNL-E949:2009dza},
NA62\,\cite{NA62:2020pwi,NA62:2021zjw},
KOTO\,\cite{KOTO:2020prk,Ferber:2023iso}, 
KTeV\,\cite{KTEV:2000ngj},
and BABAR\,\cite{BaBar:2015jvu,Winkler:2018qyg},
and at 95\,\% credibility level from
MicroBooNE\,\cite{MicroBooNE:2021usw,MicroBooNE:2022ctm,Ferber:2023iso},
L3\,\cite{L3:1996ome,Ferber:2023iso},
CHARM\,\cite{CHARM:1985anb,Winkler:2018qyg},
LHCb\,\cite{LHCb:2015nkv,LHCb:2016awg,Winkler:2018qyg},
\belletwo\,\cite{Belle-II:2023ueh},
and CMS\,\cite{CMS:2023bay}  for $\alpha_D = 0.1$, $\mchione = 1.25\,\gevcc$, and $\map = 3\,\mchione$.
The mass splitting is set to $\Delta m = 0.4\,\mchione$ (left) and $\Delta m = 1.0\,\mchione$ (right).
The lifetime of the \chitwo{} is set to$c\tau(\chi_2) = 0.1\,\cm$, which results in mixing parameters of $\epsilon=3.2\times10^{-3}$ (left) and $\epsilon=2.3\times10^{-4}$ (right).
All constraints but the one from this work do not depend on the presence of a dark photon or iDM.
}
    \label{fig:model_dependent5}
\end{figure*}

\begin{figure*}[htp!]
    \centering
    \includegraphics[width=0.45\textwidth]{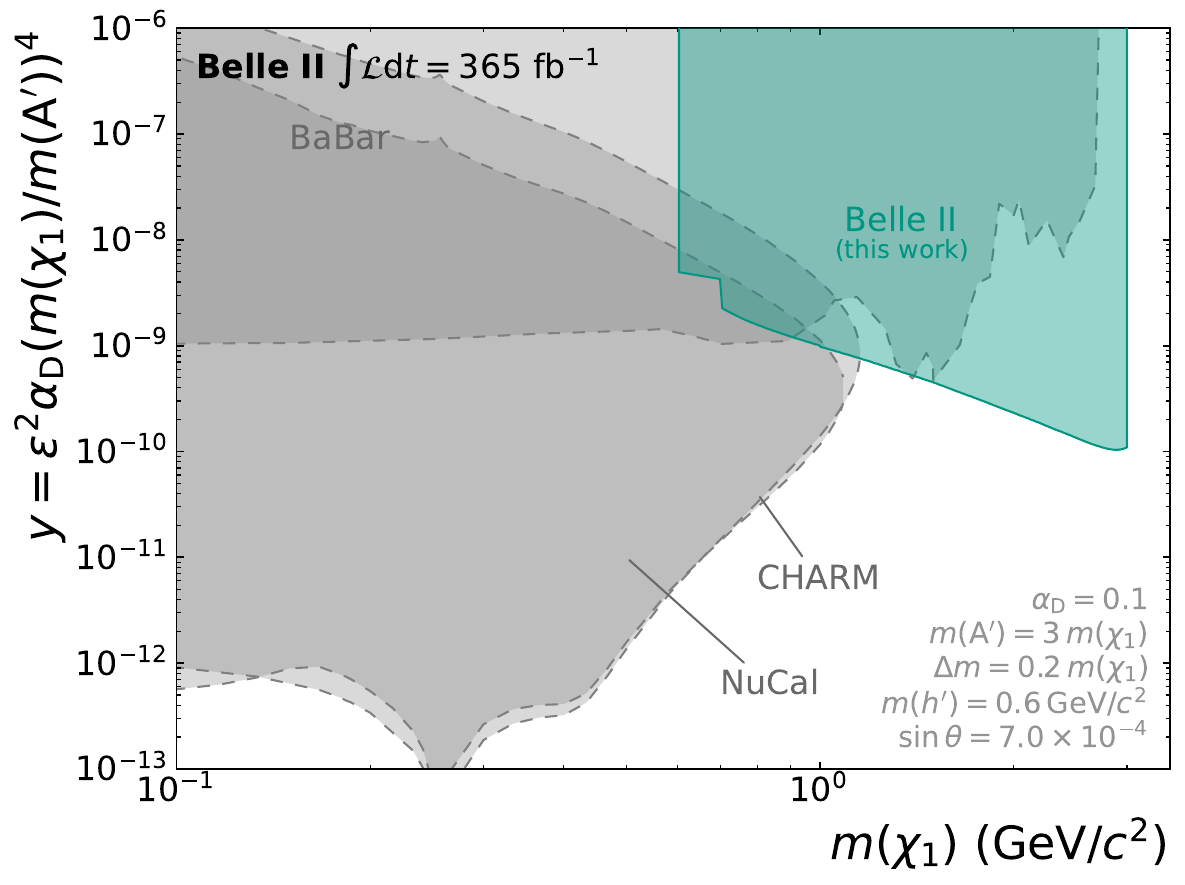}
    \includegraphics[width=0.45\textwidth]{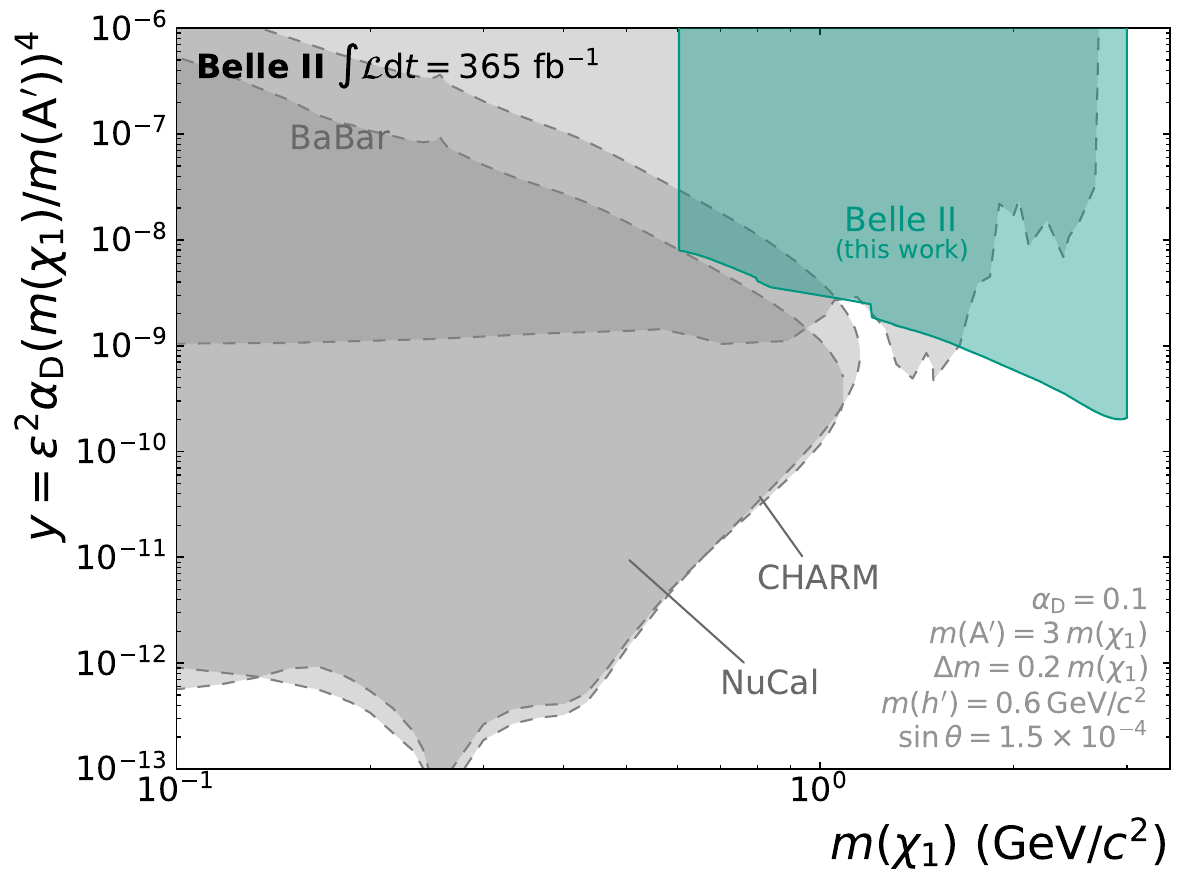}
    \includegraphics[width=0.45\textwidth]{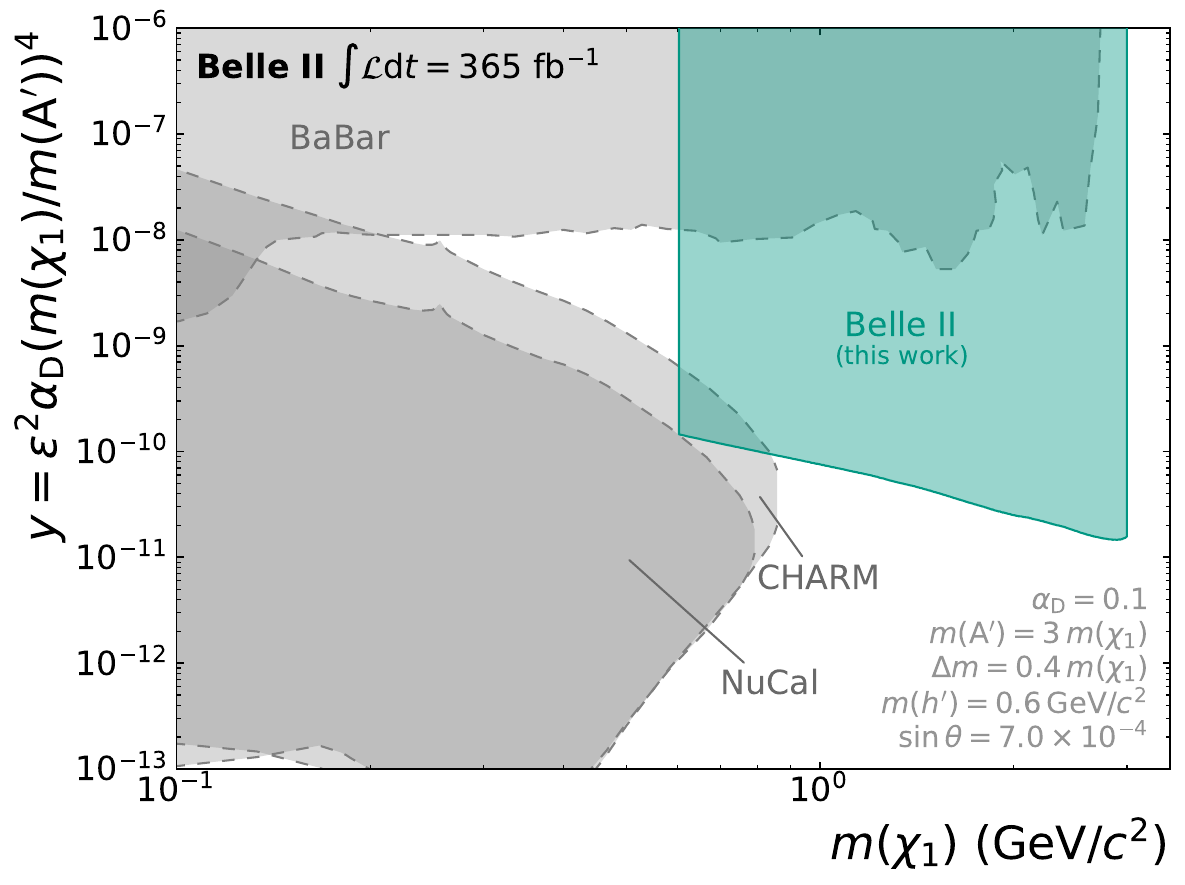}
    \includegraphics[width=0.45\textwidth]{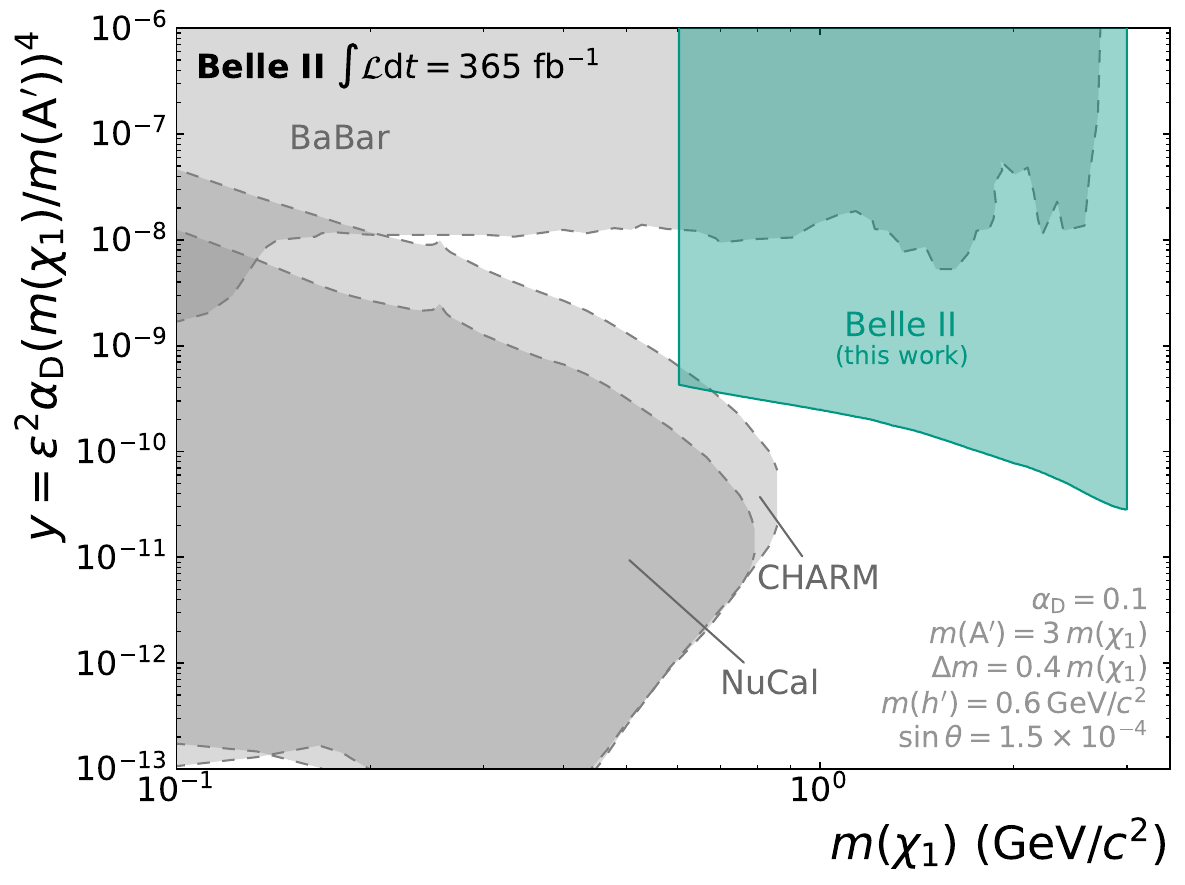}
    \caption{
Exclusion regions at 95\,\% credibility level in the plane of the dimensionless variable $y = \epsilon^2\alpha_D(m(\chi_1)/m(A'))^4$ and DM mass $\mchione$ from this work (teal) together with existing constraints at 90\,\% credibility level from
CHARM\,\cite{Gninenko:2012eq,Tsai:2019buq},
NuCal\,\cite{Blumlein:2011mv,Blumlein:2013cua,Tsai:2019buq}, and
BABAR\,\cite{BaBar:2017tiz,Duerr:2019dmv} for $\alpha_D = 0.1$, $\map = 3\,\mchione$, and $\mdh = 0.6\,\gevcc$.
The mass splitting is set to $\Delta m=0.2\,\mchione$ (top) and $\Delta m=0.4\,\mchione$ (bottom). 
The mixing angle of the dark Higgs is set to $\sin \theta = 7.0\times 10^{-4}$ ($c\tau(\dh) = 1.0\,\cm$) (left) and $\sin \theta = 1.5\times 10^{-4}$ ($c\tau(\dh) = 21.54\,\cm$) (right).
Constraints colored in gray with dashed outline are reinterpretations not performed by the experimental collaborations.
All constraints but the one from this work do not depend on the presence of a dark Higgs boson or iDM.
}
    \label{fig:model_dependent6}
\end{figure*}

\begin{figure*}[htp!]
    \centering
    \includegraphics[width=0.45\textwidth]{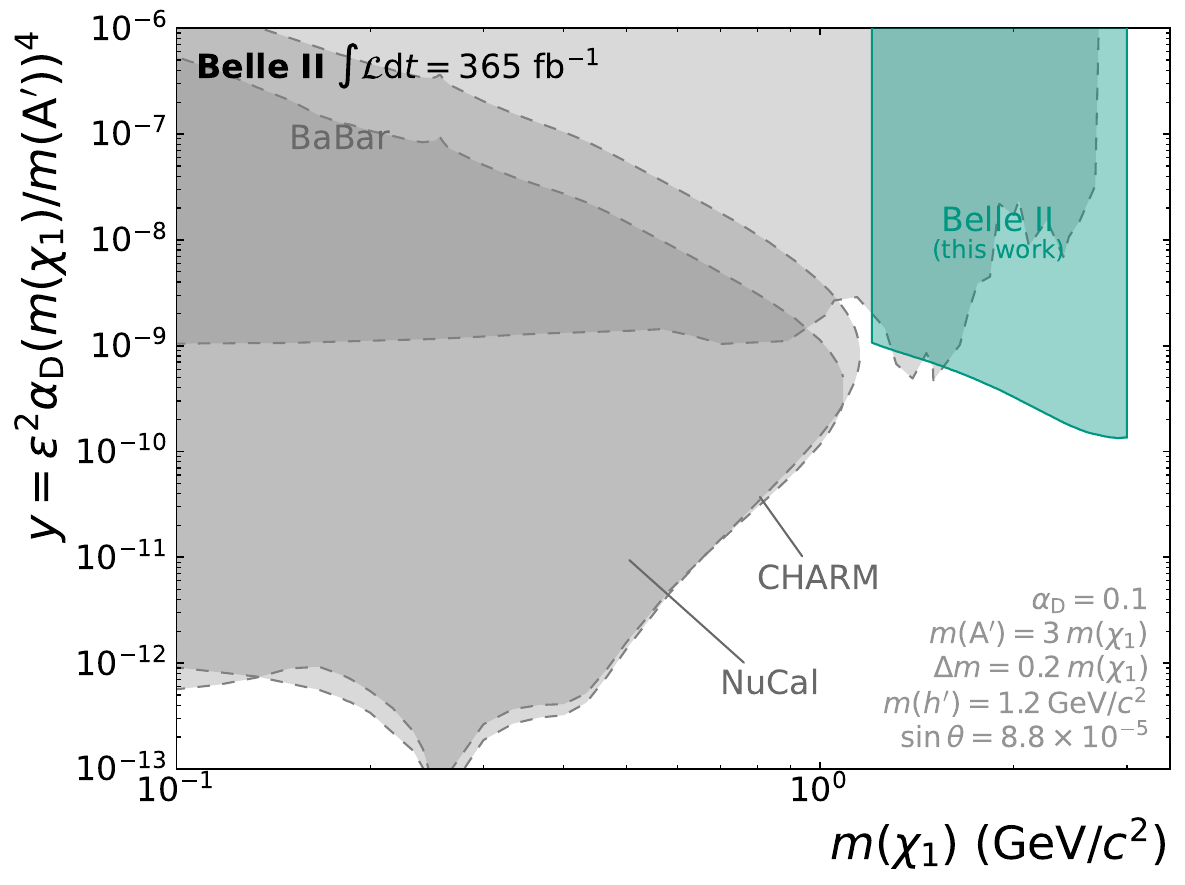}
    \includegraphics[width=0.45\textwidth]{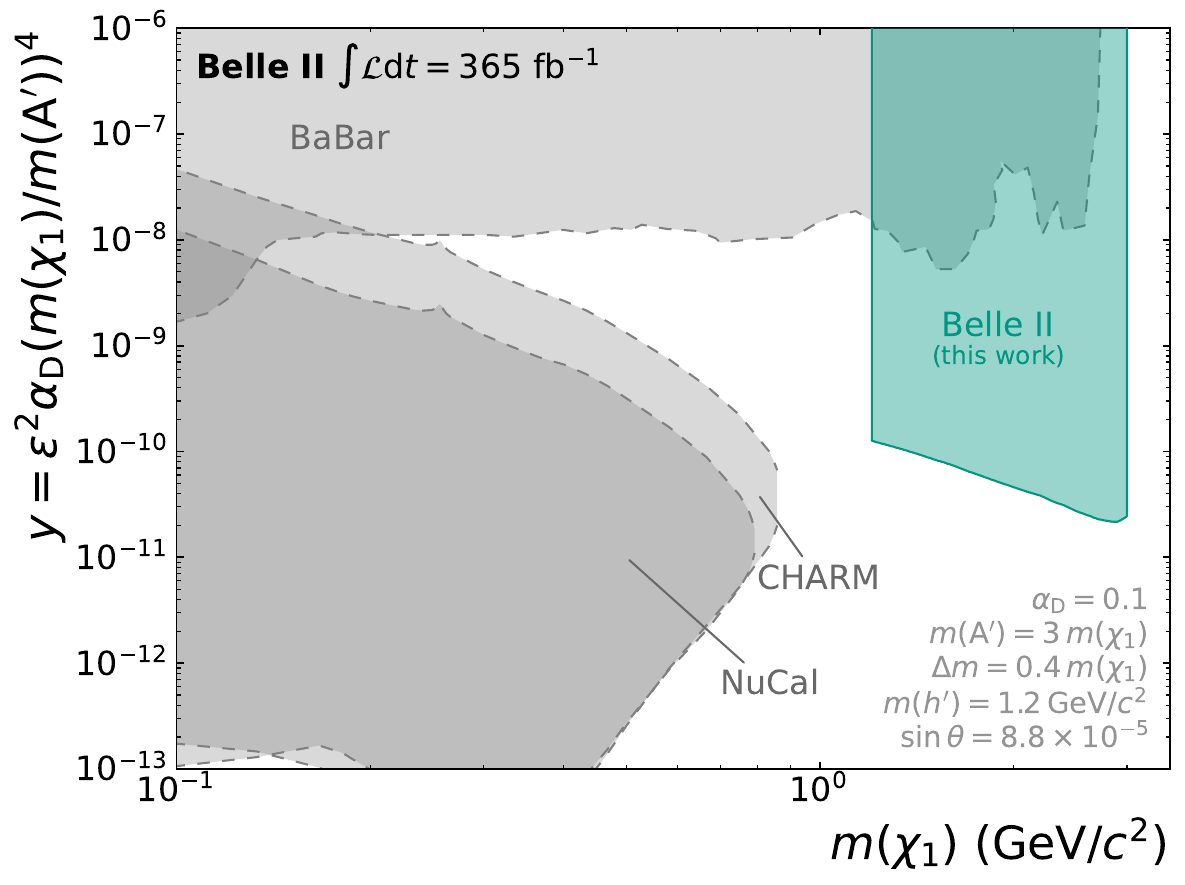}
    \caption{
Exclusion regions at 95\,\% credibility level in the plane of the dimensionless variable $y = \epsilon^2\alpha_D(m(\chi_1)/m(A'))^4$ and DM mass $\mchione$ from this work (teal) together with existing constraints at 90\,\% credibility level from
CHARM\,\cite{Gninenko:2012eq,Tsai:2019buq},
NuCal\,\cite{Blumlein:2011mv,Blumlein:2013cua,Tsai:2019buq}, and
BABAR\,\cite{BaBar:2017tiz,Duerr:2019dmv} for $\alpha_D = 0.1$, $\map = 3\,\mchione$, $\mdh = 1.2\,\gevcc$, and $\sin \theta = 8.8\times 10^{-5}$ ($c\tau(\dh) = 21.54\,\cm$).
The mass splitting is set to $\Delta m=0.2\,\mchione$ (left) and $\Delta m=0.4\,\mchione$ (right). 
Constraints colored in gray with dashed outline are reinterpretations not performed by the experimental collaborations.
All constraints but the one from this work do not depend on the presence of a dark Higgs boson or iDM.
}
    \label{fig:model_dependent7}
\end{figure*}

\begin{figure*}[htp!]
    \centering
    \includegraphics[width=0.45\textwidth]{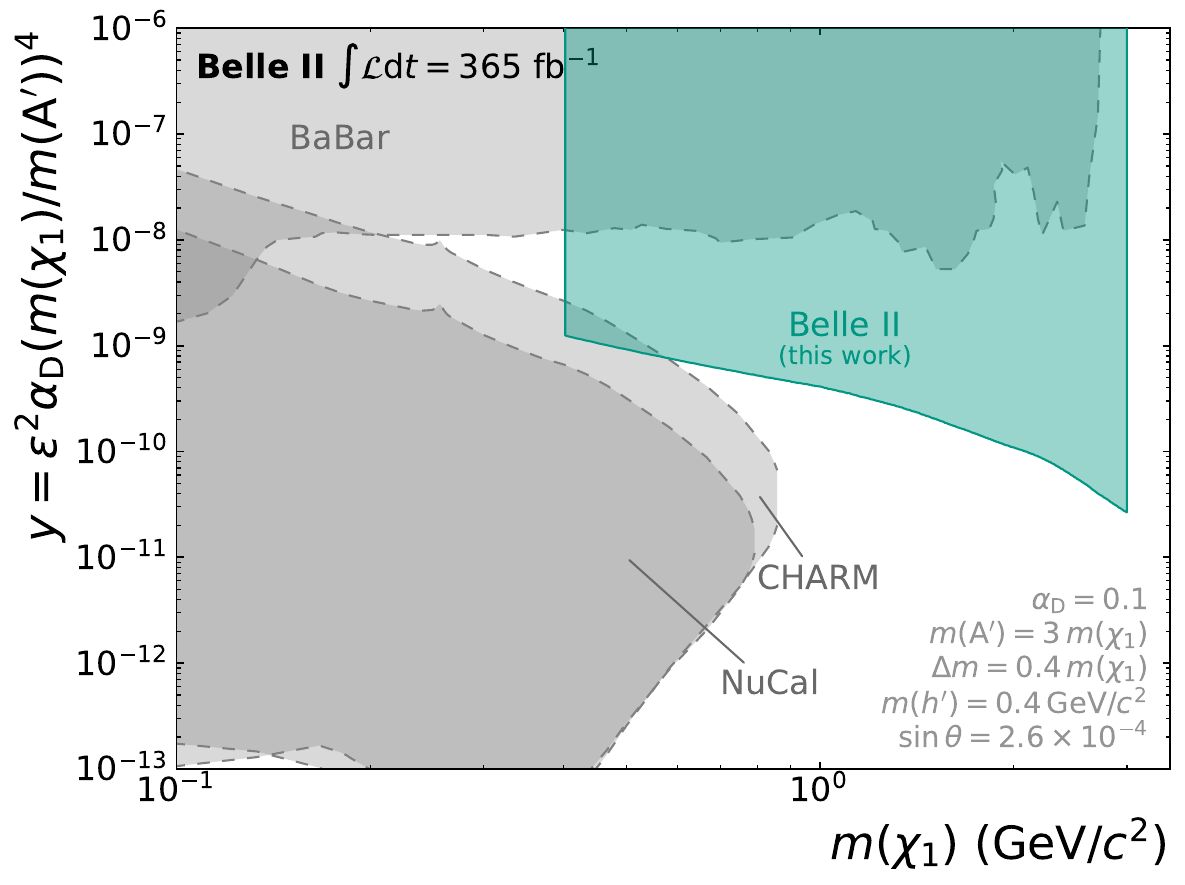}
    \includegraphics[width=0.45\textwidth]{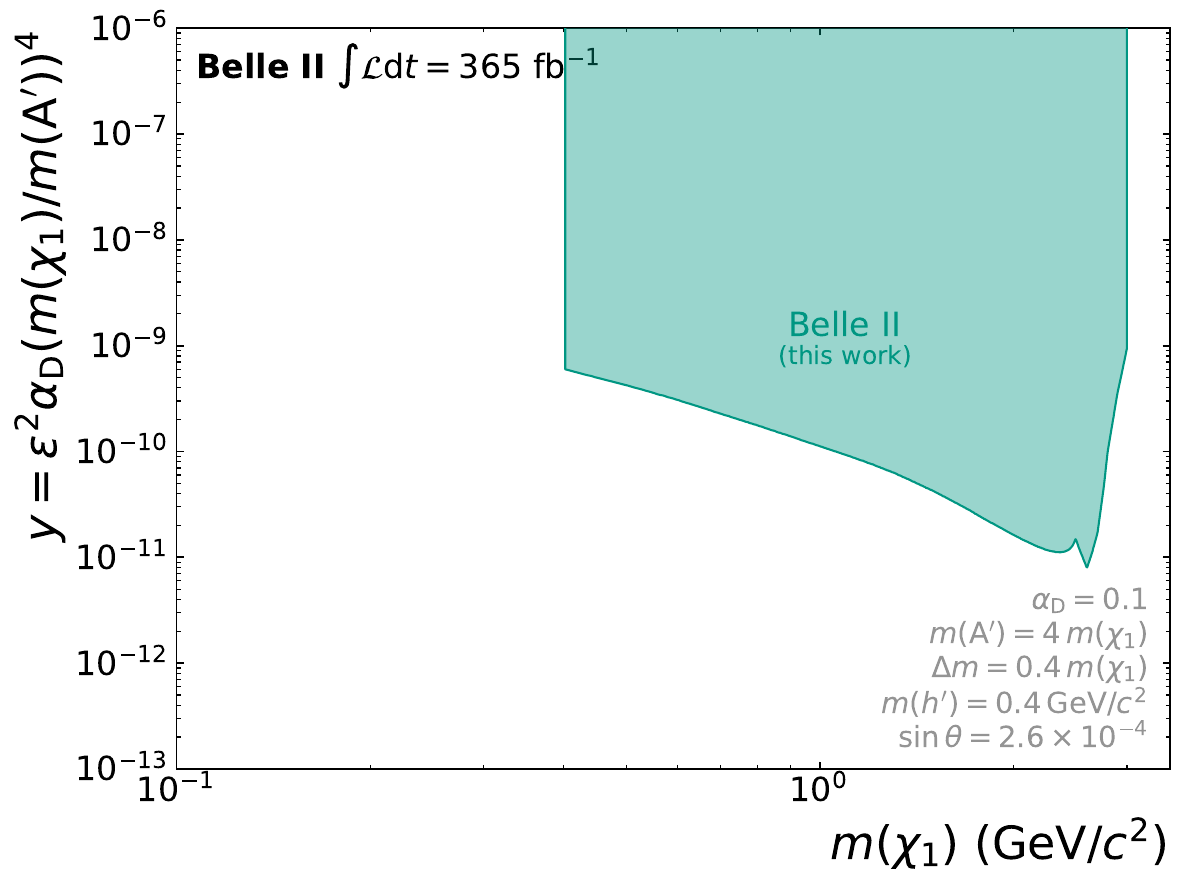}
    \caption{
Exclusion regions at 95\,\% credibility level in the plane of the dimensionless variable $y = \epsilon^2\alpha_D(m(\chi_1)/m(A'))^4$ and DM mass $\mchione$ from this work (teal) together with existing constraints at 90\,\% credibility level from
CHARM\,\cite{Gninenko:2012eq,Tsai:2019buq},
NuCal\,\cite{Blumlein:2011mv,Blumlein:2013cua,Tsai:2019buq}, and
BABAR\,\cite{BaBar:2017tiz,Duerr:2019dmv} for $\alpha_D = 0.1$, $\Delta m=0.4\,\mchione$, $\mdh = 0.4\,\gevcc$, and $\sin \theta = 2.6\times 10^{-4}$ ($c\tau(\dh) = 21.54\,\cm$).
The dark photon mass is set to $\map = 3\,\mchione$ (left) and $\map = 4\,\mchione$ (right). 
Constraints colored in gray with dashed outline are reinterpretations not performed by the experimental collaborations.
All constraints but the one from this work do not depend on the presence of a dark Higgs boson or iDM.
For $\map = 4\,\mchione$ no recasts of the BaBar, CHARM, and NuCal exclusion regions exist in the literature.
}
    \label{fig:model_dependent8}
\end{figure*}

%% file: references.bib
@article{LZ:2024zvo,
    author = "Aalbers, J. and others",
    collaboration = "LZ",
    title = "{Dark Matter Search Results from 4.2{\,}{\,}Tonne-Years of Exposure of the LUX-ZEPLIN (LZ) Experiment}",
    eprint = "2410.17036",
    archivePrefix = "arXiv",
    primaryClass = "hep-ex",
    reportNumber = "FERMILAB-PUB-24-0796-V",
    doi = "10.1103/4dyc-z8zf",
    journal = "Phys. Rev. Lett.",
    volume = "135",
    number = "1",
    pages = "011802",
    year = "2025"
}

@article{XENON:2023cxc,
    author = "Aprile, E. and others",
    collaboration = "XENON",
    title = "{First Dark Matter Search with Nuclear Recoils from the XENONnT Experiment}",
    eprint = "2303.14729",
    archivePrefix = "arXiv",
    primaryClass = "hep-ex",
    doi = "10.1103/PhysRevLett.131.041003",
    journal = "Phys. Rev. Lett.",
    volume = "131",
    number = "4",
    pages = "041003",
    year = "2023"
}

@article{PandaX:2024qfu,
    author = "Bo, Zihao and others",
    collaboration = "PandaX",
    title = "{Dark Matter Search Results from 1.54{\,}{\,}Tonne{\textperiodcentered}Year Exposure of PandaX-4T}",
    eprint = "2408.00664",
    archivePrefix = "arXiv",
    primaryClass = "hep-ex",
    doi = "10.1103/PhysRevLett.134.011805",
    journal = "Phys. Rev. Lett.",
    volume = "134",
    number = "1",
    pages = "011805",
    year = "2025"
}

@article{DarkSide-50:2022qzh,
    author = "Agnes, P. and others",
    collaboration = "DarkSide-50",
    title = "{Search for low-mass dark matter WIMPs with 12~ton-day exposure of DarkSide-50}",
    eprint = "2207.11966",
    archivePrefix = "arXiv",
    primaryClass = "hep-ex",
    reportNumber = "FERMILAB-PUB-22-589-ND-PPD-SCD",
    doi = "10.1103/PhysRevD.107.063001",
    journal = "Phys. Rev. D",
    volume = "107",
    number = "6",
    pages = "063001",
    year = "2023"
}

@article{CMS:2024zqs,
    author = "Hayrapetyan, Aram and others",
    collaboration = "CMS",
    title = "{Dark sector searches with the CMS experiment}",
    eprint = "2405.13778",
    archivePrefix = "arXiv",
    primaryClass = "hep-ex",
    reportNumber = "CMS-EXO-23-005, CERN-EP-2024-106",
    doi = "10.1016/j.physrep.2024.09.013",
    journal = "Phys. Rept.",
    volume = "1115",
    pages = "448--569",
    year = "2025"
}

@article{ATLAS:2024fdw,
    author = "Aad, Georges and others",
    collaboration = "ATLAS",
    title = "{Exploration at the high-energy frontier: ATLAS Run 2 searches investigating the exotic jungle beyond the Standard Model}",
    eprint = "2403.09292",
    archivePrefix = "arXiv",
    primaryClass = "hep-ex",
    reportNumber = "CERN-EP-2024-075",
    doi = "10.1016/j.physrep.2024.10.001",
    journal = "Phys. Rept.",
    volume = "1116",
    pages = "301--385",
    year = "2025"
}

@article{Belle-II:2023ueh,
    author = "Adachi, I. and others",
    collaboration = "Belle~II Collaboration",
    title = "{Search for a long-lived spin-0 mediator in b\textrightarrow{}s transitions at the Belle~II experiment}",
    eprint = "2306.02830",
    archivePrefix = "arXiv",
    primaryClass = "hep-ex",
    reportNumber = "Belle II Preprint 2023-009, KEK Preprint 2023-7",
    doi = "10.1103/PhysRevD.108.L111104",
    journal = "Phys. Rev. D",
    volume = "108",
    number = "11",
    pages = "L111104",
    year = "2023"
}

@article{Caldwell:2008fw,
    author = "Caldwell, Allen and Kollar, Daniel and Kroninger, Kevin",
    title = "{BAT: The Bayesian Analysis Toolkit}",
    eprint = "0808.2552",
    archivePrefix = "arXiv",
    primaryClass = "physics.data-an",
    doi = "10.1016/j.cpc.2009.06.026",
    journal = "Comput. Phys. Commun.",
    volume = "180",
    pages = "2197--2209",
    year = "2009"
}

@misc{bat-zenodo,
    author = "{F., Beaujean et al.}",
    title = "{B}{A}{T} release, version 1.0.0",
    doi = {10.5281/zenodo.1322675},
    howpublished = "\url{https://doi.org/10.5281/zenodo.1322675}"
}

@article{Kou:2018nap,
  title          = "{The Belle~II Physics Book}",
  author         = {Kou, E and others},
  journal        = {Prog. Theor. Exp. Phys.},
  volume         = {2019},
  pages          = {123C01},
  year           = {2019},
  doi            = {10.1093/ptep/ptz106},
  note           = {[Erratum: Prog. Theor. Exp. Phys. 2020, 029201 (2020)]},
  eprint         = "1808.10567",
  archivePrefix  = "arXiv",
  primaryClass   = "hep-ex"
}

@article{Abe:2010gxa,
      author         = "Abe, T. and others",
      title          = "{Belle~II technical design report}",
      collaboration  = "Belle~II Collaboration",
      year           = "2010",
      eprint         = "1011.0352",
      archivePrefix  = "arXiv",
      primaryClass   = "physics.ins-det",
      reportNumber   = "KEK-REPORT-2010-1",
      SLACcitation   = "%%CITATION = ARXIV:1011.0352;%%",
      journal = "" 
}

@article{Akai:2018mbz,
  author          = "Akai, Kazunori and Furukawa, Kazuro and Koiso, Haruyo",
  title           = "{SuperKEKB Collider}",
  collaboration   = "SuperKEKB Accelerator Team",
  journal         = "Nucl. Instrum. Methods Phys. Res. A",
  volume          = {907},
  pages           = {188},
  year            = {2018},
  doi             = "10.1016/j.nima.2018.08.017",
  eprint          = "1809.01958",
  archivePrefix   = "arXiv",
  primaryClass    = "physics.acc-ph",
  SLACcitation    = "%%CITATION = ARXIV:1809.01958;%%"
}

@article{OConnell:2006rsp,
    author = "O'Connell, D. and others",
    title = "{Minimal Extension of the Standard Model Scalar Sector}",
    eprint = "hep-ph/0611014",
    archivePrefix = "arXiv",
    reportNumber = "CALT-68-2614",
    doi = "10.1103/PhysRevD.75.037701",
    journal = "Phys. Rev. D",
    volume = "75",
    pages = "037701",
    year = "2007"
}

@article{Beacham:2019nyx,
    author = "Beacham, J. and others",
    title = "{Physics Beyond Colliders at CERN: Beyond the Standard Model Working Group Report}",
    eprint = "1901.09966",
    archivePrefix = "arXiv",
    primaryClass = "hep-ex",
    reportNumber = "CERN-PBC-REPORT-2018-007",
    doi = "10.1088/1361-6471/ab4cd2",
    journal = " J. Phys. G: Nucl. Part. Phys.",
    volume = "47",
    number = "1",
    pages = "010501",
    year = "2020"
}

@article{Tucker-Smith:2001myb,
    author = "Tucker-Smith, David and Weiner, Neal",
    title = "{Inelastic dark matter}",
    eprint = "hep-ph/0101138",
    archivePrefix = "arXiv",
    reportNumber = "UCB-PTH-00-43, LBNL-47234, UW-PT-00-17",
    doi = "10.1103/PhysRevD.64.043502",
    journal = "Phys. Rev. D",
    volume = "64",
    pages = "043502",
    year = "2001"
}

@article{CMS:2021sch,
    author = "Tumasyan, Armen and others",
    collaboration = "CMS Collaboration",
    title = "{Search for long-lived particles decaying into muon pairs in proton-proton collisions at $ \sqrt{s} $ = 13 TeV collected with a dedicated high-rate data stream}",
    eprint = "2112.13769",
    archivePrefix = "arXiv",
    primaryClass = "hep-ex",
    reportNumber = "CMS-EXO-20-014, CERN-EP-2021-266",
    doi = "10.1007/JHEP04(2022)062",
    journal = "J. High Energy Phys.",
    volume = "04",
    number = "2022",
    pages = "062",
    year = "2022"
}

@article{CMS:2023bay,
title = {Search for Inelastic Dark Matter in Events with Two Displaced Muons and Missing Transverse Momentum in Proton-Proton Collisions at $\sqrt{s}=13\text{ }\text{ }\mathrm{TeV}$},
  author = {Hayrapetyan, A. and others},
  collaboration = {CMS Collaboration},
  journal = {Phys. Rev. Lett.},
  volume = {132},
  issue = {4},
  pages = {041802},
  numpages = {21},
  year = {2024},
  doi = {10.1103/PhysRevLett.132.041802},
  eprint = "2305.11649",
  archivePrefix = "arXiv",
  primaryClass = "hep-ex",
}

@article{Mongillo:2023hbs,
    author = "Mongillo, M. and others",
    title = "{Constraining light thermal inelastic dark matter with NA64}",
    eprint = "2302.05414",
    archivePrefix = "arXiv",
    primaryClass = "hep-ph",
    reportNumber = "IRMP-CP3-23-09, FERMILAB-PUB-23-066-T",
    doi = "10.1140/epjc/s10052-023-11536-5",
    journal = "Eur. Phys. J. C",
    volume = "83",
    number = "5",
    pages = "391",
    year = "2023"
}

@article{NA64:2019auh,
  title = "{Improved limits on a hypothetical $X(16.7)$ boson and a dark photon decaying into ${e}^{+}{e}^{\ensuremath{-}}$ pairs}",
  author = {Banerjee, D. and others},
  collaboration = {NA64 Collaboration},
  journal = {Phys. Rev. D},
  volume = {101},
  issue = {7},
  pages = {071101},
  numpages = {7},
  year = {2020},
  month = {Apr},
  publisher = {American Physical Society},
  doi = {10.1103/PhysRevD.101.071101},
  eprint = "1912.11389",
  archivePrefix = "arXiv",
  primaryClass = "hep-ex",
  reportNumber = "CERN-EP-2019-284",
}

@article{Arbey:2021gdg,
    author = "Arbey, A. and Mahmoudi, F.",
    title = "{Dark matter and the early Universe: a review}",
    eprint = "2104.11488",
    archivePrefix = "arXiv",
    primaryClass = "hep-ph",
    reportNumber = "CERN-TH-2021-066",
    doi = "10.1016/j.ppnp.2021.103865",
    journal = "Prog. Part. Nucl. Phys.",
    volume = "119",
    pages = "103865",
    year = "2021"
}

@article{Planck:2018vyg,
    author = "Aghanim, N. and others",
    collaboration = "Planck Collaboration",
    title = "{Planck 2018 results. VI. Cosmological parameters}",
    eprint = "1807.06209",
    archivePrefix = "arXiv",
    primaryClass = "astro-ph.CO",
    doi = "10.1051/0004-6361/201833910",
    journal = "Astron. Astrophys.",
    volume = "641",
    pages = "A6",
    year = "2020",
    note = "[Erratum: Astron. Astrophys. 652, C4 (2021)]"
}

@article{Izaguirre:2015zva,
    author = "Izaguirre, Eder and Krnjaic, Gordan and Shuve, Brian",
    title = "{Discovering Inelastic Thermal-Relic Dark Matter at Colliders}",
    eprint = "1508.03050",
    archivePrefix = "arXiv",
    primaryClass = "hep-ph",
    doi = "10.1103/PhysRevD.93.063523",
    journal = "Phys. Rev. D",
    volume = "93",
    number = "6",
    pages = "063523",
    year = "2016"
}

@article{Izaguirre:2017bqb,
    author = "Izaguirre, Eder and others",
    eprint = "1703.06881",
    archivePrefix = "arXiv",
    primaryClass = "hep-ph",
    reportNumber = "FERMILAB-PUB-17-068-PPD, PUPT-2520",
    doi = "10.1103/PhysRevD.96.055007",
    journal = "Phys. Rev. D",
    volume = "96",
    number = "5",
    pages = "055007",
    year = "2017"
}

@article{Duerr:2019dmv,
    author = "Duerr, M. and others",
    title = "{Invisible and displaced dark matter signatures at Belle~II}",
    eprint = "1911.03176",
    archivePrefix = "arXiv",
    primaryClass = "hep-ph",
    reportNumber = "DESY-19-141, OUTP-19-10P, TTK-19-46",
    doi = "10.1007/JHEP02(2020)039",
    journal = "J. High Energy Phys.",
    volume = "02",
    number = "2020",
    pages = "039",
    year = "2020"
}

@article{Duerr:2020muu,
    author = "Duerr, M. and others",
    title = "{Long-lived dark Higgs and inelastic dark matter at Belle~II}",
    eprint = "2012.08595",
    archivePrefix = "arXiv",
    primaryClass = "hep-ph",
    doi = "10.1007/JHEP04(2021)146",
    journal = "J. High Energy Phys.",
    volume = "04",
    number = "2021",
    pages = "146",
    year = {2021}
}

@article{Tsai:2019buq,
    author = "Tsai, Yu-Dai and deNiverville, Patrick and Liu, Ming Xiong",
    title = "{Dark photon and muon $g-2$ inspired inelastic dark matter models at the high-energy intensity frontier}",
    eprint = "1908.07525",
    archivePrefix = "arXiv",
    primaryClass = "hep-ph",
    reportNumber = "FERMILAB-PUB-19-393-A-PPD",
    doi = "10.1103/PhysRevLett.126.181801",
    journal = "Phys. Rev. Lett.",
    volume = "126",
    number = "18",
    pages = "181801",
    year = "2021"
}

@article{Ferber:2023iso,
    author = "Torben Ferber and Alexander Grohsjean and Felix Kahlhoefer",
    title = "{Dark Higgs bosons at colliders}",
    eprint = "2305.16169",
    archivePrefix = "arXiv",
    primaryClass = "hep-ph",
    reportNumber = "P3H-23-034, TTP23-018",
    doi = "10.1016/j.ppnp.2024.104105",
    journal = "Prog. Part. Nucl. Phys.",
    volume = "136",
    pages = "104105",
    year = "2024"
}

@article{Alwall:2014hca,
    author = "Alwall, J. and others",
    title = "{The automated computation of tree-level and next-to-leading order differential cross sections, and their matching to parton shower simulations}",
    eprint = "1405.0301",
    archivePrefix = "arXiv",
    primaryClass = "hep-ph",
    reportNumber = "CERN-PH-TH-2014-064, CP3-14-18, LPN14-066, MCNET-14-09, ZU-TH-14-14",
    doi = "10.1007/JHEP07(2014)079",
    journal = "J. High Energy Phys.",
    volume = "07",
    number = "2014",
    pages = "079",
    year = "2014"
}

@article{Frixione:2021zdp,
    author = "Frixione, Stefano and others",
    title = "{Lepton collisions in MadGraph5\_aMC@NLO}",
    journal = "",
    eprint = "2108.10261",
    archivePrefix = "arXiv",
    primaryClass = "hep-ph",
    reportNumber = "MCNET-21-13,CP3-21-50",
    year = "2021"
}

@article{Lange:2001uf,
    author = "Lange, D. J.",
    title = "{The EvtGen particle decay simulation package}",
    doi = "10.1016/S0168-9002(01)00089-4",
    journal = "Nucl. Instrum. Methods Phys. Res. A",
    volume = "462",
    pages = "152--155",
    year = "2001"
}

@article{Jadach:1999vf,
      author         = "Jadach, S. and Ward, B. F. L. and W\c{a}s, Z.",
      title          = "{The precision Monte Carlo event generator KK for two-fermion
                        final states in $e^+e^-$ collisions}",
      journal        = "Comput. Phys. Commun.",
      volume         = "130",
      year           = "2000",
      pages          = "260",
      doi            = "10.1016/S0010-4655(00)00048-5",
      eprint         = "hep-ph/9912214",
      archivePrefix  = "arXiv",
      primaryClass   = "hep-ph",
      reportNumber   = "DESY-99-106, CERN-TH-99-235, UTHEP-99-08-01",
      SLACcitation   = "%%CITATION = HEP-PH/9912214;%%"
}

@article{Jadach:1990mz,
      author         = "Jadach, S. and K{\"u}hn, J. H. and W\c{a}s, Z.",
      title          = "{TAUOLA - a library of Monte Carlo programs to simulate
                        decays of polarized $\tau$ leptons}",
      journal        = "Comput. Phys. Commun.",
      volume         = "64",
      year           = "1990",
      pages          = "275",
      doi            = "10.1016/0010-4655(91)90038-M",
      reportNumber   = "CERN-TH-5856-90",
      SLACcitation   = "%%CITATION = CPHCB,64,275;%%"
}

@article{Berends:1984gf,
    author = "Berends, F. A. and Daverveldt, P. H. and Kleiss, R.",
    title = "{Complete lowest-order calculations for four-lepton final states in electron-positron collisions}",
    reportNumber = "Print-84-1007 (LEIDEN)",
    doi = "10.1016/0550-3213(85)90541-3",
    journal = "Nucl. Phys. B",
    volume = "253",
    pages = "441--463",
    year = "1985"
}

@article{Uehara:1996bgt,
    author = "Uehara, Sadaharu",
    title = "{TREPS: A Monte-Carlo Event Generator for Two-photon Processes at $e^+e^-$ Colliders using an Equivalent Photon Approximation}",
    journal = "",
    eprint = "1310.0157",
    archivePrefix = "arXiv",
    primaryClass = "hep-ph",
    reportNumber = "KEK-REPORT-96-11",
    month = "7",
    year = "1996"
}

@article{Balossini:2008xr,
  author = {G. Balossini and others},
  title = {Photon pair production at flavour factories with per mille accuracy},
  journal = {Phys. Lett. B},
  volume = {663},
  number = {3},
  pages = {209-213},
  year = {2008},
  doi = {https://doi.org/10.1016/j.physletb.2008.04.007},
  eprint = "0801.3360",
  archivePrefix = "arXiv",
  primaryClass = "hep-ph",
}

@article{Campanario:2019mjh,
    author = "Campanario, F. and others",
    title = "{Standard model radiative corrections in the pion form factor measurements do not explain the ${a}_{\ensuremath{\mu}}$ anomaly}",
    eprint = "1903.10197",
    archivePrefix = "arXiv",
    primaryClass = "hep-ph",
    doi = "10.1103/PhysRevD.100.076004",
    journal = "Phys. Rev. D",
    volume = "100",
    number = "7",
    pages = "076004",
    year = "2019"
}

@article{Jadach:1998gi,
    author = "Jadach, S. and others",
    title = "{Monte Carlo program KoralW 1.42 for all four-fermion final states in ${e}^{+}{e}^{-}$ collisions}",
    eprint = "hep-ph/9906277",
    archivePrefix = "arXiv",
    reportNumber = "CERN-TH-98-242, UTHEP-98-0702",
    doi = "10.1016/S0010-4655(99)00219-2",
    journal = "Comput. Phys. Commun.",
    volume = "119",
    pages = "272--311",
    year = "1999"
}

@article{Sjostrand:2014zea,
      author         = {Sj\"{o}strand, Torbj\"{o}rn and others},
      title          = "{An Introduction to PYTHIA 8.2}",
      journal        = "Comput. Phys. Commun.",
      volume         = "191",
      year           = "2015",
      pages          = "159-177",
      doi            = "10.1016/j.cpc.2015.01.024",
      eprint         = "1410.3012",
      archivePrefix  = "arXiv",
      primaryClass   = "hep-ph",
      reportNumber   = "LU-TP-14-36, MCNET-14-22, CERN-PH-TH-2014-190,
                        FERMILAB-PUB-14-316-CD, DESY-14-178, SLAC-PUB-16122",
      SLACcitation   = "%%CITATION = ARXIV:1410.3012;%%"
}

@article{Agostinelli:2002hh,
  author         = "Agostinelli, S. and others",
  title          = "{\textsc{Geant4}—a simulation toolkit}",
  collaboration  = "\textsc{Geant4} Collaboration",
  journal        = "Nucl. Instrum. Methods Phys. Res. A",
  volume         = {506},
  pages          = {250},
  year           = {2003},
  doi            = "10.1016/S0168-9002(03)01368-8",
  reportNumber   = "SLAC-PUB-9350, FERMILAB-PUB-03-339",
  SLACcitation   = "%%CITATION = NUIMA,A506,250;%%",
}

@article{Kuhr:2018lps,
      author         = "Kuhr, T. and others",
      title          = "{The Belle~II Core Software}",
      collaboration  = "Belle II Framework Software Group",
      journal        = "Comput. Softw. Big Sci.",
      volume         = "3",
      year           = "2019",
      number         = "1",
      pages          = "1",
      doi            = "10.1007/s41781-018-0017-9",
      eprint         = "1809.04299",
      archivePrefix  = "arXiv",
      primaryClass   = "physics.comp-ph",
      SLACcitation   = "%%CITATION = ARXIV:1809.04299;%%"
}

@misc{basf2-zenodo,
    collaboration = "{Belle II Collaboration}",
    title = "{Belle II Analysis Software Framework (basf2) (release-06-00-08)}",
    version      = {release-06-00-08},
    doi = {10.5281/zenodo.16267708},
    howpublished = "\url{https://doi.org/10.5281/zenodo.16267708}"
}

@article{Punzi:2003bu,
    author = "Punzi, Giovanni",
    editor = "Lyons, L. and Mount, R. P. and Reitmeyer, R.",
    title = "{Sensitivity of searches for new signals and its optimization}",
    eprint = "physics/0308063",
    archivePrefix = "arXiv",
    reportNumber = "PHYSTAT-2003-MODT002",
    journal = "eConf",
    volume = "C030908",
    pages = "MODT002",
    year = "2003"
}

@article{Winkler:2018qyg,
    author = "Winkler, Martin Wolfgang",
    title = "{Decay and detection of a light scalar boson mixing with the Higgs boson}",
    eprint = "1809.01876",
    archivePrefix = "arXiv",
    primaryClass = "hep-ph",
    reportNumber = "NORDITA-2018-087",
    doi = "10.1103/PhysRevD.99.015018",
    journal = "Phys. Rev. D",
    volume = "99",
    number = "1",
    pages = "015018",
    year = "2019"
}

@article{Gross:2010qma,
    author = "Gross, Eilam and Vitells, Ofer",
    title = "{Trial factors for the look elsewhere effect in high energy physics}",
    eprint = "1005.1891",
    archivePrefix = "arXiv",
    primaryClass = "physics.data-an",
    doi = "10.1140/epjc/s10052-010-1470-8",
    journal = "Eur. Phys. J. C",
    volume = "70",
    pages = "525--530",
    year = "2010"
}

@misc{aux:2024,
author = {},
title = {},
howpublished = {},
year = {},
note = "See Supplemental Material for additional plots and numerical results."
}

@phdthesis{Gaiser:Phd,
    author = "Gaiser, John",
    title = "{Charmonium spectroscopy from radiative decays of the $J/\psi$ and $\psi^\prime$}",
    reportNumber = "SLAC-R-255",
    school = "Stanford University",
    year = "1982"
}

@phdthesis{Skwarnicki:1986xj,
    author = "Skwarnicki, Tomasz",
    title = "{A study of the radiative CASCADE transitions between the Upsilon-Prime and Upsilon resonances}",
    reportNumber = "DESY-F31-86-02, DESY-F-31-86-02",
    school = "Cracow, INP",
    year = "1986"
}

@article{BaBar:2017tiz,
    author = "Lees, J. P. and others",
    collaboration = "BaBar Collaboration",
    title = "{Search for invisible decays of a dark photon produced in ${e}^{+}{e}^{-}$ collisions at BaBar}",
    eprint = "1702.03327",
    archivePrefix = "arXiv",
    primaryClass = "hep-ex",
    reportNumber = "BABAR-PUB-17-001, SLAC-PUB-16923",
    doi = "10.1103/PhysRevLett.119.131804",
    journal = "Phys. Rev. Lett.",
    volume = "119",
    number = "13",
    pages = "131804",
    year = "2017"
}

@article{Gninenko:2012eq,
    author = "Gninenko, S. N.",
    title = "{Constraints on sub-GeV hidden sector gauge bosons from a search for heavy neutrino decays}",
    eprint = "1204.3583",
    archivePrefix = "arXiv",
    primaryClass = "hep-ph",
    doi = "10.1016/j.physletb.2012.06.002",
    journal = "Phys. Lett. B",
    volume = "713",
    pages = "244--248",
    year = "2012"
}

@article{Blumlein:2011mv,
    author = "Bl{\"u}mlein, Johannes and Brunner, J{\"u}rgen",
    title = "{New exclusion limits for dark gauge forces from beam-dump data}",
    eprint = "1104.2747",
    archivePrefix = "arXiv",
    primaryClass = "hep-ex",
    reportNumber = "DESY-11-062, DO-TH-11-11, SFB-CPP-11-18, LPN-11-17, DESY-11--062, DO--TH-11-11, SFB-CPP--11--18, LPN-11--17",
    doi = "10.1016/j.physletb.2011.05.046",
    journal = "Phys. Lett. B",
    volume = "701",
    pages = "155--159",
    year = "2011"
}

@article{Blumlein:2013cua,
    author = {Bl\"umlein, Johannes and Brunner, J\"urgen},
    title = "{New exclusion limits on dark gauge forces from proton Bremsstrahlung in beam-dump data}",
    eprint = "1311.3870",
    archivePrefix = "arXiv",
    primaryClass = "hep-ph",
    reportNumber = "DESY-13-202, DO-TH-13-29, SFB-CPP-13-87, LPN-13-087",
    doi = "10.1016/j.physletb.2014.02.029",
    journal = "Phys. Lett. B",
    volume = "731",
    pages = "320--326",
    year = "2014"
}

@article{ref:lumi_paper,
doi = {10.1088/1674-1137/ad806c},
url = {https://dx.doi.org/10.1088/1674-1137/ad806c},
year = {2025},
month = {jan},
publisher = {Chinese Physical Society and the Institute of High Energy Physics of the Chinese Academy of Sciences and the Institute of Modern Physics of the Chinese Academy of Sciences and IOP Publishing Ltd
				},
volume = {49},
number = {1},
pages = {013001},
author = {Adachi, I. and others},
collaboration = "{Belle II Collaboration}",
title = {{Measurement of the integrated luminosity of data samples collected during 2019-2022 by the Belle II experiment}},
journal = {Chin. Phys. C}
}

@article{Barberio:1990ms,
      author         = "Barberio, Elisabetta and van Eijk, Bob and W\c{a}s, Zbigniew",
      title          = "{PHOTOS: A universal Monte Carlo for QED radiative
                        corrections in decays}",
      journal        = "Comput. Phys. Commun.",
      volume         = "66",
      year           = "1991",
      pages          = "115",
      doi            = "10.1016/0010-4655(91)90012-A",
      reportNumber   = "CERN-TH-5857-90",
      SLACcitation   = "%%CITATION = CPHCB,66,115;%%"
}

@article{Barberio:1993qi,
    author       = "Barberio, Elisabetta and W\c{a}s, Zbigniew",
    title        = "{PHOTOS: A Universal Monte Carlo for QED radiative 
                     corrections. Version 2.0}",
    reportNumber = "CERN-TH-7033-93",
    doi          = "10.1016/0010-4655(94)90074-4",
    journal      = "Comput. Phys. Commun.",
    volume       = "79",
    pages        = "291",
    year         = "1994"
}

@article{Gorbunov:2021ccu,
    author = "Gorbunov, Dmitry and Krasnov, Igor and Suvorov, Sergey",
    title = "{Constraints on light scalars from PS191 results}",
    eprint = "2105.11102",
    archivePrefix = "arXiv",
    primaryClass = "hep-ph",
    reportNumber = "INR-TH-2021-012",
    doi = "10.1016/j.physletb.2021.136524",
    journal = "Phys. Lett. B",
    volume = "820",
    pages = "136524",
    year = "2021"
}

@article{BNL-E949:2009dza,
    author = "Artamonov, A. V. and others",
    collaboration = "BNL-E949 Collaboration",
    title = "{Study of the decay $K^+\to\pi^+\nu \bar\nu$ in the momentum region $140 < P_\pi < 199$ MeV/c}",
    eprint = "0903.0030",
    archivePrefix = "arXiv",
    primaryClass = "hep-ex",
    reportNumber = "BNL-81786-2008-JA, FERMILAB-PUB-09-007-CD-T, KEK-2008-44, TRIUMF-TRI-PP-08-26, UHEP-EX-08-004",
    doi = "10.1103/PhysRevD.79.092004",
    journal = "Phys. Rev. D",
    volume = "79",
    pages = "092004",
    year = "2009"
}

@article{NA62:2021zjw,
    author = "Cortina Gil, Eduardo and others",
    collaboration = "NA62 Collaboration",
    title = "{Measurement of the very rare K$^{+}$\textrightarrow{}$ {\pi}^{+}\nu \overline{\nu} $ decay}",
    eprint = "arXiv:2103.15389",
    archivePrefix = "arXiv",
    primaryClass = "hep-ex",
    doi = "10.1007/JHEP06(2021)093",
    journal = "J. High Energy Phys.",
    volume = "06",
    number = "2021",
    pages = "093",
    year = "2021"
}

@article{NA62:2020pwi,
    author = "Cortina Gil, Eduardo and others",
    collaboration = "NA62 Collaboration",
    title = "{Search for $\pi^0$ decays to invisible particles}",
    eprint = "arXiv:2010.07644",
    archivePrefix = "arXiv",
    primaryClass = "hep-ex",
    reportNumber = "CERN-EP-2020-193",
    doi = "10.1007/JHEP02(2021)201",
    journal = "J. High Energy Phys.",
    volume = "02",
    number = "2021",
    pages = "201",
    year = "2021"
}

@article{MicroBooNE:2021usw,
    author = "Abratenko, P. and others",
    collaboration = "MicroBooNE Collaboration",
    title = "{Search for a Higgs portal scalar decaying to electron-positron pairs in the MicroBooNE detector}",
    eprint = "2106.00568",
    archivePrefix = "arXiv",
    primaryClass = "hep-ex",
    reportNumber = "FERMILAB-PUB-21-262-E",
    doi = "10.1103/PhysRevLett.127.151803",
    journal = "Phys. Rev. Lett.",
    volume = "127",
    number = "15",
    pages = "151803",
    year = "2021"
}

@article{MicroBooNE:2022ctm,
    author = "Abratenko, P. and others",
    collaboration = "MicroBooNE Collaboration",
    title = "{Search for long-lived heavy neutral leptons and Higgs portal scalars decaying in the MicroBooNE detector}",
    eprint = "2207.03840",
    archivePrefix = "arXiv",
    primaryClass = "hep-ex",
    reportNumber = "FERMILAB-PUB-22-507",
    doi = "10.1103/PhysRevD.106.092006",
    journal = "Phys. Rev. D",
    volume = "106",
    number = "9",
    pages = "092006",
    year = "2022"
}

@article{KOTO:2020prk,
    author = "Ahn, J. K. and others",
    collaboration = "KOTO Collaboration",
    title = "{Study of the $K_L \to \pi^0 \nu \bar \nu$ decay at the J-PARC KOTO experiment}",
    eprint = "2012.07571",
    archivePrefix = "arXiv",
    primaryClass = "hep-ex",
    doi = "10.1103/PhysRevLett.126.121801",
    journal = "Phys. Rev. Lett.",
    volume = "126",
    number = "12",
    pages = "121801",
    year = "2021"
}

@article{KTEV:2000ngj,
    author = "Alavi-Harati, A. and others",
    title = "{Search for the decay $K_L \to \pi^0 \mu^+ \mu^-$}",
    eprint = "hep-ex/0001006",
    archivePrefix = "arXiv",
    reportNumber = "FERMILAB-PUB-00-022-E",
    doi = "10.1103/PhysRevLett.84.5279",
    journal = "Phys. Rev. Lett.",
    volume = "84",
    pages = "5279--5282",
    year = "2000"
}

@article{L3:1996ome,
    author = "Acciarri, M. and others",
    collaboration = "L3 Collaboration",
    title = "{Search for neutral Higgs boson production through the process $e^+ e^-\to Z^* H^0$}",
    reportNumber = "CERN-PPE-96-095, CERN-PPE-96-95",
    doi = "10.1016/0370-2693(96)00987-2",
    journal = "Phys. Lett. B",
    volume = "385",
    pages = "454--470",
    year = "1996"
}

@article{CHARM:1985anb,
    author = "Bergsma, F. and others",
    collaboration = "CHARM Collaboration",
    title = "{Search for axion-like particle production in 400 GeV proton-copper interactions}",
    reportNumber = "CERN-EP-85-38",
    doi = "10.1016/0370-2693(85)90400-9",
    journal = "Phys. Lett. B",
    volume = "157",
    pages = "458--462",
    year = "1985"
}

@article{LHCb:2015nkv,
    author = "Aaij, Roel and others",
    collaboration = "LHCb Collaboration",
    title = "{Search for hidden-sector bosons in $B^0 \!\to K^{*0}\mu^+\mu^-$ decays}",
    eprint = "1508.04094",
    archivePrefix = "arXiv",
    primaryClass = "hep-ex",
    reportNumber = "CERN-PH-EP-2015-202, LHCB-PAPER-2015-036",
    doi = "10.1103/PhysRevLett.115.161802",
    journal = "Phys. Rev. Lett.",
    volume = "115",
    number = "16",
    pages = "161802",
    year = "2015"
}

@article{LHCb:2016awg,
    author = "Aaij, R. and others",
    collaboration = "LHCb Collaboration",
    title = "{Search for long-lived scalar particles in $B^+ \to K^+ \chi (\mu^+\mu^-)$ decays}",
    eprint = "1612.07818",
    archivePrefix = "arXiv",
    primaryClass = "hep-ex",
    reportNumber = "CERN-EP-2016-302, LHCB-PAPER-2016-052",
    doi = "10.1103/PhysRevD.95.071101",
    journal = "Phys. Rev. D",
    volume = "95",
    number = "7",
    pages = "071101",
    year = "2017"
}

@article{BaBar:2015jvu,
    author = "Lees, J. P. and others",
    collaboration = "BaBar Collaboration",
    title = "{Search for long-lived particles in $e^+e^-$ collisions}",
    eprint = "1502.02580",
    archivePrefix = "arXiv",
    primaryClass = "hep-ex",
    reportNumber = "BABAR-PUB-14-010, SLAC-PUB-16226",
    doi = "10.1103/PhysRevLett.114.171801",
    journal = "Phys. Rev. Lett.",
    volume = "114",
    number = "17",
    pages = "171801",
    year = "2015"
}

@article{Jeffreys:1946,
    author = "Jeffreys, H.",
    title = "{An invariant form for the prior probability in estimation problems}",
    doi = "10.1098/rspa.1946.0056",
    journal = "Proc. R. Soc. Lond. A",
    volume = "186",
    pages = "453–461",
    year = "1946"
}

@article{PhysRevD.111.032012,
  title = {{Observation of the decay ${B}^{0}\ensuremath{\rightarrow}J/\ensuremath{\psi}\ensuremath{\omega}$ at Belle II}},
  author = {Adachi, I. and others},
  collaboration = {Belle II Collaboration},
  journal = {Phys. Rev. D},
  volume = {111},
  issue = {3},
  pages = {032012},
  numpages = {8},
  year = {2025},
  month = {Feb},
  publisher = {American Physical Society},
  doi = {10.1103/PhysRevD.111.032012},
  url = {https://link.aps.org/doi/10.1103/PhysRevD.111.032012}
}
